\documentclass[12pt,twoside,a4paper]{book}
\pdfoutput=1

\usepackage{aas_macros}
\usepackage{rotating}
\usepackage{natbib}
\usepackage{psfig}
\usepackage{bm}       
\usepackage{setspace}
\usepackage{amssymb}
\usepackage{amsbsy}   
\usepackage[multiple]{footmisc}
\usepackage{ifthen}
\usepackage{multirow}
\pdfoutput=1
\newcommand{\sun}{\ensuremath{\odot}}

\newcommand{\ga}{\ensuremath{\gtrsim}}
\newcommand{\la}{\ensuremath{\lesssim}}

\newcommand{\epsscale}[1]{}

\newcommand{\HI}{\ifmmode{\rm HI}\else{\mbox{H\,{\sc{i}}}}\fi}

\newcommand{\kms}{\ensuremath{{\rm \,km\,s}^{-1}}}

\newcommand{\simgt}{\ensuremath{\ga}}
\newcommand{\simlt}{\ensuremath{\la}}

\begin{document}


\frontmatter


\begin{titlepage}

\begin{flushright}
\texttt{The Formation of Stellar Halos\\ 
in Late-Type Galaxies\\
\vspace{0.5cm}
Agostino Renda\\
\vspace{2cm}
PhD Thesis}
\end{flushright}

\cleardoublepage

\begin{center}

\vspace{8cm}
{\textsc{\Huge\bf The Formation of Stellar Halos}}\\ 
\vspace{0.5cm}
{\textsc{\Huge\bf in Late--Type Galaxies}}

\vspace{2cm}
{\Large\bf by} \\
\vspace{0.5cm}
{\Large\bf Agostino Renda} \\

\vspace{2.5cm}

{\it A Dissertation \\
Presented in fulfilment of the requirements\\
for the degree of\\
Doctor of Philosophy\\
at Swinburne University Of Technology\\
}

\vspace{1cm}
{\it May 2007}

\vspace{2cm}

\begin{figure}[h]
\centerline{\includegraphics[width=4cm]{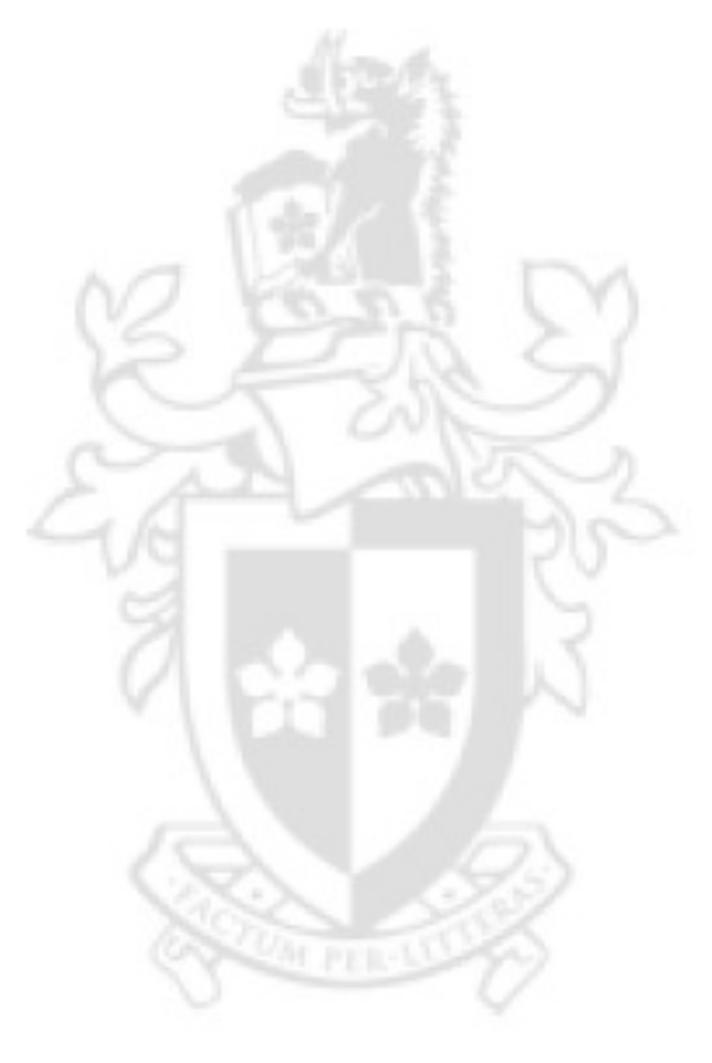}}

\end{figure}

\end{center} 
\end{titlepage}



\cleardoublepage

\onehalfspacing


\chapter*{Abstract}

\noindent Near--field observations may provide tight constraints -~i.e.~``boundary conditions''~- on any model of structure formation in the Universe. Detailed observational data have long been available for the Milky~Way (e.g.,~\citealt{FBH}) and have provided tight constraints on several Galaxy formation models (e.g.:~\citealt{Abadi}; \citealt{BC}). An implicit assumption still remains unanswered though: \textit{is the Milky~Way a ``normal'' spiral?} Searching for directions, it feels natural to look at our neighbour: Andromeda. An intriguing piece of the ``puzzle'' is provided by contrasting its stellar halo with that of our Galaxy, even more so since \cite{MM} have suggested that a correlation between stellar halo metallicity and galactic luminosity is in place and would leave the Milky~Way halo as an outlier with respect to other spirals of comparable luminosities. Further questions hence arise: \textit{is there any stellar halo--galaxy formation symbiosis?} Our first step has been to contrast the chemical evolution of the Milky~Way with that of Andromeda by means of a semi--analytic model. We have then pursued a complementary approach through the analysis of several semi--cosmological late--type galaxy simulations which sample a wide variety of merging histories. We have focused on the stellar halo properties in the simulations at redshift~zero and shown that -~at any given galaxy luminosity~- the metallicities of the stellar halos in the simulations span a range in excess of $\sim$~1~dex, a result which is strengthened by the robustness tests we have performed. We suggest that the underlying driver of the halo metallicity dispersion can be traced to the diversity of galactic mass assembly histories inherent within the hierarchical clustering paradigm.



\clearpage

\singlespacing

\chapter*{Acknowledgements}

\noindent \textit{Whether we are still on speaking terms or not, I thank all the people, dispersed all over this planet, without whom this thesis would not have been, neither completed, nor written.}

\vspace{5mm}

\noindent \textit{Among them, I thank my supervisors, Brad Gibson and Daisuke Kawata, and Mustapha Mouhcine, for their unfaltering support and encouragement through difficult times.} 



\clearpage


\chapter*{Declaration}

This thesis contains no material that has been accepted for the award
of any other degree or diploma. To the best of my knowledge, this
thesis contains no material previously published or written by another
author, except where due reference is made in the text of the thesis.
All work presented is primarily that of the author.

\vspace{4mm}

\noindent Science is a collaborative pursuit, and the studies presented in this thesis
are very much part of a team effort. The collaborators listed below have
great experience and knowledge, and were all integral in obtaining the
presented results:

\vspace{4mm}

\noindent Brad K. Gibson; Daisuke Kawata; Mustapha Mouhcine; Chris B. Brook;\\ 
Yeshe Fenner; Rodrigo A. Ibata; Amanda I. Karakas; John C. Lattanzio;\\ 
Simon C. Campbell; Alessandro Chieffi; Katia Cunha; Chris Flynn; Verne V. Smith.

\vspace{8mm}

\noindent Chapter~2 was published in

\vspace{1mm}

\noindent Renda A., Kawata D., Fenner Y., Gibson B.K.,\\
\noindent Mon.\ Not.\ R.\ Astron.\ Soc., 356, 1071 (2005)

\vspace{6mm}

\noindent Chapter~3 was partially published in

\vspace{1mm}

\noindent Renda A., Gibson B.K., Mouhcine M., Ibata R.A.,\\ 
\noindent Kawata D., Flynn C., Brook C.B.,\\
\noindent Mon.\ Not.\ R.\ Astron.\ Soc.\ Lett., 363, L16 (2005)

\vspace{6mm}

\noindent Appendix~A was published in

\vspace{1mm}

\noindent Renda A., Fenner Y., Gibson B.K., Karakas A.I.,\\
\noindent Lattanzio J.C., Campbell S., Chieffi A., Cunha K., Smith V.V.,\\
\noindent Mon.\ Not.\ R.\ Astron.\ Soc., 354, 575 (2004)

\vspace{8mm}

\noindent Minor alterations and redistribution has been made to these studies\\ in order to maintain consistency of style throughout the
thesis.

\vspace{6mm}
\noindent
Agostino Renda

\vspace{1mm}
\noindent
May 7, 2007


\clearpage

\tableofcontents
\listoffigures
\listoftables
\clearpage

\mainmatter
\onehalfspacing



\chapter{Introduction}
\label{chap:intro}

\vspace{5mm}

\begin{quotation}
\begin{flushright}
{\it{``Ne osservava le rughe,\\ 
e provava ad immaginarne la storia.''\\
  -- Anonimo}}
\end{flushright}
\end{quotation}

\vspace{15mm}

Near--field observations -~e.g.~in the Local~Group and in the nearby low redshift Universe~- of properties such as metallicity, luminosity and kinematics may provide tight constraints -~i.e.~``boundary conditions''~- on any model of structure formation in the Universe. This ``near--field'' approach to cosmology is the overarching theme of this PhD~Thesis.

The focus narrows here on stellar halos in late--type galaxies. Understanding the formation history of stellar halos is one of the challenges of galactic astrophysics. The problem is generally framed
within the context of two competing scenarios: one of ``rapid
collapse'' (Eggen, Lynden--Bell \& Sandage,~1962), in which the stellar
halo is formed by the rapid collapse of a proto--galaxy within a dynamical
timescale ($\sim$~10$^{8}$~yr), and one of ``galactic assembly''
\citep{SZ}, whereby the stellar halo is assembled over a longer timescale
($\sim$~10$^{ 9}$~yr) by the accretion of ``building--blocks'', each with
separate enrichment history. Both scenarios have their strengths and
weaknesses, and it would appear that a hybrid model is the most plausible
picture consistent with extant data (e.g.:~\citealt{CB2000}; \citealt{FBH}).

Detailed observational data have long been available for the Milky~Way (e.g.,~\citealt{FBH}) and have provided tight constraints on several Galaxy formation models (e.g.:~\citealt{Abadi}; \citealt{BC}). An implicit assumption still remains unanswered though: \textit{is the Milky~Way a ``normal'' spiral?} Searching for directions, it feels natural to look at our neighbour: Andromeda is the closest spiral to the Milky~Way, $\approx~780$~kpc (more than~2~million~light--years) away from our Galaxy. There has recently been an astonishing progress in the observations of Andromeda (e.g.,~\citealt{Ibata07}). An intriguing piece of the ``puzzle'' is provided by contrasting the stellar halo of our Galaxy with that of its
neighbour, even more so since the Hubble~Space~Telescope imaging of nearby spiral galaxy stellar halos in \cite{MM} suggests that a correlation between stellar halo metallicity and galactic luminosity is in place and would leave the Milky~Way halo as an outlier with respect to other spirals of comparable luminosities. Further questions hence arise: \textit{is there any stellar halo--galaxy formation symbiosis? Is the Milky~Way an outlier?}...\textit{Which is ``normal''?}

The layout of the PhD~Thesis we present here reflects the path we have undertaken to face the challenge these questions pose.
Our first step has been to contrast the chemical evolution of the Milky~Way with that of 
Andromeda by means of a semi--analytical model (Chapter~\ref{chap:contra}; \citealt{Renda}), which, in passing, has also led to a new understanding of the origin of Fluorine in the Milky~Way (Appendix~\ref{app:appendixA}; \citealt{RendaFluo}). We have then pursued a complementary approach through the analysis of several semi--cosmological late--type galaxy simulations which sample a wide variety of merging histories. We have first focused on the stellar halo properties in the simulations at redshift~zero (Chapter~\ref{chap:halo}, which synthesises the background which is presented in Appendices~\ref{app:appendixB}~--~\ref{app:appendixH}; \citealt{RendaHalo}), we have then traced galaxy mass assembly and metal enrichment in the simulations over the second half of the age of the Universe (Chapters~\ref{chap:mass}~and~\ref{chap:massmeta}, respectively). We finally summarise what (we think that) we see from our point of view in Chapter~\ref{chap:conclusion}.


\cleardoublepage

%
%
%


%
%
%
%




\chapter[Contrasting the Milky Way and Andromeda]{Contrasting the Milky Way and Andromeda Stellar Halos}
\label{chap:contra}

The chemical evolution history of a galaxy hides clues about how it
formed and has been changing through time. We have studied the
chemical evolution history of the Milky Way (MW) and Andromeda (M31)
to find which are common features in the chemical evolution of disc
galaxies as well as which are galaxy--dependent. We use a semi--analytic
multi--zone chemical evolution model. Such models have succeeded in
explaining the mean trends of the observed chemical properties in 
these two Local Group spiral galaxies with similar mass and
morphology. 
Our results suggest that while the evolution of the MW and M31 shares
general similarities, differences in the formation history are
required to explain the observations in detail.
In particular, we found that the observed higher metallicity in the M31 halo
can be explained by either a) a higher halo star formation efficiency 
or b) a larger reservoir of infalling halo gas 
with a longer halo formation phase.
These two different pictures
would lead to a) a higher $[$O/Fe$]$ at low metallicities or 
b) younger stellar populations in the M31 halo, respectively. 
Both pictures result in a more massive stellar halo in M31, which
suggests a possible correlation between the halo metallicity and 
its stellar mass.

\section{Introduction} 
\label{contrast:intro}

The chemical properties of galaxies hide
clues about their formation and evolution. 
Semi--analytic chemical evolution models
(\citealt{TA}; \citealt{Tinsley}) have succeeded in explaining the
mean trends of galactic systems by numerically solving a set of
equations governing the simplified evolution of the chemical elements
as they cycle through gas and stars.  One strength of these models is
that they typically have the fewest number of free parameters, making
convergence to a smaller set of solutions more likely.

Strong constraints can be placed on chemical evolution models only by
contrasting them against a comprehensive set of observed properties.
Since the most detailed observational data are generally 
available for the Milky Way (e.g.,~\citealt{FBH}), successful
agreement between model predictions and these observed properties has
been obtained by several studies in the past, which help us to
understand the formation history of the Milky Way (MW). Some models
have focused on the evolution of the chemical abundances both in the
solar neighbourhood and in the whole disc, adopting a framework in which 
the MW has been built--up inside--out by means of a single accretion event (e.g.:~\citealt{MF}; \citealt{PT}). Others have used an early infall of gas to explain the
thick disc formation, followed by a slower infall to form the thin
disc (e.g.,~Chiappini, Matteucci \& Gratton 1997). Several studies have
paid particular attention to the chemical evolution of a larger range
of elements (e.g.:~Timmes, Woosley \& Weaver 1995; \citealt{GP};
Alib\'es, Labay \& Canal 2001) or focused on the features of the
Metallicity Distribution Function (MDF) in the solar neighbourhood
(e.g.,~\citealt{FG}). However, an implicit assumption remains unanswered: 
\textit{is the MW a typical spiral?}

Andromeda (M31) is the closest spiral to our MW (e.g.,~\citealt{vanDenBergh2003}). 
Previous theoretical studies of the chemical
evolution of M31 have been done by \cite{DT} and Moll\'a, Ferrini \&
Diaz (1996), both emphasising the evolution of the M31 disc. They
concluded that M31 has a formation history and chemical evolution
similar to that of the MW. More recently, however, there has been
considerable observational progress in the study of both galaxies. 
A striking difference between M31 and the MW is that the
metallicity of the M31 halo ($\langle$[Fe/H]$\rangle\approx~-0.5$)
\footnote{Hereafter [X/Fe]$~=~{\rm log}_{10}{\rm (X/Fe)}~-~{\rm log}_{10}{\rm (X/Fe)}_{\odot}$.} 
is significantly higher than
its MW analogue ($\langle$[Fe/H]$\rangle\approx~-1.8$,
e.g.,~\citealt{RN}), as revealed by many recent studies
(e.g.:~Holland, Fahlman \& Richer 1996; Durrell, Harris, 
\& Pritchet 2001; \citealt{SvD}; \citealt{FJ}; \citealt{RG}; \citealt{FergusonETal};
\citealt{WE}; \citealt{BellazziniETal}; \citealt{BrownETal}; \citealt{ReitzelETal}; \citealt{RichETal}). 
 
It is therefore timely to attempt the construction
of a chemical evolution model for both the MW and M31, 
using the same framework. 
Such an attempt may be helpful in highlighting the 
common features in the chemical evolution of spiral galaxies 
(at least in these two spirals) 
and those which remain galaxy--dependent.

In Section~\ref{contrast:model} we describe our multi--zone
chemical evolution model, and in Section~\ref{contrast:results:MW} we 
present the results of our MW~model. In Section~\ref{contrast:results:M31} we show the results 
for M31. Finally, our results are discussed in Section~\ref{contrast:discussion}.

\section{The model}
\label{contrast:model}

In this study we use a semi--analytic multi--zone chemical evolution
model for a spiral galaxy. This model is based on \texttt{GEtool}
(\citealt{FG}; \citealt{GibsonETal}). We follow the chemical evolution of both halo and
disc of M31 and the MW under the assumption that both
galaxies have formed through two phases of gas infall. The first infall
episode corresponds to the halo build--up, and the second to the
inside--out formation of the disc. Similar formalisms have been
successful in modeling the chemical evolution in the solar
neighbourhood (e.g.:~Chiappini, Matteucci \& Gratton 1997; \citealt{ChangETal}; Alib{\' e}s, Labay \& Canal 2001). We assume that halo
stars were born in a burst induced either by the collapse of a single
proto--galactic cloud
(Eggen, Lynden--Bell \& Sandage 1962) or by the multiple merger of building--blocks
(\citealt{SZ}; \citealt{BC};
\citealt{BrookETal}). 
The disc is assumed to be built--up
inside--out by the smooth accretion of gas on a longer timescale.
Observations of HI High Velocity Clouds (HVCs) which appear to be
currently falling onto the MW (e.g.,~\citealt{PutmanETal} and
references therein) may provide evidence for such gas infall.  Recently,
HVCs have also been detected in the M31 neighbourhood, though their
interpretation as infalling clouds is debated \citep{ThilkerETal}.

We monitor the face--on projected properties of the halo and disc
components. While this geometrical simplification is suitable for
approximating the flat disc, it is less appropriate for the halo,
whose shape is roughly spherical rather than disc--like.  
However, we consider this choice acceptable in a simplified model of the 
chemical evolution of a spiral galaxy.
We follow the chemical evolution of several independent rings, $2$~kpc
wide, out to a galactocentric radius $R~=~10~R_{d}$,
where $R_{d}$ is the disc scale--length. We also
ignore the bulge component, because we are interested in the
relatively outer region ($R~>~4$~kpc). 
Each ring is a single zone onto which gas falls, 
without exchange of matter between the rings. 
We trace the chemical evolution of each zone. 
In our model, we assume that the age of the galaxy
is $t_{now}~=~13$~Gyr. The basic equations (e.g.,~\citealt{Tinsley}), 
in a zone at a radius $R$,
for the evolution of the gas surface density $\Sigma_{g,i}(R,t)$ of an
element $i$ are written as follows:
\begin{eqnarray}
\label{contrast:eq1}
&&\dot\Sigma_{g,i}(R,t)~=~\nonumber\\
&~-~&\psi(R,t)X_{i}(R,t)~+~\int_{M_{min}}^{M_{Bmin}}\psi(R,t~-~\tau_{m})\nonumber\\
&\times&Y_{i}(m,Z_{t~-~\tau_{m}})\frac{\varphi(m)}{m}dm~+~k\int_{M_{B_{min}}}^{M_{B_{max}}}\frac{\varphi(M_{B})}{M_{B}}\nonumber\\
&\times&\int_{\mu_{min}}^{0.5}f(\mu)\psi(R,t~-~\tau_{m_{2}})Y_{i}(M_{B},Z_{t~-~\tau_{m_{2}}})d\mu~dM_{B}\nonumber\\
&+& (1~-~k)\int_{M_{Bmin}}^{M_{Bmax}}\psi(R,t~-~\tau_{m})Y_{i}(m,Z_{t~-~\tau_{m}})\nonumber\\
&\times&\frac{\varphi(m)}{m}dm~+~\int_{M_{Bmax}}^{M_{max}}\psi(R,t~-~\tau_{m})Y_{i}(m,Z_{t~-~\tau_{m}})\nonumber\\
&\times&\frac{\varphi(m)}{m}dm~+~X_{i_{halo}}(t)h(R)e^{\frac{-t}{\tau_{h}}}\nonumber\\
&+&X_{i_{disc}}(t)d(R)e^{\frac{-(t~-~t_{delay})}{\tau_{d}(R)}}.\end{eqnarray}
Here, $X_{i}(R,t)~=~\frac{\Sigma_{g,i}(R,t)}{\Sigma_{g}(R,t)}$ is the
mass fraction for the element $i$; $\psi(R,t)$ is the star formation
rate (SFR);
$\varphi(m)$ is the Initial Mass Function (IMF) with mass range
$M_{min}$~-~$M_{max}$; $\tau_{m}$ is the lifetime of a star with mass
$m$; $Y_{i}(m,Z_{t~-~\tau_{m}})$ is the stellar yield of the element
from a star of mass $m$, age $\tau_{m}$ and metallicity
$Z_{t~-~\tau_m}$. The first term describes the depletion of the element $i$ which is
locked--up in newly formed stars. The second and the fourth terms show
the contribution of mass loss from low and intermediate mass
stars. The third term describes the contribution from Type~Ia~SNe~(SNe~Ia). The contribution from SNe~Ia is calculated as suggested in
\cite{GR}, where $k$, $M_{B_{min}}$, $M_{B_{max}}$, $\mu_{min}$, $\mu$, $f(\mu)$, $\tau_{m_{2}}$ are defined. 
The fifth term shows the contribution from
Type~II~SNe~(SNe~II). The sixth and the seventh terms represent the infalling halo and disc gas, respectively.

The Kroupa, Tout \& Gilmore (1993) IMF is used here. We have chosen a
lower mass limit of $M_{min}~=~0.08$~M$_{\odot}$ and imposed an upper mass
limit of $M_{max}~=~60$~M$_{\odot}$ in order to avoid the overproduction of oxygen and recover the
observed trend of $[$O/Fe$]$ at low metallicity in the solar neighbourhood in the MW. 
Such IMF upper mass limit is currently 
loosely constrained by stellar formation and evolution models.
Yet, these stellar models, and the yields they provide, are one of the most
important features in galactic chemical evolution, although
questions remain concerning the precise composition of stellar ejecta,
due to the uncertain role played by processes including mass loss,
rotation, fall--back, and the location of the mass cut, which separates
the remnant from the ejected material in SNe. 
The SNe~II
yields are from \cite{WW}. We have halved the
iron yields shown in \cite{WW}, as suggested by
\cite{TimmesETal}.  The SNe~Ia yields are from \cite{IwamotoETal}.
 The metallicity--dependent yields of \cite{RV} have
been used for stars in the mass range 1~--~8~M$_{\odot}$.
The lifetimes of stars as a function of mass and metallicity have been
taken from \cite{SchallerETal}.

\begin{table}
\begin{center}
\caption{The parameters for the~MW~and~the~M31~models.}
\begin{tabular}{@{}llllllllllll@{}}
\label{contrast:tab1}
&&&&&&&&&&\\
\hline
\hline
&&&&&&&&&&\\
     &$\alpha$&$\frac{\Sigma_{t,h}(R_{\odot},t_{now})}{\rm M_{\odot}~pc^{-2}}$&$\frac{R_{d}}{kpc}$&$\frac{\Sigma_{t,d}(R_{\odot},t_{now})}{\rm M_{\odot}~pc^{-2}}$&$\frac{\tau_{h}}{\rm Gyr}$&$\frac{t_{delay}}{\rm Gyr}$&$a_{d}$&$b_{d}$&$\nu_{h}$&$\nu_{d}$\\
&&&&&&&&&&\\
\hline
&&&&&&&&&&\\
MW&$2$&$~6$&$3.0$&$48$&$0.1$&$1.0$&$2.0$&$1.25$&$~0.125$&$0.03$\\
M31a&$2$&$~6$&$5.5$&$46$&$0.1$&$1.0$&$2.0$&$1.25$&$~0.125$&$0.03$\\
M31b&$2$&$~6$&$5.5$&$46$&$0.1$&$1.0$&$0.7$&$0.50$&$12.5~~$&$0.08$\\
M31c&$2$&$57$&$5.5$&$46$&$0.5$&$6.0$&$0.1$&$0.10$&$~0.125$&$0.08$\\
&&&&&&&&&&\\
\hline
\hline
\end{tabular}
\end{center}
\end{table}

\subsection{Infall}
\label{contrast:model:infall}

The infall rate during the halo and disc phases is simply assumed
to decline exponentially,
as seen by the adopted sixth and seventh terms in equation~\ref{contrast:eq1}. The evolution
of the total\footnote{Hereafter, by ``total'' density we mean the sum of the stellar density and of the gas density.} surface mass density,
$\Sigma_{t}(R,t)$, is described by:
\begin{eqnarray}
\label{contrast:eq2}
\frac{d\Sigma_{t}(R,t)}{dt}~=~h(R)e^{\frac{-t}{\tau_{h}}}~+~d(R)e^{\frac{-(t~-~t_{delay})}{\tau_{d}(R)}}.
\end{eqnarray}
Here, the first term describes the infall rate in the halo phase. 
The infall time--scale in the halo phase, $\tau_{h}$, is assumed to be 
independent of radius, for simplicity. The infall of disc gas starts 
with a delay of $t_{delay}$, as seen in the second term. 
The time--scale of the disc infall depends on radius as follows:
\begin{eqnarray}
\label{contrast:eq3}
\tau_{d}(R)~=~a_{d}~+~b_{d} \frac{R}{\rm kpc}~{\rm Gyr}.
\end{eqnarray}
The values for the constants $a_{d}$ and $b_{d}$ are free parameters.
The infall coefficients $h(R)$ and $d(R)$ are chosen in order to
reproduce the present--day total surface
density in the disc, $\Sigma_{t,d}(R,t_{now})$, and in the halo,
$\Sigma_{t,h}(R,t_{now})$, respectively, as follows
(e.g.,~\citealt{TimmesETal}):
\begin{eqnarray}
\label{contrast:eq4}
h(R)&=&\Sigma_{t,h}(R,t_{now})\nonumber\\
&&\times\Bigg\{\tau_{h}\Big[1~-~{\rm exp}\Big(\frac{-t_{now}}{\tau_{h}}\Big)\Big]\Bigg\}^{-1};\\
d(R)&=&\Sigma_{t,d}(R,t_{now})\nonumber\\
&&\times\Bigg\{\tau_{d}(R)\Big[1~-~{\rm exp}\Big(-\frac{t_{now}~-~t_{delay}}{\tau_{d}(R)}\Big)\Big]\Bigg\}^{-1}.
\end{eqnarray}
The infalling halo gas has been assumed to be of primordial composition.
On the other hand, it is unlikely that the accreting gas has primordial 
abundance at a later epoch, since even low density inter--galactic medium,
such as the Lyman~$\alpha$ forest, has a significant amount of metals
(e.g.,~\citealt{CS}), and it is known that the HVCs in the MW,
which could be infalling gas clouds, 
have metallicities between $0.1$ and $0.3~Z_{\odot}$
\citep{SembachETal}. Therefore, we assume that the gas accreting onto the disc is 
pre--enriched. The level of pre--enrichment can be loosely constrained
from the observed metallicity of Galactic HVCs.
We simply assume that the metallicity of the infalling disc 
material is  $Z_{infall}(R,t)~=~Z(R,t)$ if
$Z(R,t)~<~Z_{infall,max}~=~0.3~Z_{\odot}$, where $Z(R,t)$ is the
metallicity of the gas at the radius $R$ and the time $t$,
otherwise $Z_{infall}(R,t)~=~Z_{infall,max}~=~0.3~Z_{\odot}$.
The abundance pattern of the infalling disc gas 
is further unknown parameter. Following the above simple assumption,
we set the infalling disc gas, at a given galactocentric radius $R$ 
and time $t$, to have the same abundance pattern as the ISM at $R$ and $t$.
This guarantees the smooth evolution of the gas abundance and of the 
abundance patterns in each radial bin.

\subsection{Disc and halo surface density profiles}
\label{contrast:model:denpro}

We adopt the following exponential profile for the present--day 
total surface density of the disc component:
\begin{eqnarray}
\label{contrast:eq5}
\Sigma_{t,d}(R,t_{now})~=~\Sigma_{t,d}(R_{\odot},t_{now})e^{-(R~-~R_{\odot})/R_{d}}.
\end{eqnarray}
Here $R_{\odot}~=~8$~kpc, which is the galactocentric distance of the
Sun within the MW. The same definition of $R_{\odot}~=~8$~kpc 
is applied to the M31~models.

The surface density profiles of the MW and M31 disc are 
different. We have chosen 
a scale--length of $R_{d}~=~3.0$~kpc for the MW
(e.g.:~Robin, Creze \& Mohan 1992; \citealt{RuphyETal};
\citealt{Freudenreich}) and $R_{d}~=~5.5$~kpc for M31
\citep{WK}. For the MW, we assume
$\Sigma_{t,d}(R_{\odot},t_{now})~=~48$~M$_{\odot}$~pc$^{-2}$
\citep{KG}. We adopt $\Sigma_{t,d}(R_{\odot},t_{now})~=~46$~M$_{\odot}$~pc$^{-2}$ in M31, such that the mass of the M31 disc 
is similar to that of the MW disc ($\approx 10^{11}$~M$_{\odot}$ as in
\citealt{Freeman}).

We adopt a modified Hubble law for the present--day total surface
density profile of the halo \citep{BT}:
\begin{eqnarray}
\label{contrast:eq6}
\Sigma_{t,h}(R,t_{now})~=~\frac{\Sigma_{t,h_{0}}}{1~+~(R/R_{\odot})^{\alpha}}.
\end{eqnarray}
Here we set $\alpha~=~2$. 
This corresponds to a volume halo density profile:
\begin{eqnarray}
\label{contrast:eq7}
\rho_{t,h}(r,t_{now})~=~\frac{\rho_{t,h_{0}}}{\Big[1~+~\big(\frac{r}{R_{\odot}}\big)^{2}\Big]^{3/2}},
\end{eqnarray}
with $\rho_{t,h_{0}}~=~2^{3/2}\rho_{t,h}(R_{\odot},t_{now})$ and $\Sigma_{t,h_{0}}~=~2R_{\odot}\rho_{t,h_{0}}$. For the MW, we assume
$\Sigma_{t,h}(R_{\odot},t_{now})~=~6$~M$_{\odot}$~pc$^{-2}$ 
which yields a present--day stellar surface density at $R_{\odot}$
of $\Sigma_{\star,h}(R_{\odot},t_{now})~=~1.3$~M$_{\odot}$~pc$^{-2}$
for our model (see Section~\ref{contrast:results:MW}), 
which is consistent with the observed halo stellar density at the solar radius
$\rho_{\star,h}(R_{\odot},t_{now})~=~5.7\times
10^{4}$~M$_{\odot}$~kpc$^{-3}$ as estimated by Preston, Shectman \& Beers
(1991).

The assumption of $\alpha~=~2$, which implies $\rho_{\star,h}\propto r^{-3}$, 
agrees with a recent analysis
(Zibetti, White \& Brinkmann 2003) of the halo emission for 
a sample of $\approx1000$ edge--on disc galaxies within 
the Sloan Digital Sky Survey (SDSS). 
This result is similar to the density profile 
$\propto r^{-3.5}$ which has been suggested for the MW stellar halo 
(\citealt{CB2000}; \citealt{CB2001}; Sakamoto, Chiba \& Beers 2003).

\subsection{Star Formation Rate}
\label{contrast:model:SFR}

We assume that the halo star formation (SF) happens on a short
time--scale because of a rapid infall event associated with the
collapse of a single massive proto--galactic cloud \citep{EggenETal} or
with multiple mergers of building--blocks \citep{SZ}.  The disc SF is
assumed to be a more quiescent phenomenon, and likely to be driven by the
spiral arms (e.g.,~\citealt{WS}). Therefore, we adopt a different SF
law for each component.

The adopted halo SFR is described as:
\begin{equation}
\label{contrast:eq8}
\psi_{h}(R,t)~=~\nu_{h}\Big(\frac{\Sigma_{g}(R,t)}{1~M_{\odot}~pc^{-2}}\Big)^{1.5}~{\rm M_{\odot}~Gyr^{-1}~pc^{-2}},
\end{equation}
where $\nu_{h}$ is the star formation efficiency (SFE) in the
halo. Therefore, the halo SFR follows a Schmidt law with exponent
$1.5$ (e.g.,~\citealt{K}). The adopted halo SFE is $\nu_{h}~=~0.125$,
which is approximatively half of the value ($\nu_{h}~=~0.25\pm0.07$)
suggested by \cite{K} and is chosen to give the best fit to
the observed halo MDF in the solar neighbourhood (see Section~\ref{contrast:results:MW}). 
Stars born before
$t_{delay}$, when the disc phase starts (Section~\ref{contrast:model:infall}), are hereafter
labelled as ``halo stars''.

The adopted disc SFR is written as:
\begin{equation}
\label{contrast:eq9}
\psi_{d}(R,t)~=~\nu_{d}\Sigma_{g}(R,t)^{2}\,
 \frac{R_{\odot}}{R}~{\rm M_{\odot}~Gyr^{-1}~pc^{-2}},
\end{equation} 
where $\nu_{d}$ is the SFE in the disc. 
This formulation \citep{WS} reflects the
assumption that SF in the disc is triggered by the compression of the
ISM by spiral arms. 
The efficiency
factor $\nu_{d}$ is a free parameter. We have found that $\nu_{d}$
affects both the present--day gas fraction and the disc MDF. 
The value $\nu_{d}~=~0.03$ is used in our MW~model
to reproduce the observed gas density profile of the MW
disc and the observed MDF in the solar neighbourhood. 

\section{The Milky Way model}
\label{contrast:results:MW}

Using the multi--zone model described in the previous section, 
we construct a model which closely reproduces the known
observational properties of the MW. 
The adopted parameters are summarised in Table~\ref{contrast:tab1} (MW~model).

\begin{figure}
\begin{center}
\includegraphics[width=1.0\textwidth]{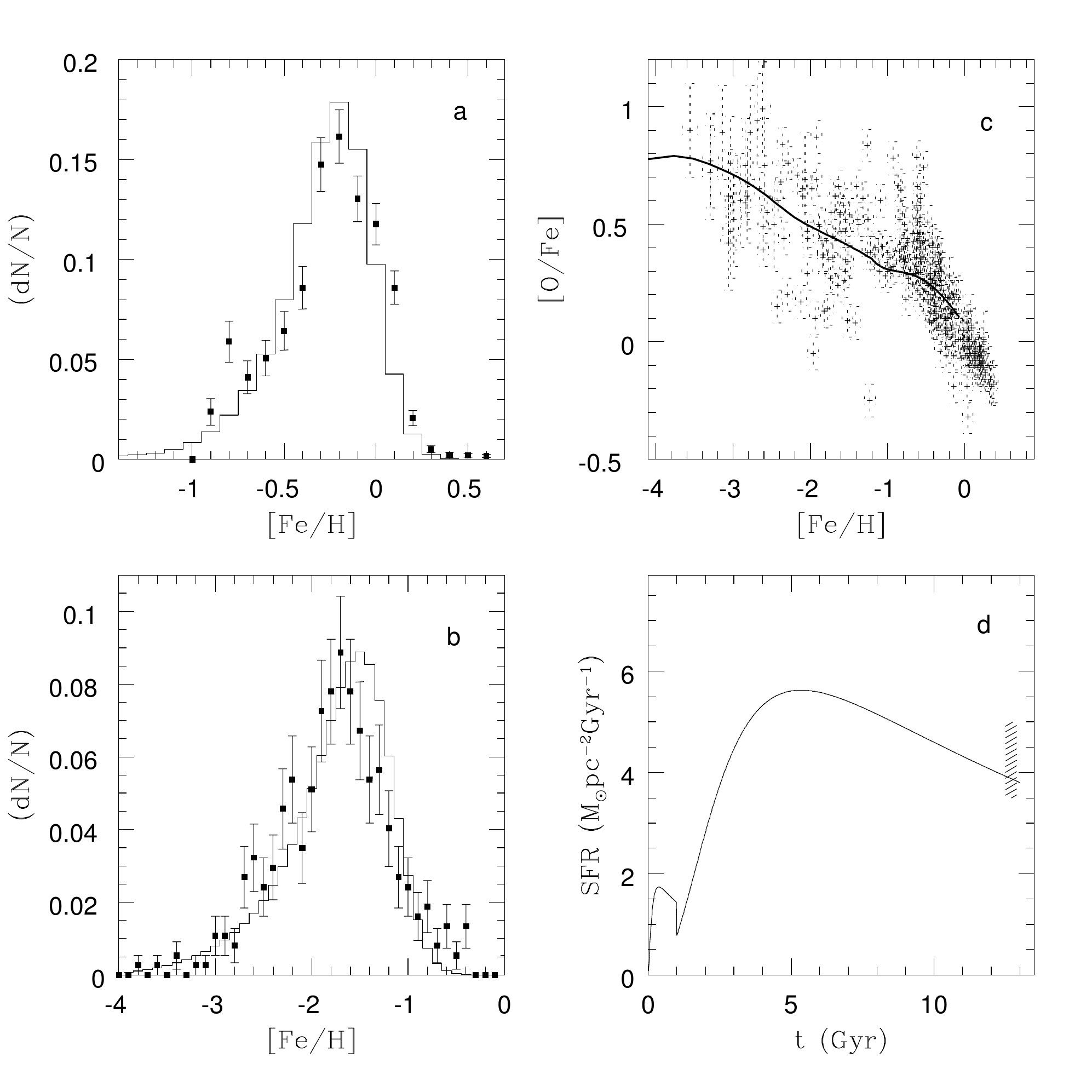}
\caption{Comparison between the results of the MW~model (solid lines) 
and the observations of: ``a'' the MDF in the MW in the solar
neighbourhood (closed boxes with error--bars, Kotoneva~et~al.~2002);
``b'' the halo MDF at $R~=~R_{\odot}$ (Ryan~\&~Norris~1991, closed boxes with
statistical Poissonian error--bars); ``c''
[O/Fe] and [Fe/H] for stars observed in the
solar neighbourhood (Carretta~et~al.~2000, who included reanalysis of
Sneden~et~al.~1991, Tomkin~et~al.~1992, Kraft~et~al.~1992, Edvardsson~et~al.~1993; 
Gratton~et~al.~2003; Bensby,~Feltzing~\&~Lundstr\"om~2003; Cayrel~et~al.~2003); ``d''
the present--day SFR in the solar neighbourhood 
as summarised in Rana~(1991, the shaded region).
}
\label{contrast:fig1}            
\end{center}
\end{figure}

\begin{figure}
\begin{center}
\includegraphics[width=1.0\textwidth]{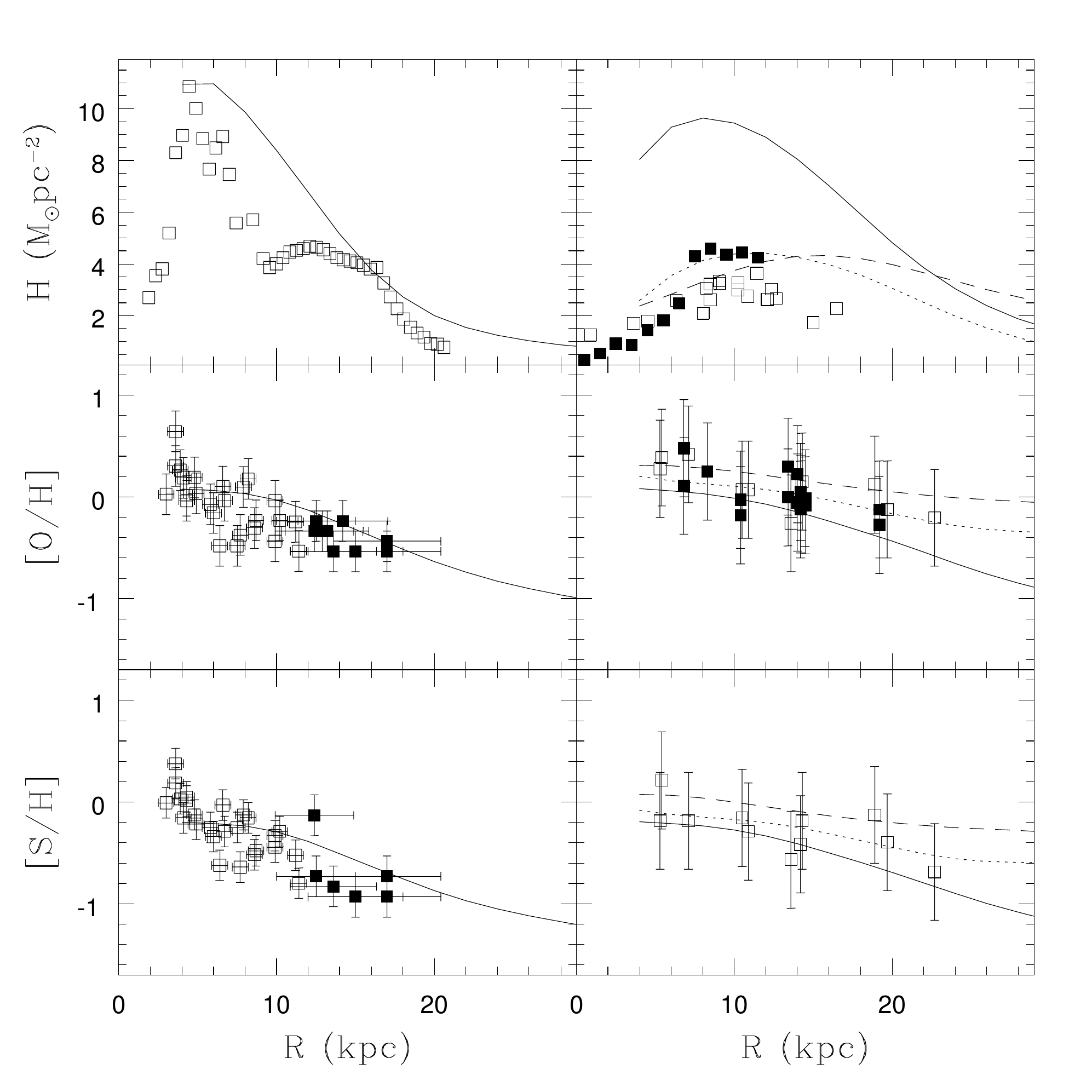}
\caption{Radial distributions of the present--day hydrogen surface density 
(upper) and of the present--day oxygen (middle) and sulfur (bottom) abundances 
for the MW~model (left) and the M31~models (right).
In the left panels, the solid lines are for the MW~model results.
In the right panels, solid, dotted and dashed lines display the~M31a, M31b and M31c~models, respectively.
The observed distributions of the total surface density of hydrogen
for the MW are from Dame~(1993, open boxes), and those for M31
are from Dame~et~al.~(1993, open boxes) and 
Loinard~et~al.~(1999, closed boxes).
Observations of abundances in HII regions in the MW, from Vilchez~\&~Esteban~(1996) 
and Afflerbach~et~al.~(1997) (closed and open boxes,
respectively), are shown in the left panels.
Those in M31, from Dennefeld~\&~Kunth~(1981) and Blair~et~al.~(1982) 
(closed and open boxes, respectively), 
are shown in the right panels.
}
\label{contrast:fig2}
\end{center}            
\end{figure}

Fig.~\ref{contrast:fig1}~``a'' compares the model~MDF in the solar
neighbourhood with the observed MDF of K dwarfs
\citep{KotonevaETal}. Here we have assumed a K dwarf
mass range of 0.8~--~1.4~M$_{\odot}$ and convolved the model~MDF 
with a Gaussian error function with $\sigma~=~0.15$~dex, 
consistent with the known empirical uncertainties of the observational
data. There is good agreement between the model and the data.
In Fig.~\ref{contrast:fig1}~``b'', the predicted halo MDF at the solar radius is shown. 
Our MDF includes all stars 
still on the main--sequence at the present--time. The MDF has been
convolved with a Gaussian error function with $\sigma~=~0.25$~dex (e.g.,~\citealt{A}). 
Our model is in agreement with the observed halo MDF in the
solar neighbourhood \citep{RN}.
The evolution of $[$O/Fe$]$ as a function of $[$Fe/H$]$ for
the gas component in the solar neighbourhood is shown in Fig.~\ref{contrast:fig1}~``c''. The model result is
consistent with the trend observed in local
stars. Fig.~\ref{contrast:fig1}~``d'' displays the predicted SF history in the solar neighbourhood.
The SF history has been estimated by several authors (e.g.:~\citealt{BN}; \citealt{HernandezETal}).
Unfortunately, such observational estimates are not defined well enough
to provide useful constraints on a chemical evolution model.
Nevertheless, Fig.~\ref{contrast:fig1}~``d'' demonstrates that our MW~model is consistent with 
the broad range of the estimated SFR as summarised in \cite{Rana}.

The left panels of Fig.~\ref{contrast:fig2} show the predicted radial distribution 
of the present--day hydrogen column density (upper) and of the present--day oxygen (middle)
and sulfur (bottom) abundances of the gas, 
and compare them to the observations.
The observed hydrogen surface density is 
obtained by summing the surface densities of H$_{2}$ and HI 
\citep{D}\footnote{The HI surface density profile has been recently 
confirmed by \cite{NakaS}.}. The result of our MW~model is 
compatible with the observed hydrogen distribution at the inner radii 
within the uncertainties ($\approx 50\%$) of the observed values
\footnote{The contribution of HII should also be
considered. However, so far no reliable measurement has been achieved.}.
However, the hydrogen surface density at the outer radii 
is overestimated when compared with the observations.

The predicted present--day radial abundance profiles of oxygen and sulfur also reproduce
the abundances observed in HII regions 
(\citealt{VE}; \citealt{AfflerbachETal}), although at the outer radii
the model overestimates the sulfur abundances.

Our semi--analytic model, whose parameter values are summarised in Table~\ref{contrast:tab1},
satisfies the general MW observational constraints.
We now use the same framework to study the chemical evolution of M31.

\section{The M31 models}
\label{contrast:results:M31}

The M31~models employ for the disc a different surface density profile  
from that of the MW~model. Since the halo density profile of M31 is unknown,
we adopt the same halo profile used in the MW~model 
(see Section~\ref{contrast:results:MW}).
The SF history of halo and disc is 
described by the parameters listed in Table~\ref{contrast:tab1}.
In Section~\ref{contrast:results:M31:MW}, we first show a M31~model which adopts
the same parameter set of our MW~model. This model 
fails to reproduce some crucial features observed in M31.
Therefore, in Sections~\ref{contrast:results:M31:M31b}~and~\ref{contrast:results:M31:M31c}, we present two models 
able to better explain the key observations.

\subsection{M31a~model: MW analogue}
\label{contrast:results:M31:MW}

First, we construct a M31~model (M31a) 
with the same parameter values of 
our MW~model. These parameters
are summarised in Table~\ref{contrast:tab1}.
The right panels of Fig.~\ref{contrast:fig2} show the radial distributions of 
the present--day hydrogen surface density and
the radial profiles of the present--day oxygen and sulfur abundances of the gas phase. 
The radial profiles of oxygen and sulfur abundances are reproduced within
the observational errors. The M31a~model results in too 
high hydrogen surface density when compared with the data.
This is because the observed hydrogen surface density in M31
is smaller than in the MW.

\begin{figure}
\begin{center}
\includegraphics[width=1.0\textwidth]{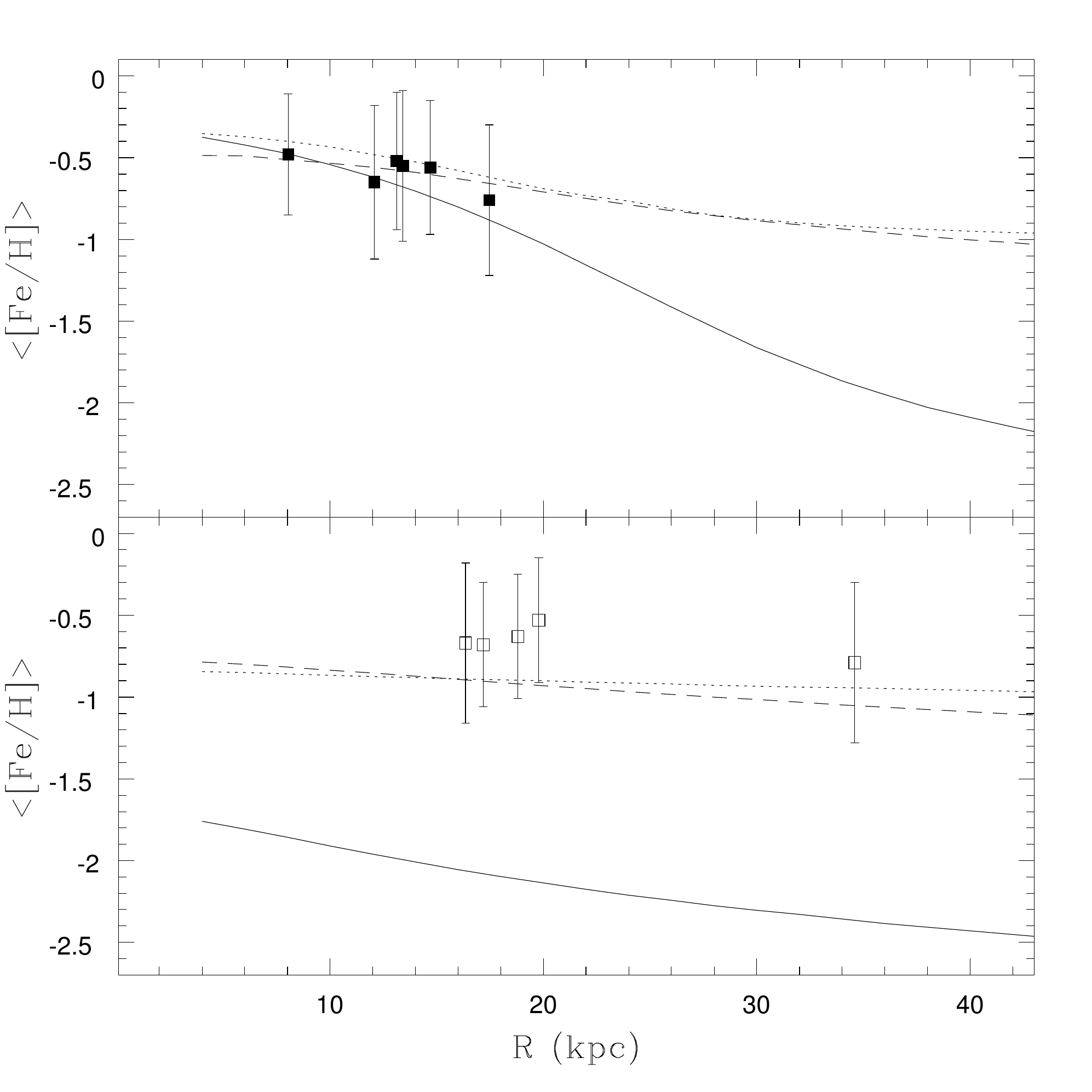}
\caption{Radial profile of the present--day mean stellar 
metallicity of both disc (upper panel) and halo (lower panel).
Solid, dotted and dashed lines represent the results of the~M31a,
M31b and M31c~models, respectively.
The mean metallicities from Bellazzini~et~al.~(2003) are also
shown (closed boxes with 1~$\sigma$ dispersion of their MDF).}
\label{contrast:fig3}
\end{center}
\end{figure}

Fig.~\ref{contrast:fig3} shows the radial profile for the mean $[$Fe/H$]$ 
of main--sequence (MS) stars for the M31a~model.
Hereafter we simply call $\langle[$Fe/H$]\rangle$ the "mean metallicity".
Model results are compared with
observations by \cite{BellazziniETal}\footnote{The observed MDFs 
are derived from red giant branch (RGB) stars, whereas
the model produces the MDFs of MS stars, 
since it is difficult to construct MDFs of RGB stars within our framework.
We assume that this inconsistency does not significantly affect our comparison.}.
We have chosen as reference fields those which 
are mostly disc-- (or halo--) dominated, with estimated halo-- (or disc--) 
contamination around or lower than 10\% 
(see Table~1 in \citealt{BellazziniETal}):
the disc--dominated fields (G287, G119, G33, G76, G322 and G272) 
lie at deprojected\footnote{The inclination angle of the M31
disc is $i_{M31}\approx 12.5^{\circ}$.} galactocentric distances
$R\approx 8$~kpc, $R\approx 12$~kpc, $R\approx 13$~kpc, $R\approx
14$~kpc, $R\approx 15$~kpc and $R\approx 18$~kpc, respectively; the
halo--dominated fields (G319, G11, G351, G219 and G1) lie at 
projected galactocentric distances $R\approx 16$~kpc, $R\approx
17$~kpc, $R\approx 19$~kpc, $R\approx 20$~kpc and $R\approx 34$~kpc,
respectively.
Fig.~\ref{contrast:fig3} shows that M31a is in broad agreement 
with the observed mean metallicity in the disc--dominated fields, 
though the metallicity gradient is 
slightly steeper than the observed one (Table~\ref{contrast:tab2}).
On the other hand, the mean metallicity of the halo in 
M31a is too low, compared to the data. 

To overcome the failure of M31a, we explored the parameter space of the
M31~models which could explain the observational properties of M31,
and found two viable solutions.
In the following, we present these two models.

\begin{figure}
\begin{center}
\includegraphics[width=1.0\textwidth]{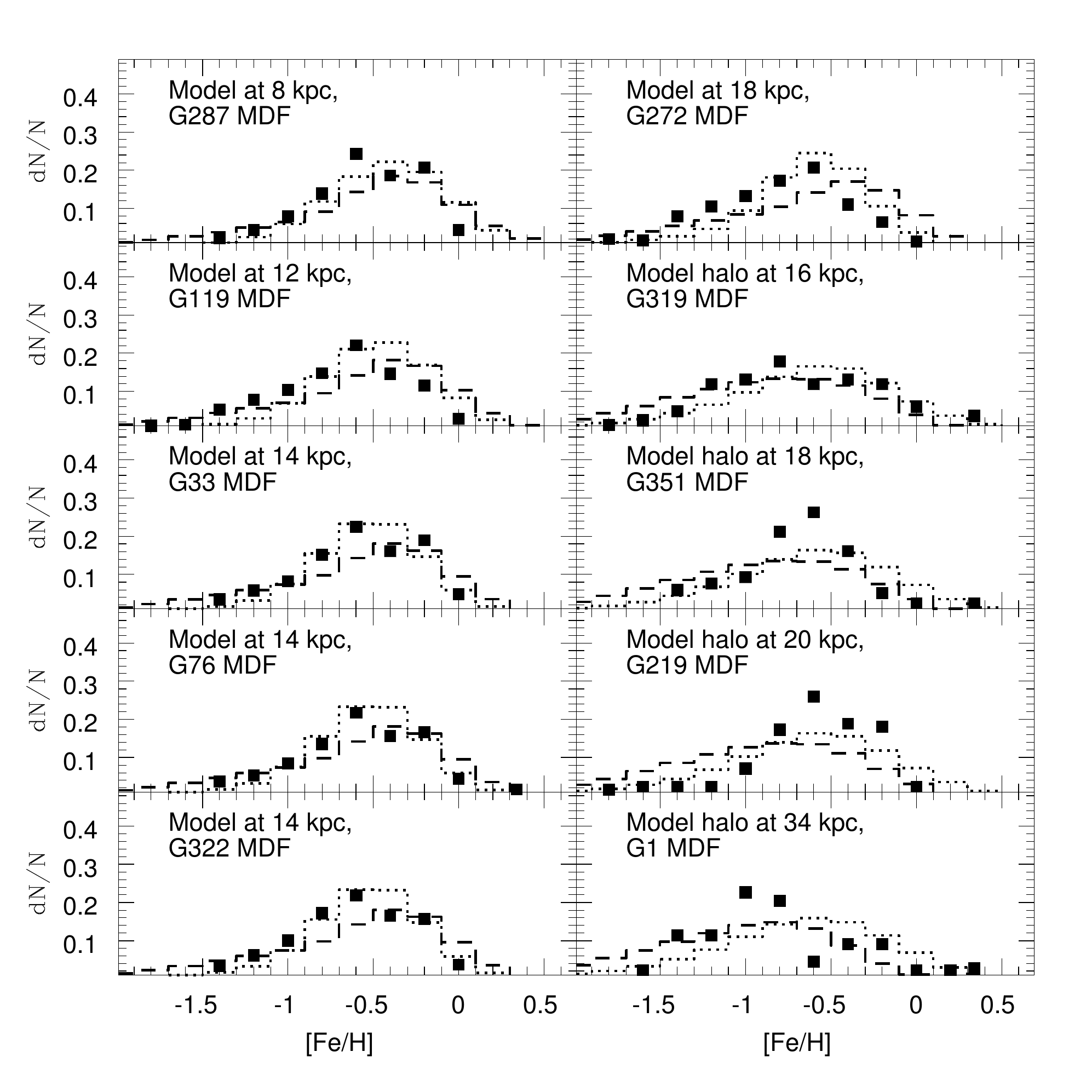}
\caption{The MDFs of the~M31b and M31c~models (dotted and dashed lines, 
respectively).
The MDFs observed from Bellazzini~et~al.~(2003) are also shown (closed boxes).}
\label{contrast:fig4}
\end{center}
\end{figure}

\subsection{M31b~model: our best model}
\label{contrast:results:M31:M31b}

The model parameters of M31b are summarised in Table~\ref{contrast:tab1}.
Compared with the M31a~model, i.e. the MW analogue, this model has
a shorter time--scale for the infall of disc gas (i.e.
smaller $a_d$ and $b_d$) and higher disc and halo SFE.
We found that a combination of higher disc SFE and 
shorter time--scale for the disc infall
leads to better agreement with the observed 
radial profile of the present--day hydrogen surface density.
In addition, a higher SFE in the halo phase leads to 
a more metal--rich halo, and we adopt a SFE~100~times higher in the halo
phase than in the M31a~model.

In M31b the radial profile 
of the present--day oxygen and sulfur abundances reproduces the observational data 
to roughly the same degree of the M31a~model
(Fig.~\ref{contrast:fig2}). On the other hand, the higher disc SFE leads to 
stronger SF and therefore
larger depletion of gas, and lower present--day hydrogen 
surface density than in M31a. 
Consequently, the result of the M31b~model
is in better agreement with the observed radial profile of 
the hydrogen surface density (Fig.~\ref{contrast:fig2}). 
The hydrogen surface density profiles
in both the M31b~model and the observations peak at a radius
of $\approx~8$~kpc with a value of $\approx$~4.5~M$_{\odot}$~pc$^{-2}$ \citep{LoinardETal}.
However, and similarly in the MW~model, the hydrogen surface density at the outer radii 
is overestimated in M31b when compared with observations from \cite{DameETal}. 
We have tried to
reproduce the observed hydrogen surface density at the outer radii by
changing various parameters. However, we are unable to find a better parameter set.
This might suggest that, to explain the low gas surface density at the
outer radii in M31, another mechanism which is not included in our semi--analytic
model is required.

The additional benefit of the shorter time--scale of the disc infall 
in M31b is its smaller SFR at the present--day.
Since M31b has a SFR which peaks at an earlier epoch than in M31a, 
M31b predicts a lower present--day SFR 
(~$\approx$~1.5~${\rm M}_{\rm \sun}~{\rm yr}^{-1}$) in the disc (4~--~20~kpc)
than in M31a (~$\approx$~2.4~${\rm M}_{\rm \sun}~{\rm yr}^{-1}$).
The observational estimates of the present--day SFR in M31 still 
have a large uncertainty. For example, \cite{W2002}
estimated the SFR for the last~$\approx 5$~Gyr roughly between~$\approx~2$ and~$\approx~20$~M$_{\odot}$~yr$^{-1}$, while \cite{W2003} 
inferred that the total SFR for the
M31 disc has been~$\approx~1$~M$_{\odot}$~yr$^{-1}$ over the past $60$~Myr. 
This value is higher than the SFR of~$\approx~0.2$~M$_{\odot}$~yr$^{-1}$ 
estimated from H~$\alpha$ and Far~Infra--Red luminosities 
\citep{DevereuxETal}. 
In addition, all these values could be affected by systematic errors 
(e.g.,~\citealt{Bell}). Nevertherless, these observations suggest that 
M31 has a lower SFR than in the MW, and M31b is consistent with 
this trend.

Fig.~\ref{contrast:fig3} shows improved agreement between the results of the
M31b~model and the observations of the present--day mean metallicity for both disc-- 
and halo--dominated fields. In the disc, M31b has
a shallower metallicity gradient than in M31a, and better reproduces
the observed gradient (Table~\ref{contrast:tab2}). This is mainly due to the small
$b_d$ we adopted (Table~\ref{contrast:tab1}). 

Fig.~\ref{contrast:fig4} directly compares the MDFs of M31b with 
the MDF obtained in \cite{BellazziniETal} in different fields.
The model MDFs are derived from main--sequence stars,
and convolved with a Gaussian error function with $\sigma~=~0.25$~dex. 
There is qualitative agreement between the MDFs of the M31b~model and 
those observed, especially in the disc--dominated fields. 
The main exception is the G1 halo--dominated field, 
whose observed MDF is more metal--poor and bimodal. 
The MDFs of the G351 and G219 halo--dominated fields
show narrower peaks than those
of the model. Differences in the shape of the observed halo MDFs among
the different fields suggest possible inhomogeneities 
in the M31 halo (see also \citealt{BellazziniETal}).

In the framework of our semi--analytic models,
M31b gives the best fit for the observational constraints
available in M31. In Section~\ref{contrast:discussion}, 
we discuss the formation process of M31 which is implied by these results. 

\begin{table}
\caption{Radial gradients~(dex/kpc) for the present--day mean stellar metallicity of the MDFs
in the M31~models. The results for the MW~model and for the M31~fields observed in Bellazzini~et~al.~(2003) are also presented as reference. $\left(\Delta\langle[Fe/H]\rangle/\Delta R\right)$ refers to the range 4~--~16~kpc for the MW~model and 4~--~20~kpc for the M31~models. $\left(\Delta\langle[Fe/H]\rangle/\Delta R\right)_{halo}$ refers to the range 4~--~30~kpc in the halo
for both the MW~and~the~M31~models.}
\begin{center}
\begin{tabular}{@{}lll@{}}
\label{contrast:tab2}
&&\\
\hline
\hline
&&\\
&$\left(\Delta\langle[Fe/H]\rangle/\Delta R\right)$&$\left(\Delta\langle[Fe/H]\rangle/\Delta R\right)_{halo}$\\
&&\\
\hline
&&\\
  MW & $-0.058$ & $-0.021~$\\
M31$_{\rm obs}$ & $-0.026$ & $-0.008$\\
M31a & $-0.041$ & $-0.021~$\\
M31b & $-0.021$ & $-0.003~$\\
M31c & $-0.014$ & $-0.009~$\\
&&\\
\hline
\hline
\end{tabular}
\end{center}
\end{table}

\subsection{M31c~model: another possible model}
\label{contrast:results:M31:M31c}

The results of the M31b~model
suggest that a stronger halo SF
can produce a metal--rich halo.
However, applying the higher SFE is not
the only way to induce stronger SF.
This can also be achieved by employing a
larger reservoir of infalling halo gas, 
without increasing the SFE.
Here we discuss the M31c~model,
which assumes a much larger present--day halo surface density
and a longer (6~Gyr) halo phase (Table~\ref{contrast:tab1}).
Again, to explain the observed hydrogen surface density, M31c uses
a shorter disc infall time--scale, and a weaker radial dependence,
which also leads to a present--day SFR similar to that of the M31b~model.

Fig.~\ref{contrast:fig2} shows the predicted present--day radial distributions of oxygen and sulfur 
abundances which are in agreement with the observational data to roughly 
the same degree of the other models. The present--day hydrogen surface
density profile of the M31c~model is consistent with \cite{DameETal}
in the inner region of the disc at $R~<~12$~kpc. However, M31c 
has a broader peak at a larger radius than in M31b.
Consequently, the model overestimates the surface density 
in the outer region at $R~>~12$~kpc.  
Again, this result might suggest that additional physical mechanisms
operate, 
to explain the observed low hydrogen surface density at the outer radii.

Fig.~\ref{contrast:fig3} shows that M31c also reproduces the observed
present--day radial distributions of the metallicity, similarly to the M31b~model.
The metallicity gradient in M31c is flatter than in M31b at the inner radii 
as a consequence of the weaker radial dependence
of the disc infall time--scale (i.e. a smaller $b_d$). 
In the halo fields, M31c also has systematically higher metallicity 
than in M31a and a steeper gradient than in M31b. 
Although the gradient is in better agreement with 
the observations than in M31b (Table~\ref{contrast:tab2}), M31c has lower mean metallicities than observed, 
especially at the outer radii.
In Fig.~\ref{contrast:fig4}, the predicted MDFs at different radii are 
displayed. The MDFs of M31c are also in good agreement
with those observed, especially in the disc. 
On the other hand, the halo MDFs show a more pronounced
metal--poor tail than in M31b. 

Thus, the M31c~model also leads to acceptable results, although 
M31b is in better agreement with the observations. 
The next section discusses the formation history of M31 
implied by the M31b~and~M31c~models.

\section{Discussion}
\label{contrast:discussion}

The results in the previous section have shown 
that the main trends of the chemical evolution
history of both the MW and M31 can be described using the same
framework. 
We have also found that to
explain the observations in more detail, different sets of
parameter values for the formation history are required
for the MW and M31, respectively. 
In this section, we discuss how these different parameters values
relate to different galaxy formation histories.

The previous sections have shown that 
the main differences
between the MW and M31 include 
the present--day hydrogen surface density in the disc and, more significantly, 
the present--day halo metallicity.
The smaller hydrogen surface density in M31
can be explained by a combination of
the shorter disc infall time--scale and the higher disc SFE.

The most striking difference is that 
the mean metallicity observed in the M31 halo is about ten
times higher than that observed in the MW halo, though both
galaxies have similar mass and morphology.

\textit{Which is a ``typical''
halo?} Metal--rich halos seem more common than metal--poor ones, as 
pointed out by \cite{ZibettiETal} who analysed a sample
of about 1000 edge--on disc galaxies within the SDSS. 
\cite{HH} have also shown that the NGC5128 halo MDF
closely resembles 
that in the M31 halo.
This similarity
is perhaps suggestive of a common history in the halo formation of both
galaxies, despite their different Hubble--type.

A straightforward way to obtain a more metal--rich halo would 
be by requiring homogeneous pre--enrichment of the infalling halo gas 
with a metallicity $Z~\approx~0.1~Z_{\odot}$. 
However, this metallicity seems too high, 
if such pre--enrichment occurred homogeneously in the whole universe.
For example, although this value is close to the mean metallicity
of Damped Lyman $\alpha$ systems (DLAs), 
many DLAs have much lower metallicity
(\citealt{Pettini03}; \citealt{ProchaskaETal}). 
If DLAs are more evolved systems than the infalling halo gas, 
their metallicity should be higher than that induced by pre--enrichment. 
Therefore, a pre--enrichment of $Z~\approx 0.1~Z_{\odot}$ may not be tenable. 

We found that the observed metal--rich halo of M31 could be explained 
by two scenarios without any pre--enrichment: a) a higher halo SFE; 
b) a larger reservoir of infalling halo gas with a longer halo phase. 
Scenario~a) might be explained in a hierarchical clustering regime, 
where the halo is formed by accretion of building--blocks. 
Theory predicts that gas removal by
supernova--driven winds should be more effective in lower mass
systems, leading to a consequent suppression of the SFR and
a low SFE (e.g.:~\citealt{DS}; \citealt{E}). 
In addition, observations based on the SDSS also
suggest that the SFE decreases with decreasing stellar mass in low mass
galaxies with a stellar mass M$_{\star}~<~3\times10^{10}$~M$_{\odot}$
\citep{KauffmannETal}. Therefore, higher SFE in the M31 halo
phase would be explained if the building--blocks of the M31 halo
are systematically higher mass systems than those of the MW halo
(see also \citealt{HH}).
This result again supports the notion of 
diversity in the formation of spiral galaxies which are apparently
similar in mass and morphology and belong to the same Local Group. 
Admittedly, our
model does not adhere to the hierarchical--clustering
scenario in a self--consistent manner, however, 
our chemical evolution models can be
interpreted as the ``mean'' SF and chemical evolution
history of the stars which end up at each radius.

As an observational consequence of our models,
we found that both scenarios~a)~and~b) predict 
a more massive stellar halo in M31, 
respectively about 6 and 9 times more massive
than in the MW, 
whose stellar halo mass is $\approx 10^{9}$~M$_{\odot}$.
This result suggests 
that there might be a correlation between
the halo metallicity and its stellar mass. 
Using dynamical simulations, Bekki, Harris \& Harris (2003) show that
the stellar halo comes from the outer part of
the progenitor discs when the bulge is formed by a major merger of
two spiral galaxies. Based on this, they also predict the correlation 
of the metallicity of the stellar halos and the mass of the bulges which
were formed by major mergers, since larger bulges have the larger 
progenitors, and progenitor spiral galaxies should follow the observed 
mass--metallicity relation. 
Although they do not mention a correlation between
the masses of the bulge and halo, it would naturally be expected.
Thus, a major merger scenario could explain our conclusion.
Future observational
surveys will better quantify the correlation between the halo
metallicity and its stellar mass (e.g.,~\citealt{MM}).

It is possible to distinguish scenarios~a)~and~b) through observation.
First, due to a longer halo phase, scenario~b) predicts 
intermediate--age population in the M31 halo. 
This picture could explain the
recent evidence of intermediate--age population
by deep imaging of the M31 halo \citep{BrownETal}.

It is worth mentioning that the metallicity gradient 
in the stellar halo is sensitive to the assumed present--day stellar halo density profile, 
especially in scenario~b); a steeper density profile leads to a steeper 
metallicity gradient.
\cite{PvdB1994} suggested that the outer halo of M31 can be modeled 
by a power law surface brightness profile of $I\propto R^{-4}$, 
which is much steeper than what we assumed ($\propto R^{-2}$). 
We found that such a steep profile rules out scenario~b) to
reproduce the flat metallicity gradient observed in the M31 halo.
However, it is still difficult to accurately measure the
halo surface brightness of M31 (e.g.,~\citealt{ZibettiETal}).
Thus, more observational estimates of the M31 halo 
surface brightness profile would be an important test for this scenario.

We also found that the higher halo SFE in scenario~a) leads to 
about 0.2~dex higher [O/Fe] when [Fe/H]$<$~-1~dex,
due to intense halo SF. Although measuring [O/Fe] is a hard challenge
for the current available instruments, this task could be
accomplished by the next--generation large--aperture optical telescope.

\subsection{Prospect}
\label{contrast:discussion:prospe}

The framework we have used can explain the main trends in the chemical
properties of both the MW and M31. However, recent observations
of stellar streams both in the MW 
(e.g.:~\citealt{HWdZ99}; \citealt{CB2000}; \citealt{ILIC02}; \citealt{BrookETal}; 
\citealt{NHF04}; \citealt{MKL04}) 
and M31 (\citealt{IbataETal};
\citealt{McConnachieETal}; \citealt{MerrettETal};
\citealt{ZuckerETal}; \citealt{LewisETal}) 
clearly identify inhomogeneities in the chemical and
dynamical history of both galaxies, which could suggest that 
a significant
fraction of the halo stars results from late accretion of satellite
galaxies. 
In light of these recent observations, it will be
important to study both the MW and M31 in more detail by employing 
a self--consistent
chemo--dynamical model (e.g.,~\citealt{BrookETal2004}) to trace their interrelated chemical and
dynamical histories. 





\chapter[The Halo Metallicity--Luminosity Relationship]{\LARGE Stellar Halo\\ Metallicity--Luminosity Relationship\\ for Late--Type Galaxies}
\label{chap:halo}

The stellar halos of late--type galaxies bear important chemo--dynamical
signatures of galaxy formation. We present here the analysis of 89
semi--cosmological late--type galaxy simulations, spanning $\sim$~4~magnitudes
in total galactic luminosity. These simulations sample a wide variety of
merging histories and show significant dispersion in halo metallicity at a
given total luminosity~-~more than a factor of~10 in metallicity. 
Our analysis suggests that galaxies
with a more extended merging history possess halos which have younger and
more metal rich stellar populations than the stellar halos associated with
galaxies with a more abbreviated assembly. A correlation between halo
metallicity and its surface brightness has also been found, reflecting the
correlation between halo metallicity and its stellar mass. Our
simulations are compared with recent Hubble~Space~Telescope observations
of resolved stellar halos in nearby spirals.

\section{Introduction} 
\label{melu:intro}

Understanding the formation history of stellar halos is one of the
classical pursuits of galactic astronomy. The problem is generally framed
within the context of two competing scenarios: one of ``rapid
collapse'' (Eggen, Lynden--Bell \& Sandage,~1962), in which the stellar
halo is formed by the rapid collapse of a proto--galaxy within a dynamical
timescale ($\sim$~10$^{8}$~yr), and one of ``galactic assembly'' 
\citep{SZ}, whereby the stellar halo is assembled on a longer timescale
($\sim$~10$^{ 9}$~yr) by the accretion of ``building--blocks'', each with
separate enrichment history. Both scenarios have their strengths and
weaknesses, and it would appear that a hybrid model is the most plausible
option consistent with extant data (e.g.:~\citealt{CB2000}; \citealt{FBH}). 

An intriguing piece of the halo formation ``puzzle'' is provided by 
comparing the stellar halo of our own Milky Way with that of its
neighbour, M31. First, despite their comparable total galactic luminosities,
the stellar halo of M31 is {\it apparently} much more metal--rich than that of
the Milky Way (e.g.:~\citealt{RN}; \citealt{MK}; Durrell, Harris \& Pritchet~2001; 
\citealt{FergusonETal}; \citealt{BrownETal}; \citealt{BellazziniETal}; \citealt{Durrell04}; \citealt{Ferguson05}; \citealt{Irwin}). In fact, the halo of M31 bears a closer resemblance to that of NGC~5128
(e.g.,~\citealt{HH}), despite their differing morphological classifications.

The stellar halo--galaxy formation symbiosis has been further brought to
light by the recent work of Mouhcine~et~al. (see also Tikhonov, Galazutdinova \&
Drozdovsky~2005). The deep Hubble~Space~Telescope imaging of nearby spiral galaxy stellar
halos in \cite{MM} suggests a correlation between stellar 
halo metallicity and total galactic luminosity. On the surface, this
correlation appears to leave our own Milky Way halo as an
outlier with respect to other spirals, with a stellar halo metallicity $\sim$~1~dex
lower than spirals of comparable luminosities (such as M31, as alluded to
earlier). 

However, it must be noted that the metallicity of the Galactic 
halo in the Solar Neighbourhood (e.g.,~in Ryan \& Norris 1991) comes 
from spectroscopic metallicities in a kinematically--selected 
sample, whereas that of the M31 halo, as well as those of the 
stellar halos of nearby spiral galaxies \citep{MM}, 
have primarily been derived from photometric metallicities in 
topographically--selected samples. Since it is now becoming 
possible to obtain spectroscopic metallicities for significant 
samples of kinematically--selected giants in the M31 halo (e.g.:~\citealt{Ibata04}; \citealt{Kalirai}; \citealt{Chapman}; \citealt{Ibata07}), 
it will be crucial to assess the consistency of the M31 stellar halo metallicity as
derived from kinematically-- and from topographically--selected samples, respectively.

A grasp of the true scatter around the general trend in the halo metallicity--luminosity relation awaits a larger observational data set, however any theory which attempts to explain such a relation needs to simultaneously account for the {\it apparent} metallicity discrepancy between the Milky Way and the Andromeda stellar halos. The question arises as to what is driving the scatter of halo metallicity for galaxies of comparable luminosities?

In what follows, we investigate whether differences in the stellar halo
assembly history can explain the diversity seen in halo metallicities.
Using chemo--dynamical simulations,
we have constructed a sample 
of 89 late--type galaxies, spanning
$\sim$~4~magnitudes in luminosity, and sampling a wide range of assembly histories
at a given luminosity (or mass). We contrast the stellar halo
metallicity--galactic luminosity relation in the simulations with the recent empirical
determination of \cite{MM}. The framework in which
the simulations have been conducted is described in Section~\ref{melu:model}, while
Sections~\ref{melu:results}~and~\ref{melu:discussion} present the results and the related discussion, respectively.

\section{Simulations}
\label{melu:model}

The simulations employed here are patterned after the semi--cosmological
adiabatic feedback model of \cite{BrookETal2004}, and were constructed using
the chemo--dynamical code {\tt GCD+} \citep{KG03}. {\tt GCD+}
self--consistently treats
the effects of gravity, gas dynamics, radiative cooling, star
formation, and chemical enrichment (including Type~II~and~Ia~Supernovae and 
relaxing the instantaneous recycling approximation). 
In this framework, gas within the SPH smoothing kernel of Type~II~Supernova explosions is prevented from cooling for the lifetime of the lowest mass star which ends as a Type~II~Supernova (\citealt{BrookETal2004} and references therein). Details of both {\tt GCD+} and the feedback model can be found in \cite{KG03} and \cite{BrookETal2004}, respectively.

This semi--cosmological version of \texttt{GCD+} is based on the
seminal work of \cite{KG91}. The initial condition for a given
simulation is an
isolated sphere of dark matter and gas. This ``top--hat'' overdensity has an
amplitude $\delta_{\rm i}$ at initial redshift z$_{\rm i}$, which is approximately
related to the collapse redshift z$_{\rm c}$ by z$_{\rm c}~=~0.36\delta_{\rm i}(1~+~{\rm z}_{\rm i})~-~1$ (e.g.,~\citealt{Padmanabhan}). We set z$_{\rm c}~=~2.0$, which
determines $\delta_{\rm i}$ at z$_{\rm i}~=~40$. Small--scale density fluctuations
based on a CDM power spectrum are superimposed on the sphere using
\texttt{COSMICS} \citep{Bertschinger}, and the amplitude of the fluctuations is
parameterised by $\sigma_{8}$. These fluctuations are the seeds for local
collapse and subsequent star formation. Solid--body rotation corresponding
to a spin parameter $\lambda$ 
is imparted to the initial sphere to incorporate the
effects of longer wavelength fluctuations. For the flat CDM model
described here, the relevant parameters include $\Omega_{0}~=~1$, baryon
fraction $\Omega_{\rm b}~=~0.1$, H$_{0}~=~50$~km~s$^{-1}$~Mpc$^{-1}$, spin
parameter $\lambda~=~0.06$, and $\sigma_{8}~=~0.5$.
We employed 14147~dark~matter and 14147~gas/star~particles.

A series of 89 simulations -~all with the \textit{same} number of particles, collapse redshift z$_{\rm c}$, and spin
parameter $\lambda$~- were completed spanning
a factor of 50 in mass in four separate mass ``bins'', 
each with a related initial comoving radius R$^{\rm i}_{40}$ 
at z$_{\rm i}~=$~40:
M$_{\rm tot}$~=~10$^{11}$~M$_{\odot}$ (R$^{\rm i}_{40}~=$~0.7~Mpc);
5$\times$10$^{11}$~M$_{\odot}$ (1.1~Mpc);
10$^{12}$~M$_{\odot}$ (1.4~Mpc);
5$\times$10$^{12}$~M$_{\odot}$ (2.4~Mpc).

For each total
mass, we run simulations with different patterns of small--scale density
fluctuations which lead to different hierarchical assembly histories. 
This was controlled by setting different random seeds
for the Gaussian perturbation generator in {\tt COSMICS}.

\section{Results}
\label{melu:results}

\begin{figure}
\begin{center}
\includegraphics[width=1.0\textwidth]{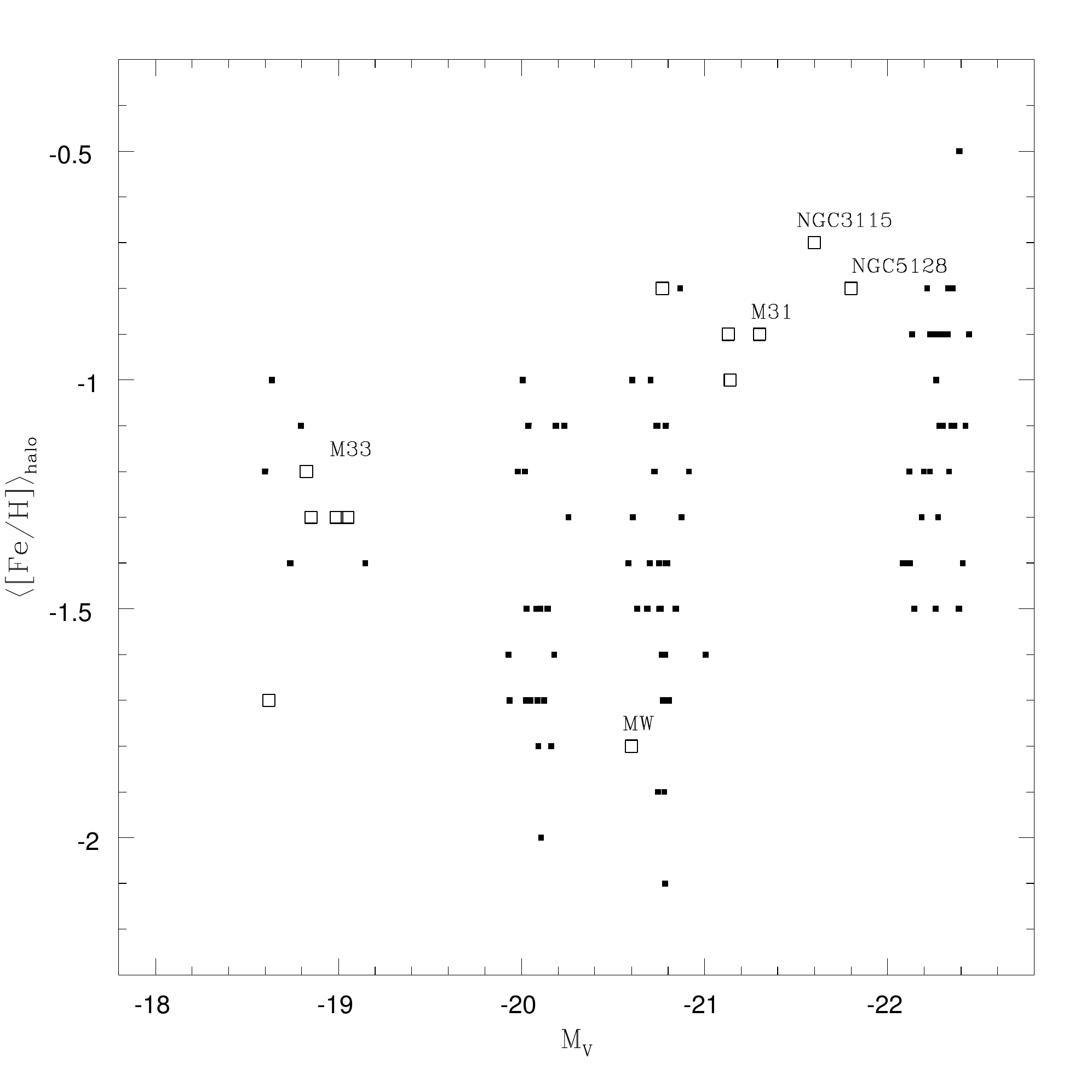}
\caption{The stellar halo metallicity--total galactic V--band luminosity relation ($\langle{\rm [Fe/H]}\rangle_{\rm halo}$~--~M$_{\rm V}$).
A filled box corresponds to the peak of the MDF for a simulation. The related MDFs are displayed in Appendix~\ref{app:appendixB}.
Open (unlabelled) boxes represent the data of Mouhcine~et~al.~(2005); labelled boxes represent
additional data taken from the literature (see the text for details).
68\% of the stars in the simulated (observed) MDFs are typically enclosed
within $\pm$0.7~dex ($\pm$0.35~dex) of the peak of the related MDF.}
\label{melu:fig1}
\end{center}
\end{figure}

Our grid of 89 
simulations was employed to populate the redshift z~=~0
stellar halo metallicity--luminosity plane 
(i.e.~$\langle [$Fe/H$]\rangle_{\rm halo}$~--~M$_{\rm V}$)
shown in Fig.~\ref{melu:fig1} (filled symbols). We have been conservative in our
topographical definition of ``stellar halo'', adopting a projected cut--off radius of 15~kpc
to delineate the halo from (possible) disc and bulge contaminants. Such a choice should also allow an easier comparison with the observations of halo fields in nearby spiral galaxies. 

The stellar
metallicity distribution (MDF) was then generated using all the stellar particles from the ``halo region'', convolved with a Gaussian with 
$\sigma_{\rm [Fe/H]}$~=~0.15~dex, representing the typical observational
uncertainties (e.g.,~\citealt{BellazziniETal}). Only those simulations with~$>$~100 halo stellar particles are included in the analysis here. The related MDFs are displayed in Appendix~\ref{app:appendixB}.

For each MDF, the peak is displayed in Fig.~\ref{melu:fig1}; 
68\% of the stellar particles within a given (simulated)
MDF\footnote{Each MDF has been fitted by an univariate skew--normal distribution (e.g.:~\citealt{Azzalini}; {\tt http://azzalini.stat.unipd.it/SN/Intro/intro.html}).} typically span 1.4~dex in metallicity. We
confirmed that the halo MDFs are
insensitive to projection effects, even in the most massive simulations, 
suggesting that our cut--off for the ``halo region'' is
sufficient for minimising disc and bulge contamination. The optical
properties of our simulated stellar populations were derived using the 
population synthesis models of \cite{ML}, taking into account the age 
and the metallicity of each stellar particle. We note that the V--band
luminosity shown along the abscissa of Fig.~\ref{melu:fig1} is the {\it total} galactic
luminosity (i.e.~halo + bulge + disc).

Fig.~\ref{melu:fig1} also shows the observed halo
metallicity--luminosity values for 13 nearby spirals, primarily taken from 
\cite{MM}, supplemented with data from the literature
(\citealt{BM}; \citealt{Elson}; Harris, Harris \& Poole~1999; 
Brooks,~Wilson~\&~Harris~2004)\footnote{Note that the identification of the stellar halo in M33 is still debated (e.g.,~Tiede, Sarajedini \& Barker~2004; see also \citealt{McConnachie}).}. For the observational
datasets we located the MDF peak exactly as we did
for the simulated datasets. The related MDFs are displayed in Figures~\ref{appB:data:fig1}~and~\ref{appB:data:fig2}; 68\% of the stars in the observed MDFs are typically enclosed within $\pm$~0.35~dex of the MDF peak, a factor of $\sim$~2 narrower than the
simulations\footnote{Our current model does not take into account any pre--enrichment scenario due to extremely metal--poor stars (Pop~III hereafter) whose detailed physics is still much debated (e.g.:~Woosley, Heger \& Weaver~2002; \citealt{Larson}). An early and homogeneous pre--enrichment of the proto--galactic masses due to Pop~III could lead to narrower MDFs at lower redshift.}.

What is readily apparent from Fig.~\ref{melu:fig1} is that significant dispersion in halo
metallicity ($\simgt$~1~dex) exists at any given total galactic luminosity in our
simulations. For a set of simulations with the same initial total mass,
the \textit{only} difference among these runs can be traced
to the random pattern of the initial small--scale density fluctuations;
this translates directly into differing hierarchical assembly histories.
Qualitiatively, it would appear that the assembly history alone
may account for the diversity in halo metallicity at a given galactic luminosity,
and thus account for the apparent outliers in the observed trend. The dispersion in halo metallicity at any given total galactic luminosity is summarised by the histograms in Fig.~\ref{melu:extra:fig1}, where the distribution of the MDF peaks for the semi--cosmological simulations in \cite{RendaHalo} is shown, although such histograms should be taken as plain summaries, since for each simulation the pattern of the initial small--scale density fluctuations - and the related hierarchical assembly history - has been arbitrarily chosen rather than drawn from a hypothetical a priori distribution of patterns.

\begin{figure}
\begin{center}
\includegraphics[width=1.0\textwidth]{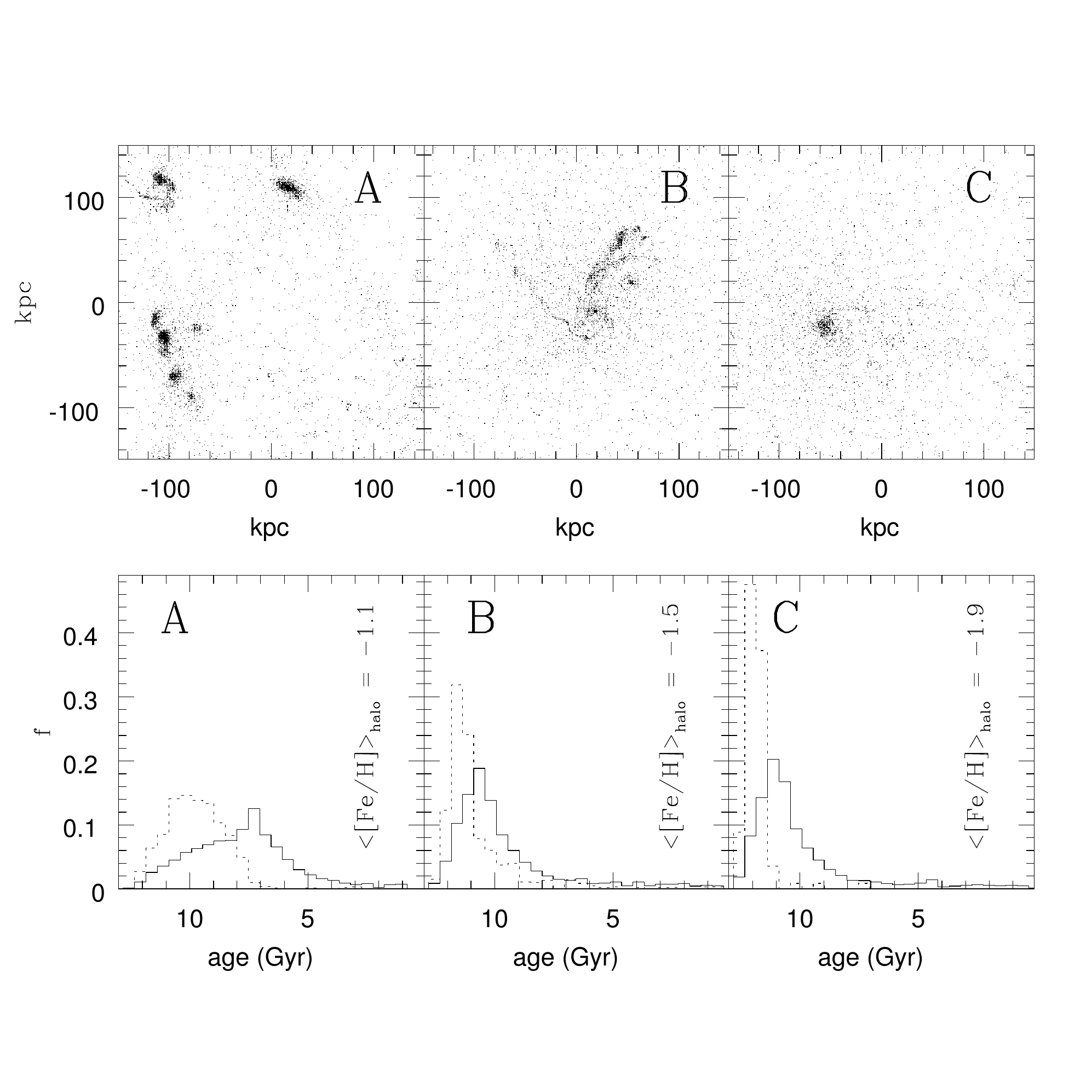}
\caption{Upper panels: Projected distribution of the gas particles at
redshift z~=~1.5 for simulations with M$_{\rm tot}$~=~10$^{12}$~M$_\odot$.
Lower panels: The associated stellar age distributions at
z~=~0 for the galaxies in the upper panels.
The solid (dotted) histogram corresponds to the
stellar age distribution for the entire galaxy (stellar halo).
The corresponding z~=~0~halo metallicities are denoted in each panel.}
\label{melu:fig2}
\end{center}
\end{figure}

\begin{figure}
\begin{center}
\includegraphics[width=1.0\textwidth]{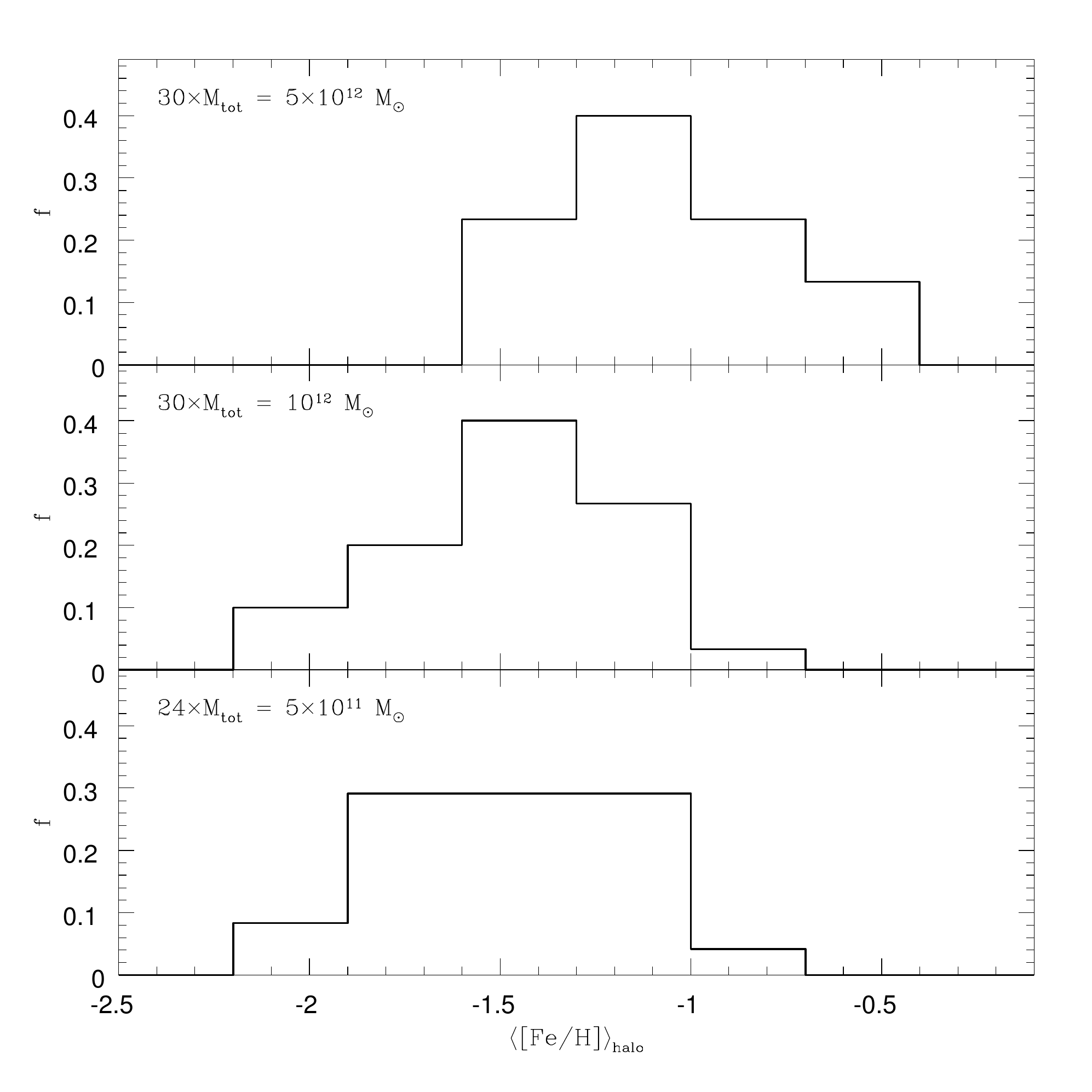}
\caption{Distributions of the z~=~0~MDF peaks for the semi--cosmological simulations in Renda~et~al.~(2005b) with M$_{\rm tot}$~=~:~5$\times$10$^{11}$~M$_\odot$; 10$^{12}$~M$_\odot$; 5$\times$10$^{12}$~M$_\odot$.}
\label{melu:extra:fig1}
\end{center}
\end{figure}

\begin{figure}
\begin{center}
\includegraphics[width=1.0\textwidth]{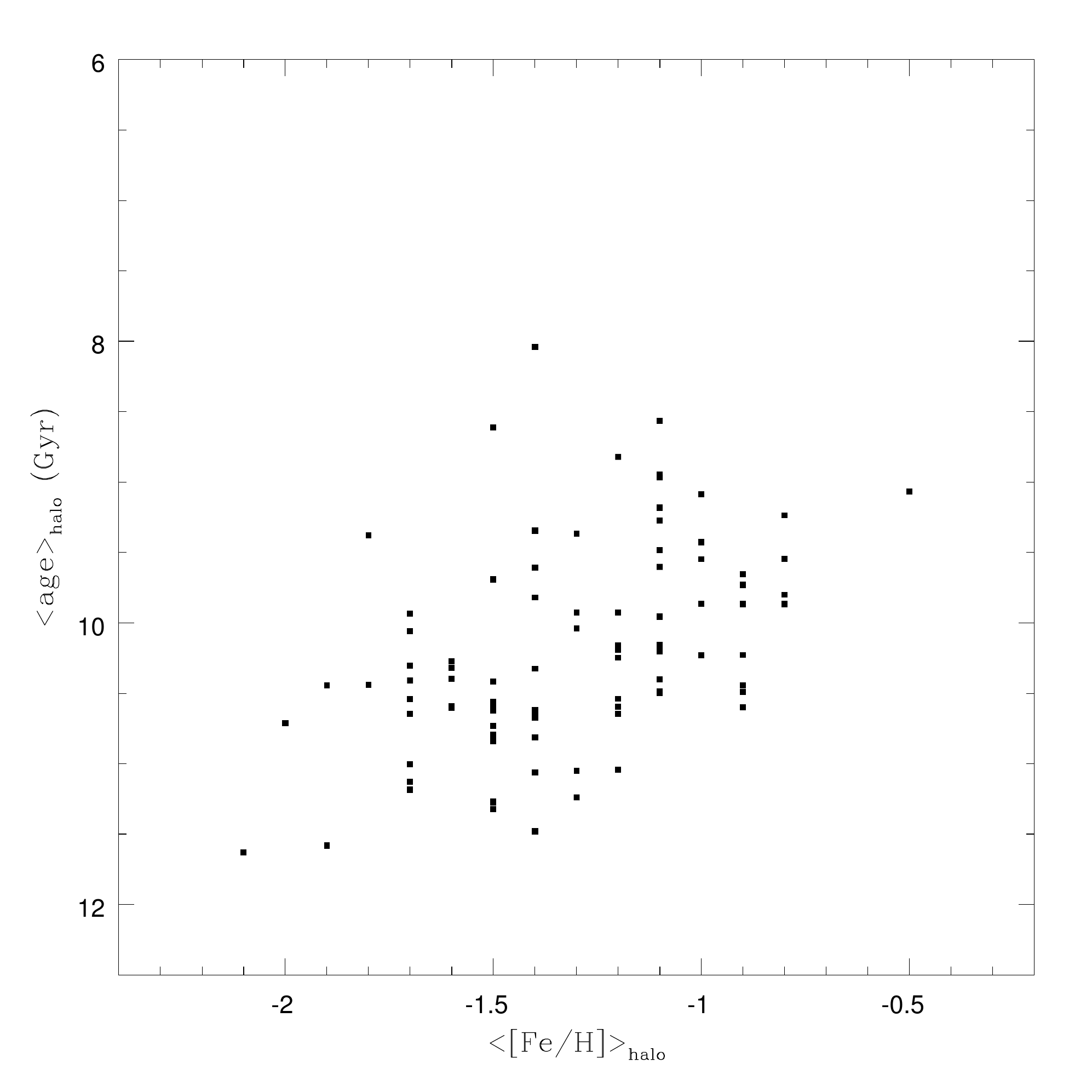}
\caption{The relation between the mean stellar halo age and the peak of the halo MDF in our simulations. The halo ADFs are displayed in Appendix~\ref{app:appendixD}.}
\label{melu:extra:fig2}
\end{center}
\end{figure}

\begin{figure}
\begin{center}
\includegraphics[width=1.0\textwidth]{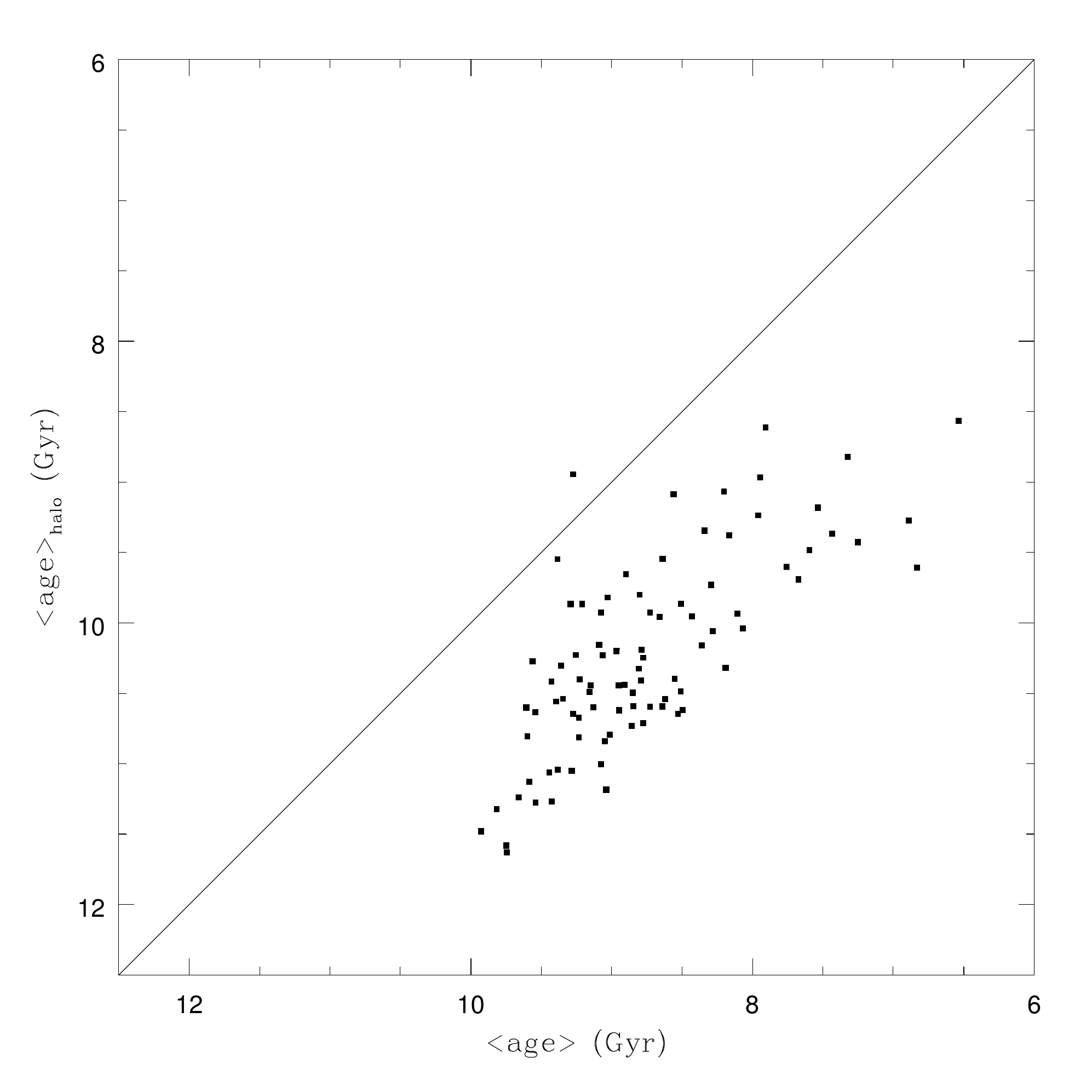}
\caption{The relation between the mean halo age and the mean age for the entire galaxy in our simulations. Both the halo and the galaxy ADFs for each simulation are displayed in Appendix~\ref{app:appendixD}.}
\label{melu:extra:fig3}
\end{center}
\end{figure}


\begin{figure}
\begin{center}
\includegraphics[width=1.0\textwidth]{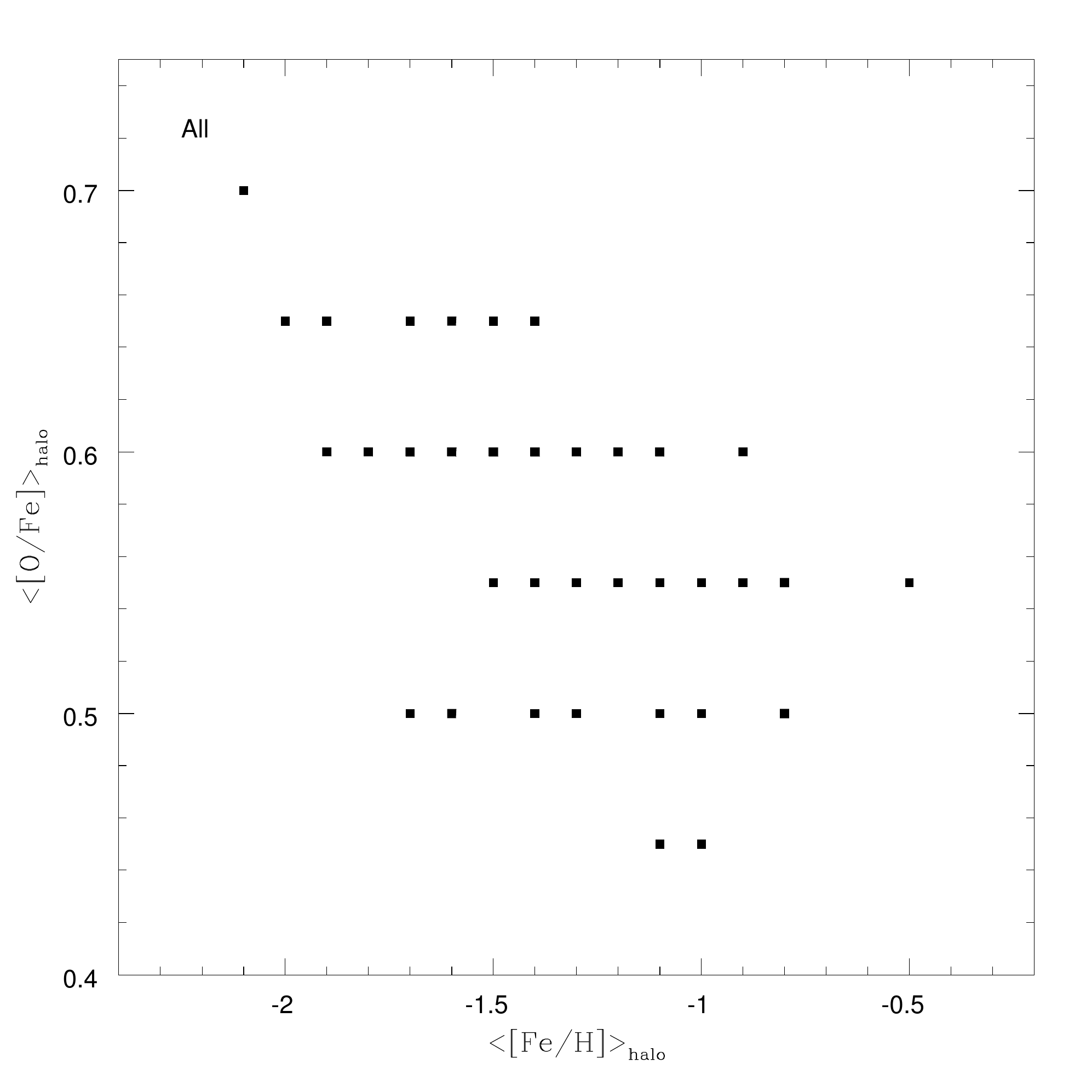}
\caption{The relation between the peak of the halo [O/Fe] distribution and the peak of the halo MDF in our simulations. The related MDFs are displayed in Appendix~\ref{app:appendixB}. The halo~$[$O/Fe$]$~distributions are displayed in Appendix~\ref{app:appendixC}.}
\label{melu:fig3}
\end{center}
\end{figure}

\begin{figure}
\begin{center}
\includegraphics[width=1.0\textwidth]{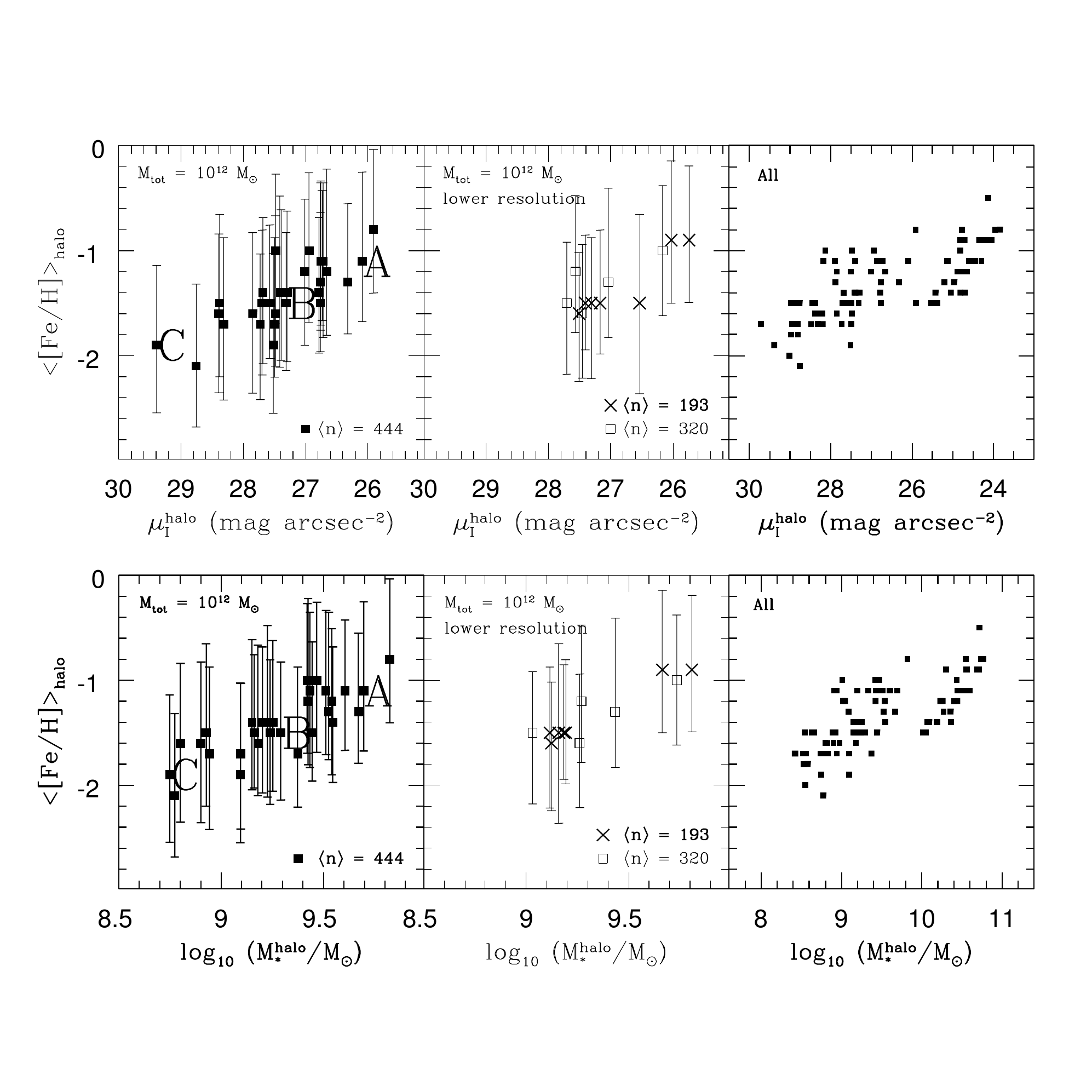}
\caption{Upper panels: Stellar halo metallicity--halo I--band surface brightness relation
($\langle [$Fe/H$]\rangle_{\rm halo}$~--~$\mu_{\rm I}^{\rm halo}$). 
In the left panel, our fiducial simulations with
M$_{\rm tot}$~=~10$^{12}$~M$_\odot$ and 14147$\times$2 particles are shown,
while the middle panel shows the corresponding lower--resolution
simulations with 5575$\times$2 (crosses) and 9171$\times$2 (open boxes)
particles. The error bars show the 68\% Confidence Level around each MDF peak.
The average number of halo particles is shown in the bottom
right corner of each panel. Simulations with fewer than~100~stellar halo
particles (at R$~>~$15~kpc) are not included here. Simulations A~--~C of Fig.~\ref{melu:fig2}
are displayed in the left panel. The right panel shows
the $\langle [$Fe/H$]\rangle_{\rm halo}$~--~$\mu_{\rm I}^{\rm halo}$ relation
for all the simulations.
Lower panels: Halo metallicity--stellar mass relation
($\langle [$Fe/H$]\rangle_{\rm halo}$~--~M$_{*}^{\rm halo}$). 
In the left panel, our fiducial simulations with
M$_{\rm tot}$ = 10$^{12}$~M$_\odot$, while the middle panel shows the corresponding lower--resolution
simulations. The right panel shows the $\langle [$Fe/H$]\rangle_{\rm halo}$~--~M$_{*}^{\rm halo}$ relation
for all the simulations.}
\label{melu:fig4}
\end{center}
\end{figure}

Fig.~\ref{melu:fig2} shows snapshots of the gas particles at redshift z~=~1.5 and the 
associated z~=~0 stellar Age Distribution Function (ADF) for
three of the simulations with M$_{\rm tot}$~=~10$^{12}$~M$_\odot$. 
These representative runs show that simulations with 
more metal--rich halos are assembled over a longer timescale,
and thus possess a broader ADF (in both the halo and the associated galaxy).
Such a scenario is consistent with the evidence, presented by \cite{BrownETal},
of a substantial intermediate--age metal--rich population in a topographically--selected halo field
of M31.
Conversely, simulations with more metal--poor halos
are assembled earlier through more of a monolithic process,
with a consequently narrower ADF. The relationship between the mean stellar halo age and the peak of the halo MDF is summarised in Fig.~\ref{melu:extra:fig2}. We note that the mean halo age in the simulations broadly correlates with the mean galaxy age as shown in Fig.~\ref{melu:extra:fig3}, thus suggesting that the assembly history of the halo ``knows'' of the
formation history of the other galaxy structural components (e.g.,~bulge
and disc). Both the halo and the galaxy ADFs for each simulation are displayed in Appendix~\ref{app:appendixD}.

Further, the most metal--poor stellar halos (which, recall,
formed preferentially via more of a monolithic collapse) possess $\alpha$--elements to
iron ratios a factor of $\sim$~2 higher than the most metal--rich halos (which
formed preferentially over a more extended period of hierarchical clustering),
as shown in Fig.~\ref{melu:fig3}. Such a trend is expected if the different amounts of $\alpha$--elements and Iron released
by Type~II and Ia Supernovae over different timescales is taken into account
(e.g.:~Timmes, Woosley \& Weaver~1995; \citealt{WHW}). The halo~$[$O/Fe$]$~distributions are displayed in Appendix~\ref{app:appendixC}.

Future analysis of higher resolution simulations will clarify whether or not the relationships Figures~\ref{melu:extra:fig2},~\ref{melu:extra:fig3}~and~\ref{melu:fig3} point to are broadened in our sample because of the current resolution - thus mimicking tighter correlations.

The upper panels of Fig.~\ref{melu:fig4} show the stellar halo metallicity--I--band surface
brightness relation ($\langle{\rm [Fe/H]}\rangle_{\rm halo}$~--~$\mu_{\rm I}^{\rm halo}$) for simulations with
M$_{\rm tot}$~=~10$^{12}$~M$_\odot$; the surface brightness was measured
at a projected distance of 20~kpc from the dynamical centre of each simulation.
The halo metallicity--stellar mass relation ($\langle [$Fe/H$]\rangle_{\rm halo}$~--~M$_{*}^{\rm halo}$)
is shown in the lower panels of Fig.~\ref{melu:fig4}. The three galaxies
presented in Fig.~\ref{melu:fig2} are also displayed in Fig.~\ref{melu:fig4}.
An immediate correlation is apparent with more massive halos possessing higher
surface brightness and also higher metallicity. This is consistent
with a picture in which galaxies that experienced more extended
assembly histories have more massive stellar halos, with both higher halo metallicity and halo
surface brightness~-~i.e.~higher stellar halo density.

\subsection{Robustness}
\label{melu:results:robu}
\subsubsection{Resolution}
\label{melu:results:robu:numeri}

The issue of model convergence (resolution) is always a concern when
interpreting cosmological simulations (particularly when including
baryons). To test this, we conducted a series of simulations for
M$_{\rm tot}$~=~10$^{12}$~M$_\odot$, with 5575$\times$2 and 9171$\times$2
particles, to supplement the default grid (which used 14147$\times$2
particles); the middle panels of Fig.~\ref{melu:fig4} show where these lower--resolution
simulations sit in the halo metallicity--surface brightness plane and in the halo metallicity--stellar mass plane, respectively.

The consistency between higher (left panels) and lower (middle panels)
resolution simulations is reassuring\footnote{Since we limit our analysis to simulations with~$>$~100 halo stellar particles, the $\mu_{\rm I}^{\rm 
halo}~>~28$~mag~arcsec$^{-2}$ region of the halo metallicity--surface
brightness plane and the M$_{*}^{\rm halo}~<~10^{9}$~M$_\odot$ region of the halo metallicity--stellar mass plane 
are underpopulated by the lower--resolution runs (middle
panels of Fig.~\ref{melu:fig4}).}.

\subsubsection{Halo Semantics}
\label{melu:results:robu:sema}

\begin{figure}
\begin{center}
\includegraphics[width=1.0\textwidth]{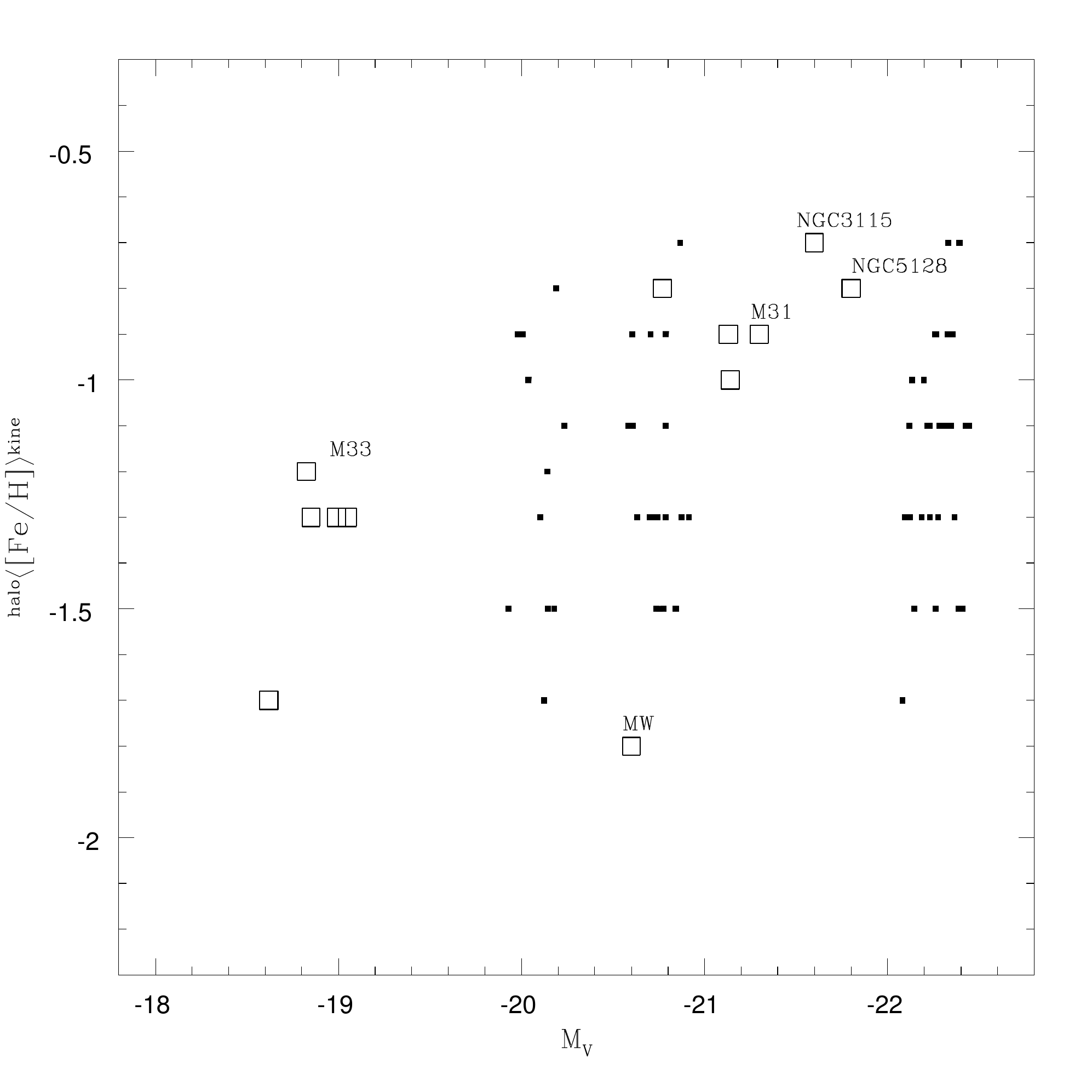}
\caption{The stellar halo metallicity--total galactic V--band luminosity relation ($^{\rm halo}\langle [$Fe/H$]\rangle^{\rm kine}$~--~M$_{\rm V}$). A filled box corresponds to the peak of the MDF which refers, for a simulation, to the ensemble of counter--rotating (v$_{\theta}~<~0$) stellar particles within each topographical halo (at a projected radius R$~>~$15~kpc). The related MDFs are displayed in Appendix~\ref{app:appendixF}. The observational data label--code is as in Fig.~\ref{melu:fig1}.}
\label{melu:kine:fig1}
\end{center}
\end{figure}

Up to now we have dealt with a topographical definition of stellar halo: we have labelled as ``halo'' the ensemble of stellar particles in a simulation -~at~z~=~0~- at a projected radius R$~>~$15~kpc. What if the ``halo'' is defined in a different way? That's to say:  given a different ``halo'' labelling, how robust is the dispersion in halo metallicity at a given total luminosity? As a robustness test, we have tried to include kinematical information in our halo definition -~in a simple way. We have relabelled as ``halo'' the ensemble of stellar particles in a simulation -~at~z~=~0~- at a projected radius R$~>~$15~kpc \textit{and} counter--rotating i.e.~with circular velocity v$_{\theta}~<~0$. The galactic circular velocity profiles of the sample we have analysed are displayed in Appendix~\ref{app:appendixE}. Here we focus on the stellar halo\footnote{Clearly, the galactic circular velocity profiles show a range of disc features among the late--type galaxy simulations we have analysed. Such features and their relations with both the framework we have chosen at the current resolution and the patterns of initial density fluctuations i.e.~the formation histories we have arbitrarily sampled will be the topic of future studies.}. Only those simulations with~$>$~100 stellar particles which are counter--rotating with v$_{\theta}~<~0$ at a projected radius R$~>~$15~kpc are included in the robustness test here. The related MDFs are displayed in Appendix~\ref{app:appendixF}. The MDF peaks are shown in Fig.~\ref{melu:kine:fig1}.

The dispersion in metallicity at a given total luminosity we found among the topographical halos is not wiped out when we include kinematical information in our halo definition, as shown in Fig.~\ref{melu:kine:fig1} -~to be compared with Fig.~\ref{melu:fig1}. 

\begin{figure}
\begin{center}
\includegraphics[width=1.0\textwidth]{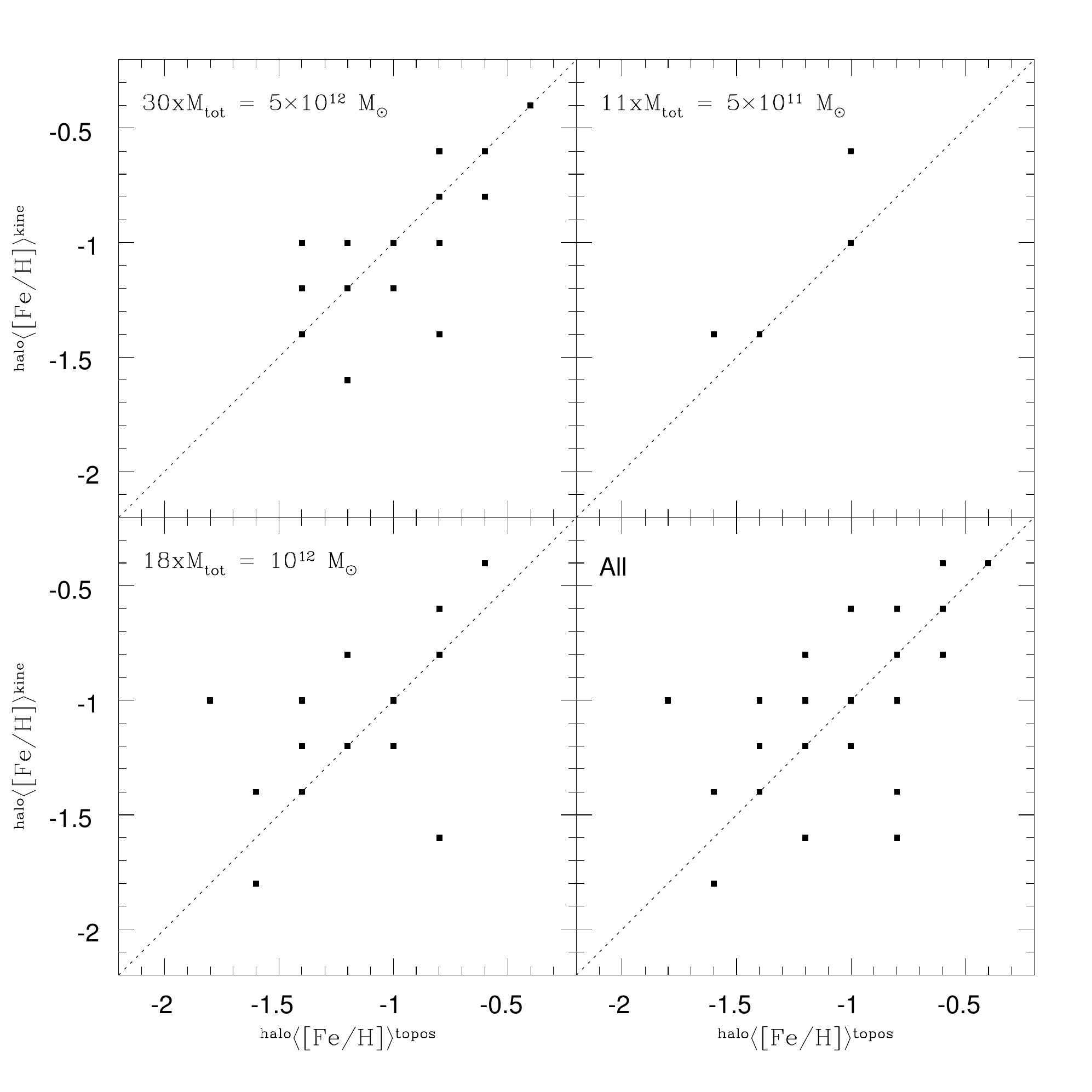}
\caption{The relationship between $^{\rm halo}\langle [$Fe/H$]\rangle^{\rm topos}$ (the ``topographical'' halo metallicity) and $^{\rm halo}\langle [$Fe/H$]\rangle^{\rm kine}$ (the metallicity of the ensemble of counter--rotating stellar particles within each topographical halo).}
\label{melu:kine:fig2}
\end{center}
\end{figure}

\begin{figure}
\begin{center}
\includegraphics[width=1.0\textwidth]{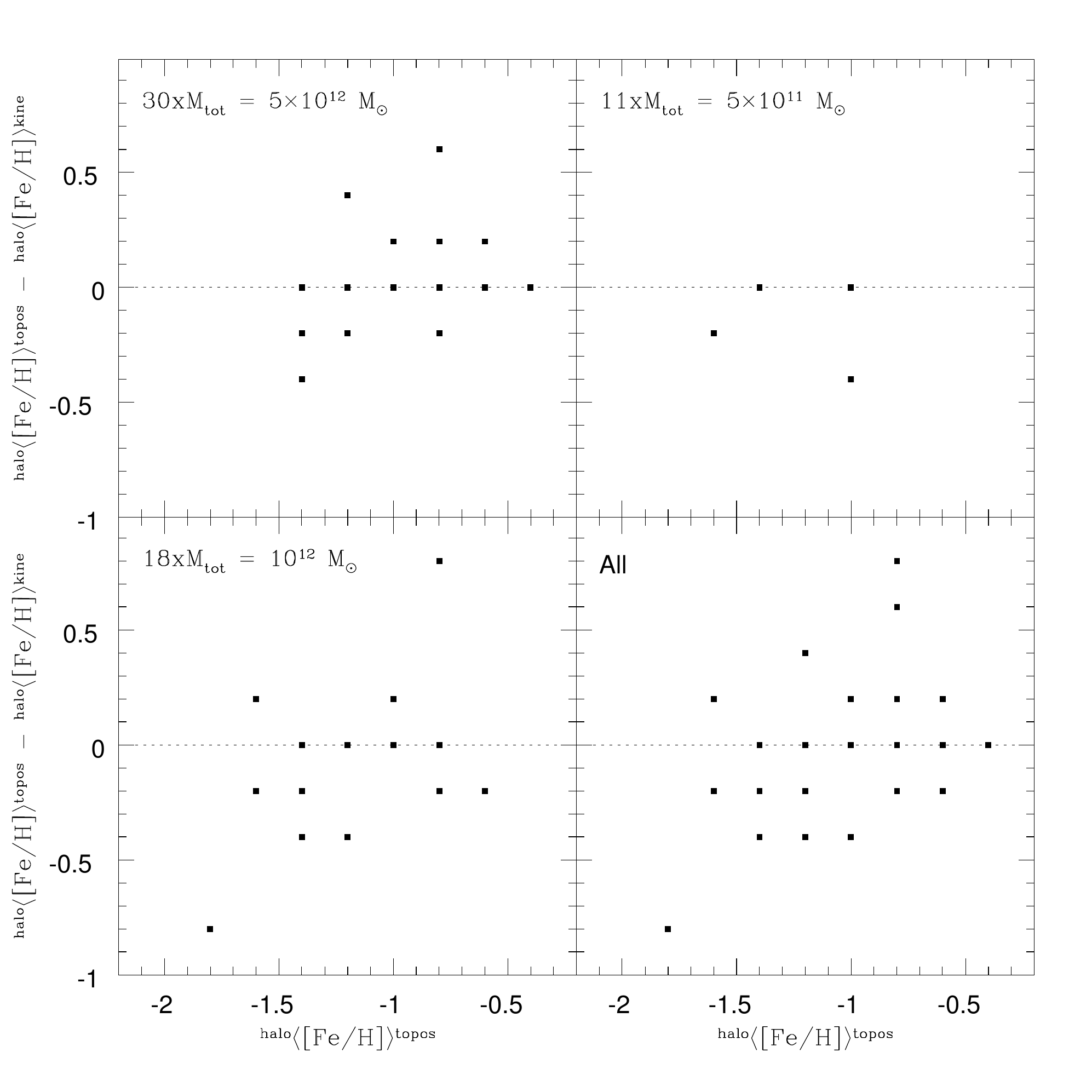}
\caption{The discrepancy between $^{\rm halo}\langle [$Fe/H$]\rangle^{\rm topos}$ (the ``topographical'' halo metallicity) and $^{\rm halo}\langle [$Fe/H$]\rangle^{\rm kine}$ (the metallicity of the ensemble of counter--rotating stellar particles within each topographical halo) as a function of the ``topographical'' halo metallicity.}
\label{melu:kine:fig3}
\end{center}
\end{figure}

Given a simulation, the relationship between $^{\rm halo}\langle [$Fe/H$]\rangle^{\rm topos}$ (the ``topographical'' halo metallicity) and $^{\rm halo}\langle [$Fe/H$]\rangle^{\rm kine}$ (the metallicity of the ensemble of counter--rotating stellar particles within each topographical halo) is shown in Fig.~\ref{melu:kine:fig2} whereas the discrepancy in metallicity between the two ``halo'' labels is shown in Fig.~\ref{melu:kine:fig3} as a function of the ``topographical'' halo metallicity. 

\begin{figure}
\begin{center}
\includegraphics[width=1.0\textwidth]{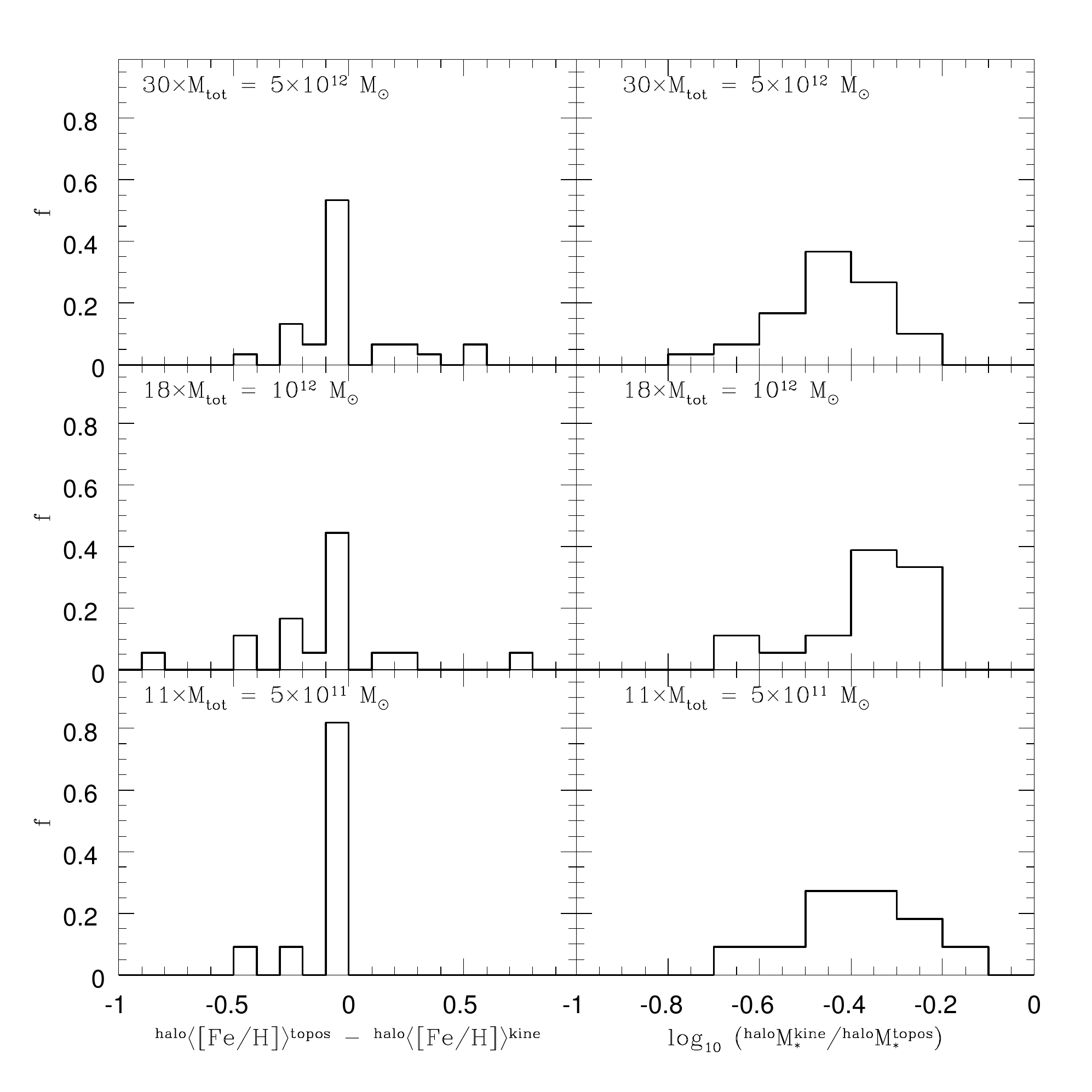}
\caption{Left panels: For each subset, the distribution of the discrepancy between $^{\rm halo}\langle [$Fe/H$]\rangle^{\rm topos}$ (the ``topographical'' halo metallicity) and $^{\rm halo}\langle [$Fe/H$]\rangle^{\rm kine}$ (the metallicity of the ensemble of counter--rotating stellar particles within each topographical halo). Right panels: For each subset, the distribution of the ``topographical'' halo mass fraction the ensemble of its counter--rotating stellar particles amounts to.}
\label{melu:kine:fig4}
\end{center}
\end{figure}

For each subset, the distribution of the ``topographical'' halo mass fraction the ensemble of its counter--rotating stellar particles amounts to is shown in the right panels of Fig.~\ref{melu:kine:fig4}. 

Both Figures~\ref{melu:kine:fig2}~and~\ref{melu:kine:fig3} support what is summarised by the histograms in the left panels of Fig.~\ref{melu:kine:fig4} i.e.~the discrepancy between $^{\rm halo}\langle [$Fe/H$]\rangle^{\rm topos}$ and $^{\rm halo}\langle [$Fe/H$]\rangle^{\rm kine}$ is distributed around $\approx$~0~dex within a range of $\approx~\pm$~0.5~dex -~although the significance of such a distribution is undermined by both the small size and the current resolution of the sample of simulations which have been analysed. A larger sample size and higher resolution in the simulations will allow a stronger statistical analysis and more complex kinematical selections in the future.

\subsubsection{Halo Metallicity}
\label{melu:results:robu:cmd}

Up to now, each halo MDF we have analysed and the related metallicity peak have referred to the ensemble of stellar particles within what has been labelled as ``halo'' in a simulation, either the ensemble of stellar particles at a projected radius R$~>~$15~kpc (the ``topographical'' halo) or the ensemble of counter--rotating (v$_{\theta}~<~0$) stellar particles within the ``topographical'' halo. 

As a robustness test, we generate the simulated Colour--Magnitude--Diagram (CMD hereafter) for each ``topographical'' halo. Next, we approach each simulated CMD the same way an observed CMD is approached in \cite{MMc},~i.e., through the same pipeline, a metallicity--colour relationship is constructed out of the fiducial Red~Giant~Branch (RGB hereafter) tracks to derive the metallicity of each (generated) star in the RGB of the (simulated) CMD.

\begin{figure}
\begin{center}
\includegraphics[width=1.0\textwidth]{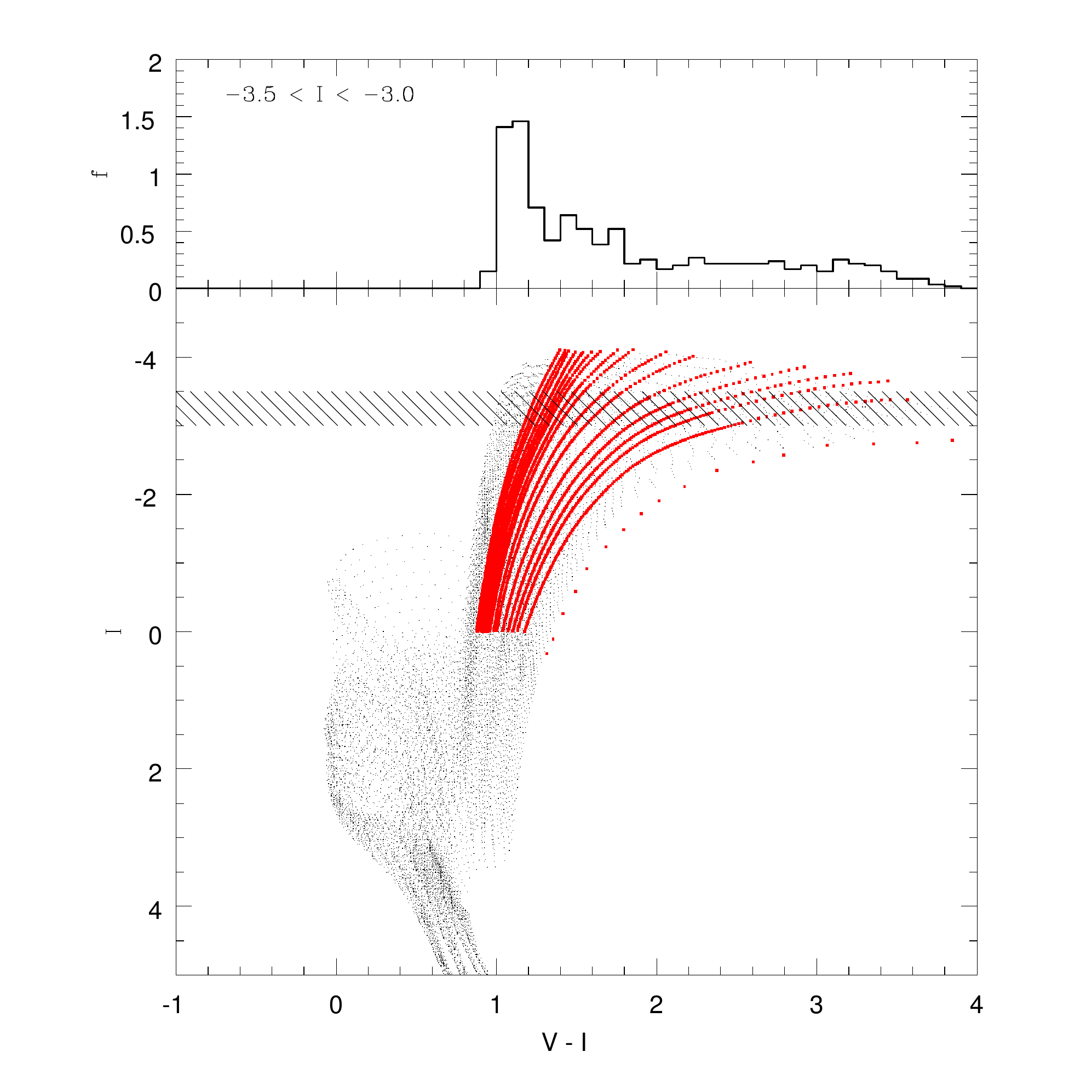}
\caption{The Padova theoretical isochrones (Girardi et al., 2002) up to the RGB~tip in the metallicity range of 0.0001$~<~$Z$~<~$0.030 which have been used to generate the simulated CMD out of the ensemble of stellar particles in the ``topographical'' halo for each galaxy simulation. They are shown as black dots. Red dots display the fiducial RGB~tracks (i.e.~0.8~M$_\odot$~VandenBerg~et~al.~2000~RGB~tracks and the empirical NGC~6791~RGB~track) which are used, as in Mouhcine~et~al.~(2005), to construct a metallicity--colour relationship to derive $[$Fe/H$]$ out of the (V~-~I) colour of each (generated) star in the RGB of the (simulated) CMD at -3.5$~<~$I$~<~$-3.0 (i.e.~the shaded region). The histogram in the top panel displays the (V~-~I) colour distribution for the Padova theoretical isochrones at -3.5$~<~$I$~<~$-3.0.}
\label{melu:CMD:iso}
\end{center}
\end{figure}

\begin{figure}
\begin{center}
\includegraphics[width=1.0\textwidth]{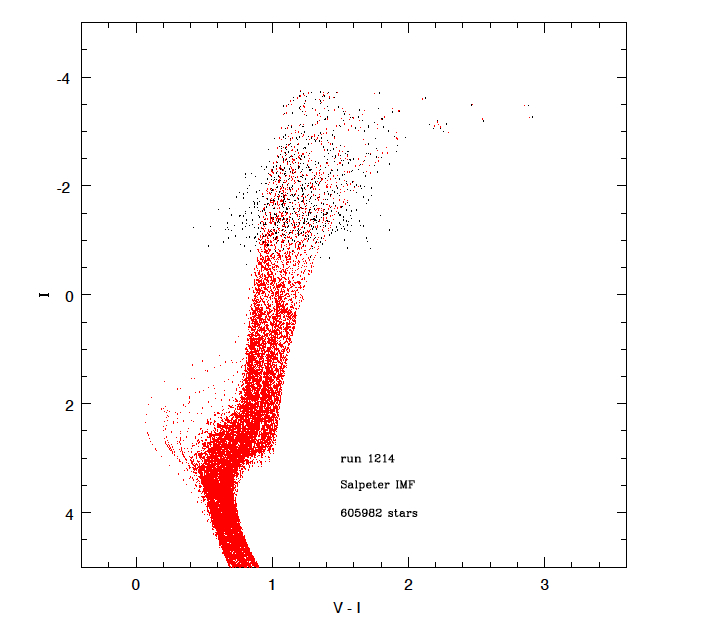}
\caption{Topographical halo simulated CMD for the label~``A'' semi--cosmological simulation (Figures~\ref{melu:fig2}~and~\ref{melu:fig4}) at M$_{\rm tot}$~10$^{12}$~M$_\odot$. The observational errors are taken into account (e.g.,~Secker~\&~Harris 1993) and are randomly drawn for I~$<$~-1 from a Normal distribution with $\sigma_{\rm I}~=~\sigma_{\rm V}~=~{\rm exp}({\rm I}~-~0.6)$ and $\sigma_{\rm V~-~I}~=~\sqrt{\sigma_{\rm I}^{2}~+~\sigma_{\rm V}^{2}}$ which implies $\sigma_{\rm I}~\approx~0.01$ at I~=~-4.0 and $\sigma_{\rm I}~\approx~0.2$ at I~=~-1 whereas at I~$>$~-1 $\sigma_{\rm I}~=~\sigma_{\rm V}~=~0$. Red dots display the simulated CMD without taking observational errors into account. Black dots display the simulated CMD when taking observational errors into account.}
\label{melu:CMD:CMD1214}
\end{center}
\end{figure}

\begin{figure}
\begin{center}
\includegraphics[width=1.0\textwidth]{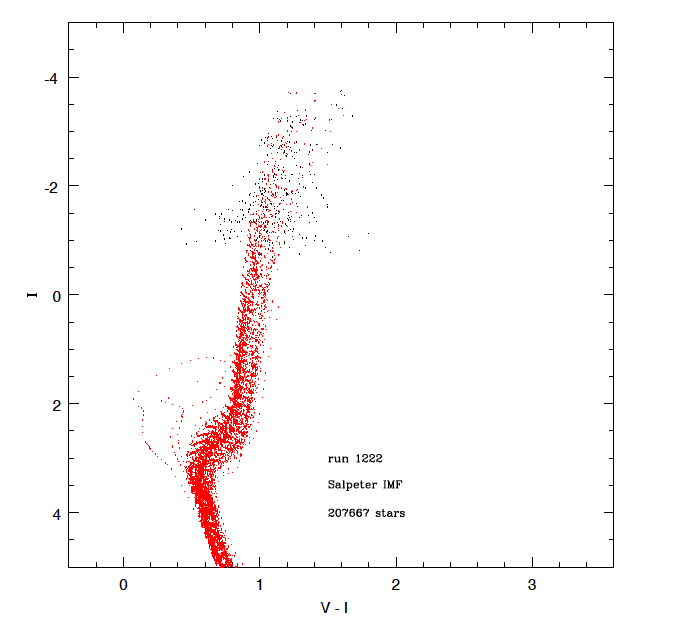}
\caption{Topographical halo simulated CMD for the label~``B'' semi--cosmological simulation (Figures~\ref{melu:fig2}~and~\ref{melu:fig4}) at M$_{\rm tot}$~10$^{12}$~M$_\odot$. The observational errors are taken into account (e.g.,~Secker~\&~Harris 1993) and are randomly drawn for I~$<$~-1 from a Normal distribution with $\sigma_{\rm I}~=~\sigma_{\rm V}~=~{\rm exp}({\rm I}~-~0.6)$ and $\sigma_{\rm V~-~I}~=~\sqrt{\sigma_{\rm I}^{2}~+~\sigma_{\rm V}^{2}}$ which implies $\sigma_{\rm I}~\approx~0.01$ at I~=~-4.0 and $\sigma_{\rm I}~\approx~0.2$ at I~=~-1 whereas at I~$>$~-1 $\sigma_{\rm I}~=~\sigma_{\rm V}~=~0$. Red dots display the simulated CMD without taking observational errors into account. Black dots display the simulated CMD when taking observational errors into account.}
\label{melu:CMD:CMD1222}
\end{center}
\end{figure}

\begin{figure}
\begin{center}
\includegraphics[width=1.0\textwidth]{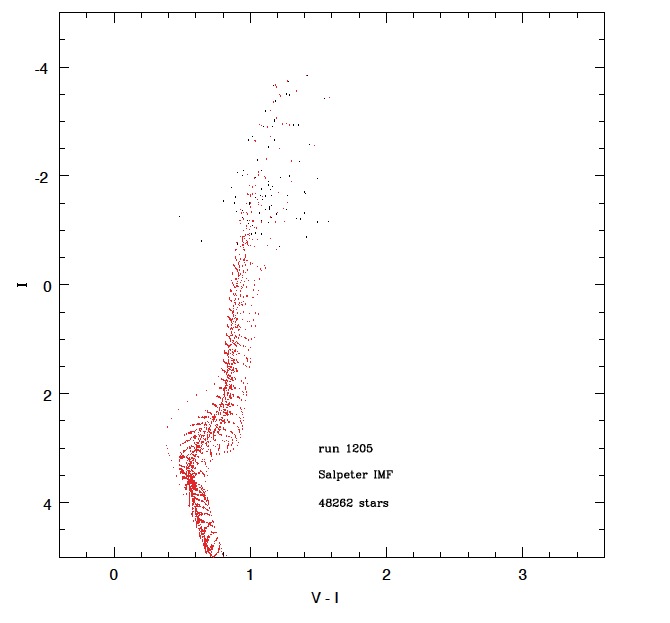}
\caption{Topographical halo simulated CMD for the label~``C'' semi--cosmological simulation (Figures~\ref{melu:fig2}~and~\ref{melu:fig4}) at M$_{\rm tot}$~10$^{12}$~M$_\odot$. The observational errors are taken into account (e.g.,~Secker~\&~Harris 1993) and are randomly drawn for I~$<$~-1 from a Normal distribution with $\sigma_{\rm I}~=~\sigma_{\rm V}~=~{\rm exp}({\rm I}~-~0.6)$ and $\sigma_{\rm V~-~I}~=~\sqrt{\sigma_{\rm I}^{2}~+~\sigma_{\rm V}^{2}}$ which implies $\sigma_{\rm I}~\approx~0.01$ at I~=~-4.0 and $\sigma_{\rm I}~\approx~0.2$ at I~=~-1 whereas at I~$>$~-1 $\sigma_{\rm I}~=~\sigma_{\rm V}~=~0$. Red dots display the simulated CMD without taking observational errors into account. Black dots display the simulated CMD when taking observational errors into account.}
\label{melu:CMD:CMD1205}
\end{center}
\end{figure}

The Padova\footnote{\texttt{http://pleiadi.pd.astro.it}} theoretical isochrones \citep{Girardi} in the metallicity range of 0.0001$~<~$Z$~<~$0.030 have been used to generate the simulated CMD out of the ensemble of stellar particles in the ``topographical'' halo for each simulation. The theoretical isochrones are taken into account up to the RGB~tip and are shown in Fig.~\ref{melu:CMD:iso} as black dots. Data storage has constrained us to choose 5$\%$~halo~mass resolution i.e.~for each stellar particle a Simple Stellar Population, whose total mass amounts to 5$\%$ of the stellar particle mass, is generated, with a Salpeter \citep{Salpeter} Initial~Mass~Function (IMF hereafter) over a mass range of 0.1$~<~$m$_{*}$/M$_\odot$$~<~$m$_{\rm top}$ where m$_{\rm top}~=~$m$_{\rm RGB}^{\rm tip}$ is the largest stellar mass in the theoretical isochrones -~at the given stellar particle age and metallicity~- for a star behind the RGB~tip. As a conservative choice, neither age nor metallicity extrapolation is performed. Given the IMF shape and both the age and the metallicity range of the ensemble of stellar particles the simulated CMD is generated out of, we consider the information--loss such IMF~mass~range implies not a significant one. The observational errors are taken into account (e.g.,~\citealt{SH}) and are randomly drawn for I~$<$~-1 from a Normal distribution with $\sigma_{\rm I}~=~\sigma_{\rm V}~=~{\rm exp}({\rm I}~-~0.6)$ and $\sigma_{\rm V~-~I}~=~\sqrt{\sigma_{\rm I}^{2}~+~\sigma_{\rm V}^{2}}$ which implies $\sigma_{\rm I}~\approx~0.01$ at I~=~-4.0 and $\sigma_{\rm I}~\approx~0.2$ at I~=~-1 whereas (since our focus is on the upper~RGB) at I~$>$~-1 we set $\sigma_{\rm I}~=~\sigma_{\rm V}~=~0$. The simulated CMDs for the simulations labelled as A~--~C in Figures~\ref{melu:fig2}~and~\ref{melu:fig4} are shown in Figures~\ref{melu:CMD:CMD1214}~--~\ref{melu:CMD:CMD1205} at a resolution (0.5$\%$) lower than that (5$\%$) of the corresponding simulated CMDs because of image storage constraints.

\begin{figure}
\begin{center}
\includegraphics[width=1.0\textwidth]{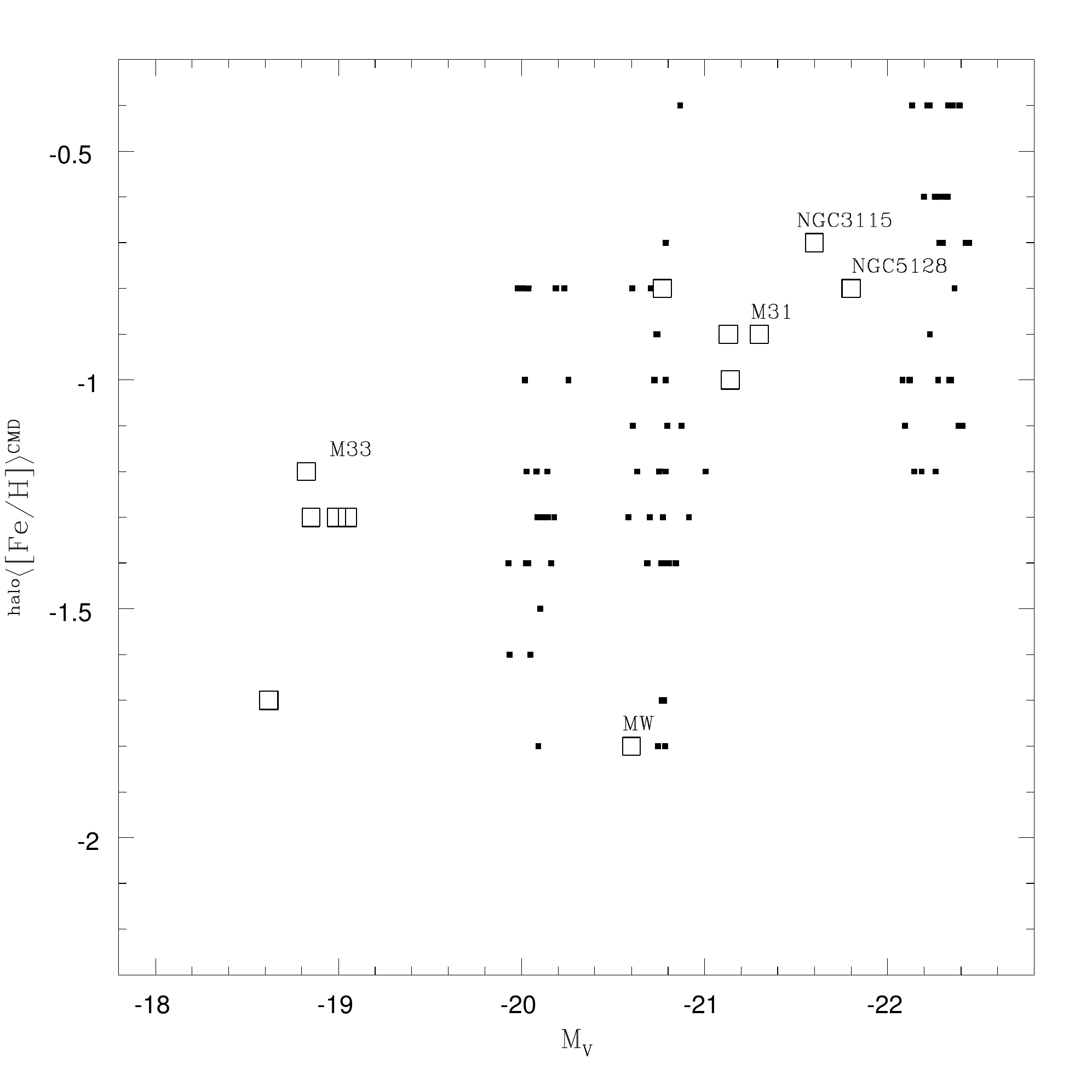}
\caption{The stellar halo metallicity--total galactic V--band luminosity relation ($^{\rm halo}\langle [$Fe/H$]\rangle^{\rm CMD}$~--~M$_{\rm V}$). A filled box corresponds to the peak of the MDF which is derived out of the (V~-~I) colour distribution at -3.5$~<~$I$~<~$-3.0 via metallicity--colour relationship through the same pipeline as in Mouhcine~et~al.~(2005). The metallicity of each artificial star of the ensemble the MDF refers to is derived via metallicity--colour relationship out of the (V~-~I) colour of the generated star in the RGB of the simulated ``topographical'' halo CMD at -3.5$~<~$I$~<~$-3.0. The related (V~-~I) colour distributions are displayed in Appendix~\ref{app:appendixG} whereas the related MDFs derived via metallicity--colour relationship are shown in Appendix~\ref{app:appendixH}. The observational data label--code is as in Fig.~\ref{melu:fig1}.}
\label{melu:CMD:fig1}
\end{center}
\end{figure}

Once the simulated CMD is generated, both the same fiducial RGB~tracks (i.e.~0.8~M$_\odot$ \citealt{VdBrgb}~RGB~tracks and the empirical NGC~6791~RGB~track) and the same pipeline as in \cite{MMc} are used to construct a metallicity--colour relationship to derive $[$Fe/H$]$ out of the (V~-~I) colour of each (generated) star in the RGB of the (simulated) CMD at -3.5$~<~$I$~<~$-3.0. The fiducial RGB~tracks are shown in Fig.~\ref{melu:CMD:iso} as red dots. The related (V~-~I) colour distributions for each simulated CMD at -3.5$~<~$I$~<~$-3.0 are displayed in Appendix~\ref{app:appendixG} whereas the related MDFs derived via metallicity--colour relationship (through the same pipeline as in \citealt{MMc}) are shown in Appendix~\ref{app:appendixH}. The MDF peaks are shown in Fig.~\ref{melu:CMD:fig1}.

The dispersion in metallicity at a given total luminosity we found among the topographical halos is not wiped out when the halo metallicity is derived from the halo simulated CMD via metallicity--colour relationship, as shown in Fig.~\ref{melu:CMD:fig1} -~to be compared with Fig.~\ref{melu:fig1}. 

\begin{figure}
\begin{center}
\includegraphics[width=1.0\textwidth]{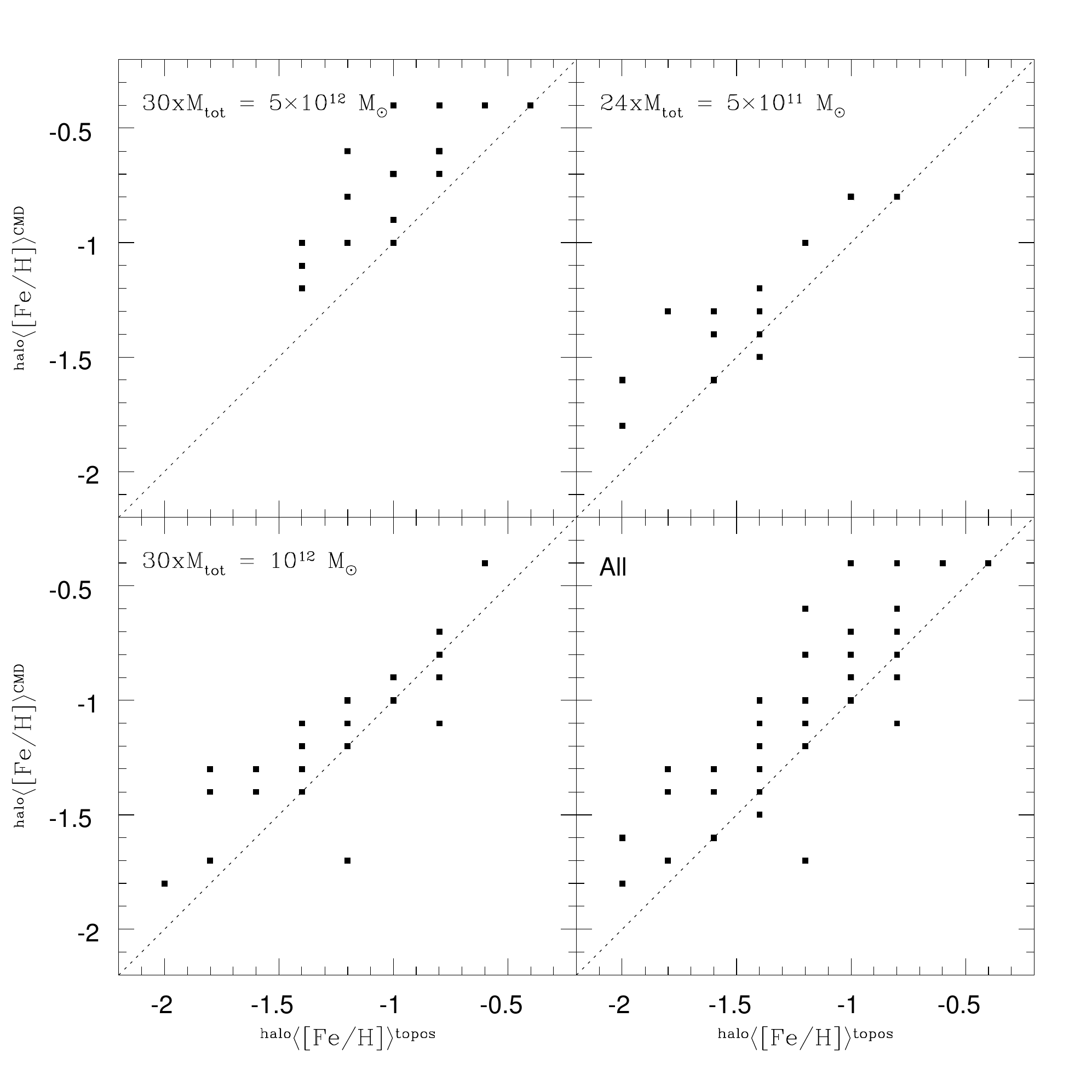}
\caption{The relationship between $^{\rm halo}\langle [$Fe/H$]\rangle^{\rm topos}$ (the ``topographical'' halo metallicity) and $^{\rm halo}\langle [$Fe/H$]\rangle^{\rm CMD}$ (the ``topographical'' halo metallicity as derived from the halo simulated CMD via metallicity--colour relationship).}
\label{melu:CMD:fig2}
\end{center}
\end{figure}

\begin{figure}
\begin{center}
\includegraphics[width=1.0\textwidth]{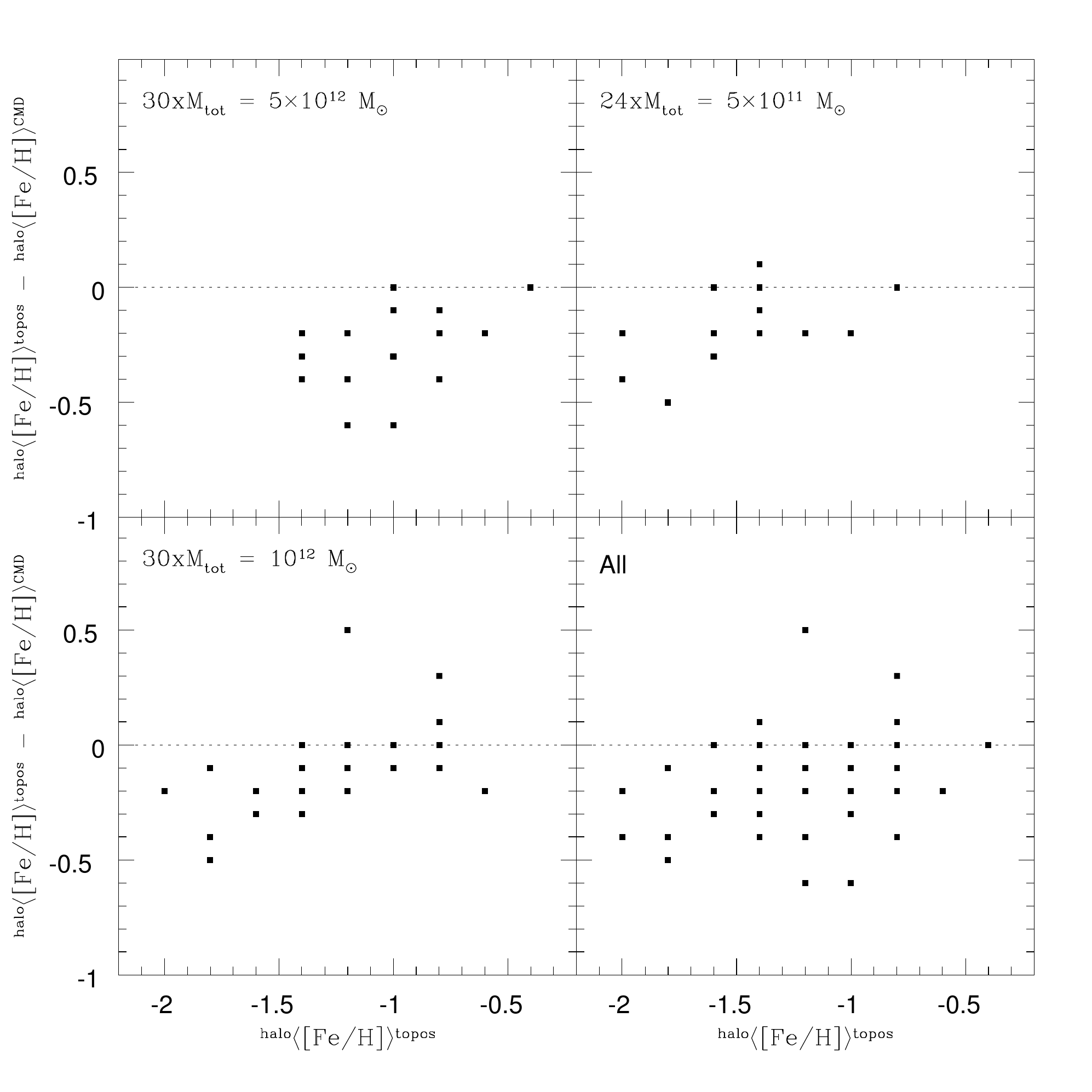}
\caption{The discrepancy in metallicity between $^{\rm halo}\langle [$Fe/H$]\rangle^{\rm topos}$ (the ``topographical'' halo metallicity) and $^{\rm halo}\langle [$Fe/H$]\rangle^{\rm CMD}$ (the ``topographical'' halo metallicity as derived from the halo simulated CMD via metallicity--colour relationship) as a function of the ``topographical'' halo metallicity.}
\label{melu:CMD:fig3}
\end{center}
\end{figure}

Given a simulation, the relationship between $^{\rm halo}\langle [$Fe/H$]\rangle^{\rm topos}$ (the ``topographical'' halo metallicity) and $^{\rm halo}\langle [$Fe/H$]\rangle^{\rm CMD}$ (the ``topographical'' halo metallicity as derived from the halo simulated CMD via metallicity--colour relationship) is shown in Fig.~\ref{melu:CMD:fig2} whereas the discrepancy in metallicity between $^{\rm halo}\langle [$Fe/H$]\rangle^{\rm topos}$ and $^{\rm halo}\langle [$Fe/H$]\rangle^{\rm CMD}$ is shown in Fig.~\ref{melu:CMD:fig3} as a function of $^{\rm halo}\langle [$Fe/H$]\rangle^{\rm topos}$. 

\begin{figure}
\begin{center}
\includegraphics[width=1.0\textwidth]{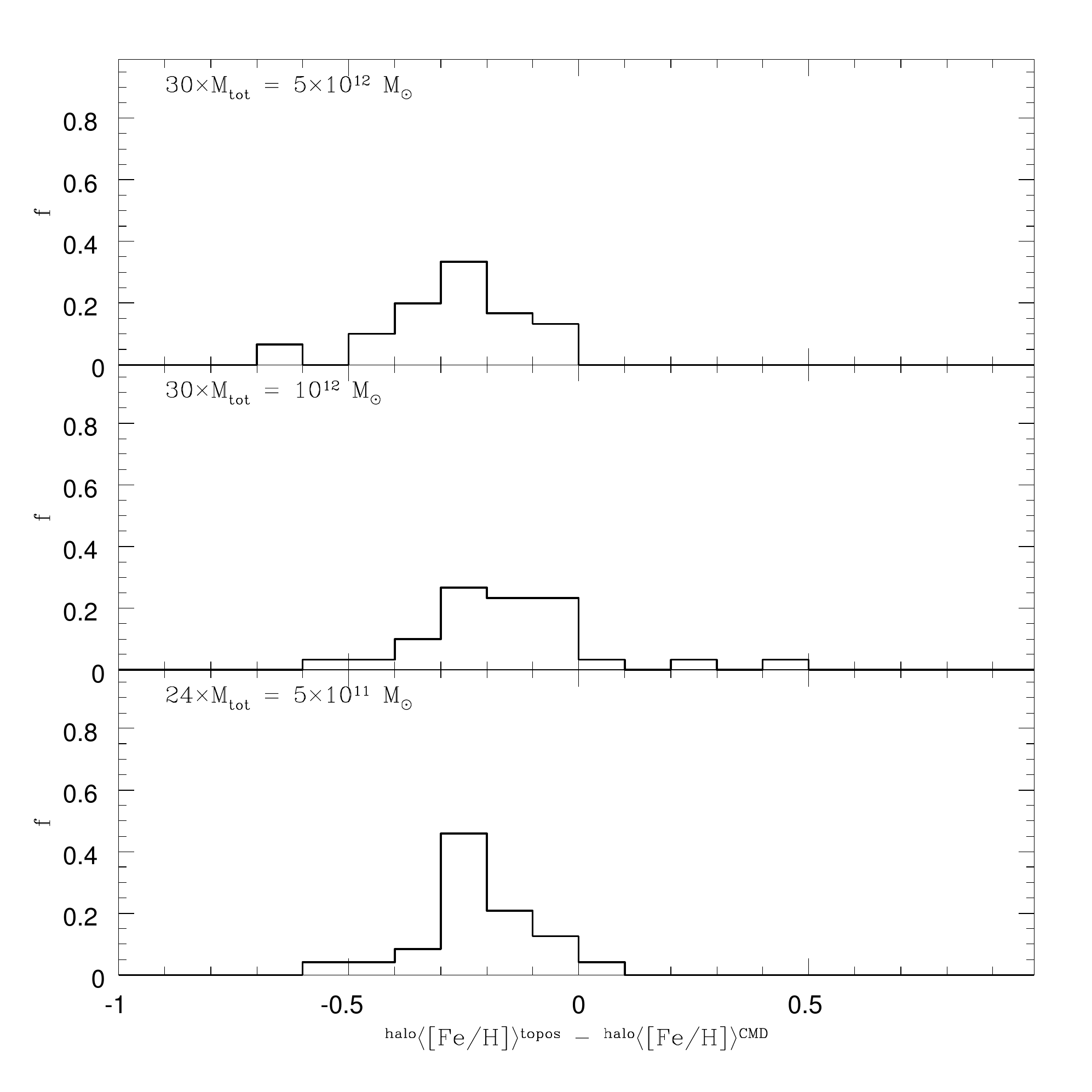}
\caption{For each subset, the distribution of the discrepancy between $^{\rm halo}\langle [$Fe/H$]\rangle^{\rm topos}$ (the ``topographical'' halo metallicity) and $^{\rm halo}\langle [$Fe/H$]\rangle^{\rm CMD}$ (the ``topographical'' halo metallicity as derived from the halo simulated CMD via metallicity--colour relationship).}
\label{melu:CMD:fig4}
\end{center}
\end{figure}

The histograms in Fig.~\ref{melu:CMD:fig4} display the distribution of the discrepancy between $^{\rm halo}\langle [$Fe/H$]\rangle^{\rm topos}$ and $^{\rm halo}\langle [$Fe/H$]\rangle^{\rm CMD}$, for each subset, and show that $^{\rm halo}\langle [$Fe/H$]\rangle^{\rm CMD}$ is generally biased towards higher metallicities ($\lesssim$~0.3~dex more metal--rich as shown in Fig.~\ref{melu:CMD:fig4}) because the fiducial~RGB~tracks do not extend towards colours as blue as those the Padova theoretical isochrones extend towards, as shown in Fig.~\ref{melu:CMD:iso}, which (given that, as a conservative choice, no extrapolation has been performed beyond the range covered by the fiducial tracks) amounts to narrowing the metallicity range which is sampled at low metallicities, once the metallicity is derived out of the metallicity--colour relationship.

Interestingly, since the fiducial~RGB~tracks have globular cluster age, the robustness test above allows us to evaluate which bias could be brought in by such uniformly old halo age assumption when the halo metallicity is derived out of the metallicity--colour relationship by interpolating among fiducial~RGB~tracks (e.g.,~as in \citealt{MMc}). Such bias can be quantified by the histograms in Fig.~\ref{melu:CMD:fig4}, even if, admittedly, the construction of each simulated CMD out of the corresponding ``topographical'' halo and its subsequent analysis via fiducial~RGB~tracks have been constrained by our conservative choice not to perform any extrapolation beyond the range covered by either the theoretical isochrones we used to generate the simulated CMDs or the fiducial~RGB~tracks we used to analyse them. The discrepancy between $^{\rm halo}\langle [$Fe/H$]\rangle^{\rm topos}$ (i.e.~the metallicity of the ``topographical'' halo as derived by the analysis of the MDF of the stellar particles in the simulation which constitute the ``topographical'' stellar halo ensemble -~the stellar particles in the simulation at a projected radius R$~>~$15~kpc) and $^{\rm halo}\langle [$Fe/H$]\rangle^{\rm CMD}$ (i.e.~the metallicity of the ``topographical'' halo as derived from the halo simulated CMD via metallicity--colour relationship built out of the fiducial~RGB~tracks with globular cluster age) amounts to $\lesssim$~0.3~dex. Future analysis of higher resolution simulations would further improve such comparison.

\section{Discussion}
\label{melu:discussion}

We have presented here an analysis of the characteristics of late--type galaxy 
stellar halos formed within a grid of 89 simulations, with 
particular emphasis placed 
upon the relationship between stellar halo metallicity
and the associated galactic luminosity. It helps to stress here that ``Halo Semantics'' is currently a controversial topic (e.g.,~\citealt{Ibata07} and references therein): we have labelled as ``topographical halo'' the ensemble of stellar particles in a simulation at a projected radius R$~>~$15~kpc, at~z~=~0. Although future analysis of simulations at resolutions higher than the current would further improve the study we have undertaken, 
we have shown that -~at any given total luminosity or conversely total dynamical mass~- the stellar halo metallicities in the simulations 
span a range in excess of $\sim$~1~dex, a result which is strengthened by the robustness tests we have performed. 

We suggest that the 
underlying driver of this metallicity dispersion can be traced to the 
diversity of galactic mass assembly histories inherent within the 
hierarchical clustering paradigm. Galaxies with a more protracted 
assembly history possess more metal--rich and younger stellar halos, with
an associated greater dispersion in age, than galaxies which experience
more of a monolithic collapse.

For a given total luminosity (or dynamical mass), those galaxies with more
extended assembly histories also possess more massive stellar halos, which 
in turn leads to a direct correlation between the stellar halo
metallicity and its surface brightness (as anticipated by earlier 
semi--analytical models~-~e.g.,~\citealt{Renda}). By extension, such a 
correlation may prove to be a useful diagnostic tool for disentangling the
formation history of late--type galaxies.

Recently, \cite{MM} have presented an observed correlation between
stellar halo metallicity and total galactic luminosity, as shown in Fig.~\ref{melu:fig1}. 
The observed dispersion in the mean halo metallicity at a given galactic
luminosity is {\it smaller} than what we find in our simulations. {\it However}
the latter can account for the outliers in the observed trend. Since our motivation has been to study
which is the effect of the pattern of the initial density fluctuations {\it alone}
on the stellar halo features at redshift z~=~0 in late--type galaxy simulations, it helps to note that galaxy formation, as it is observed,
is an ongoing process which is the result of the interplay among different parameters,
of which the pattern of initial density fluctuations (thus the merging history) is one.
We have shown that the merging history {\it alone} may be held responsible of the dispersion
in halo metallicity at comparable total galactic luminosities,
as {\it apparently} observed for example in our Milky Way and in Andromeda (see Section~\ref{melu:intro}).

This begs the question...\textit{Which is normal}? 
Our study suggests that if the stellar halo was (primarily) assembled 
through more of a monolithic collapse, such a low metallicity is indeed what should
be expected; conversely, the fact that the M31 stellar halo is significantly metal--rich is suggestive of a more
protracted assembly history (see also \citealt{Hammer}).
An observational consequence of these differing formation histories is
the prediction that the M31 stellar halo should possess a high
surface brightness;
observations tentatively support this prediction (Reitzel, Guhathakurta \& Gould~1998; \citealt{Irwin}).

Further observations are needed to tighten our grasp of the strength and scatter of the stellar halo metallicity--luminosity relation, which we have shown to be a useful diagnostic tool for disentangling the formation history of late--type galaxies.




\chapter{Galaxy Mass Assembly since z~$\sim~1$}
\label{chap:mass}


We analyse 112 N--body/hydrodynamical late--type galaxy simulations 
to study the stellar mass assembly and the relationships between
stellar, gaseous and total mass in galaxies 
since z~$\sim~1$ down to the present day, with an emphasis 
on the effects of the merging histories on the 
z~$=~0$ galaxy properties.

The redshift evolution of the stellar mass in simulations
with comparable luminosities at z~$=~0$ shows a significant 
dispersion, which lies in their differing merging histories. 
This contrasts with the redshift evolution of the total 
mass within the central 100~kpc, which is already 
assembled by z~$\sim~1.5$. 
The stellar--to--total mass ratio in the simulations broadly agrees with the observed (e.g.,~\citealt{Conselice}). We find that massive simulations are generally more evolved than their lower mass counterparts.

\section{Introduction}

The hierarchical structure 
assembly has become a well established paradigm in 
the last decade. The properties of the cosmic microwave 
background and the large scale structure 
(e.g.: \citealt{WMAP}; \citealt{Bahcall}) support a picture 
where a large fraction of matter is non--baryonic, 
gravitationally interacting only, and assembled in 
a hierarchical fashion. Galaxy formation in such a 
scenario is a continuous process where galaxy 
properties are the results of its merging history, 
mass and environment. 

Observationally, much 
progress has been made in studying the global 
star formation history and the build--up of the stellar 
mass over a significant fraction of the Hubble time 
(e.g.: \citealt{Lilly96}; \citealt{Madau96}). 
However, many details are still missing.
The analysis of galaxy stellar and total masses 
at different cosmic epochs can thus assist in 
constraining the likely scenarios of galaxy evolution.

\cite{Conselice} have recently found a lack of evolution in 
the galaxy stellar--to--total mass ratio over the redshift 
range ${\rm 0.2 \la z \la 1.2}$ (see also \citealt{BZ06}), 
thus implying
that if galaxies continue to build up over the second 
half of the Hubble time stellar and dark components 
are then growing together. 

Here, we focus on galaxy mass assembly 
over the last half of the age of the Universe.
The results of the simulations, the comparison
with the data, and the implications of our results
are described in Sect.~\ref{assemb:resu}. We summarise our
conclusions in Sect.~\ref{assemb:concl}.

\section{Results}
\label{assemb:resu}

We have 
constructed an ensemble of 112 late--type galaxy simulations 
in a chemo--dynamical framework (Section~\ref{melu:model}), 
spanning a factor of~$\sim~$50 in total mass, and sampling 
a range of assembly histories at a given total mass. 

This sample includes the same 89~semi--cosmological \texttt{GCD+}~simulations analysed in \cite{RendaHalo} and 23 more simulations, completed within the same framework, with the same parameters described in Section~\ref{melu:model}, and distributed among the same mass ``bins''. These latter simulations sample more patterns of small--scale density fluctuations, which lead to different hierarchical assembly histories. A subset of 9~simulations (with ${\rm M_{tot}~=~10^{11} M_{\odot}}$, collapse redshift z$_{\rm c}~=~1.5$ and spin parameter $\lambda~=~0.054$) samples more density fluctuations patterns.

As a benchmark for the semi--cosmological framework, 
we have also analysed 2~among the disc galaxy simulations presented 
in \cite{Bailin}, which are fully cosmological \texttt{GCD+} runs 
and adopt a $\Lambda$--dominated~CDM 
cosmology (\citealt{Bailin} and references therein). The photometric properties of a stellar particle in a simulation are modeled 
as those of a Simple~Stellar~Population (SSP hereafter) with the same 
age and metallicity. The evolution of the SSP 
properties as a function of age and metallicity 
is taken from \cite{ML}. 

In the following, the galaxy 
total mass, including both dark and baryonic (i.e.~gaseous and stellar) matter, is 
measured for the region within 100~kpc from 
the centre of the stellar mass distribution, at each redshift, 
whereas the stellar 
and the gaseous mass are measured within 15~kpc from the centre of the stellar mass, at each redshift.

\begin{figure}
\centering
\includegraphics[width=1.0\textwidth]{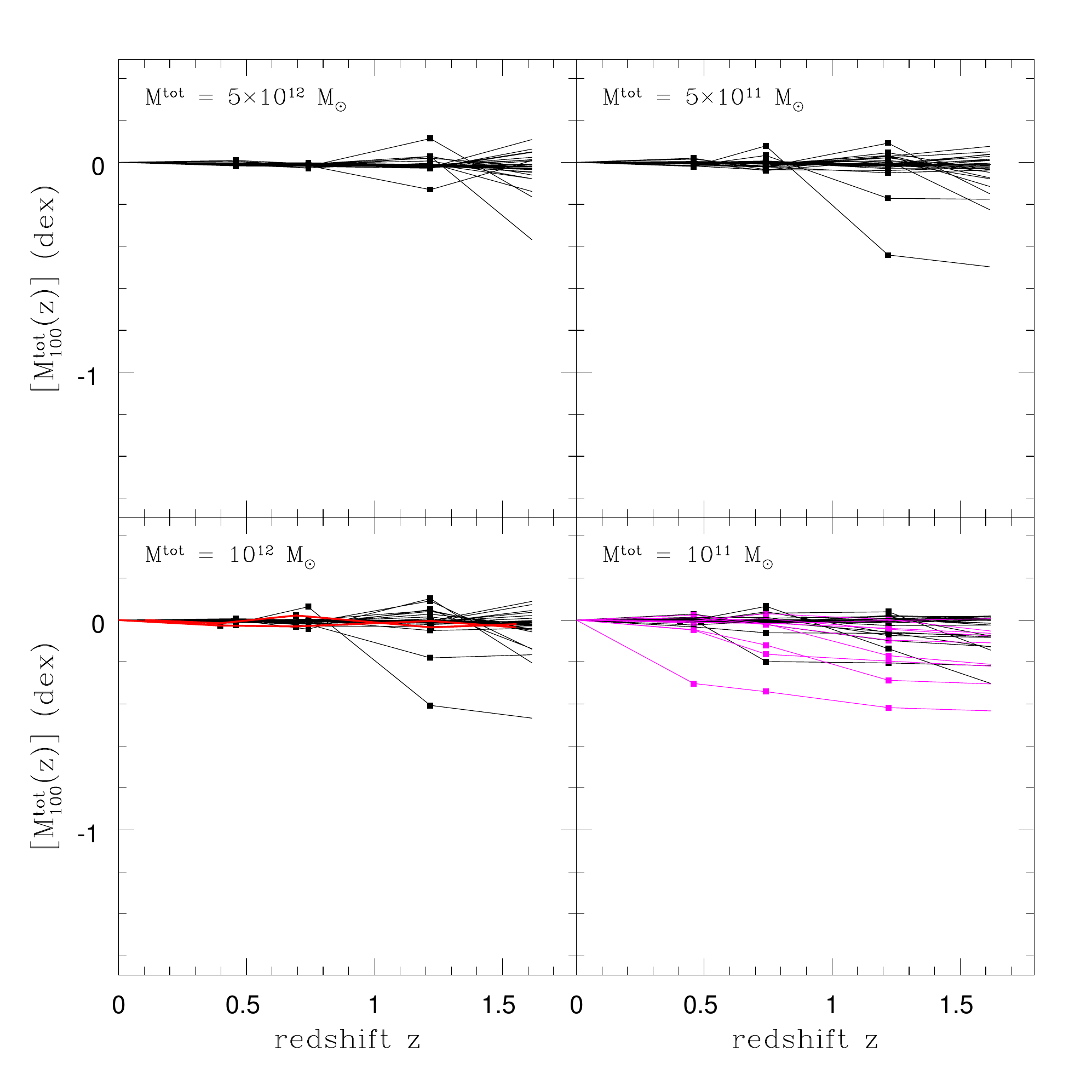}
\caption{The redshift evolution of the total
mass within the 100~kpc central region, normalised
to the z~$=~0$ value for each simulation.
The semi--cosmological simulations with collapse
redshift z$_{\rm c}~=~2$ are shown as black lines. 
Magenta for the collapse redshift z$_{\rm c}~=~1.5$ runs. 
The cosmological simulations
are shown in red.}
\label{fig:mass_assembly}
\end{figure}

\begin{figure}
\centering
\includegraphics[width=1.0\textwidth]{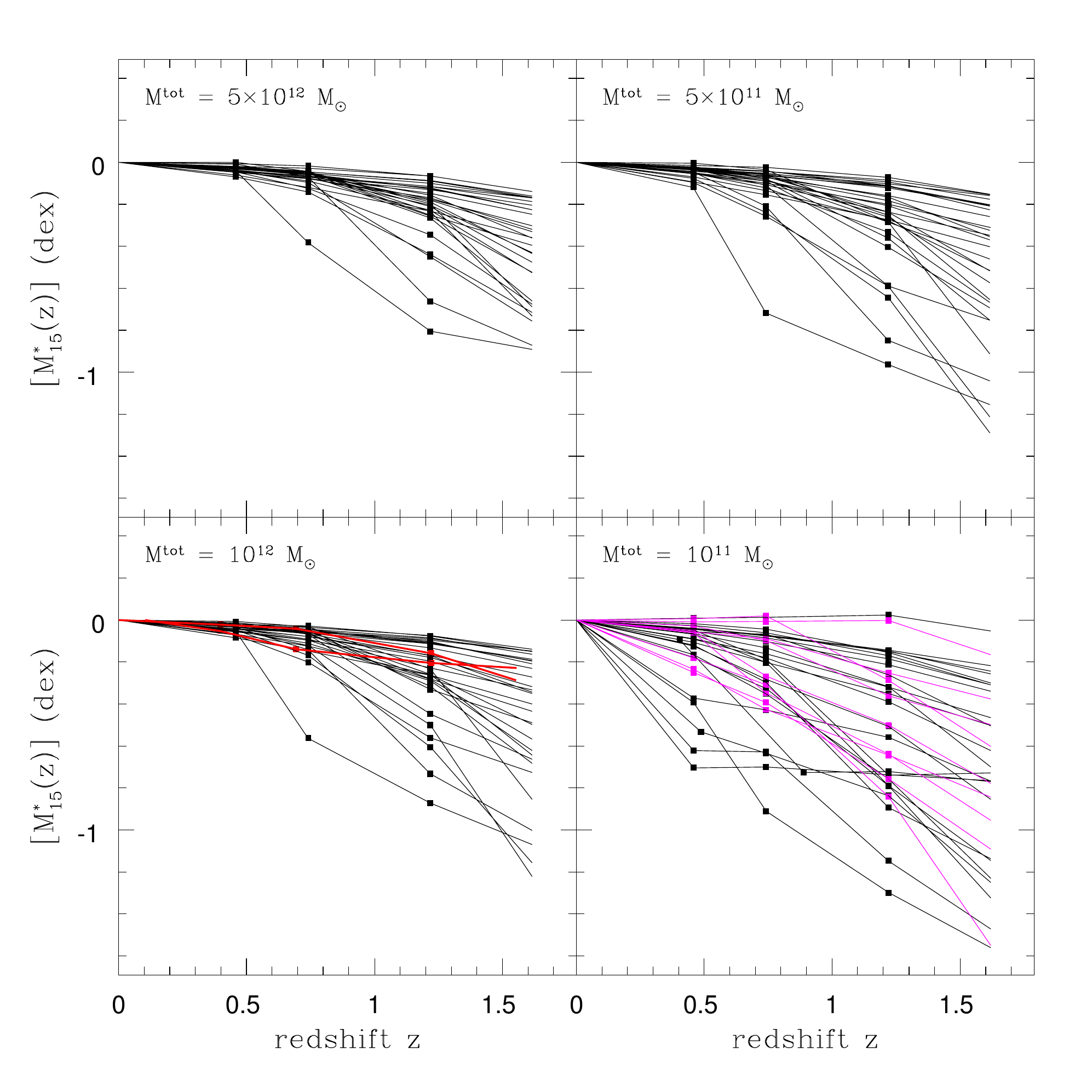}
\caption{The redshift evolution of the stellar
mass within the 15~kpc central region, normalised
to the z~$=~0$ value for each simulation.
The semi--cosmological simulations with collapse
redshift z$_{\rm c}~=~2$ are shown as black lines.
Magenta for the collapse redshift z$_{\rm c}~=~1.5$ runs.
The cosmological simulations
are shown in red.}
\label{fig:stellar_mass_assembly}
\end{figure}

\begin{figure}
\centering
\includegraphics[width=1.0\textwidth]{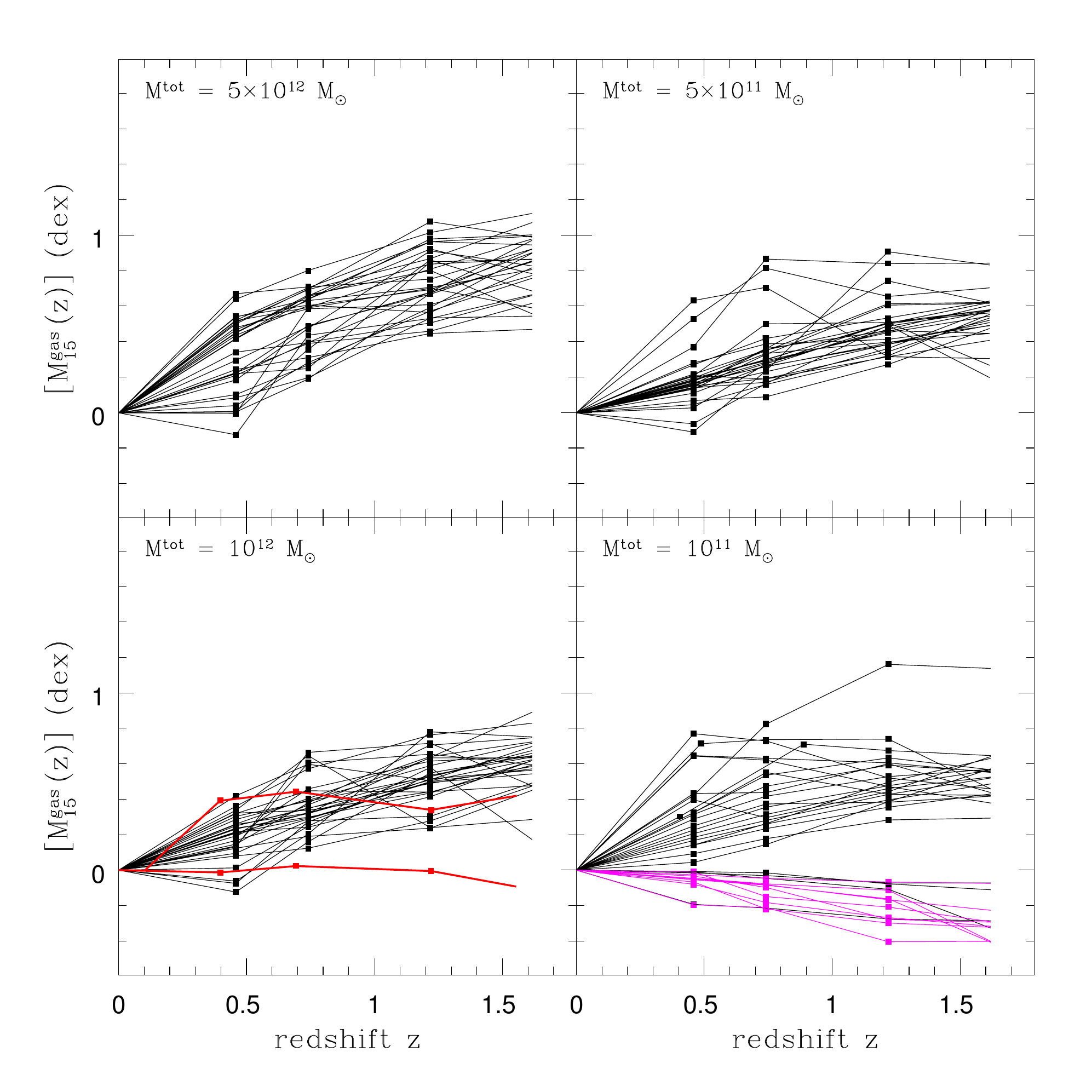}
\caption{The redshift evolution of the gaseous
mass within the 15~kpc central region, normalised
to the z~$=~0$ value for each simulation.
The semi--cosmological simulations with collapse
redshift z$_{\rm c}~=~2$ are shown as black lines.
Magenta for the collapse redshift z$_{\rm c}~=~1.5$ runs.
The cosmological simulations
are shown in red.}
\label{fig:gas_mass_assembly}
\end{figure}

\begin{figure}
\centering
\includegraphics[width=1.0\textwidth]{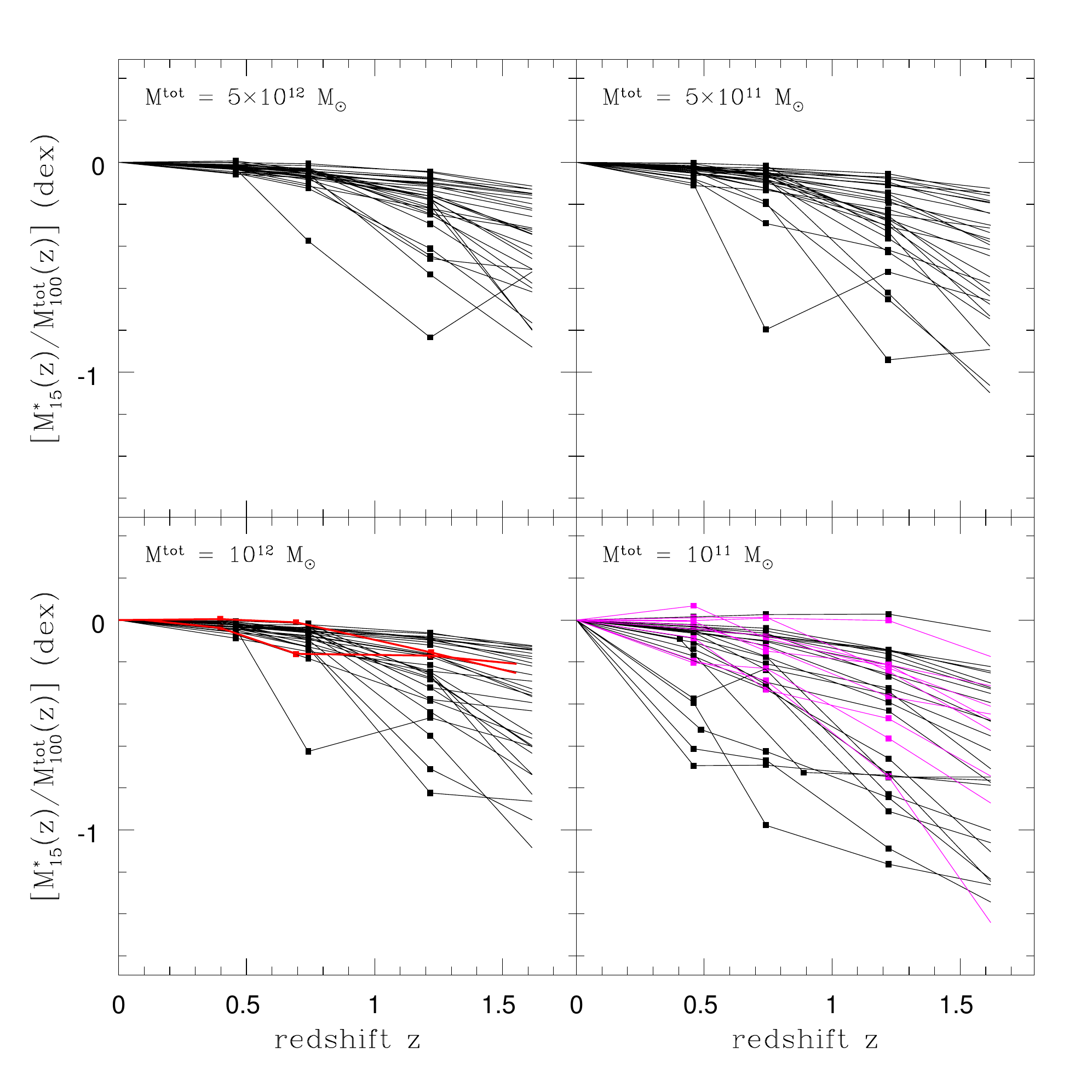}
\caption{The redshift evolution of the stellar 
(within the 15~kpc central region) to total
(within the 100~kpc central region) mass ratio, normalised
to the z~$=~0$ value for each simulation.
The semi--cosmological simulations with collapse
redshift z$_{\rm c}~=~2$ are shown as black lines.
Magenta for the collapse redshift z$_{\rm c}~=~1.5$ runs.
The cosmological simulations
are shown in red.}
\label{fig:star2tot_assembly}
\end{figure}

\begin{figure}
\centering
\includegraphics[width=1.0\textwidth]{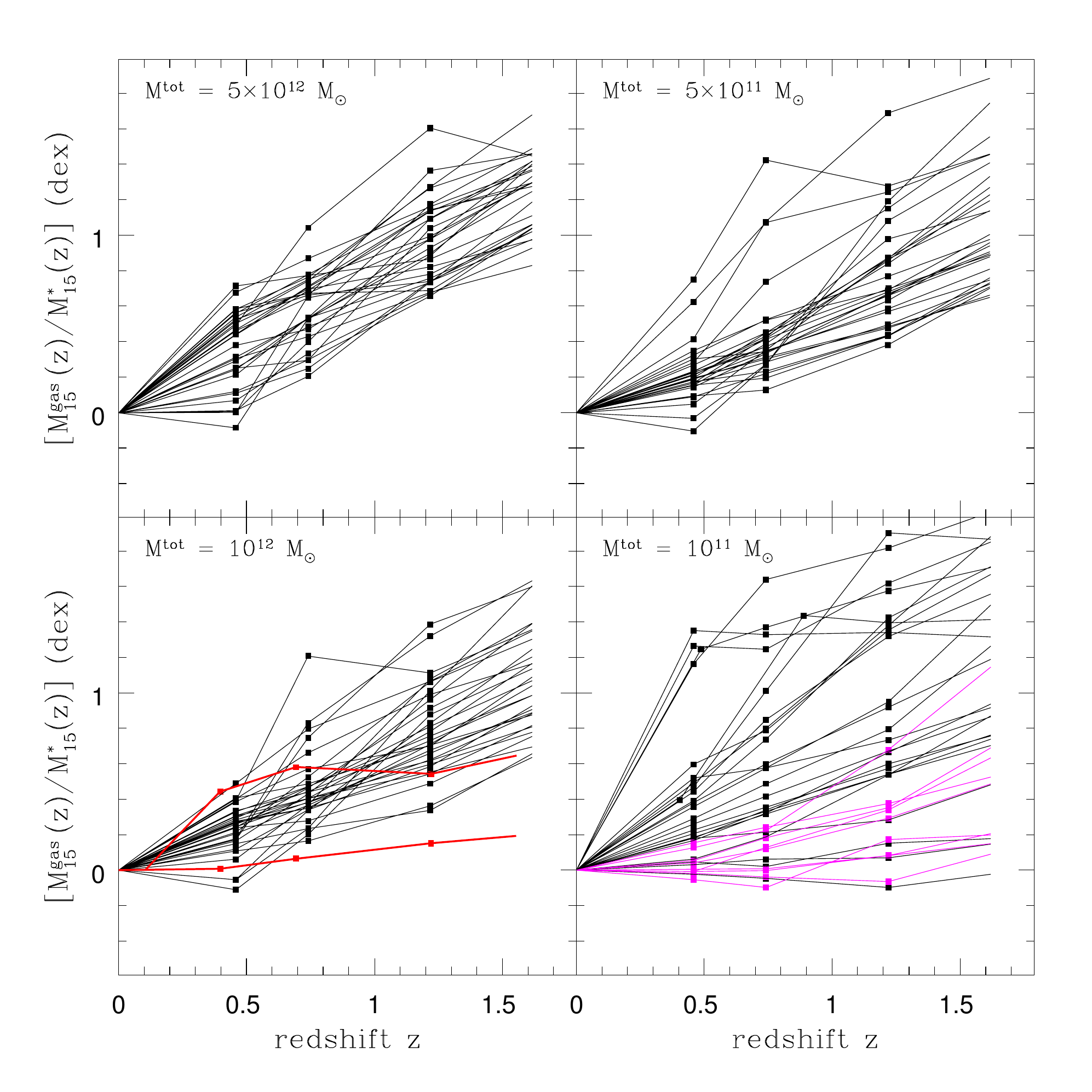}
\caption{The redshift evolution of the gaseous--to--stellar mass ratio
(within the 15~kpc central region), 
normalised to the z~$=~0$ value for each simulation.
The semi--cosmological simulations with collapse
redshift z$_{\rm c}~=~2$ are shown as black lines.
Magenta for the collapse redshift z$_{\rm c}~=~1.5$ runs.
The cosmological simulations
are shown in red.}
\label{fig:gas2star_assembly}
\end{figure}

Figures~\ref{fig:mass_assembly},~\ref{fig:stellar_mass_assembly} and~\ref{fig:gas_mass_assembly} display the redshift evolution of the total, stellar and gaseous mass normalised to the z~$=~0$~value for each simulation. Fig.~\ref{fig:star2tot_assembly} displays the redshift evolution of the stellar--to--total mass ratio whereas Fig.~\ref{fig:gas2star_assembly} displays the redshift evolution of the gaseous--to--stellar mass ratio. Black (red) lines show the
semi--cosmological (cosmological) runs. 
Magenta for the semi--cosmological simulations 
with collapse redshift z$_{\rm c}~=~1.5$. 

The consistency between the assembly histories of total, stellar and gaseous mass in both semi--cosmological and cosmological simulations is reassuring as a robustness test for the semi--cosmological framework.

The assembly of the total mass within the inner 100~kpc in the simulations 
is $\sim$~completed by z~$\ga~1.5$,
independently of their total mass. On the other hand, 
the stellar mass assembly histories within the central 15~kpc are 
more extended, i.e. at z~$\sim~$1, 
$\sim$~half of the baryonic mass in the inner region is 
gaseous, with a large variety of assembly patterns.
Massive simulations have assembled 
their stellar content early on (typically by 
z~$\sim~$1.5) and have burnt a significant fraction 
of their gas supply by now, others 
continue to form stars to much later epochs, whereas the 
simulations with collapse redshift z$_{\rm c}~=~1.5$ 
have their baryonic content dominated by the gaseous 
component at all redshifts. Galaxies with comparable 
stellar masses can have significantly different assembly 
histories of their stellar content, and diverse gaseous 
contents. This suggests that galaxy stellar mass 
should be considered \textit{neither} a robust tracer of the 
merging history \textit{nor} a good tracer of the total baryonic 
galaxy mass.

\begin{figure}
\centering
\includegraphics[width=1.0\textwidth]{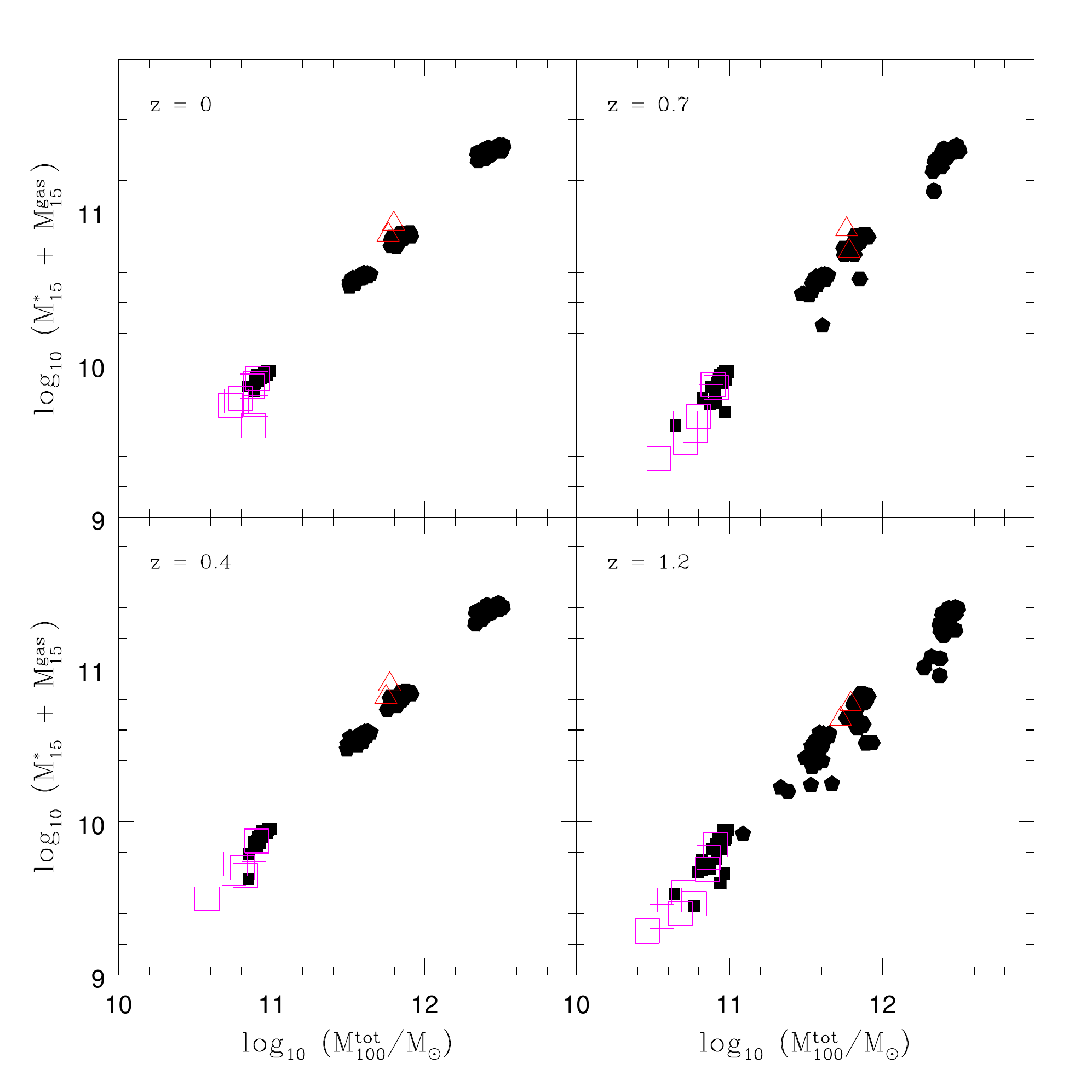}
\caption{The relationship between 
baryonic mass (i.e.~gaseous~and~stellar mass 
within the 15~kpc central region)
and total mass (within the 100~kpc central region), 
at different redshifts.
The semi--cosmological simulations with collapse redshift
z$_{\rm c}~=~2$ are shown as filled symbols (heptagons for
M$^{\rm tot}~=~5\times10^{12}$~M$_{\odot}$, hexagons
for M$^{\rm tot}~=~10^{12}$~M$_{\odot}$, pentagons for
M$^{\rm tot}~=~5\times10^{11}$~M$_{\odot}$, boxes for
M$^{\rm tot}~=~10^{11}$~M$_{\odot}$). Larger
empty boxes for the 
collapse redshift z$_{\rm c}~=~1.5$ runs. 
The cosmological simulations are shown as
large open triangles.}
\label{fig:baryo2tot}
\end{figure}

\begin{figure}
\centering
\includegraphics[width=1.0\textwidth]{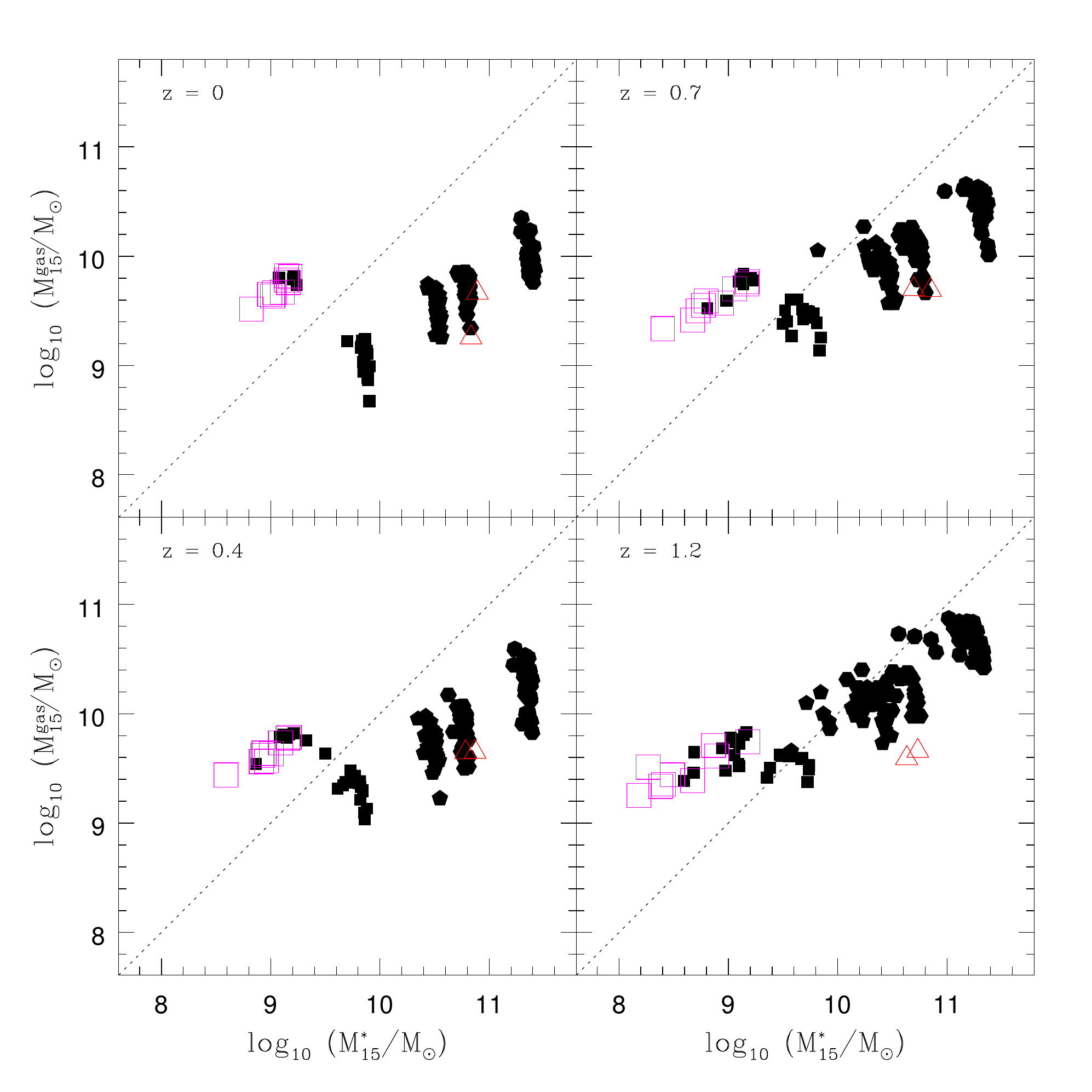}
\caption{The relationship between
gaseous mass and stellar mass (within the 15~kpc central region),
at different redshifts.
Symbols are the same as in Fig.~\ref{fig:baryo2tot}.}
\label{fig:gas2star}
\end{figure}

\begin{figure}
\centering
\includegraphics[width=1.0\textwidth]{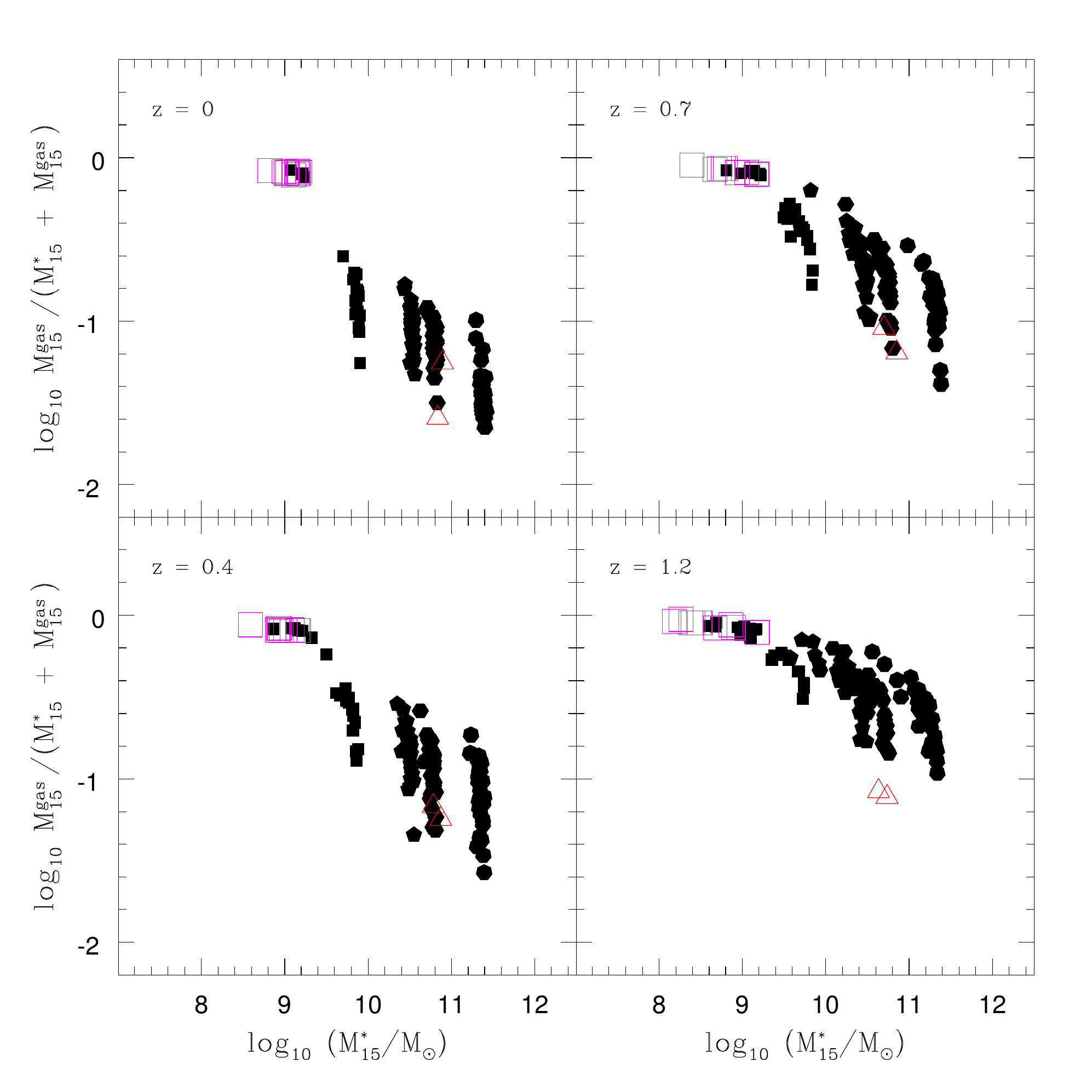}
\caption{The relationship between
baryonic gas fraction
and stellar mass,
at different redshifts.
Symbols are the same as in Fig.~\ref{fig:baryo2tot}.}
\label{fig:baryogas2star}
\end{figure}

Although the slope of the baryonic--to--total mass ratio does not significantly evolve over the redshift range which is analysed (as displayed in Fig.~\ref{fig:baryo2tot}), 
Figures~\ref{fig:gas2star}~and~\ref{fig:baryogas2star} show that, 
in our sample, galaxies of any given 
stellar mass, and at any given redshift, span a wide 
range of gaseous--to--stellar mass ratios. 
Star formation does not uniformly proceed in all galaxies: 
some galaxies have converted a significant fraction 
of their gas reservoirs into stars, whereas in others 
the baryonic content is still dominated by the gaseous component. 

\begin{figure}
\centering
\includegraphics[width=1.0\textwidth]{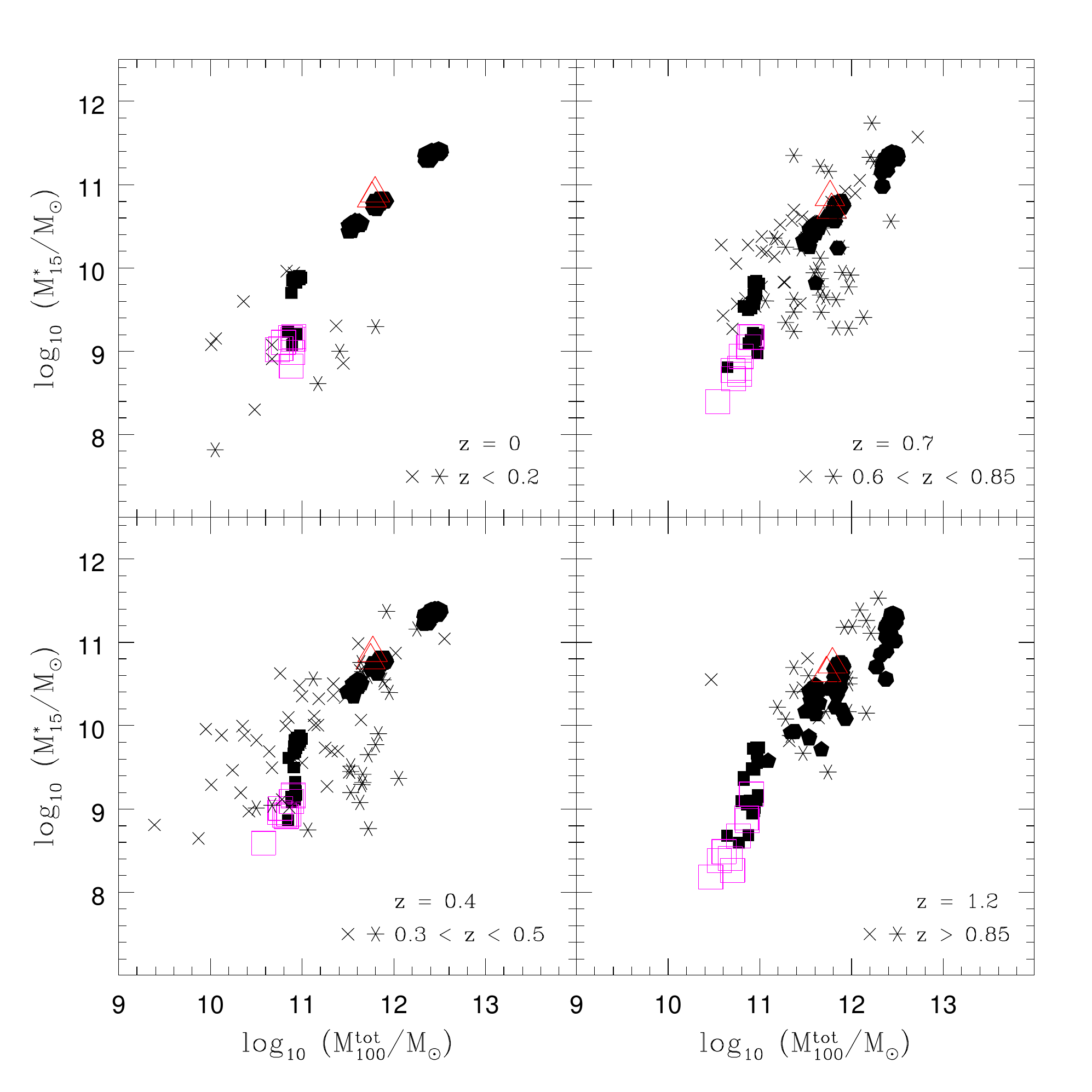}
\caption{Stellar mass (within the 15~kpc central region) 
against total mass (within the 100~kpc central region), 
at different redshifts. 
Symbols are the same as in Fig.~\ref{fig:baryo2tot}. 
The stellar--to--total mass ratio in the simulations is 
compared with the ratio drawn from the observations in \cite{Conselice} and \cite{BZ06}, displayed as six and four~vertices stars, 
respectively.} 
\label{fig:stellar2tot_ratio}
\end{figure}

The relationship between
stellar mass (within the 15~kpc central region) 
and total mass (within the 100~kpc central region) 
in the simulations is shown in 
Fig.~\ref{fig:stellar2tot_ratio}, at different redshifts. 
Stellar and total mass in the simulations 
are correlated since z~$\sim~1$ down to the present day, 
thus suggesting that galaxy stellar and dark masses grow 
together continuously, similarly to the results 
drawn from the observations in \cite{Conselice} and \cite{BZ06}.

The scatter around the stellar--to--total mass ratio in the simulations increases with redshift. However, it is smaller than the dispersion around the observed relationship. The simulations analysed here are sampling neither a full range nor an a~priori distribution of parameters (Section~\ref{melu:model}). 
Varying such parameters might increase the scatter in the simulations.
Also, the dispersion around the observed relationship could be partially due
to systematics in the observations. In fact, it is difficult to accurately 
estimate the dynamical mass for an intermediate 
redshift galaxy. 
The dynamical masses of these objects have been usually estimated 
using circular velocities 
obtained by long slit spectroscopy. However, \cite{Flores} 
have recently pointed out that long slit spectroscopy 
reveals overdispersed galaxy properties, 
because it tends to bias the circular velocity in galaxies with 
disturbed velocity fields. 

As previously shown in Fig.~\ref{fig:baryo2tot}, a tighter correlation exists between total mass and baryonic mass. Also, the baryonic--to--total mass ratio is less dispersed than the stellar--to--total mass ratio, similarly to what is observed for z~$~\sim~2$ galaxies \citep{erb}, thus reinforcing that the galaxy stellar mass should be considered neither a robust tracer of the
galaxy merging history nor a good tracer of the total baryonic
galaxy mass.

\section{Conclusions}
\label{assemb:concl}

We have presented the analysis of a sample of semi--cosmological late--type galaxy simulations with stellar mass larger than $\sim~10^9$~M$_{\odot}$,
to explore whether hierarchical formation scenarios may 
account for the constraints on galaxy mass assembly 
over the second half of the age of the Universe. 

The main results can be summarised as follows:

\begin{itemize}

\item[-] Stellar, gaseous, and total masses are found to be 
correlated in the simulations since z~$\sim~1$ down to z~$=~0$, 
in broad agreement with the scaling relations observed in local and intermediate redshift galaxies.

\item[-] Galaxies with comparable
stellar masses in the simulations can have significantly different assembly
histories of their stellar content, and diverse gaseous
contents. This suggests that the galaxy stellar mass
should be considered neither a robust tracer of the
galaxy merging history nor a good tracer of the total baryonic
galaxy mass.

\end{itemize}




\chapter{The Mass--Metallicity Relation since z~$\sim~1$.}
\label{chap:massmeta}

\newcommand{\doh}{\mbox{${\rm 12+\log(O/H)}$}}

\def\kms{km\,s$^{-1}$}
\def\spose#1{\hbox to 0pt{#1\hss}}
\def\simlt{\mathrel{\spose{\lower 3pt\hbox{$\mathchar"218$}} 
\raise 2.0pt\hbox{$\mathchar"13C$}}}

\def\simgt{\mathrel{\spose{\lower 3pt\hbox{$\mathchar"218$}}   
\raise 2.0pt\hbox{$\mathchar"13E$}}}



We analyse 112 N--body/hydrodynamical late--type galaxy simulations
to study the metal enrichment of the stellar component and the interstellar medium 
since z~$\sim~1$, down to z~$=~0$.
The relationships between stellar mass and metallicity for both stellar and gaseous components in the simulations at z~$=~0$ are in broad agreement with the relationships locally observed. The dispersion around these relationships in the simulations, which lies in their diverse merging histories, broadly agrees with the observed. We find that the integrated stellar populations in the simulations are dominated by stars as old as 4~--~10~Gyr. For massive simulations, the age range of the integrated stellar populations agrees with the observations. On the contrary, simulations with stellar mass $\sim~10^{9}$~M$_{\odot}$ at z~$=~0$ tend to be older than locally observed galaxies with comparable stellar masses.

\section{Introduction} 
\label{mmr:intro}
 
The correlation between galaxy metallicity and luminosity 
in the local universe is one of the most significant 
observational results in galaxy evolution. 
\cite{lequeux} first revealed that Oxygen abundance 
increases with total mass for irregular galaxies. 
The luminosity--metallicity relation for irregulars was 
later confirmed by \cite{skillman89} among others. 
Subsequent studies have extended the relation 
to spiral galaxies (e.g.:~\citealt{GS87}; \citealt{Z94}; \citealt{G97}), 
and to elliptical galaxies \citep{BH91}. 
More recently, large samples of star--forming galaxies 
drawn from galaxy redshift surveys, e.g.~the~2dF~Galaxy 
Redshift~Survey and the Sloan Digital Sky Survey (SDSS 
hereafter), have been used to confirm 
the luminosity--metallicity relation over a broader 
range (\citealt{lam04}; \citealt{Tremonti}). 
The stellar populations of local 
galaxies with low stellar mass are found to be generally 
young and metal--poor, 
whereas massive galaxies are found to be old and metal--rich
 (e.g.,~\citealt{Gallazzi}). 
Different groups have explored the properties of the interstellar gas at earlier epochs in intermediate--~(0~$\lesssim$~z~$\lesssim$~1:~\citealt{H01}; \citealt{Lilly}; 
\citealt{Liang}; \citealt{maier04}; \citealt{KK04}; 
\citealt{maier05}; \citealt{M06b}; 
\citealt{lam06}) and high--redshift 
galaxies (1.5~$\lesssim$~z~$\lesssim$~4:~\citealt{Pettini}; \citealt{KK00}; 
\citealt{mehlert};  \citealt{lemoine03}; \citealt{erb}; \citealt{maier06}). 

Here, we aim to 
discuss the metal enrichment histories of galactic 
systems in a hierarchical clustering scenario. 
The results of the simulations, 
the comparison with the data, and the implications of 
our results are described in Sect.~\ref{mmr:resu}. 
We summarise our conclusions in Sect.~\ref{mmr:concl}.

\section{Results}
\label{mmr:resu}

We have
constructed an ensemble of 112 simulated late--type galaxies
in a chemo--dynamical framework (Section~\ref{melu:model}),
spanning a factor of~$\sim~$50 in total mass, and sampling
a range of assembly histories at a given total mass.

This sample includes the same 89~semi--cosmological \texttt{GCD+}~simulations analysed in \cite{RendaHalo} and 23 more simulations, completed within the same framework, with the same parameters described in Section~\ref{melu:model}, and distributed among the same mass ``bins''. These latter simulations sample more patterns of small--scale density fluctuations, which lead to different hierarchical assembly histories. A subset of 9~simulations (with ${\rm M_{tot}~=~10^{11} M_{\odot}}$, collapse redshift z$_{\rm c}~=~1.5$ and spin parameter $\lambda~=~0.054$) samples more density fluctuations patterns.

As a benchmark for the semi--cosmological framework,
we have also analysed 2~among the disc galaxy simulations presented
in \cite{Bailin}, which are fully cosmological \texttt{GCD+} runs 
and adopt a $\Lambda$--dominated~CDM
cosmology (\citealt{Bailin} and references therein). 
The photometric properties of a stellar particle in a simulation are modeled
as those of a Simple~Stellar~Population (SSP hereafter) with the same
age and metallicity. The evolution of the SSP
properties as a function of age and metallicity
is taken from \cite{ML}. Optical properties have been converted into the SDSS photometric system as in \cite{Fukugita}.

In the following, the stellar mass 
is measured within the 15~kpc central region, at each redshift.
Metallicity and age of the integrated stellar populations, 
and the gas--phase metallicity, are all measured within the 
10~kpc central region, at each redshift. 

We estimate the Oxygen abundance as {\doh}. 
Caution should be taken when comparing 
the estimated abundances for simulated galaxies with 
the observed, since the latter, based on 
integrated spectra, may suffer from systematics 
(e.g.,~\citealt{mk06}).

\begin{figure}
\begin{center}
\includegraphics[width=1.0\textwidth]{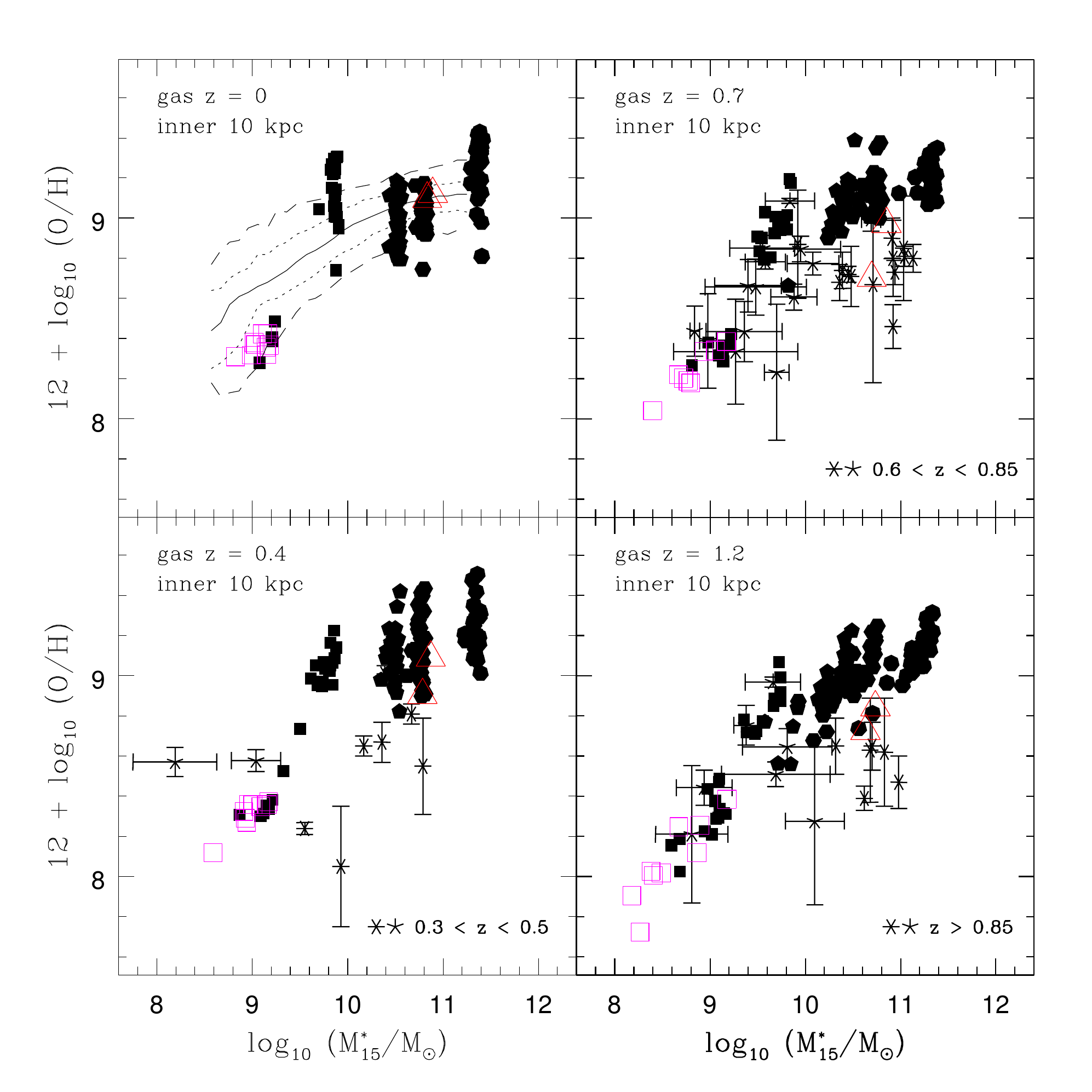}
\caption{Gas--phase Oxygen abundance 
against integrated stellar mass in the simulations, 
at different redshifts.
The semi--cosmological simulations with collapse redshift
z$_{\rm c}~=~2$ are shown as filled symbols (heptagons
for M$^{\rm tot}~=~5\times10^{12}$~M$_{\odot}$, hexagons
for M$^{\rm tot}~=~10^{12}$~M$_{\odot}$, pentagons for
M$^{\rm tot}~=~5\times10^{11}$~M$_{\odot}$, boxes for
M$^{\rm tot}~=~10^{11}$~M$_{\odot}$). 
Larger empty boxes for collapse redshift z$_{\rm c}~=~1.5$ runs. 
The cosmological simulations are shown as
large empty triangles. The z~$=~0$~gas--phase Oxygen abundance
in the simulations is compared with z~$\sim~0.1$~galaxies in the SDSS \citep{Tremonti}:
solid line for the median; dotted line for 16 and
84 percentile; dashed line for 2.5 and 97.5
percentile. The results at low and intermediate
redshifts are compared with the samples observed in 
\cite{savaglio05} and \cite{Liang06}, 
shown as five and six vertices stars, respectively.}
\label{mmr:oh_ms_gas}
\end{center}
\end{figure}

\begin{figure}
\begin{center}
\includegraphics[width=1.0\textwidth]{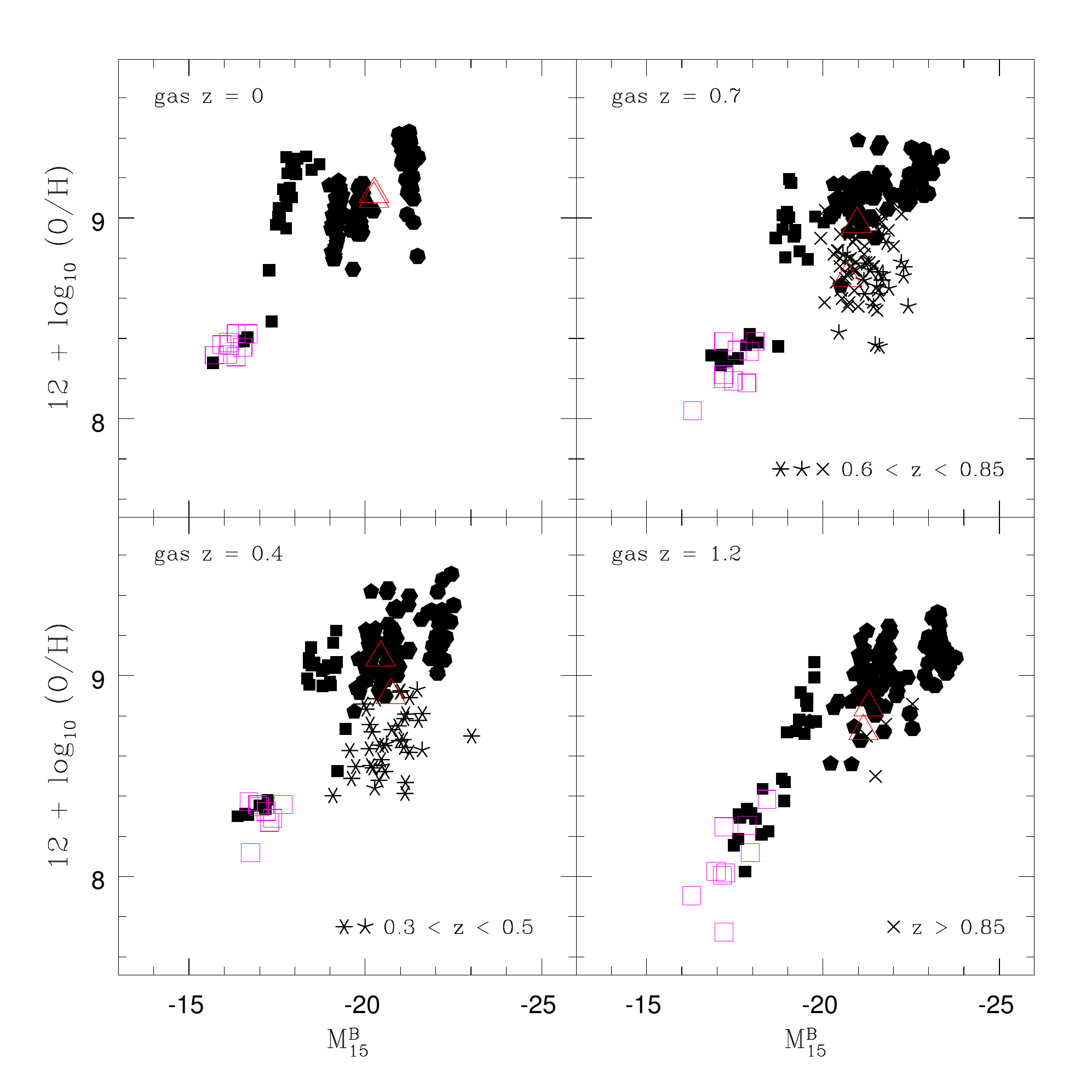}
\caption{Gas--phase Oxygen abundance against B--band
luminosity. Symbols are the same as in
Fig.~\ref{mmr:oh_ms_gas}. The redshift evolution in the simulations 
is compared with the relationship between the observed luminosity and the gas--phase
metallicity at intermediate redshifts in
\cite{Lilly}, \cite{Liang}, and \cite{M06b}, shown
as four, five and six vertices stars, respectively.}
\label{mmr:oh_mb_gas}
\end{center}
\end{figure}

\begin{figure}
\begin{center}
\includegraphics[width=1.0\textwidth]{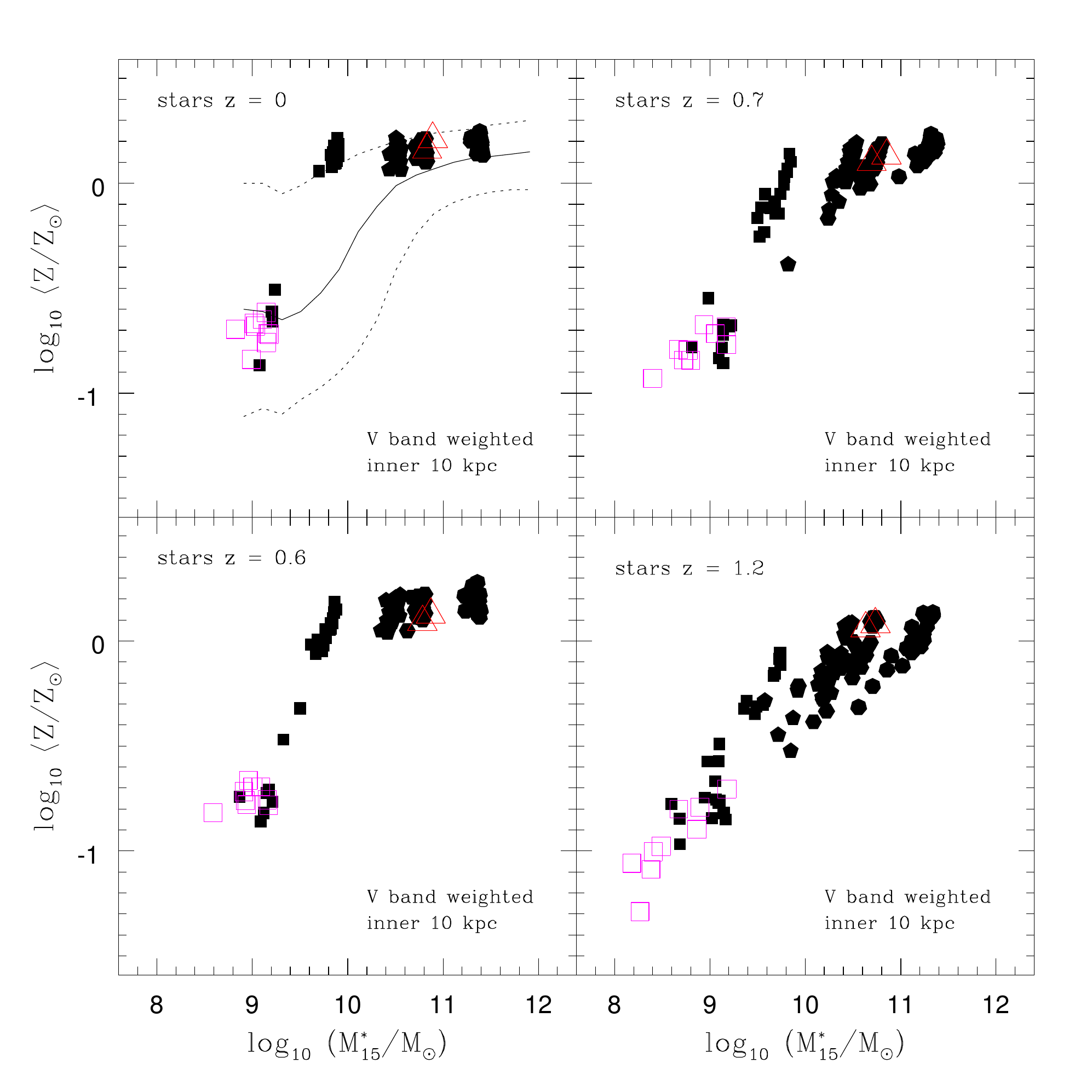}
\caption{The V--band luminosity--weighted metallicity
of the integrated stellar populations against the integrated
stellar mass, at different cosmic epochs. Symbols are
the same as in Fig.~\ref{mmr:oh_ms_gas}. The results at z~$=~0$
are compared with the z~$\sim~0.1$~relation between stellar mass and
metallicity in the SDSS \citep{Gallazzi}: solid line for the median; dotted line
for 16 and 84 percentile.}
\label{mmr:zs_mstar}
\end{center}
\end{figure}

\begin{figure}
\begin{center}
\includegraphics[width=1.0\textwidth]{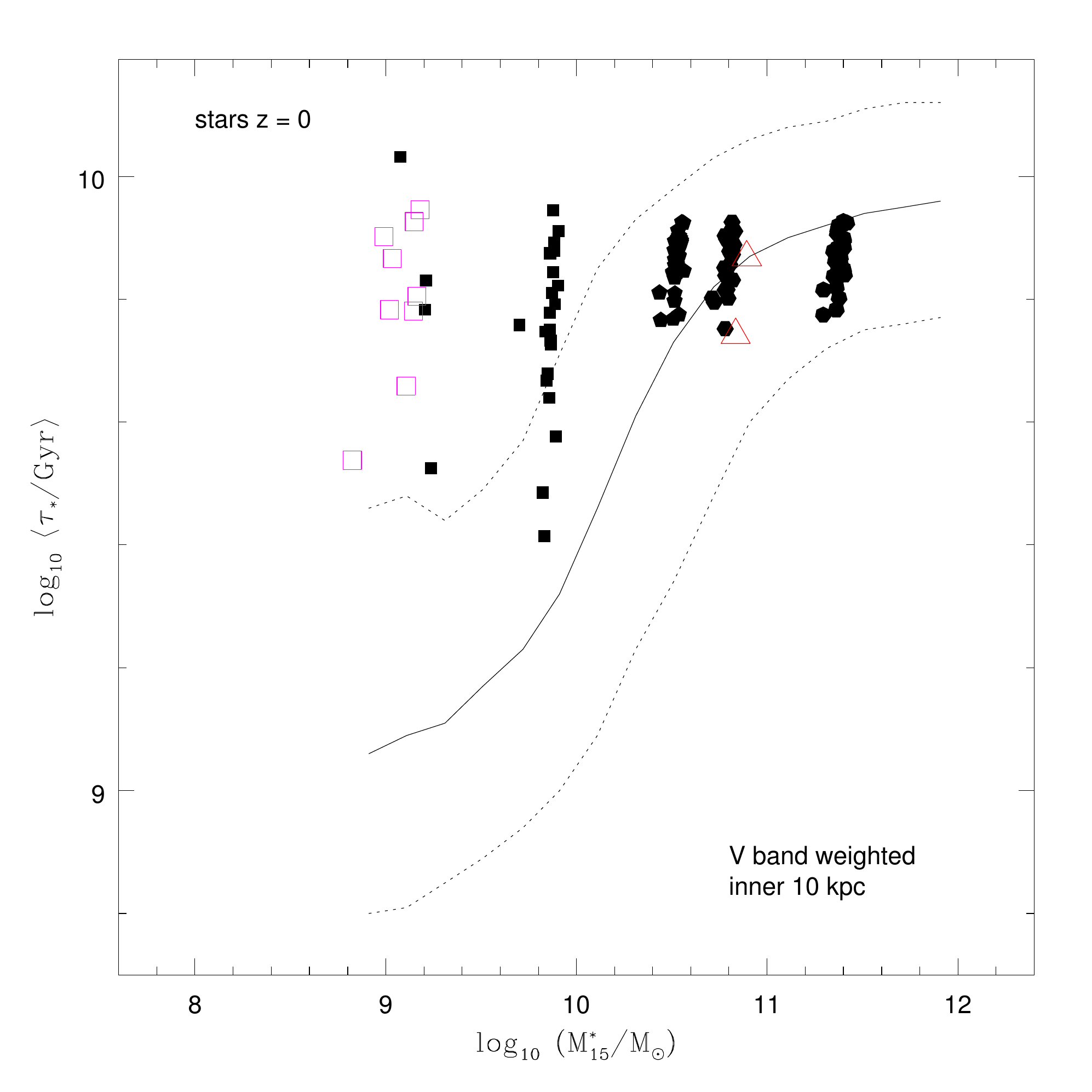}
\caption{The V--band luminosity--weighted age of the
integrated stellar populations, against the stellar
mass at redshift z~$=~0$. Symbols are the same as in
Fig.~\ref{mmr:oh_ms_gas}. The results are compared with the
z~$\sim~0.1$~relation between stellar mass and age in the SDSS 
\citep{Gallazzi}: solid line for
the median; dotted line for 16 and 84 percentile.}
\label{mmr:agestar_mstar}
\end{center}
\end{figure}

\begin{figure}
\begin{center}
\includegraphics[width=1.0\textwidth]{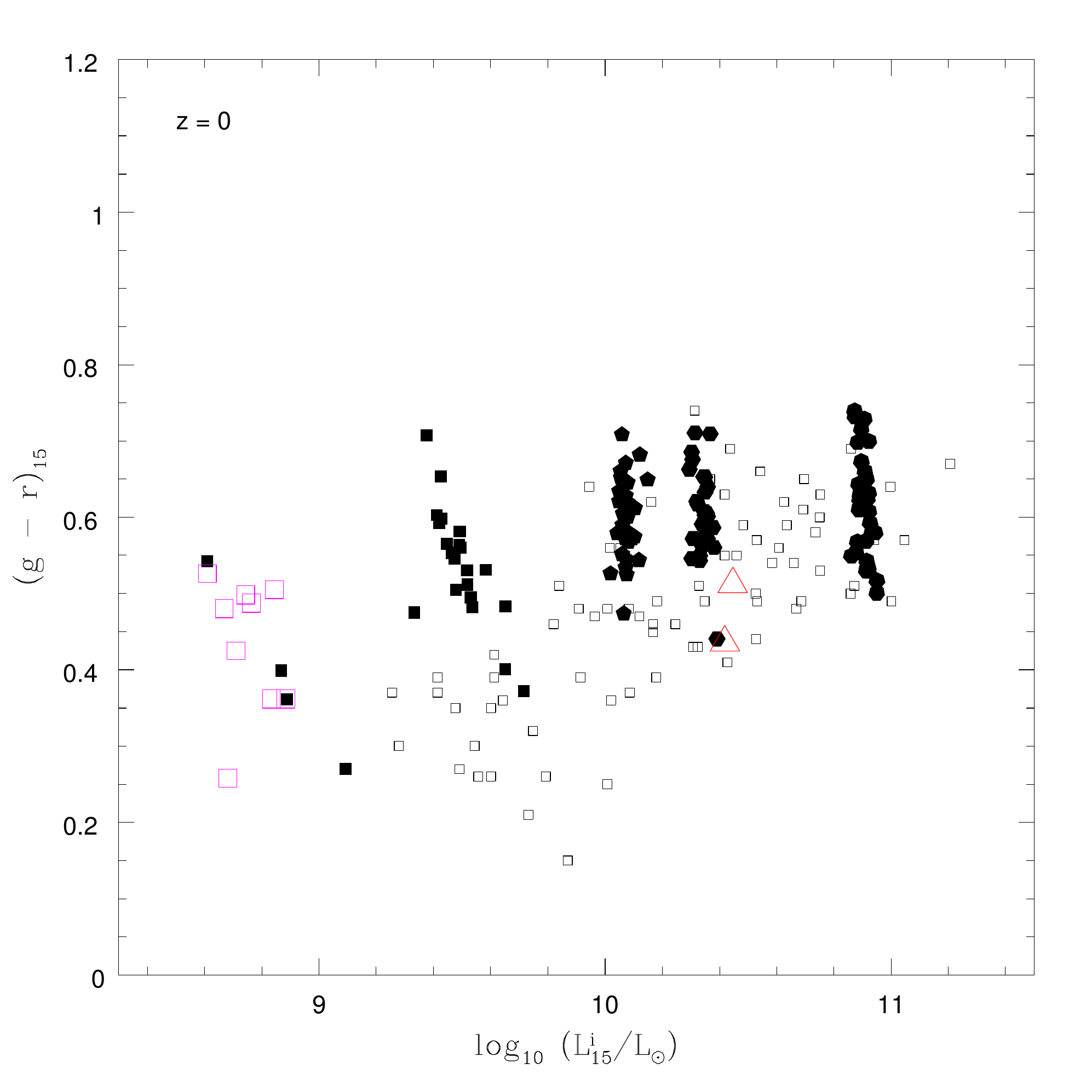}
\caption{The z~$=~0$~Colour--Magnitude relation in the simulations. 
Symbols are the same as in Fig.~\ref{mmr:oh_ms_gas}. The CMR in the simulations 
is compared with the local CMR for a sample of disc--dominated SDSS galaxies 
selected by \cite{Pizagno}, shown as empty boxes.}
\label{mmr:cmr}
\end{center}
\end{figure}

\subsection{Gas--phase Oxygen abundance vs. stellar mass}

Fig.~\ref{mmr:oh_ms_gas} shows a
correlation between gas--phase Oxygen abundance
and stellar mass in the simulations, at different redshifts.
The consistency between semi--cosmological and cosmological simulations (see also Section~\ref{assemb:resu}) is reassuring as a robustness test for the semi--cosmological framework. 

Massive galaxies with comparable stellar masses span
a range of gas--phase abundances, due to their
differing assembly histories, thus suggesting
that stellar mass should not be considered
a good tracer of gas--phase metallicity.

The results at redshift z~$=~0$ 
are compared and in broad agreement 
with the observed relation between stellar mass 
and gas--phase Oxygen abundance for a sample of 
SDSS galaxies at z~$\sim~0.1$ \citep{Tremonti}. 
The dispersion of the gas--phase metallicity at
a given stellar mass at z~$=~0$ is similar to
what is locally observed in the SDSS. 
This is a remarkable result, given that simulations and observations 
adopt different pipelines to estimate the gas--phase metallicity: for example, 
the SDSS fibre spectra preferentially sample the
inner $\sim~25\%$ of a galaxy
(e.g.,~\citealt{Tremonti} and references therein) 
whereas we are considering the Oxygen abundance
within the inner 10~kpc in the simulations.

The collapse--redshift z$_{\rm c}~=~1.5$ runs at M$_{\rm tot}~=~10^{11}$~M$_{\odot}$ are broadly as evolved as the most metal--poor z$_{\rm c}~=~2.0$ simulations at the same M$_{\rm tot}$. If the semi--cosmological simulations were constrained to agree with the observations at redshift z~$\sim~0$, the collapse--redshift range at M$_{\rm tot}~=~10^{11}$~M$_{\odot}$ would then be $1.5~\lesssim~{\rm z_{c}}~\lesssim~2.0$, thus suggesting that the observed low stellar mass galaxies may have collapsed out of density peaks with smaller values of $\delta_{\rm i}$ (see also Section~\ref{melu:model}).

Fig.~\ref{mmr:oh_ms_gas} shows that the redshift 
evolution of the mass--metallicity relationship 
depends on galaxy mass. 
Gas--phase abundances do not significantly evolve 
in simulations with stellar 
mass larger than $\sim~10^{10}$~M$_{\odot}$,
and the mass--metallicity relationship 
remains nearly as flat as it is at z~$\sim~$1. 
On the contrary, there is a systematic metal enrichment
for low stellar mass simulations, 
which increase their stellar contents by a large factor,
migrating from stellar masses of 
$\sim~10^{9}$~M$_{\odot}$ to 
$\sim~10^{10}$~M$_{\odot}$. 
Such behaviour is consistent with a scenario where massive 
galaxies have burnt a significant fraction of 
their gas reservoirs and are dominated by stars 
early on, thus leaving their metal enrichment 
almost unchanged since z~$\sim~1$. Low stellar mass 
galaxies would rather form stars in a more extended fashion, 
and a sizeable fraction of their baryons would still be 
gaseous at the present epoch. 

For intermediate redshifts, the mass--metallicity relation in the simulations is compared and in broad agreement with that reported at $0.4~\lesssim$~z~$\lesssim~1$ in \cite{savaglio05} and \cite{Liang06}, which too support a flat mass--metallicity relation for massive galaxies since z~$\sim~1$ down to z~$\sim~0$.

The gas--phase Oxygen abundances for low stellar masses 
in the simulations agree with the observed values, 
whereas the metallicity for more massive simulations 
tend to be systematically higher than the observed abundances.
Part of the offset could be explained by systematics
affecting the measurements, especially at the high--metallicity end 
(e.g.:~\citealt{EK05}; \citealt{Kennicutt03}).

\subsection{Gas--phase Oxygen abundance vs. galaxy luminosity}

As a reflection of the relationship between 
stellar mass and gas--phase metallicity, the 
relationship between gas metallicity and galaxy luminosity
is displayed in Fig.~\ref{mmr:oh_mb_gas}. 

The luminosity--metallicity relationship is compared 
and in broad agreement with observations 
at intermediate redshift: \cite{Lilly} at 
0.5~$\lesssim$~z~$\lesssim$~0.9; \cite{Liang} 
at 0~$\lesssim$~z~$\lesssim$~1.2; \cite{M06b} at 
0.2~$\lesssim$~z~$\lesssim$~0.8. As previously noted, this is a remarkable result, 
given that simulations and observations adopt different pipelines 
to estimate the gas--phase metallicity.

The Oxygen 
abundances of luminous galaxies in the simulations, 
i.e. B--band magnitude brighter than -20, 
are larger than what is observed at 0.4~$\lesssim$~z~$\lesssim$~0.7. 
Systematics in the observations may play a role here too, 
as discussed above in connection with the mass--metallicity relationship for massive galaxies.

\subsection{Stellar metallicity/age vs. stellar mass}

Age and metallicity of integrated stellar 
populations are powerful tracers of galaxy star 
formation and metal enrichment histories.
Fig.~\ref{mmr:zs_mstar} shows the relationship between stellar mass and 
V--band luminosity--weighted metallicity, at different redshifts. 

The general behaviour is similar 
to that of the gas--phase Oxygen abundance as 
a function of stellar mass (Fig.~\ref{mmr:oh_ms_gas}), 
although stellar metallicities are less dispersed than 
gas--phase Oxygen abundances at a given stellar mass. 

The results at redshift z~$=~0$ in the simulations are 
compared with the observed relation for a large sample of SDSS galaxies 
at 0.005~$\lesssim$~z~$\lesssim~0.22$ as reported in \cite{Gallazzi}, 
whose sample includes both quiescent early--type and actively star--forming 
galaxies, which might explain the smaller stellar metallicity 
dispersion at a given stellar mass in the simulations 
(which have late--type morphologies) compared with the dispersion 
observed in the more heterogeneous SDSS sample.

Fig.~\ref{mmr:agestar_mstar} shows the z~$=~0$~relationship between stellar mass 
and V--band luminosity--weighted age of 
the integrated stellar populations in the simulations, 
compared with the observed relationship. 
We find no relation between 
age and stellar mass. 
Massive galaxies with stellar mass larger than 
$\sim~10^{10}$~M$_{\odot}$ have age ranging from 
$\sim~6.5$~Gyr to $\sim~9$~Gyr, in broad agreement with the age
derived for SDSS galaxies with comparable stellar masses. 
For less massive galaxies in the simulations, 
the age range is broader, 
going from $\sim~3$ to $\sim~10$~Gyr, thus older than in \cite{Gallazzi}, whose sample is however much less populated and thus less robust at stellar masses M$_{*} < 10^{10}$~M$_{\odot}$ (see Fig.~6 in~\citealt{Gallazzi}).

The age issue in the simulations as outlined above signals that feedback should more effectively 
prevent gas from collapsing and cooling in the 
early stages of the hierarchy in order to ensure that 
most of the stellar content in low stellar mass galaxies is not formed early on. 
This has been already invoked in previous studies (e.g.:~\citealt{SN99}). 

\subsection{The Colour--Magnitude Relation}

Another signature of feedback--related problems in our sample, Fig.~\ref{mmr:cmr} 
shows the z~$=~0$ relationship between the (g~-~r) colour and 
the i--band luminosity in the simulations, 
compared\footnote{Since dust--correction has 
not been implemented in our code, we have compared 
the colour--magnitude relation (CMR hereafter) with the sample of \cite{Pizagno}, whose 
internal dust reddening was corrected for 
as in \cite{Tully}, rather than comparing the CMR in the simulations 
with the low redshift 
New York University Value--Added Galaxy Catalogue of 
\cite{Blanton}, which do not take into account extinction but the Galactic extinction as in \cite{Schlegel}.} with a sample of disc--dominated local galaxies in the SDSS,
selected by \cite{Pizagno}.

The integrated stellar population colours 
of luminous galaxies in the simulations are in good agreement
with the observed colours at comparable luminosities, 
which is expected due to the agreement, in massive simulations, 
of age and stellar metallicity with the observations.
The colour dispersion in the simulations 
-~which reflects the diverse formation 
histories which have been sampled~- agrees with the observed dispersion. 
This is a remarkable result, given that the observed sample has explicitly been selected to include disc--dominated galaxies with a disc--to--total luminosity ratio f$_{\rm d}$~$\geq$~0.9 (see \citealt{Pizagno}), whereas the simulated sample consists of late--type galaxy simulations which have not been morphologically selected.

Although the colour dispersion for less massive simulations is comparable with the observed dispersion, fainter simulations show integrated (g~-~r) colours generally redder than 
the observed, i.e. the simulated CMR is shallower than the observed. The discrepancy 
between observed and simulated colours for faint galaxies is due to 
the age discrepancy for faint galaxies between observations and simulations.
Although faint simulations in our sample 
agree with stellar and gaseous metallicities for locally observed galaxies with comparable luminosities and/or stellar masses, the former are older, resulting in redder colours. 

\section{Conclusions}
\label{mmr:concl}

We have presented the analysis of a sample of semi--cosmological late--type galaxy simulations
with stellar mass larger than $\sim~10^9$~M$_{\odot}$, 
to explore galaxy metal enrichment over the last half of the age of the Universe. 
A particular emphasis has been placed 
upon the relationship between stellar mass and metallicity 
of both gaseous and stellar components.

The main results can be summarised as follows. 

\begin{itemize}

\item[-] A correlation between gas--phase Oxygen abundance and 
 stellar mass is present at all redshifts since z~$\sim~1$ down to z~$=~0$.

\item[-] The z~$=~0$~relationship between stellar mass and gas--phase metallicity 
in the simulations is in broad agreement with the z~$\sim~0.1$~correlation observed 
in SDSS galaxies.

\item[-] The z~$=~0$~dispersion of the gas--phase Oxygen abundances for simulations with comparable masses and diverse merging histories is in broad agreement with the z~$\sim~0.1$~dispersion observed in SDSS galaxies. This suggests that the latter could be due to the differing formation histories of galaxies with comparable masses.

\item[-] The mass--metallicity relationship for massive simulations is $\sim~$flat since z~$\sim~1$ down to z~$=~0$, thus suggesting that they experience substantial metal enrichment early on.

\item[-] No relationship is found in the simulations between age and stellar mass. The age range for massive simulations
is similar to that observed for comparable mass z~$\sim~0$ galaxies. 
However, the z~$=~0$~stellar populations of less massive simulations tend to be dominated by 
stars older than the observed. A mechanism to delay star formation seems necessary to solve the discrepancy, as also suggested by the z~$=~0$ Colour--Magnitude Relation in the simulations, which is shallower than the observed.

\end{itemize}



\cleardoublepage


\chapter{Conclusions and Future Directions}
\label{chap:conclusion}

\textit{Which is ``normal''?}...\textit{Is the Milky~Way an outlier? Is there any stellar halo--galaxy formation symbiosis?} These driving questions in mind, we have tried to face the challenge they pose and undertaken the path which has led to this PhD~Thesis. Here we summarise what (we think that) we see from our point of view.

We have presented in Chapter~\ref{chap:halo} (which synthesises the background which is in Appendices~\ref{app:appendixB}~--~\ref{app:appendixH}; see also \citealt{RendaHalo}) an analysis of several late--type galaxy stellar halos within a grid of semi--cosmological simulations, with particular emphasis placed
upon the relationship between stellar halo metallicity
and galactic luminosity. It helps to stress here that ``Halo Semantics'' is currently a controversial topic (e.g.,~\citealt{Ibata07} and references therein): we have labelled as ``topographical halo'' the ensemble of stellar particles in a simulation at a projected radius R$~>~$15~kpc, at~z~=~0. Although future analysis of simulations at resolutions higher than the current would further improve the study we have undertaken, we have shown that -~at any given galaxy luminosity or, conversely, galaxy dynamical mass~- halo metallicities in the simulations span a range in excess of $\sim$~1~dex, a result which is strengthened by the robustness tests we have performed. 

We suggest that the
underlying driver of the halo metallicity dispersion can be traced to the
diversity of galactic mass assembly histories inherent within the
hierarchical clustering paradigm. Galaxies with a more protracted
assembly history possess more metal--rich and younger stellar halos, with
an associated greater dispersion in age, than galaxies which experience
more of a monolithic collapse.

For a given galaxy luminosity (or dynamical mass), those galaxies with more
extended assembly histories also possess more massive stellar halos, which
in turn leads to a direct correlation between the stellar halo
metallicity and its surface brightness (as anticipated by earlier
semi--analytical models~-~e.g.:~Chapter~\ref{chap:contra}; \citealt{Renda}). By extension, such a correlation may prove to be a useful diagnostic tool for disentangling the formation history of late--type galaxies.

Recently, \cite{MM} have presented an observed correlation between
stellar halo metallicity and galactic luminosity.
The observed halo metallicity dispersion at a given galactic
luminosity is {\it smaller} than what we find in our simulations. {\it However}
the latter can account for the outliers in the observed trend. Since our motivation has been to study
which is the effect of the pattern of the initial density fluctuations {\it alone}
on the stellar halo features at redshift z~=~0 in late--type galaxy simulations,
it helps to note that galaxy formation, as it is observed,
is an ongoing process which is the result of the interplay among different parameters,
of which the pattern of the initial density fluctuations (thus the merging history) is one.
We have shown that the merging history {\it alone} may be held responsible of the dispersion
in halo metallicity at comparable galactic luminosities.

We have then analysed the stellar mass assembly,
and the relationships between
stellar, gaseous and total mass 
since z~$\sim~1$ down to z~$=~0$ in the simulations,
to explore whether hierarchical formation scenarios may
account for the constraints on galaxy mass assembly
over the second half of the age of the Universe (Chapter~\ref{chap:mass}).

The redshift evolution of the stellar mass in simulations
with comparable luminosities at z~$=~0$ shows a large
dispersion, which lies in their differing merging histories.
This contrasts with the redshift evolution of the total
mass within the central 100~kpc, which is already
assembled by z~$\sim~1.5$. The stellar--to--total mass ratio in the simulations is in broad agreement with the observed (e.g.,~\citealt{Conselice}). 
We find that massive simulations are generally more
evolved than their lower mass counterparts. Galaxies with comparable luminosities/stellar masses at z~$=~0$ in the simulations can have significantly different assembly histories of their stellar content, and diverse gaseous contents. This suggests that stellar mass should not be considered neither a robust tracer of the galaxy merging history nor a good tracer of the total baryonic galaxy mass.

We have finally analysed the metal enrichment of the stellar and gaseous components in the simulations (Chapter~\ref{chap:massmeta}). A correlation between gas--phase Oxygen abundance and stellar mass is present since z~$\sim~1$ down to z~$=~0$. The z~$=~0$~relationship between stellar mass and gas--phase metallicity in the simulations is in broad agreement with the z~$\sim~0.1$~correlation observed in SDSS galaxies \citep{Tremonti}. The z~$=~0$~dispersion of the gas--phase Oxygen abundances for simulations with comparable masses and diverse merging histories is in broad agreement with the dispersion observed at z~$\sim~0.1$ in SDSS galaxies. This suggests that the latter could be due to the differing formation histories of galaxies with comparable masses. The mass--metallicity relationship for massive simulations is $\sim~$flat since z~$\sim~1$ down to z~$=~0$, thus suggesting that they experience substantial metal enrichment early on. No relationship is found in the simulations between age and stellar mass. The age range for massive simulations is similar to that observed for comparable mass z~$\sim~0$ galaxies.
However, the z~$=~0$~stellar populations of less massive simulations tend to be dominated by stars older than the observed. A mechanism to delay star formation seems necessary to solve the discrepancy, as also suggested by the z~$=~0$ Colour--Magnitude Relation in the simulations, which is shallower than the observed.

\textit{Where do we go from here?} However promising they are, chemo--dynamical simulations still suffer from feedback related issues (e.g.:~\citealt{SN99}; \citealt{TP}; \citealt{Abadi}; \citealt{Robertson}; \citealt{Stinson}; \citealt{Kaufmann}) which are reflected here in the mass--metallicity relationship and in the 
Colour--Magnitude Relation (Chapter~\ref{chap:massmeta}). It is reassuring though -~as a robustness test for the semi--cosmological framework~- that there is consistency between the results of semi--cosmological and cosmological simulations, as shown in Chapters~\ref{chap:mass}~and~\ref{chap:massmeta}. 

We have focused here in this PhD~Thesis on the formation of stellar halos in late--type galaxies. We have shown that the diverse galaxy merging histories may be held responsible of the observed dispersion
in halo metallicity at comparable galactic luminosities (Chapter~\ref{chap:halo}). Higher resolution simulations, a deeper understanding of the role of feedback processes in galaxy formation, and further observations are all together needed to take the next step on this research path.


\clearpage

\part*{Appendices}

\appendix



\chapter[The origin of fluorine in the Milky Way]{On the origin of fluorine\\ in the Milky Way}
\label{app:appendixA}

The main astrophysical factories of fluorine ($^{19}$F) are thought
to be Type~II supernovae, Wolf--Rayet stars, and the Asymptotic Giant 
Branch (AGB)
of intermediate mass stars. We present a model for the chemical evolution of 
fluorine in the Milky Way in a semi--analytic multi--zone chemical evolution 
framework. For the first time, we quantitatively demonstrate the impact of 
fluorine nucleosynthesis in Wolf--Rayet and AGB stars. 
The inclusion of these latter two fluorine production sites provides a possible
solution to the long--standing discrepancy between model predictions and the
fluorine abundances observed in Milky Way giants. Finally, fluorine is 
discussed as a possible probe of the role of supernovae and intermediate
mass stars in the chemical evolution history of the globular cluster
$\omega$~Centauri.

\section{Introduction} 
\label{app:appendixA:intro}

The three primary astrophysical factories for fluorine ($^{19}$F) production 
have long been thought to be 
Type~II Supernovae (SNe~II), Wolf--Rayet (WR) stars, and Asymptotic Giant
Branch (AGB) stars (e.g., respectively:~\citealt{WW}; \citealt{MA}; \citealt{ForestiniETal}; 
Mowlavi, Jorissen \&
Arnould 1998). 
Previous attempts to model the Galactic production and evolution of 
$^{19}$F have been restricted to explore the
role of SNe~II alone (e.g.:~Timmes, Woosley \& Weaver 1995; Alib\'es,
Labay \& Canal 2001).

The above problem has now been ameliorated by 
the release of the first detailed yield predictions
for fluorine production from WR and AGB stars. We are now in a position to
incorporate these yields into a Galactic chemical evolution framework, in
order to assess the respective contributions of the three putative 
fluorine production sites. To do so, we will make use of {\tt GEtool}, a
semi--analytical multi--zone
Galactic chemical evolution package which has been 
calibrated with extant observational data for the Milky Way
(\citealt{FG}; \citealt{GibsonETal}).

Specifically, in what follows, we compare the model fluorine distribution in
the Milky Way with the abundances observed by Jorissen,
Smith \& Lambert (1992) in near--solar metallicity giants. Further, our 
model predictions are contrasted with new  
fluorine determinations for
giants in the Large Magellanic Cloud (LMC) and $\omega$~Centauri 
\citep{CunhaETal}. In addition, new results for more $\omega$ Centauri giants 
from \cite{SmithETal} are included. 
The latter two systems are likely to have had {\it very}
different star formation and chemical evolution histories from those of 
the Milky Way, but despite these obvious differences, a comparison against
these new data can be valuable. In Section~\ref{app:appendixA:nucleosynth}, we provide a cursory overview of
the three traditional $^{19}$F nucleosynthesis sites; the chemical evolution
code in which the nucleosynthesis products from these factories have been
implemented is described in Section~\ref{app:appendixA:model}. Our results are then presented and
summarised in Sections~\ref{app:appendixA:results} and \ref{app:appendixA:discussion}, respectively.

\section{Nucleosynthesis of $^{19}$F}
\label{app:appendixA:nucleosynth}

\subsection{Type~II Supernovae}
\label{app:appendixA:nucleosynth:SNeII}

The massive star progenitors to SNe~II produce fluorine primarily as the result
of spallation of $^{20}$Ne by $\mu$ and $\tau$ neutrinos near the collapsed
core (\citealt{WH}; \citealt{WoosleyETal}). 
A fraction of the $^{19}$F thus created is destroyed by the subsequent
shock but most is returned to the ambient Interstellar Medium (ISM). The fluorine yields by neutrino 
spallation are very sensitive to the assumed spectra of $\mu$ and $\tau$ 
neutrinos \citep{WHW}, 
which could be nonthermal and deficient on their high--energy tails, 
lowering the equivalent temperature of the neutrinos 
in the supernova model \citep{MB}. 
An additional source of $^{19}$F derives from pre--explosive CNO 
burning in helium shell. 
However, fluorine production by neutrino spallation is largely dominant, 
as evident by comparing the models in \cite{WW}, 
and recent models which do not include neutrino nucleosynthesis 
of fluorine \citep{LC}.
Most recently, \cite{HegerETal} suggest that the relevant
neutrino cross sections need to be revised downwards; if 
confirmed, the associated SNe~II $^{19}$F yield
would decrease by $\sim$~50\%.  In light of the preliminary nature of 
the Heger et~al. claim, we retain the conservative choice offered by 
the \cite{WW} compilation.

\subsection{Asymptotic Giant Branch stars}
\label{app:appendixA:nucleosynth:AGB}

The nucleosynthesis pathways for fluorine production within AGB
stars involve both helium burning and combined hydrogen--helium burning phases 
(e.g.:~\citealt{ForestiniETal}; \citealt{JorissenETal}; 
\citealt{MowlaviETal})
and are companions for the nucleosynthesis by slow neutron accretion 
(s--process) 
\citep{MowlaviETal}.
Provided a suitable source of protons is available, fluorine can be synthesised
via $^{14}$N($\alpha$,$\gamma$)$^{18}$F\,($\beta^{+}$)\,
$^{18}$O(p,$\alpha$)\,$^{15}$N($\alpha$,$\gamma$)\,$^{19}$F. 
Primary sources of uncertainty in predicting fluorine nucleosynthesis 
in AGB stars relate to the adopted reaction rates, 
especially $^{14}$C$(\alpha,\gamma)^{18}$O and $^{19}$F$(\alpha,{\rm p})^{22}$Ne, 
and the treatment of the nucleosynthesis occurring during the convective 
thermal pulses. Nucleosynthesis during the interpulse periods can also 
be important if protons from the envelope are partially mixed 
in the top layers of the He intershell (partial mixing zone), as
\cite{LugaroETal} have recently demonstrated. Nucleosynthesis in this zone 
may result in a significant increase in the predicted $^{19}$F yields.
The magnitude of these 
systematic uncertainties for stellar models
with mass $\sim$~3~M$_{\odot}$ and metallicities Z~=~0.004~--~0.02 are
$\sim$~50~$\%$, while for stellar models
with mass M~=~5~M$_{\odot}$ and metallicity Z~=~0.02 the uncertainty
is a factor of $\sim$~5, 
due to the uncertain $^{19}$F$(\alpha,{\rm p})^{22}$Ne reaction rate.
Characterising the mass-- and metallicity-- dependence of 
the partial mixing zone--$^{19}$F relationship needs 
to be completed before
we can assess its behaviour self--consistently within our chemical evolution
model of the Milky Way. For the present study, we have adopted the yields
presented in \cite{RendaFluo}, based upon the Karakas \& Lattanzio (2003, and
references therein) models, which themselves do not include 
$^{19}$F nucleosynthesis via partial mixing. This choice is 
a conservative one, and thus should be considered as a lower limit to the
production of $^{19}$F from AGB stars.

For stars more massive than $\approx$~4~M$_{\odot}$, the convective
envelope is so deep that it penetrates into the top of the
hydrogen--burning shell so that nucleosynthesis actually occurs in the
envelope of the star. Such ``hot--bottom--burning'' acts to destroy $^{19}$F, 
and should be treated self--consistently within the AGB
models considered. 

\subsection{Wolf--Rayet stars}
\label{app:appendixA:nucleosynth:WR}

Fluorine production in WR stars is tied to its nucleosynthesis during the 
helium--burning phase. At the end of this phase though, significant 
fluorine destruction occurs via
$^{19}$F$(\alpha,{\rm p})^{22}$Ne. Any earlier synthesised $^{19}$F
must be removed from the stellar interior in order to avoid
destruction. For massive stars to be significant contributors to net
fluorine production, they must experience mass loss on a timescale that
allows the removal of $^{19}$F before its destruction. This requirement is 
met by WR stars. 

Recently, \cite{MA} studied the role that such stars can play in the
chemical evolution of fluorine by adopting updated reaction rates coupled
with extreme mass--loss rates in not--rotating stellar models. 
They pointed out that WR mass--loss is strongly metallicity--dependent, 
and that the number of WR stars at low
metallicities is very small. Their WR yields reflect such
metallicity--dependence, with minimal fluorine returned to the ISM
at low metallicities, but significant $^{19}$F returned at solar and
super--solar metallicities. The WR yields are sensitive to the adopted reaction 
and mass--loss rates, while rotating models could favour an early entrance 
into the WR phase for a given mass, 
decrease the minimum initial mass for a star to go through a WR phase 
at a given metallicity, and open more nucleosynthetic channels 
because of the mixing induced by rotation. 
Therefore, after \cite{MA}, 
we consider the aforementioned WR yields as lower limits.

\section{The model}
\label{app:appendixA:model}

In this study we employ {\tt GEtool}, our 
semi--analytical multi--zone chemical evolution package to model a sample
Milky Way--like disk galaxy (\citealt{FG};
\citealt{GibsonETal}). A dual--infall framework is constructed in which
the first infall episode corresponds to the formation of the halo,
and the second to the inside--out formation of the disk. 

A Kroupa, Tout \& Gilmore (1993) initial mass function (IMF) has 
been assumed, with lower and upper
mass limits of 0.08~M$_{\odot}$ and 120~M$_{\odot}$, respectively. 
Stellar yields are one of the most
important features in galactic chemical evolution models, yet
questions remain concerning the precise composition of stellar ejecta,
due to the uncertain role played by processes including mass loss,
rotation, fall--back, and the location of the mass cut, which separates
the remnant from the ejected material in SNe. The SNe~II
yields are from \cite{WW}; for stars more massive than
60~M$_{\odot}$, the yields are assumed to be mass--independent. 
Such assumption is made to avoid extreme extrapolation 
from the most massive star in the Woosley \& Weaver models (40~M$_{\odot}$) 
to the upper end of the IMF (120~M$_{\odot}$), and has 
negligible effect on the results, given the shape of the adopted IMF.

We have halved the iron yields shown in \cite{WW}, as suggested by
\cite{TimmesETal}.
The Type~Ia (SNe~Ia) yields of
\cite{IwamotoETal} were also employed. We adopted the 
metallicity--dependent yields of \cite{RV} for single stars in the mass range~1~--~8~M$_{\odot}$. For the purposes of this work, which focuses on fluorine, 
the choice of the Renzini \& Voli yields set does not affect the results.
Metalllicity--dependent stellar lifetimes 
have been taken from
\cite{SchallerETal}. 

We have constructed three Milky Way (MW) model variants
that differ only in their respective
treatments of $^{19}$F production:  1) MWa assumes that SNe~II are the only
source of $^{19}$F; 2) MWb includes yields from both SNe~II {\it and}
WR stars; 3) MWc includes all three sources of fluorine -
SNe~II, WR, and AGB stars. 

We end by noting that within our adopted dual--infall framework
for the chemical evolution 
of the Milky Way, our model is constrained by an array of observational
boundary conditions, including the present--day star and gas distributions
(both in density and metallicity), abundance ratio patterns, age--metallicity
relation, and G--dwarf distribution \citep{GibsonETal}.  While the modification
of any individual ingredient within the model framework will have an
impact, to some degree, upon the predicted chemical evolution, this can 
only eventuate at the expense of one or more of the aforementioned 
boundary conditions that we require our model to adhere to. 
Within our framework, yield uncertainties will dominate the systematic
uncertainties for the predicted evolution of $^{19}$F.

\subsection{Fluorine yields}
\label{app:appendixA:yields}

\begin{figure}
\begin{center}
\includegraphics[width=1.0\textwidth]{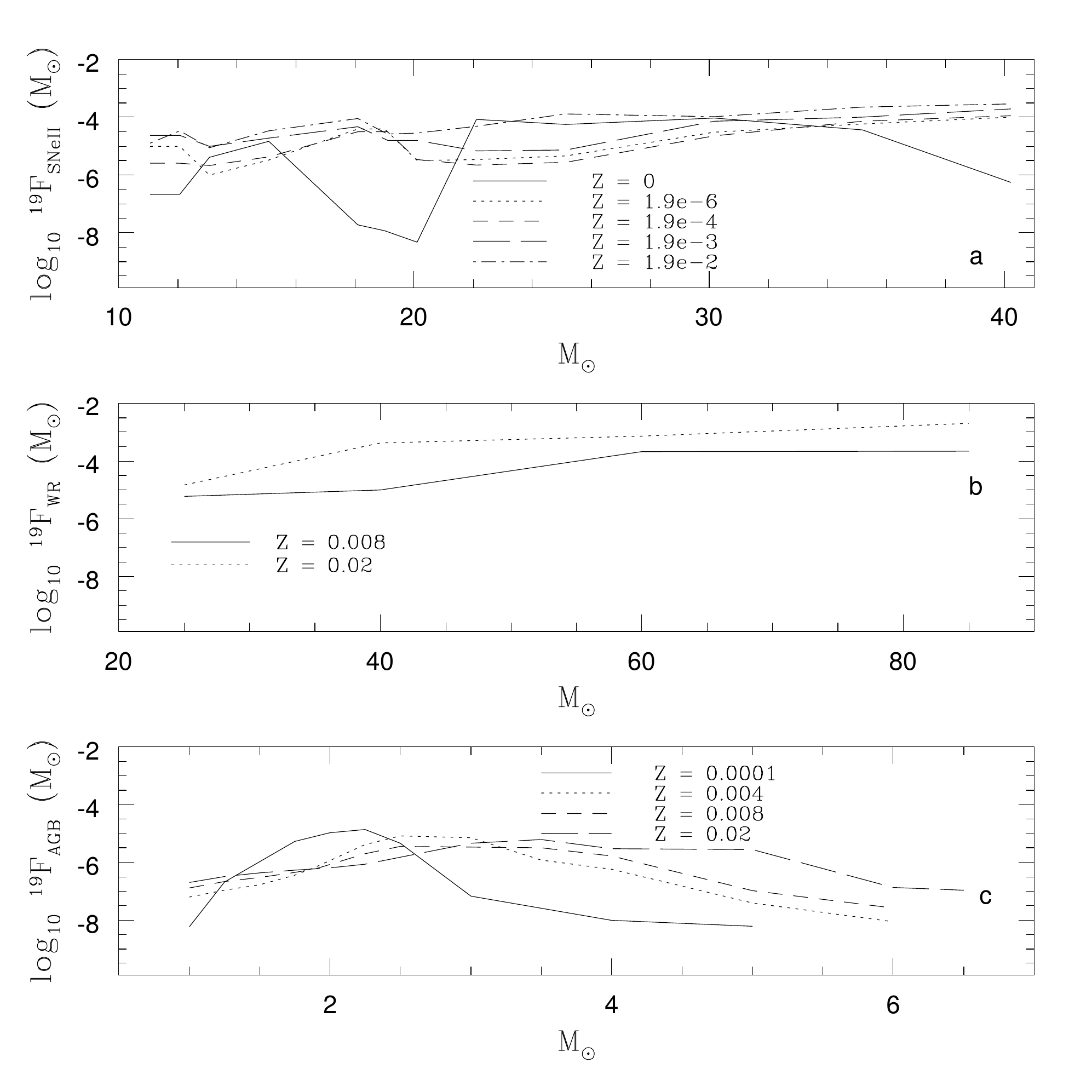}
\caption{Fluorine yields from: ``a'', SNe~II (Woosley \& Weaver 1995);
``b'', WR (Meynet \& Arnould 2000); ``c'', AGB stars.}
\label{app:appendixA:fig1}            
\end{center}
\end{figure}

We now summarise the $^{19}$F yields
employed in our three ``Milky Way'' models.

1) SNe~II $^{19}$F yields are taken from \cite{WW} and assumed to be 
mass--independent for stellar masses in excess of 60~M$_{\odot}$. 

2) WR $^{19}$F yields are taken 
from \cite{MA} for stellar masses in the range~25~--~120~M$_{\odot}$: each star within this range is assumed to evolve
through the WR stage. Such simplifying assumption 
could overestimate the WR contribution to fluorine, 
even though the adopted WR yields are themselves lower limits (Section~\ref{app:appendixA:nucleosynth:WR}).
The WR fluorine contribution has been added to the corresponding
SNe~II contribution (which comes from a different stage of the stellar
evolution).

3) AGB $^{19}$F and oxygen yields in the mass range~1~--~6.5~M$_{\odot}$ have
been derived from stellar models constructed with the 
Mount Stromlo Stellar Structure Code (\citealt{FL}; \citealt{KarakasETal}), 
and are presented in \cite{RendaFluo}. 
The post--processing nucleosynthesis models 
with 74 species and time--dependent diffusive
convective mixing are described in detail in \cite{FrostETal} and \cite{KL}.

To ensure internal consistency, we have also employed the AGB 
oxygen yields \it in lieu \rm of those of Renzini \& Voli (1981),
within this mass range.

\begin{figure}
\begin{center}
\includegraphics[width=1.0\textwidth]{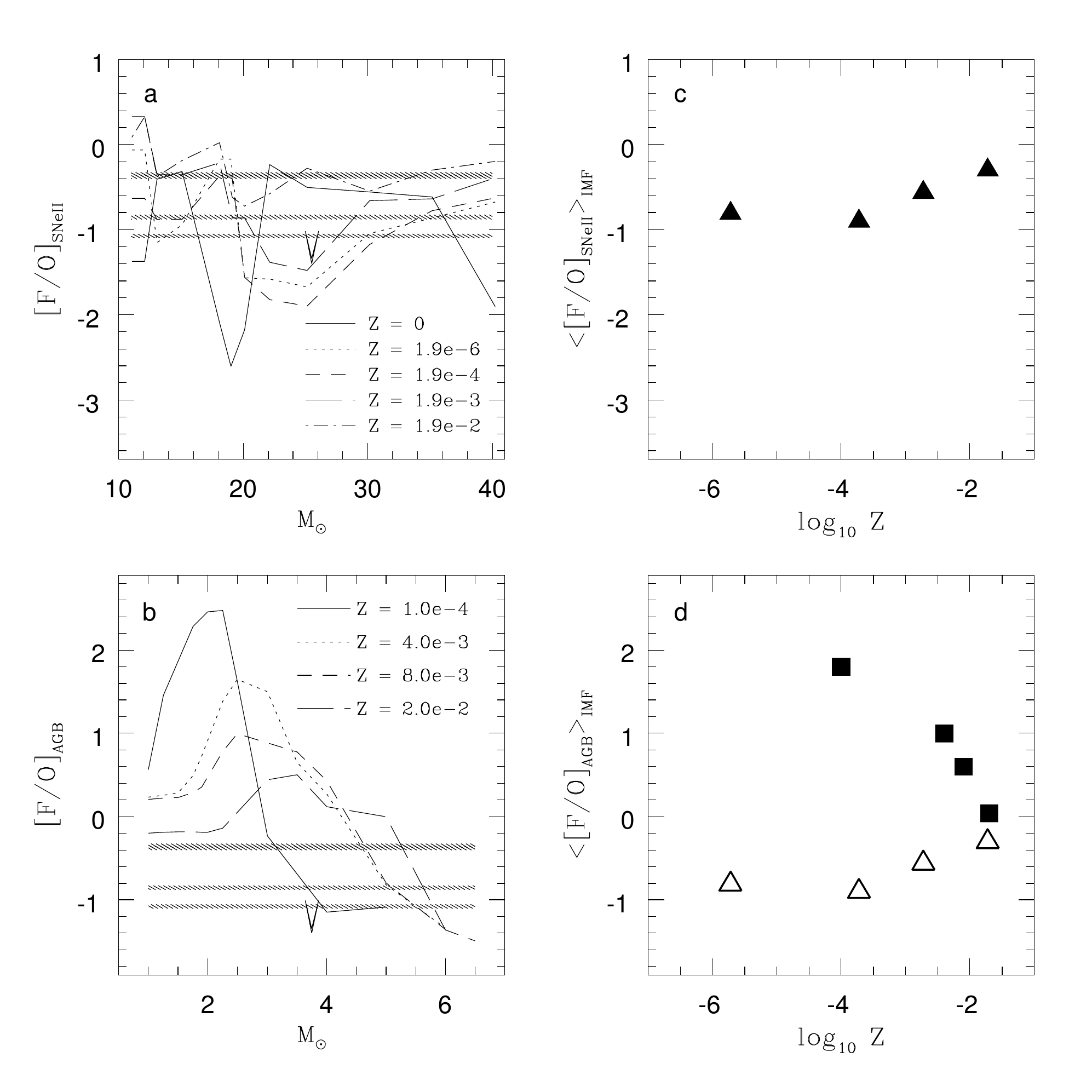}
\caption{[F/O] and $\langle$[F/O]$\rangle_{\rm IMF}$ for SNe~II and AGB yields 
(upper and lower panels, respectively). Here, A($^{19}$F)$_{\odot} = 
4.55$ (see discussion in Cunha et al. 2003) and A(O)$_{\odot} = 8.736$ (e.g.,~Holweger 2001). 
The shaded regions in panels~``a'' and ``b'' show the observed [F/O] 
in $\omega$~Cen giants (Cunha et al. 2003). The $\langle$[F/O]$\rangle_{\rm IMF}$ 
are weighted by the IMF over the SNe~II~(11~--~40~M$_{\odot}$) and 
AGB~(1~--~6.5~M$_{\odot}$) mass range, respectively. In panel~``d'',
both $\langle$[F/O]$_{\rm AGB}\rangle_{\rm IMF}$ and 
$\langle$[F/O]$_{\rm SNe~II}\rangle_{\rm IMF}$ are shown (closed boxes and open 
triangles, respectively).}
\label{app:appendixA:fig2}            
\end{center}
\end{figure}

The above fluorine yields are shown in Figure~\ref{app:appendixA:fig1}. In Figure~\ref{app:appendixA:fig2}, the yields are
expressed as [F/O]\footnote{Hereafter,
[X/Y]~=~${\rm log}_{10}$(X/Y)$~-~{\rm log}_{10}$(X/Y)$_{\odot}$ and 
A(X)~=~12~+~${\rm log}_{10}$(n$_{\rm X}$/n$_{\rm H}$).
An accurate determination of photospheric solar abundances 
requires detailed modeling of the solar granulation 
and accounting for departures from local thermodynamical equilibrium 
(e.g.,~Allende Prieto, Lambert \& Asplund 2001).
We adopt the solar fluorine abundance suggested by Cunha et~al. (2003),
and the solar iron and oxygen abundances from Holweger (2001).} 
and $\langle$[F/O]$\rangle_{IMF}$, the latter corresponding to
the mean [F/O] yields for
SNe~II and AGB stars, weighted by the IMF over the SNe~II and AGB mass
range, respectively. We have not shown a comparable entry for the WR stars as
a self--consistent treatment of the oxygen production was not included in Meynet
\& Arnould (2000). Here, oxygen has been used as the normalisation 
to make easier the comparison with the observations, 
especially in $\omega$ Centauri, though oxygen can be synthesised in various 
stellar sites, and its yields can be affected by different reaction rates 
and modeling of helium cores, semi--convection, convective boundary layers,
and mass--loss (e.g.:~\citealt{WHW}; \citealt{DrayETal}). 

\section{Results}
\label{app:appendixA:results}

\begin{figure}
\begin{center}
\includegraphics[width=1.0\textwidth]{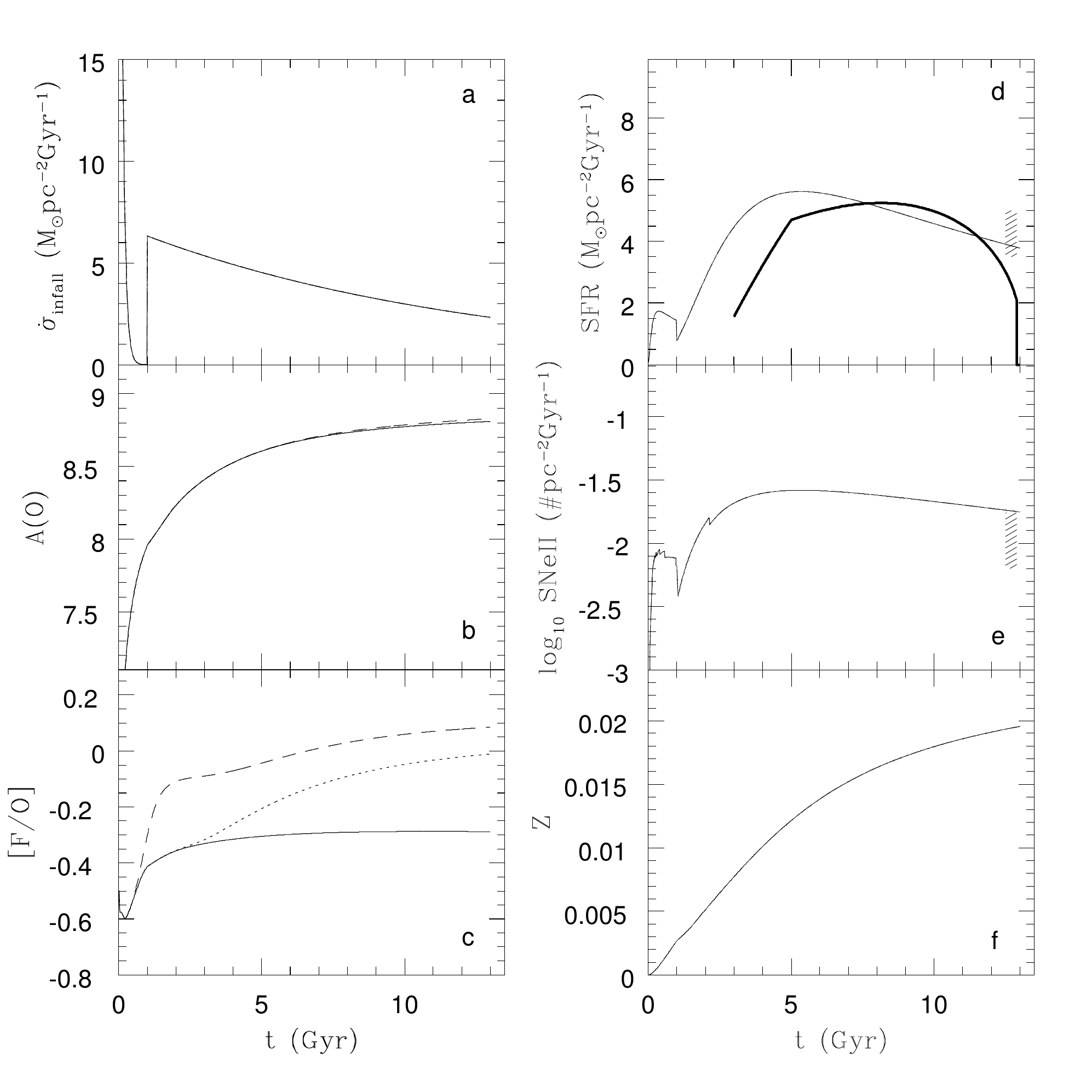}
\caption{Predicted evolution in the solar neighbourhood of: ``a'', the gas infall
rate $\dot{\sigma}_{infall}$; ``b'', A(O); ``c'', [F/O]; ``d'', the star formation rate (SFR); ``e'', the SNe~II rate; ``f'', the metallicity Z (MWa, solid line; MWb, dotted; MWc,
short--dashed). The SFR history at the solar neighbourhood obtained by 
Bertelli \& Nasi (2001) is also shown as a thick solid line in panel~``d'',
while the shaded region shows the range of values suggested by Rana (1991). 
A range of values corresponding to the estimated
SNe~II rate is shown in panel~``e'' (Cappellaro et~al. 1999).
}
\label{app:appendixA:fig3}            
\end{center}
\end{figure}

\begin{figure}
\begin{center}
\includegraphics[width=1.0\textwidth]{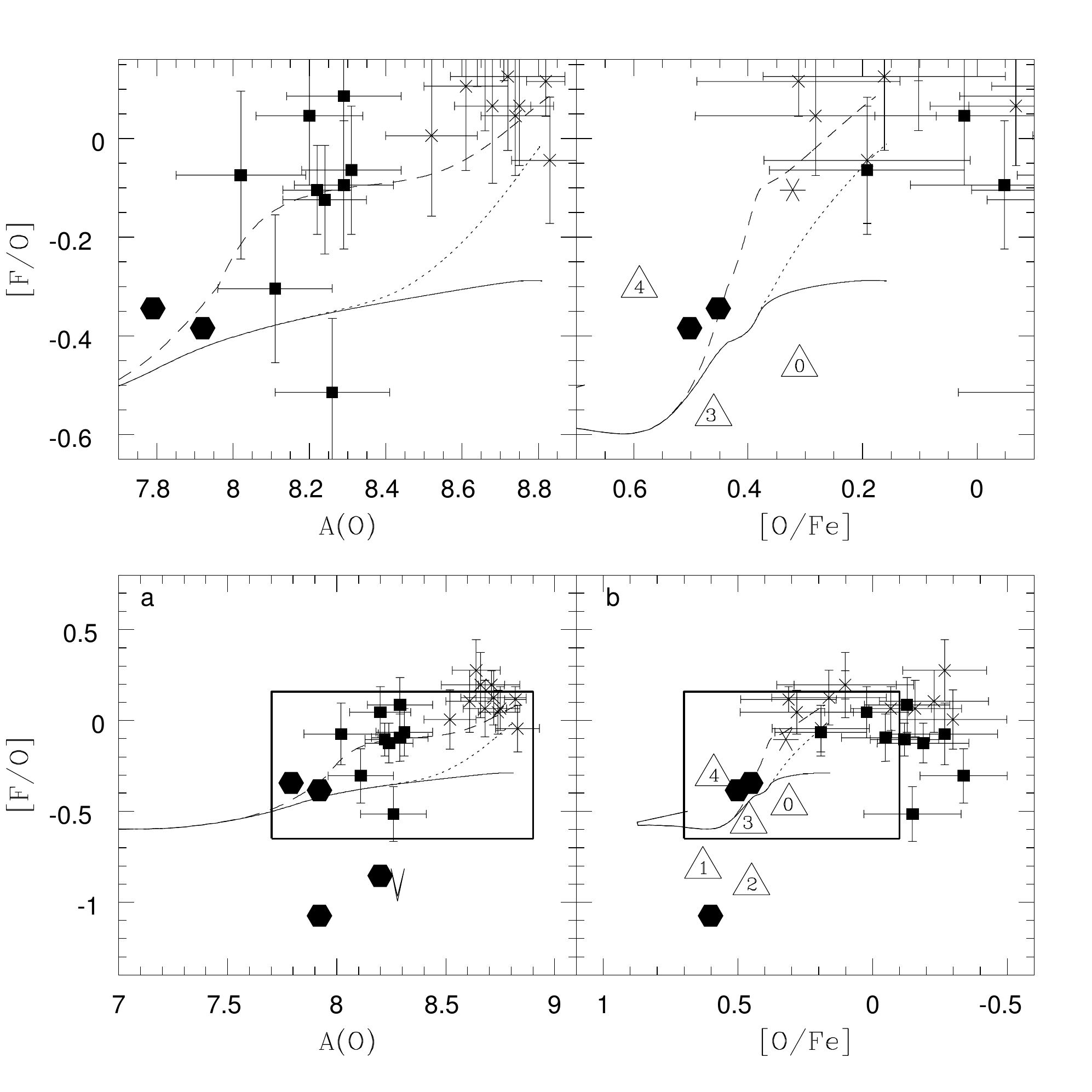}
\caption{In panel~``a'', [F/O] as a function of A(O) for the MW models (MWa, 
solid line; MWb, dotted; MWc, short--dashed). Also shown are the
values observed in Milky Way, LMC, and $\omega$~Cen giants (crosses,
boxes and hexagons, respectively).
In panel~``b'': [F/O] as a function of [O/Fe],
compared with the IMF--weighted $\langle$[F/O]$_{\rm SNe~II}\rangle_{\rm IMF}$  and  $\langle$[O/Fe]$_{\rm SNe~II}\rangle_{\rm IMF}$ yields for SNe~II
(open triangles). Within the open triangles: ``0'' corresponds to Z=0; ``1'', 
to Z=1.9$\times$10$^{-6}$; ``2'', to Z=1.9$\times$10$^{-4}$;
``3'', to Z=1.9$\times$10$^{-3}$; ``4'', to Z=1.9$\times$10$^{-2}$.
The upper panels represent enlargements of the framed regions delineated in
the corresponding bottom panels.
}
\label{app:appendixA:fig4}            
\end{center}
\end{figure}

In Figure~\ref{app:appendixA:fig3}, the evolution of [F/O], A(O), the gas infall rate
$\dot{\sigma}_{infall}$, the star formation rate (SFR), the SNe~II rate 
and the gas--phase global metallicity $Z$
of the three models at the solar neighbourhood are summarised. 
The empirical SFR history derived by \cite{BN} is shown as a thick solid
line in panel~``d'' of Figure~\ref{app:appendixA:fig3}, while the shaded region corresponds to the range of values
suggested by \cite{Rana}. A conservative range of estimated SNe~II rates
is also shown in panel~``e'' (Cappellaro, Evans \& Turatto 1999).\footnote{The
range of values shown in panel~``e'' of Figure~\ref{app:appendixA:fig3} is derived from the sample
of S0a~--~Sb galaxies in Cappellaro et al. (1999) - $0.42\pm0.19$~SNu, where
$1 $~SNu$=1$~SN($100$~yr)$^{-1}$($10^{10}$~L$_{\odot}^{B}$)$^{-1}$, assuming
L$_{MW}^{B}=2\times10^{10}$~L$_{\odot}^{B}$ and a galactic radial extent of
15~kpc. Given these assumptions, the estimated SNe~II
rate at the solar neighbourhood is necessarily uncertain.}. 
Figure~\ref{app:appendixA:fig4} then shows the
the evolution of [F/O] versus A(O)  (panel~``a''), and the
evolution of [F/O] versus [O/Fe] (panel~``b''), compared against
the IMF--weighted SNe~II yields (recall Figure~\ref{app:appendixA:fig2}). 

The MWa model provides a satisfactory reproduction of the estimated star
formation history and SNe~II rate in
the solar neighbourhood (Figure~\ref{app:appendixA:fig3}, panels~``d'' and ``e''). This model, whose only
fluorine source is SNe~II, underproduces fluorine with respect to the
abundances measured in K and M Milky Way giants observed by \cite{JorissenETal} 
and reanalysed by \cite{CunhaETal} (Fig.~\ref{app:appendixA:fig4}, panel~``a''). Fig.~\ref{app:appendixA:fig4}, panel~``a'' does not show the s--process 
enriched AGB stars of spectral types MS, S, or C in \cite{JorissenETal}, 
where freshly synthesised fluorine could be mixed to the stellar surface. 
Such inclusion of self--polluted $^{19}$F--rich stars could obscure any metallicity trend.
The results of the MWa and MWb models show that the
additional contribution from WR stars 
increases [F/O] by up to factor of 2 by the present--day, but 
it is negligible
in excess of $\sim$~9~Gyr ago (Figure~\ref{app:appendixA:fig3}, panel~``c'').

The addition of both
WR {\it and} AGB sources within the MWc model 
leads to a present--day [F/O] that is
$\sim$~0.4~dex greater than in the MWa case. Further, and perhaps more
important, AGB stars are now shown to deliver
significant amounts of fluorine to the ISM during the early epochs of the
Milky Way's evolution. Such a result is entirely consistent (and expected)
given the metallicity--dependence of the AGB yields; said yields possess
[F/O] ratios which are greater at lower metallicities (recall Figure~\ref{app:appendixA:fig2}). 
We can conclude that it is only the addition of {\it both} the WR and the AGB
contributions which allow for a significant improvement in the 
comparison between galactic models incorporating fluorine evolution and
the observational data.

\section{Discussion}
\label{app:appendixA:discussion}

We have studied the Galactic chemical evolution of fluorine, for the first
time using new grids of stellar models which provide self--consistent
predictions of fluorine nucleosynthesis for stars in both the WR and AGB
phases of stellar evolution. We have shown that the WR contribution is 
significant at solar and super--solar metallicities because of the adopted
metallicity--dependent mass--loss prescription employed in the stellar models.
In contrast, the contribution of AGB stars to fluorine production
peaks during the early epochs of the Galaxy's evolution (again due to the
metallicity--dependent behaviour of the AGB models). In combination,
the addition of the WR and AGB yields 
leads to a significant improvement in the galactic chemical evolution models
when compared against observations. 

The comparison between our MW models and the fluorine abundances in LMC and
$\omega$~Cen giants \citep{CunhaETal} is not straightforward, as the
latter two have star formation (and therefore chemical evolution)
histories different from that of the MW. However, it is interesting to 
speculate on the possible origin of
fluorine in $\omega$~Cen, given the unique nature of this ``globular cluster''
(e.g.,~\citealt{S}). Specifically, $\omega$~Cen 
is the most massive Galactic cluster, and unlike most globulars, possesses a
significant spread in metallicity 
($\sim$~1.5~dex) amongst its stellar population. It has been suggested that
$\omega$~Cen is actually the remnant core of a tidally--disrupted 
dwarf galaxy \citep{BF}. Such a scenario could naturally drive radial
gas inflows to the dwarf nucleus, potentially triggering starbursts.

Interestingly, SNe~II ejecta are characterised by low
$\langle$[F/O]$\rangle_{IMF}$ (Figure~\ref{app:appendixA:fig2} panel~''c'') and high
$\langle$[O/Fe]$\rangle_{IMF}$ (Figure~\ref{app:appendixA:fig4} panel~''b''), whereas AGB ejecta have higher
$\langle$[F/O]$\rangle_{IMF}$ (Figure~\ref{app:appendixA:fig2} panel~``d''). The observed $\omega$~Cen giants
have primarily low [F/O] (Figures:~\ref{app:appendixA:fig2}, panels~``a''~--~``b''; \ref{app:appendixA:fig4}, panels~``a''~--~``b\
'') and high [O/Fe]
(Figure~\ref{app:appendixA:fig4}, panel~``b'') values, consistent with a picture in which their interiors have
been polluted by the ejecta of an earlier generation of 
SNe~II, but not from a comparable generation of AGB.
Given that the observed oxygen
abundance in such $\omega$~Cen giants is similar to that seen in comparable
LMC and MW giants, as in Figure~\ref{app:appendixA:fig4}, panel~``a'', this would suggest that 
the chemical enrichment of $\omega$~Cen proceeded on a short timescale
(to avoid pollution from the lower mass progenitors to the AGB stars) and
in an inhomogeneous manner (given the significant scatter in observed
fluorine abundances), as previously discussed by \cite{CunhaETal}. 
Should the (downward) revised neutrino cross sections alluded
to in Section~\ref{app:appendixA:nucleosynth:SNeII} be confirmed (Heger et~al. 2004), the concurrent 
factor of $\sim$~2 reduction in SNe~II $^{19}$F production
would improve the agreement of the model with the observed [F/O] ratio
in $\omega$~Cen giants.  This would consequently
strengthen our conclusions which already support a picture 
whereby these giants have been 
polluted by earlier generations of SNe~II ejecta.





\newpage

\begin{center}
\chapter{Stellar Halo Metallicity Distributions}
\label{app:appendixB}
\end{center}

\begin{figure}
\begin{center}
\includegraphics[width=1.0\textwidth]{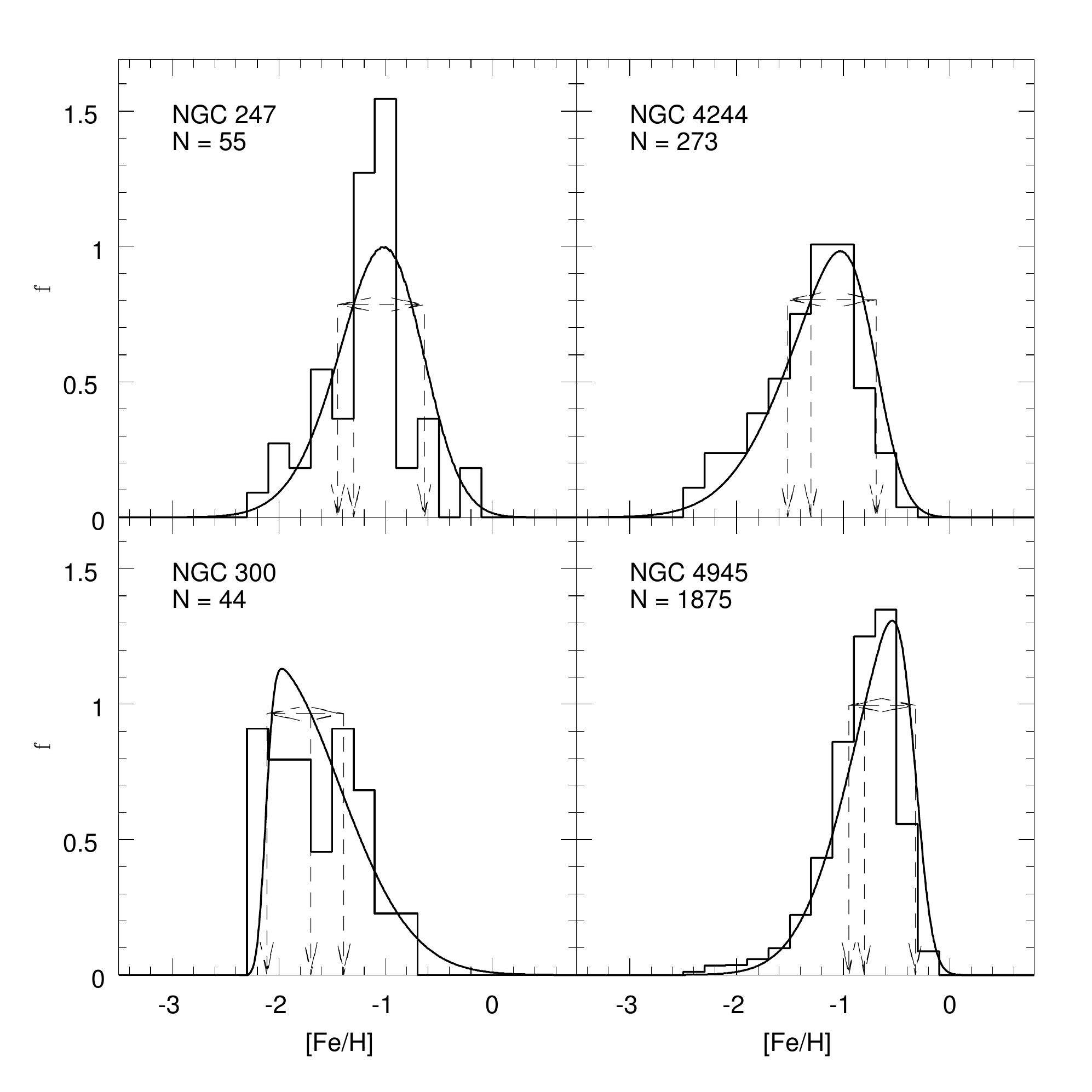}
\caption{MDFs for the halo fields observed in Mouhcine~et~al.~(2005), reanalysed by the same pipeline as in Renda~et~al.~(2005b). The 68\%~Confidence~Level range and the number of stellar particles each MDF refers to are also shown.}
\label{appB:data:fig1}            
\end{center}
\end{figure}

\begin{figure}
\begin{center}
\includegraphics[width=1.0\textwidth]{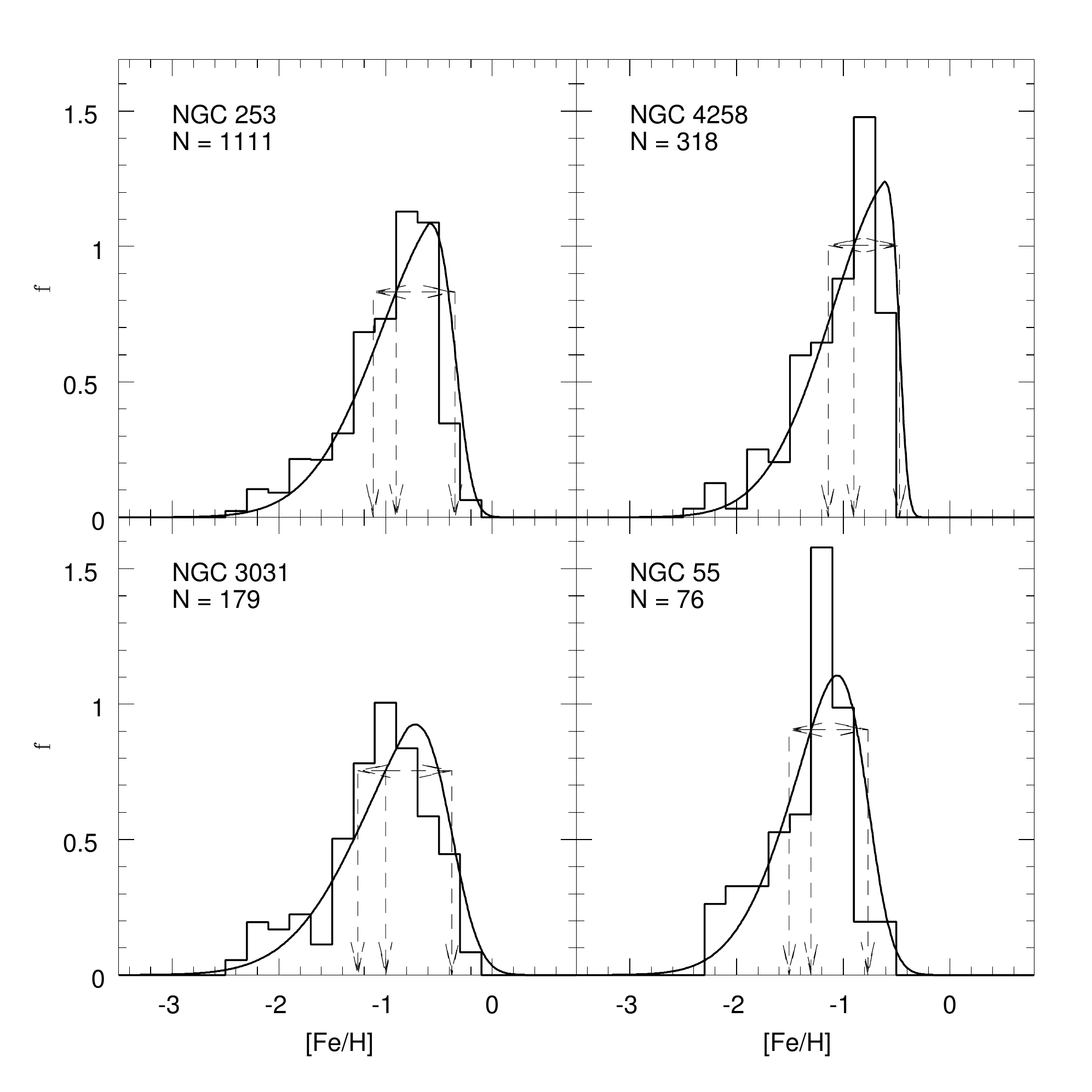}
\caption{(continued) MDFs for the halo fields observed in Mouhcine~et~al.~(2005), reanalysed by the same pipeline as in Renda~et~al.~(2005b). The 68\%~Confidence~Level range and the number of stellar particles each MDF refers to are also shown.}
\label{appB:data:fig2}
\end{center}
\end{figure}

\begin{figure}
\begin{center}
\includegraphics[width=1.0\textwidth]{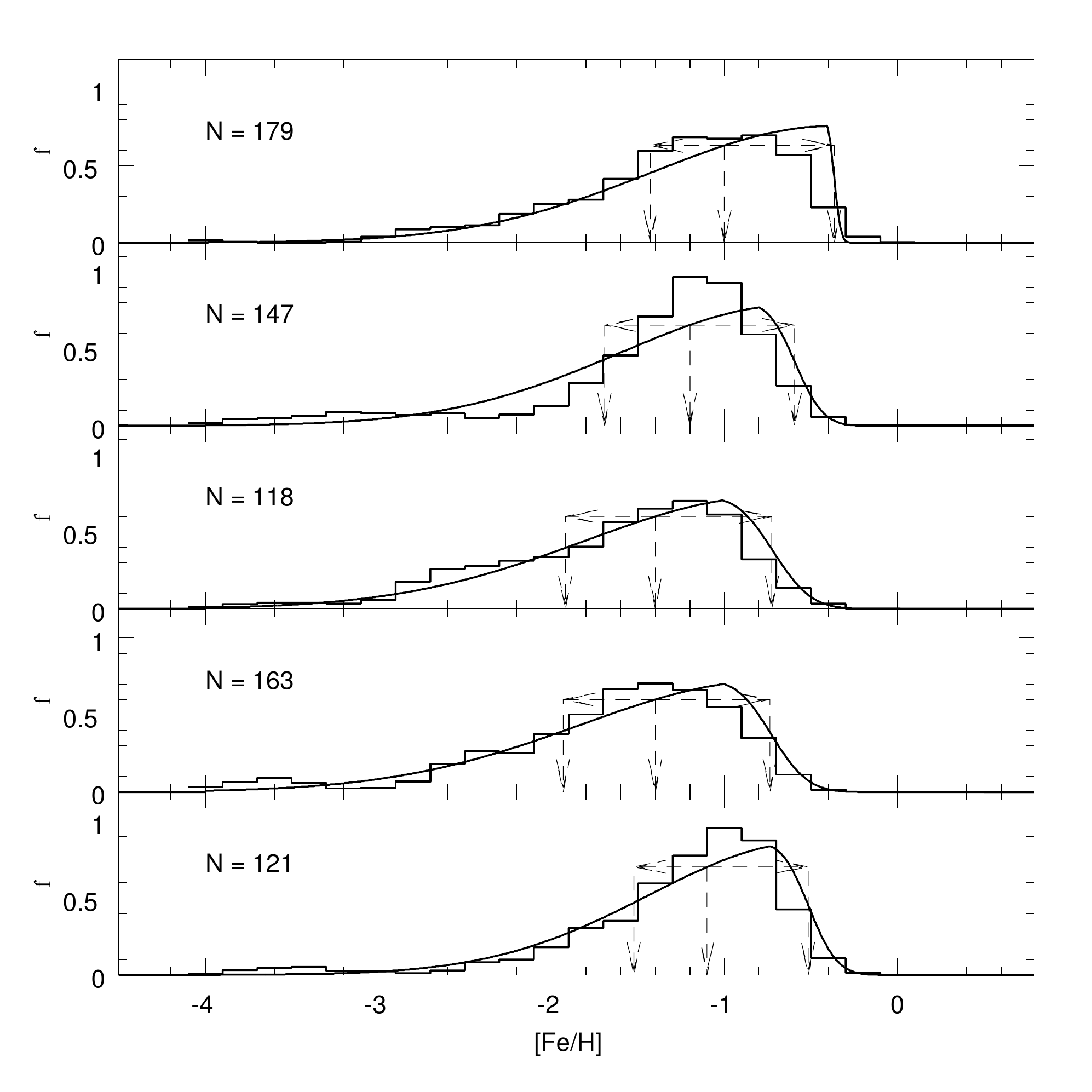}
\caption{Halo stellar particle MDFs for the M$_{\rm tot}$~=~10$^{11}$~M$_\odot$ semi--cosmological simulations in Renda~et~al.~(2005b). The 68\%~Confidence~Level range and the number of stellar particles each MDF refers to are also shown.}
\label{appB:sim1e11:fig}
\end{center}
\end{figure}

\begin{figure}
\begin{center}
\includegraphics[width=1.0\textwidth]{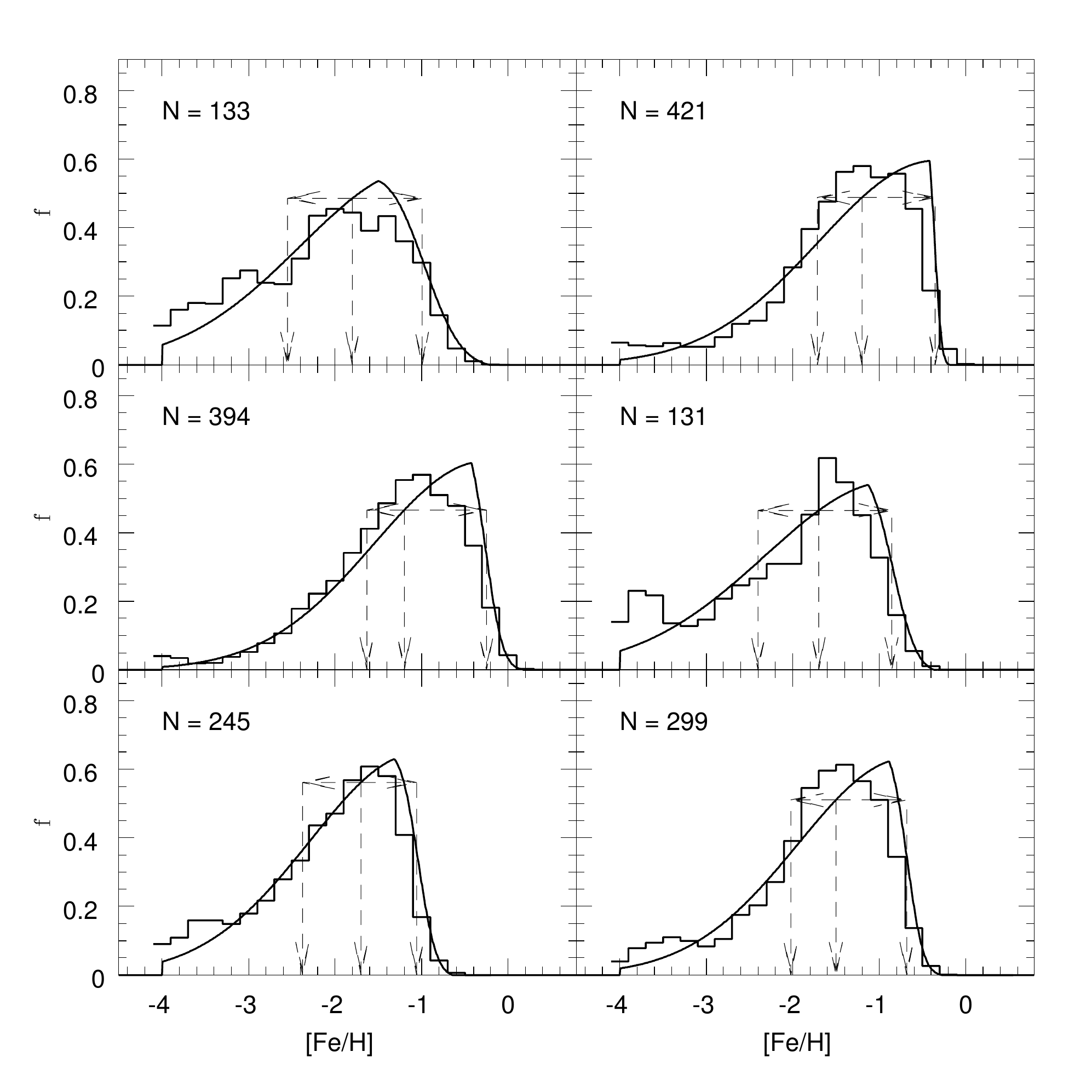}
\caption{Halo stellar particle MDFs for the M$_{\rm tot}$~=~5$\times$10$^{11}$~M$_\odot$ semi--cosmological simulations in Renda~et~al.~(2005b). The 68\%~Confidence~Level range and the number of stellar particles each MDF refers to are also shown.}
\label{appB:sim5e11:fig1}
\end{center}
\end{figure}

\begin{figure}
\begin{center}
\includegraphics[width=1.0\textwidth]{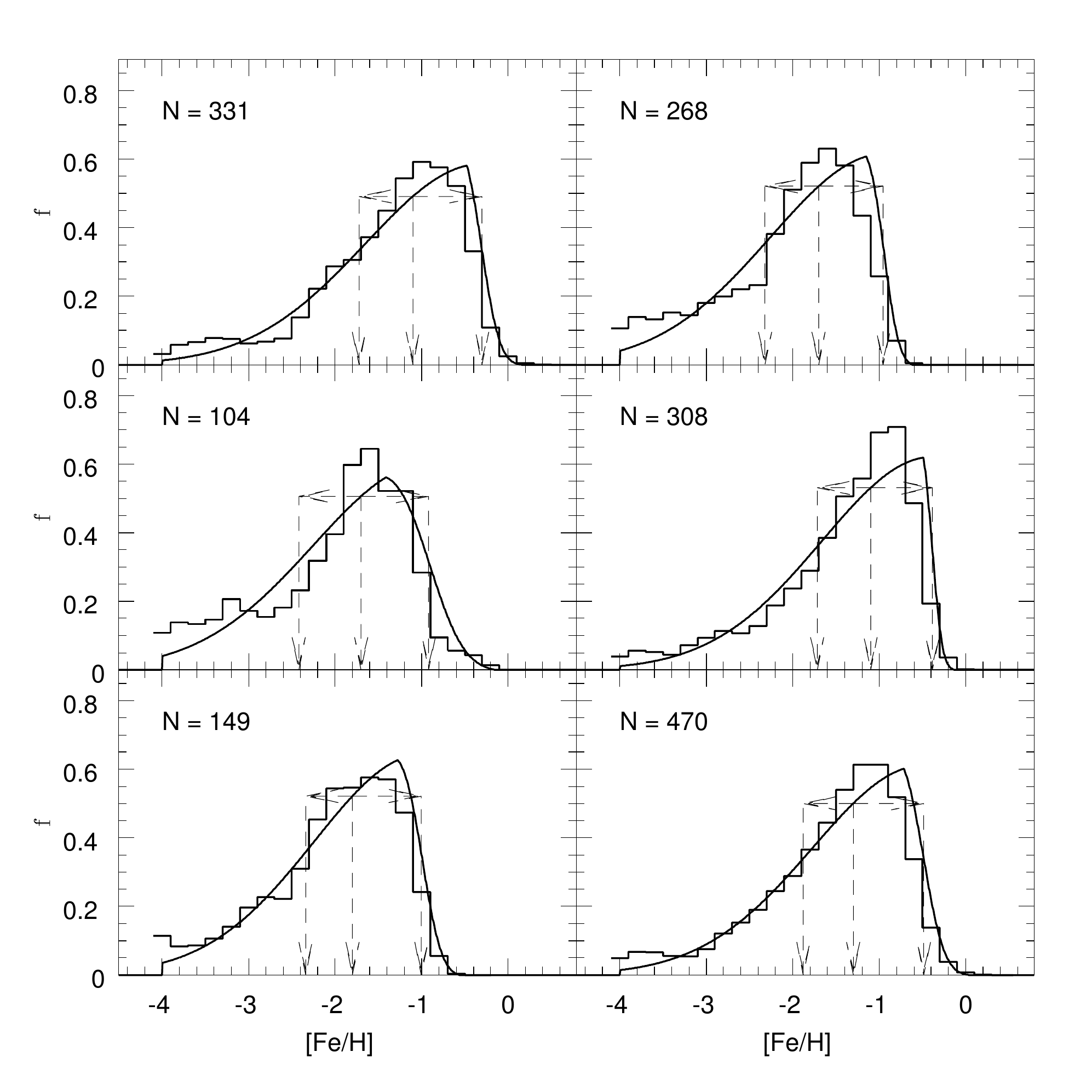}
\caption{(continued) Halo stellar particle MDFs for the M$_{\rm tot}$~=~5$\times$10$^{11}$~M$_\odot$ semi--cosmological simulations in Renda~et~al.~(2005b). The 68\%~Confidence~Level range and the number of stellar particles each MDF refers to are also shown.}
\label{appB:sim5e11:fig2}
\end{center}
\end{figure}

\begin{figure}
\begin{center}
\includegraphics[width=1.0\textwidth]{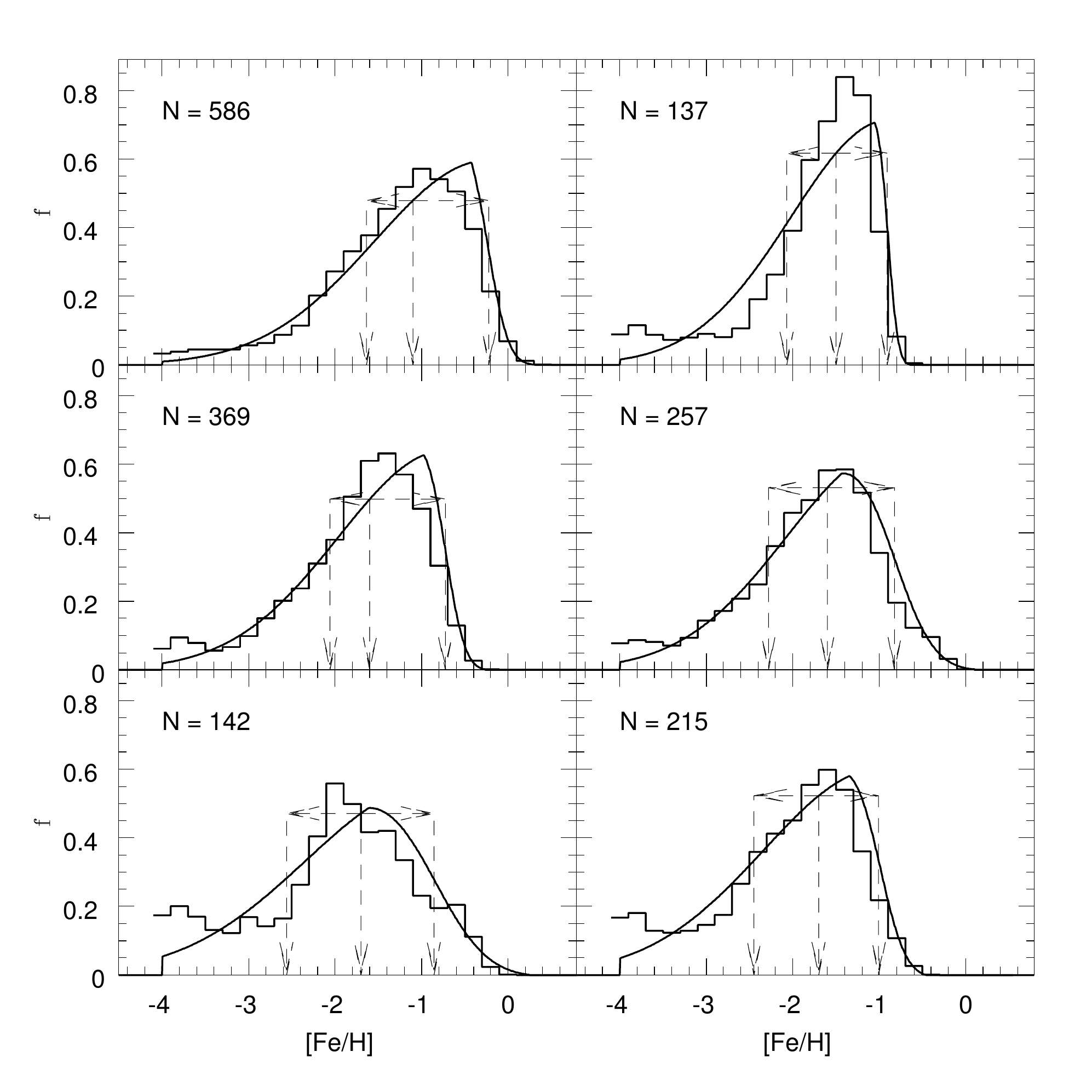}
\caption{(continued) Halo stellar particle MDFs for the M$_{\rm tot}$~=~5$\times$10$^{11}$~M$_\odot$ semi--cosmological simulations in Renda~et~al.~(2005b). The 68\%~Confidence~Level range and the number of stellar particles each MDF refers to are also shown.}
\label{appB:sim5e11:fig3}
\end{center}
\end{figure}

\begin{figure}
\begin{center}
\includegraphics[width=1.0\textwidth]{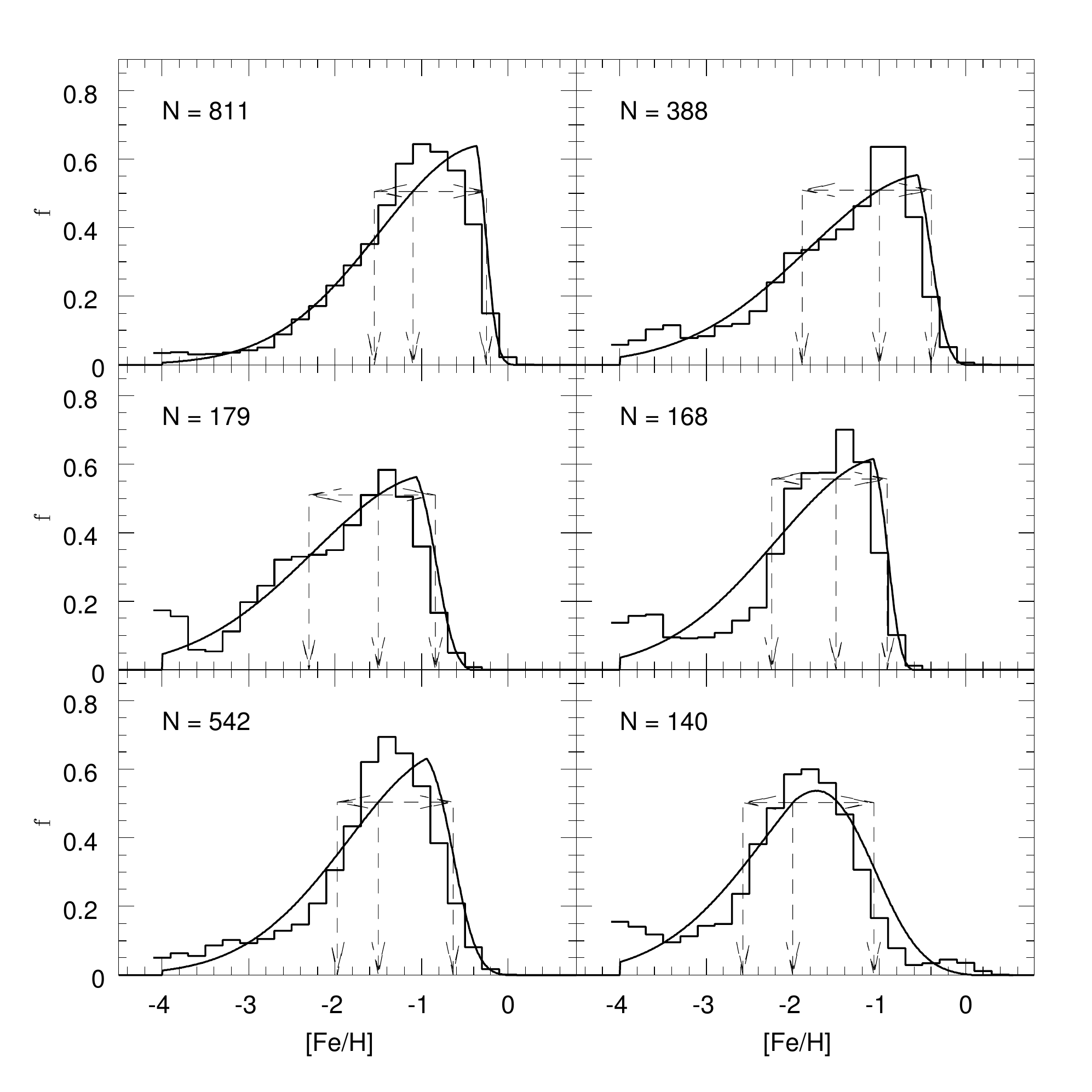}
\caption{(continued) Halo stellar particle MDFs for the M$_{\rm tot}$~=~5$\times$10$^{11}$~M$_\odot$ semi--cosmological simulations in Renda~et~al.~(2005b). The 68\%~Confidence~Level range and the number of stellar particles each MDF refers to are also shown.}
\label{appB:sim5e11:fig4}
\end{center}
\end{figure}

\begin{figure}
\begin{center}
\includegraphics[width=1.0\textwidth]{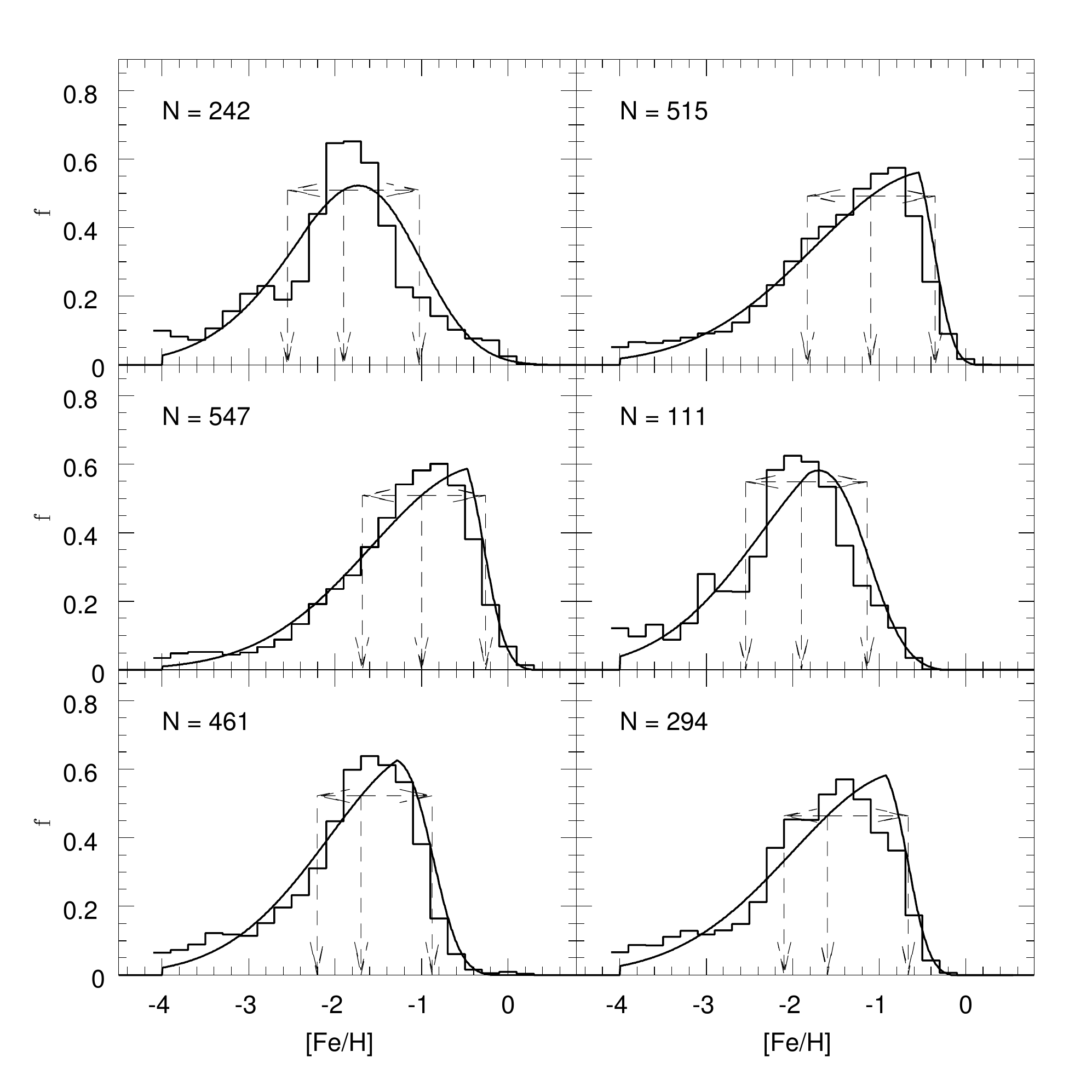}
\caption{Halo stellar particle MDFs for the M$_{\rm tot}$~=~10$^{12}$~M$_\odot$ semi--cosmological simulations in Renda~et~al.~(2005b). The 68\%~Confidence~Level range and the number of stellar particles each MDF refers to are also shown.}
\label{appB:sim1e12:fig1}
\end{center}
\end{figure}

\begin{figure}
\begin{center}
\includegraphics[width=1.0\textwidth]{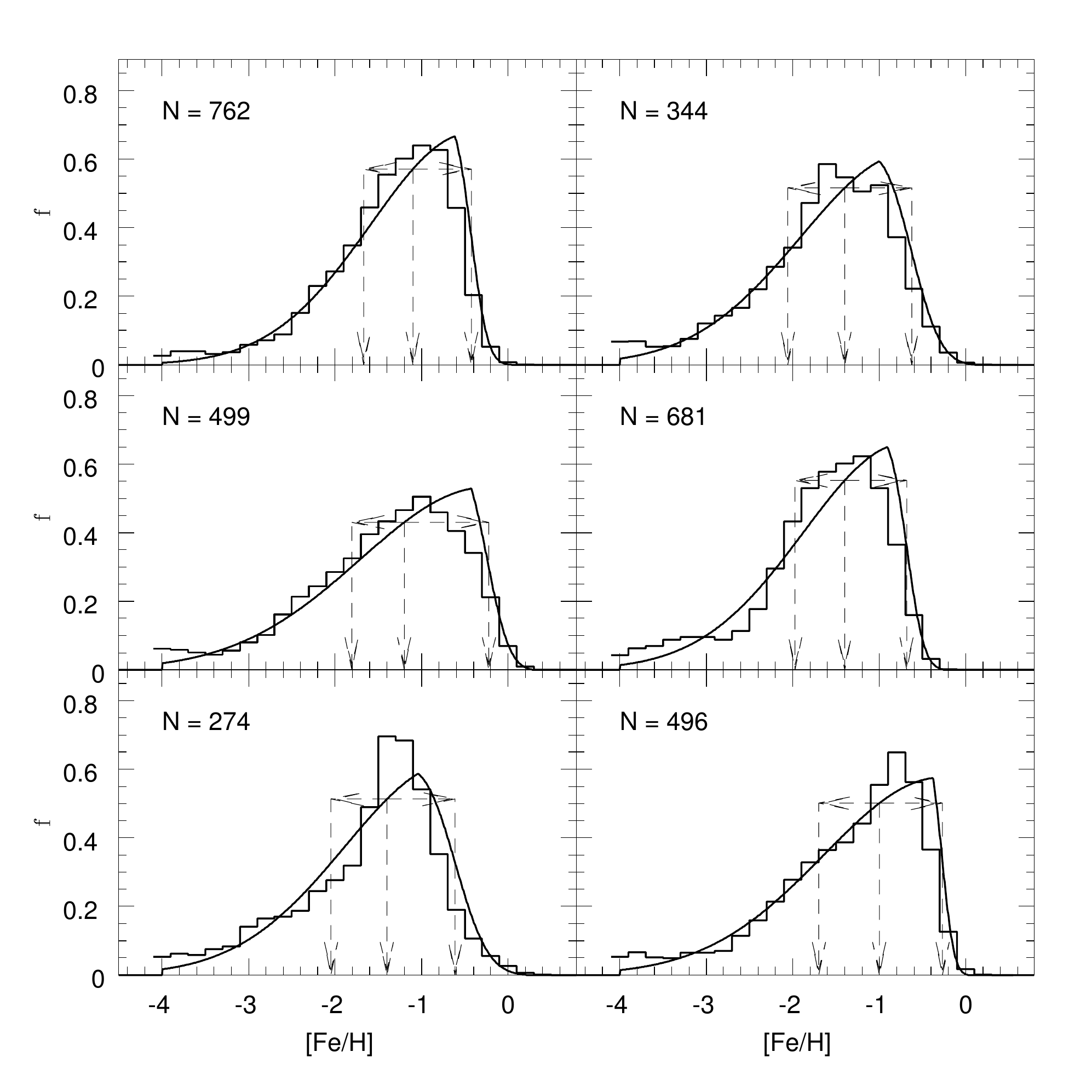}
\caption{(continued) Halo stellar particle MDFs for the M$_{\rm tot}$~=~10$^{12}$~M$_\odot$ semi--cosmological simulations in Renda~et~al.~(2005b). The 68\%~Confidence~Level range and the number of stellar particles each MDF refers to are also shown.}
\label{appB:sim1e12:fig2}
\end{center}
\end{figure}

\begin{figure}
\begin{center}
\includegraphics[width=1.0\textwidth]{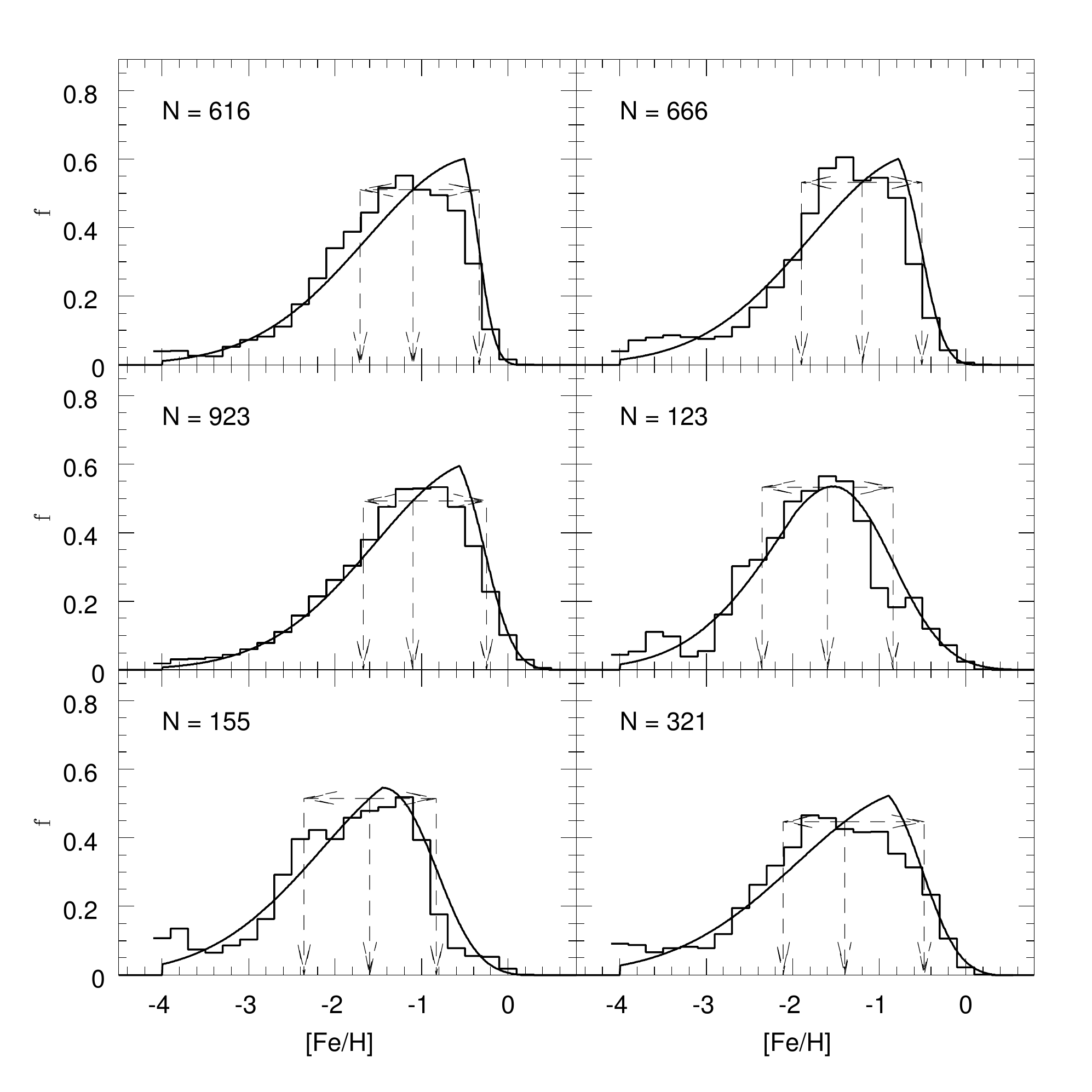}
\caption{(continued) Halo stellar particle MDFs for the M$_{\rm tot}$~=~10$^{12}$~M$_\odot$ semi--cosmological simulations in Renda~et~al.~(2005b). The 68\%~Confidence~Level range and the number of stellar particles each MDF refers to are also shown.}
\label{appB:sim1e12:fig3}
\end{center}
\end{figure}

\begin{figure}
\begin{center}
\includegraphics[width=1.0\textwidth]{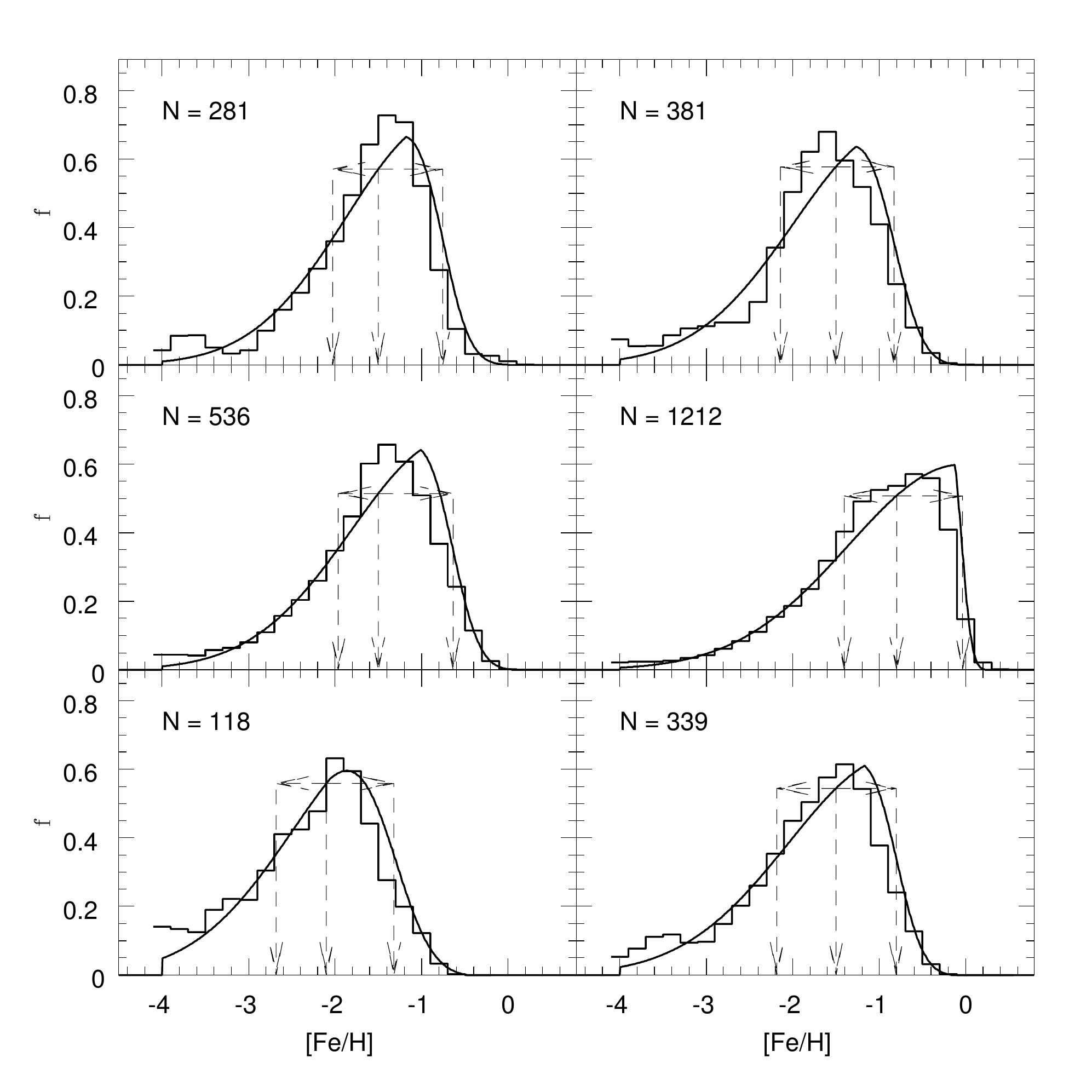}
\caption{(continued) Halo stellar particle MDFs for the M$_{\rm tot}$~=~10$^{12}$~M$_\odot$ semi--cosmological simulations in Renda~et~al.~(2005b). The 68\%~Confidence~Level range and the number of stellar particles each MDF refers to are also shown.}
\label{appB:sim1e12:fig4}
\end{center}
\end{figure}

\begin{figure}
\begin{center}
\includegraphics[width=1.0\textwidth]{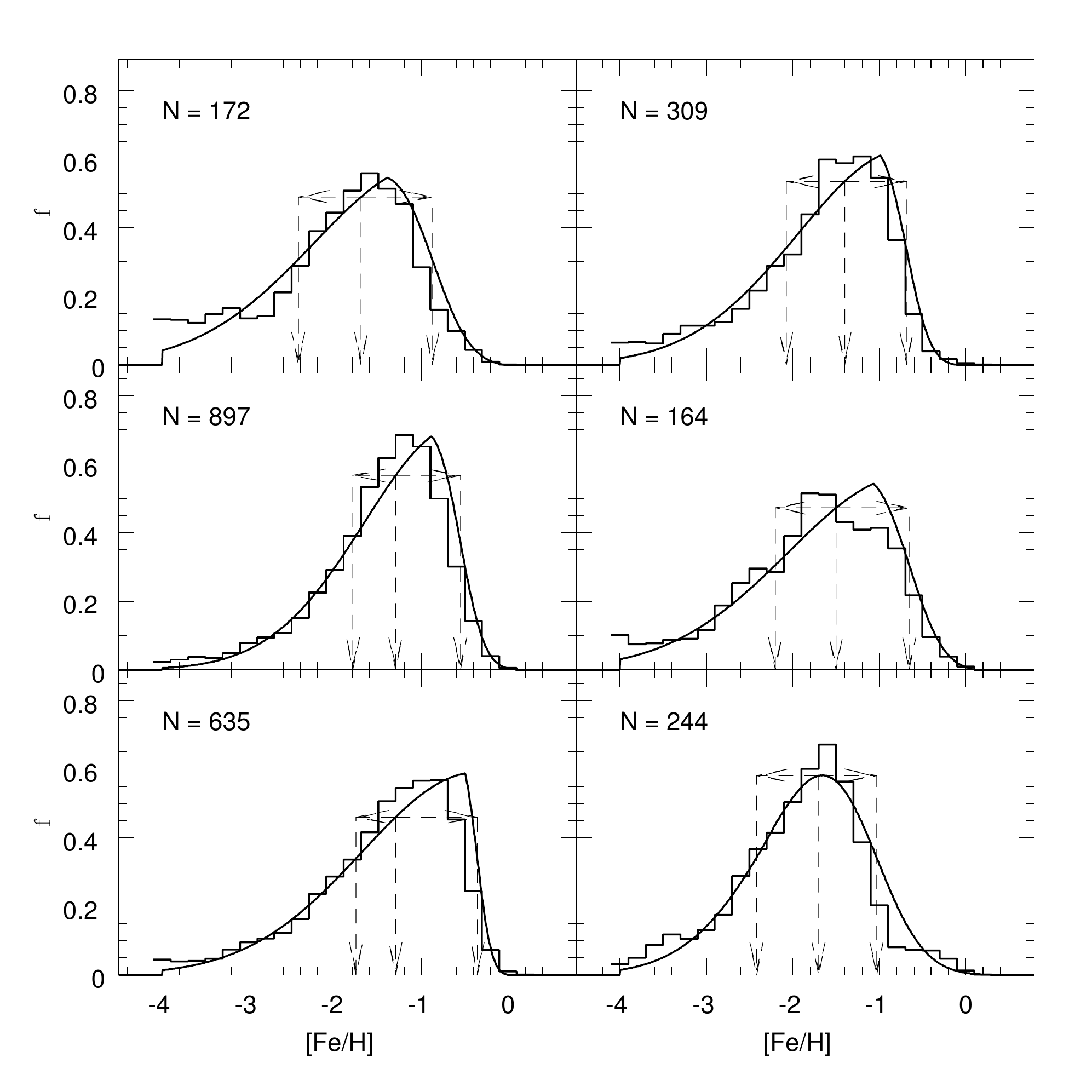}
\caption{(continued) Halo stellar particle MDFs for the M$_{\rm tot}$~=~10$^{12}$~M$_\odot$ semi--cosmological simulations in Renda~et~al.~(2005b). The 68\%~Confidence~Level range and the number of stellar particles each MDF refers to are also shown.}
\label{appB:sim1e12:fig5}
\end{center}
\end{figure}

\begin{figure}
\begin{center}
\includegraphics[width=1.0\textwidth]{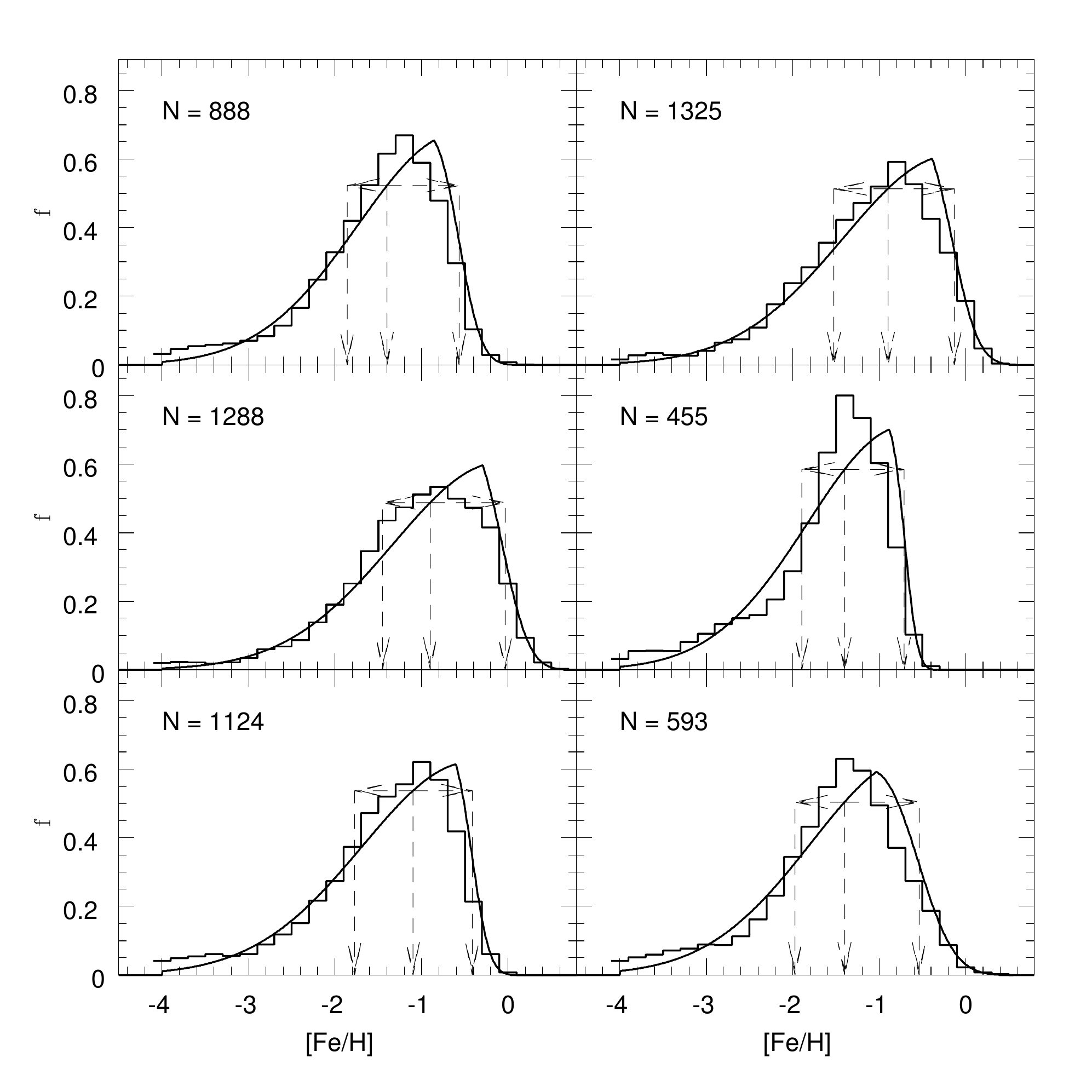}
\caption{Halo stellar particle MDFs for the M$_{\rm tot}$~=~5$\times$10$^{12}$~M$_\odot$ semi--cosmological simulations in Renda~et~al.~(2005b). The 68\%~Confidence~Level range and the number of stellar particles each MDF refers to are also shown.}
\label{appB:sim5e12:fig1}
\end{center}
\end{figure}

\begin{figure}
\begin{center}
\includegraphics[width=1.0\textwidth]{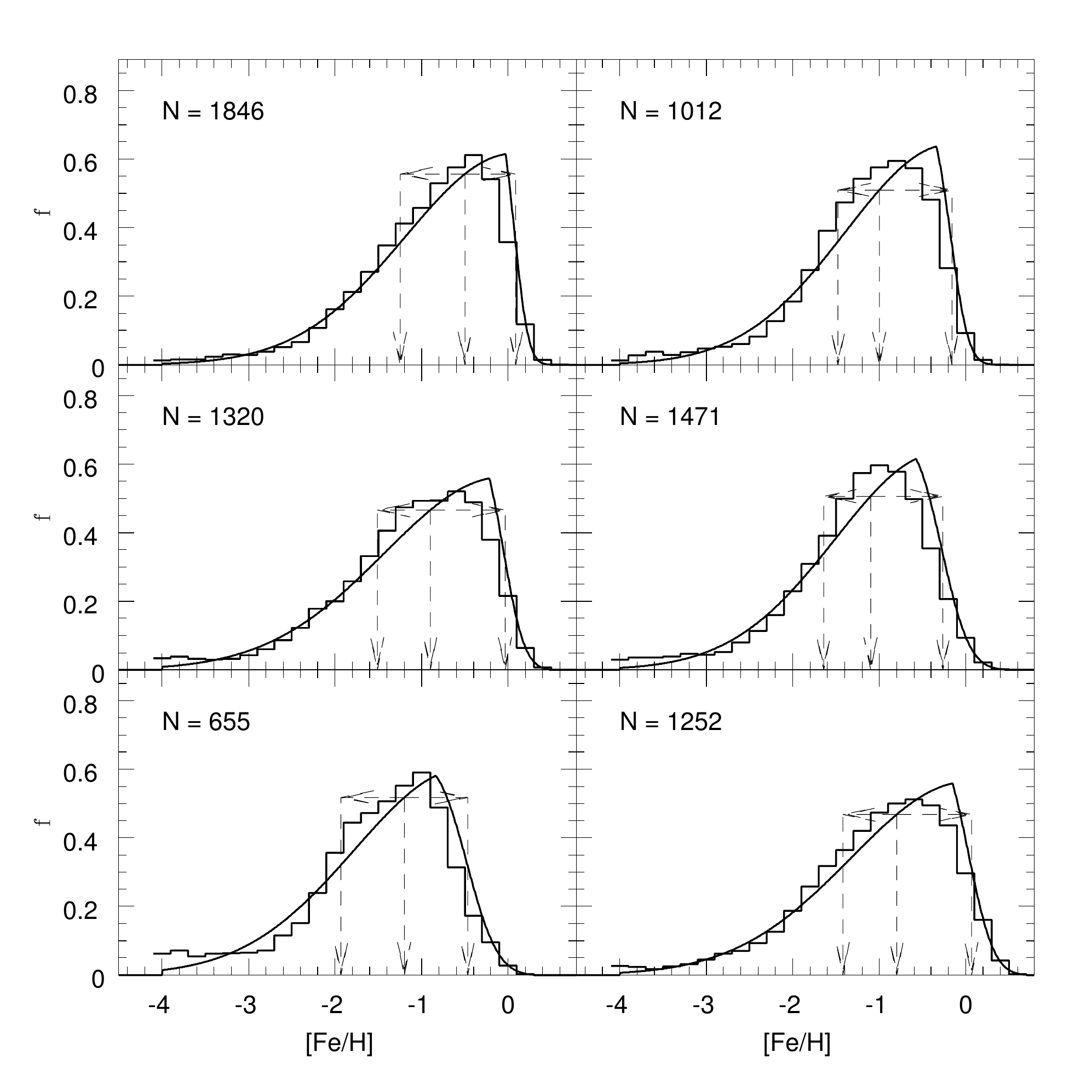}
\caption{(continued) Halo stellar particle MDFs for the M$_{\rm tot}$~=~5$\times$10$^{12}$~M$_\odot$ semi--cosmological simulations in Renda~et~al.~(2005b). The 68\%~Confidence~Level range and the number of stellar particles each MDF refers to are also shown.}
\label{appB:sim5e12:fig2}
\end{center}
\end{figure}

\begin{figure}
\begin{center}
\includegraphics[width=1.0\textwidth]{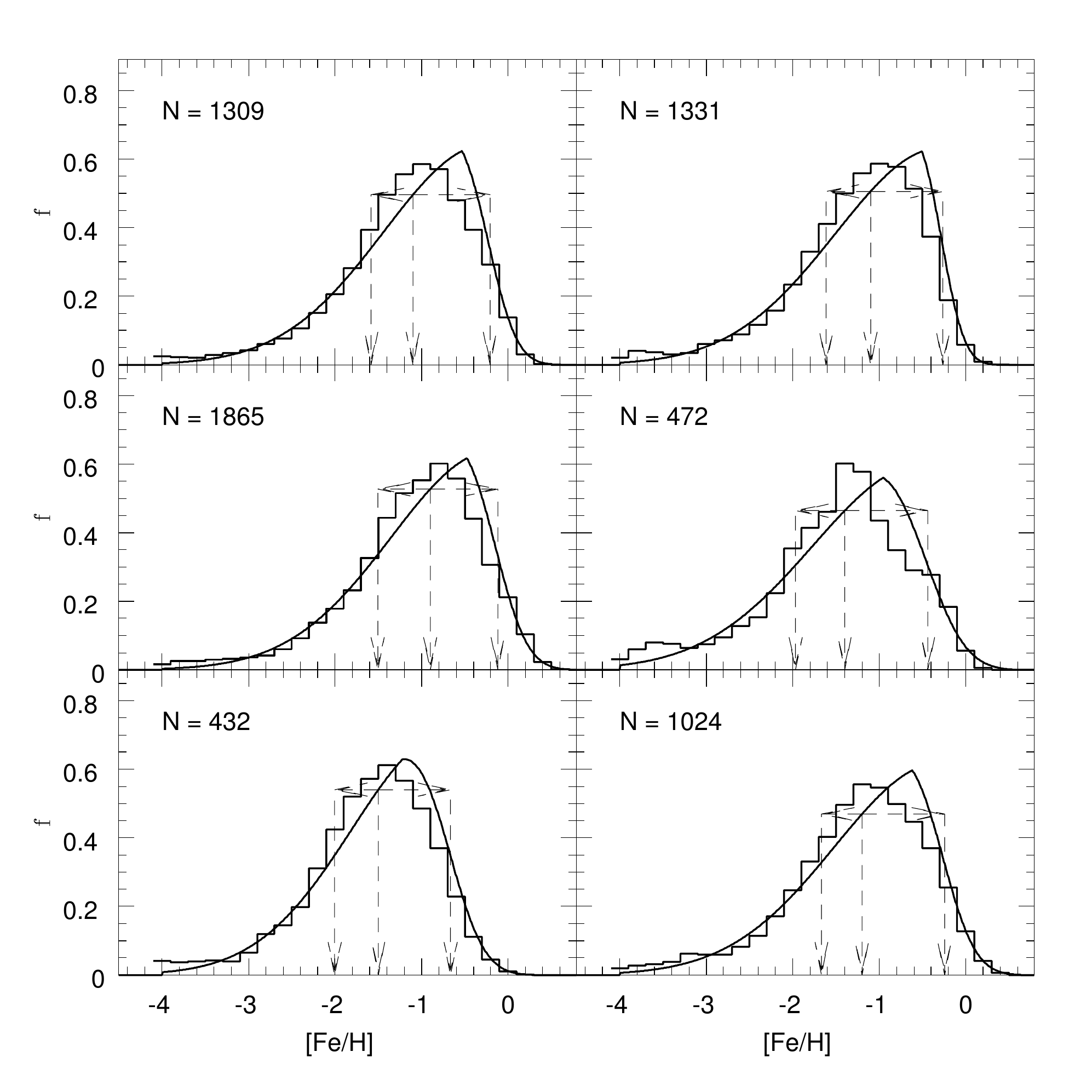}
\caption{(continued) Halo stellar particle MDFs for the M$_{\rm tot}$~=~5$\times$10$^{12}$~M$_\odot$ semi--cosmological simulations in Renda~et~al.~(2005b). The 68\%~Confidence~Level range and the number of stellar particles each MDF refers to are also shown.}
\label{appB:sim5e12:fig3}
\end{center}
\end{figure}

\begin{figure}
\begin{center}
\includegraphics[width=1.0\textwidth]{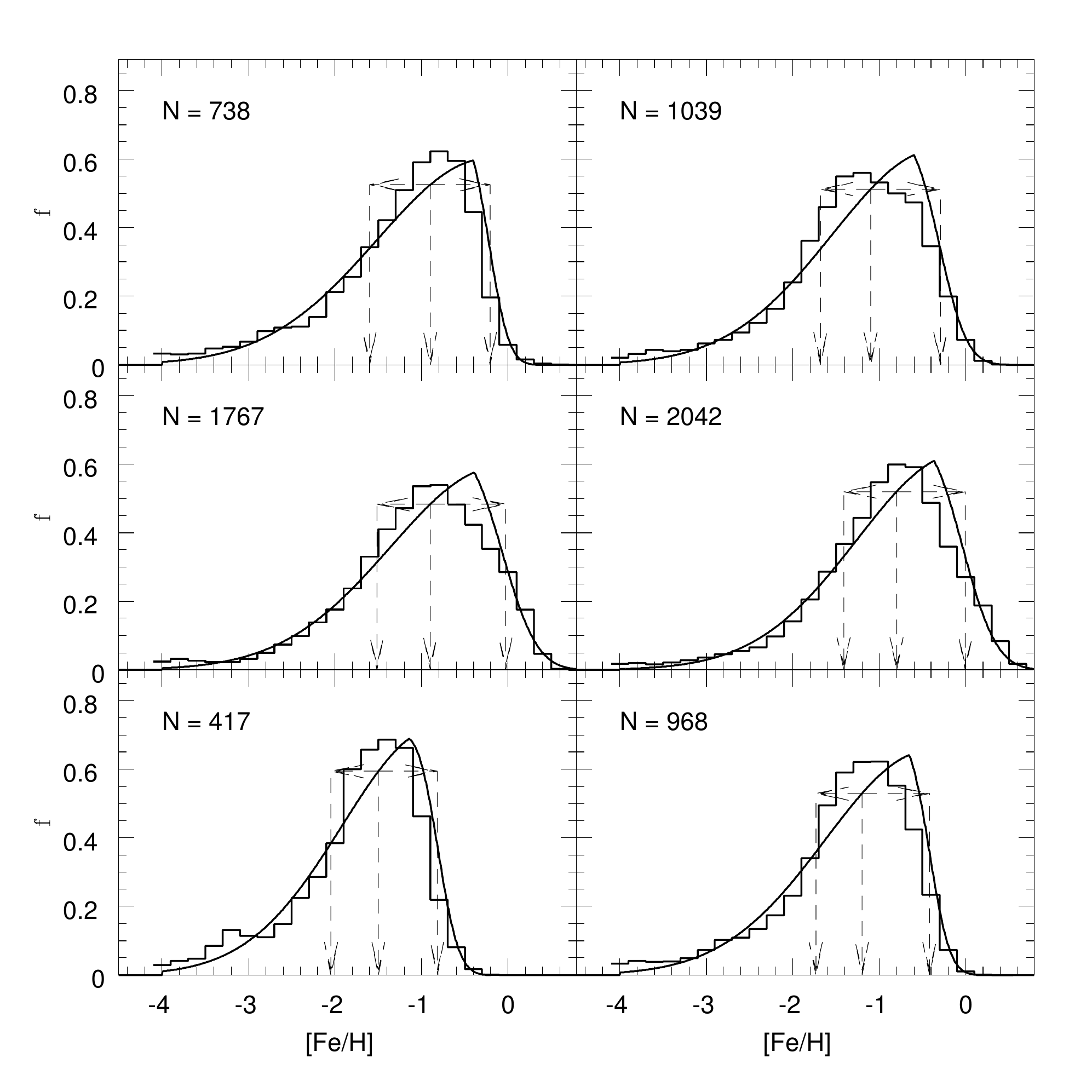}
\caption{(continued) Halo stellar particle MDFs for the M$_{\rm tot}$~=~5$\times$10$^{12}$~M$_\odot$ semi--cosmological simulations in Renda~et~al.~(2005b). The 68\%~Confidence~Level range and the number of stellar particles each MDF refers to are also shown.}
\label{appB:sim5e12:fig4}
\end{center}
\end{figure}

\begin{figure}
\begin{center}
\includegraphics[width=1.0\textwidth]{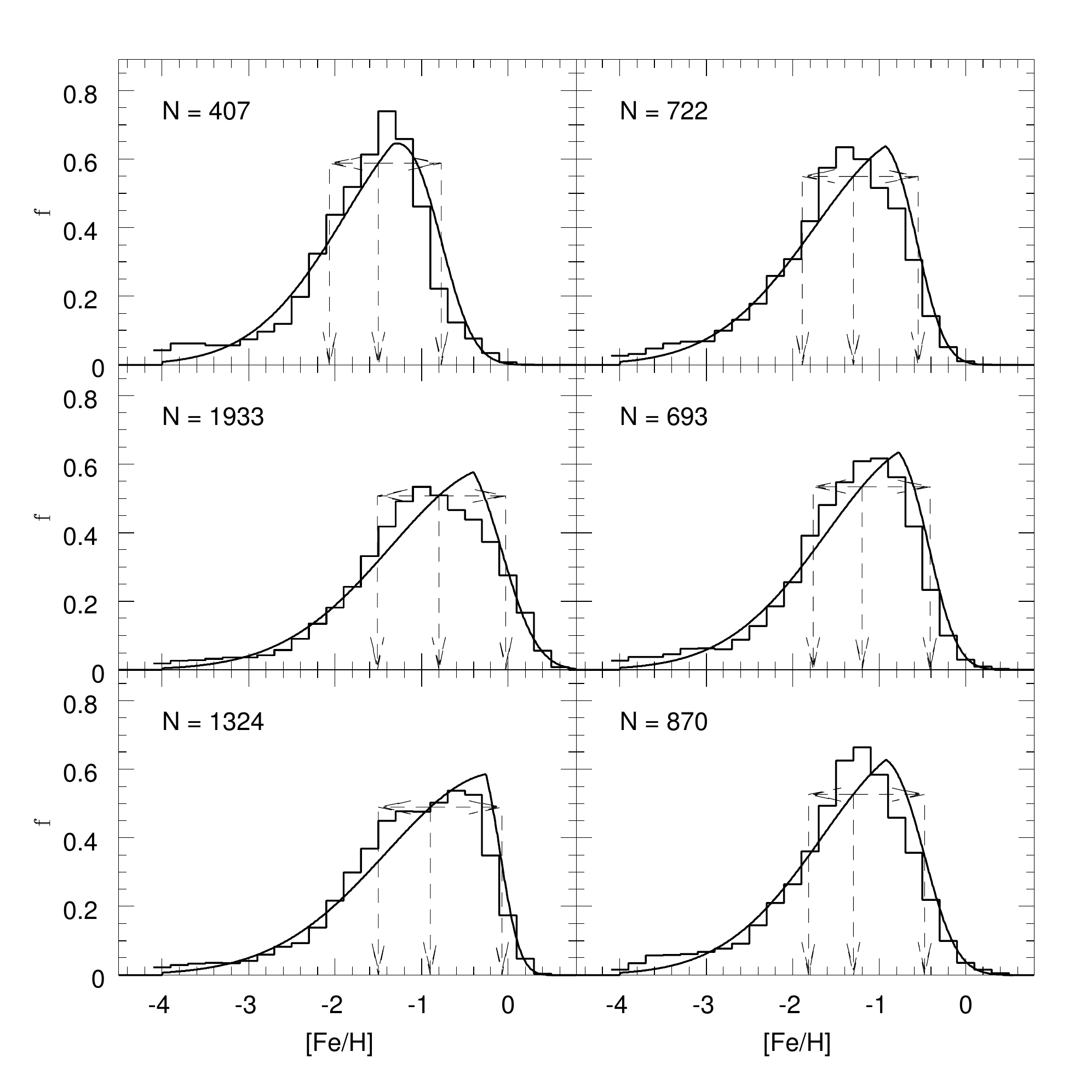}
\caption{(continued) Halo stellar particle MDFs for the M$_{\rm tot}$~=~5$\times$10$^{12}$~M$_\odot$ semi--cosmological simulations in Renda~et~al.~(2005b). The 68\%~Confidence~Level range and the number of stellar particles each MDF refers to are also shown.}
\label{appB:sim5e12:fig5}
\end{center}
\end{figure}




\newpage 

\begin{center}
\chapter{Stellar Halo $[$O/Fe$]$ Distributions}
\label{app:appendixC}
\end{center}

\begin{figure}
\begin{center}
\includegraphics[width=1.0\textwidth]{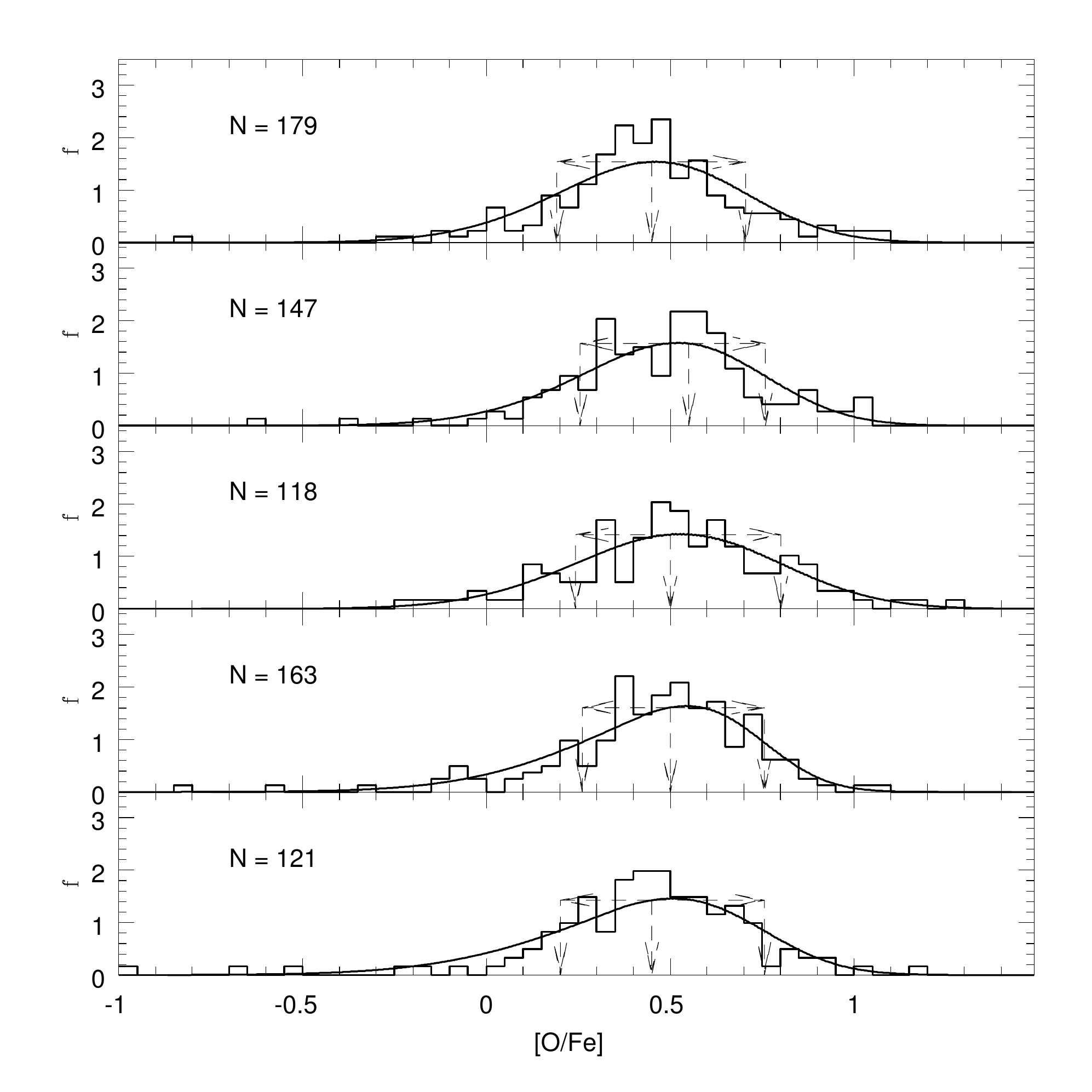}
\caption{Halo stellar particle $[$O/Fe$]$ distribution for the M$_{\rm tot}$~=~10$^{11}$~M$_\odot$ semi--cosmological simulations in Renda~et~al.~(2005b). The 68\%~Confidence~Level range and the number of stellar particles each distribution refers to are also shown.}
\label{appC:sim1e11:fig}
\end{center}
\end{figure}

\begin{figure}
\begin{center}
\includegraphics[width=1.0\textwidth]{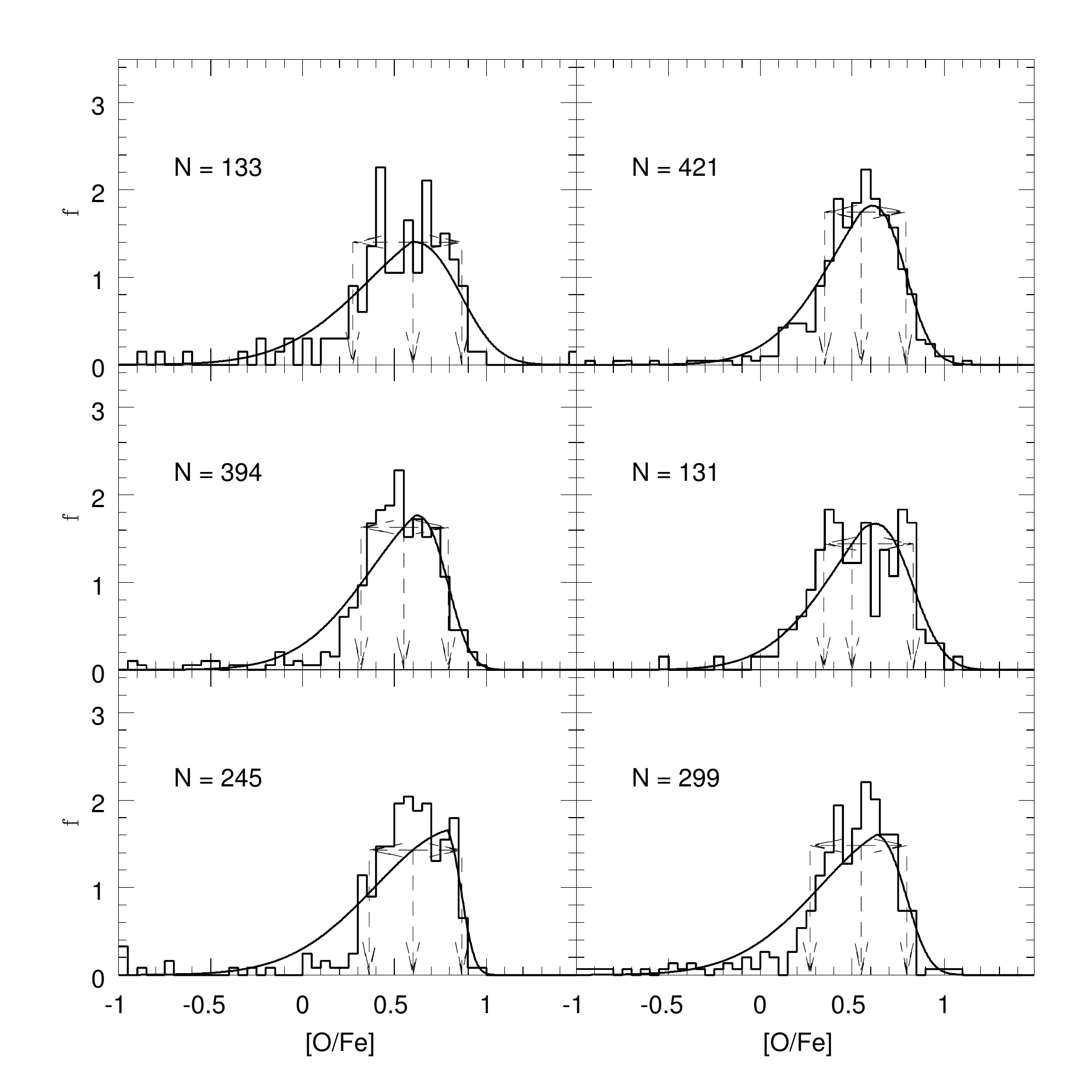}
\caption{Halo stellar particle $[$O/Fe$]$ distribution for the M$_{\rm tot}$~=~5$\times$10$^{11}$~M$_\odot$ semi--cosmological simulations in Renda~et~al.~(2005b). The 68\%~Confidence~Level range and the number of stellar particles each distribution refers to are also shown.}
\label{appC:sim5e11:fig1}
\end{center}
\end{figure}

\begin{figure}
\begin{center}
\includegraphics[width=1.0\textwidth]{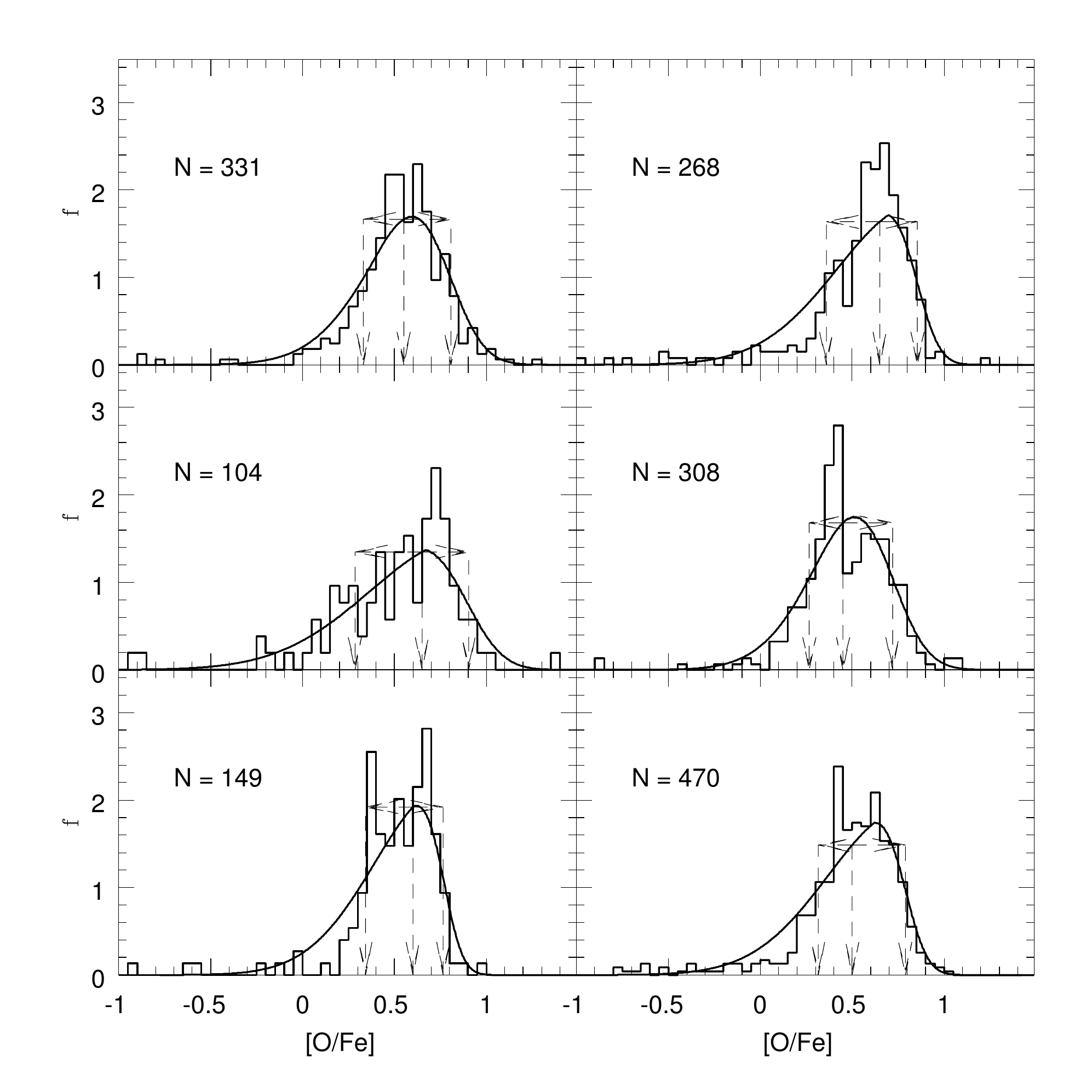}
\caption{(continued) Halo stellar particle $[$O/Fe$]$ distribution for the M$_{\rm tot}$~=~5$\times$10$^{11}$~M$_\odot$ semi--cosmological simulations in Renda~et~al.~(2005b). The 68\%~Confidence~Level range and the number of stellar particles each distribution refers to are also shown.}
\label{appC:sim5e11:fig2}
\end{center}
\end{figure}

\begin{figure}
\begin{center}
\includegraphics[width=1.0\textwidth]{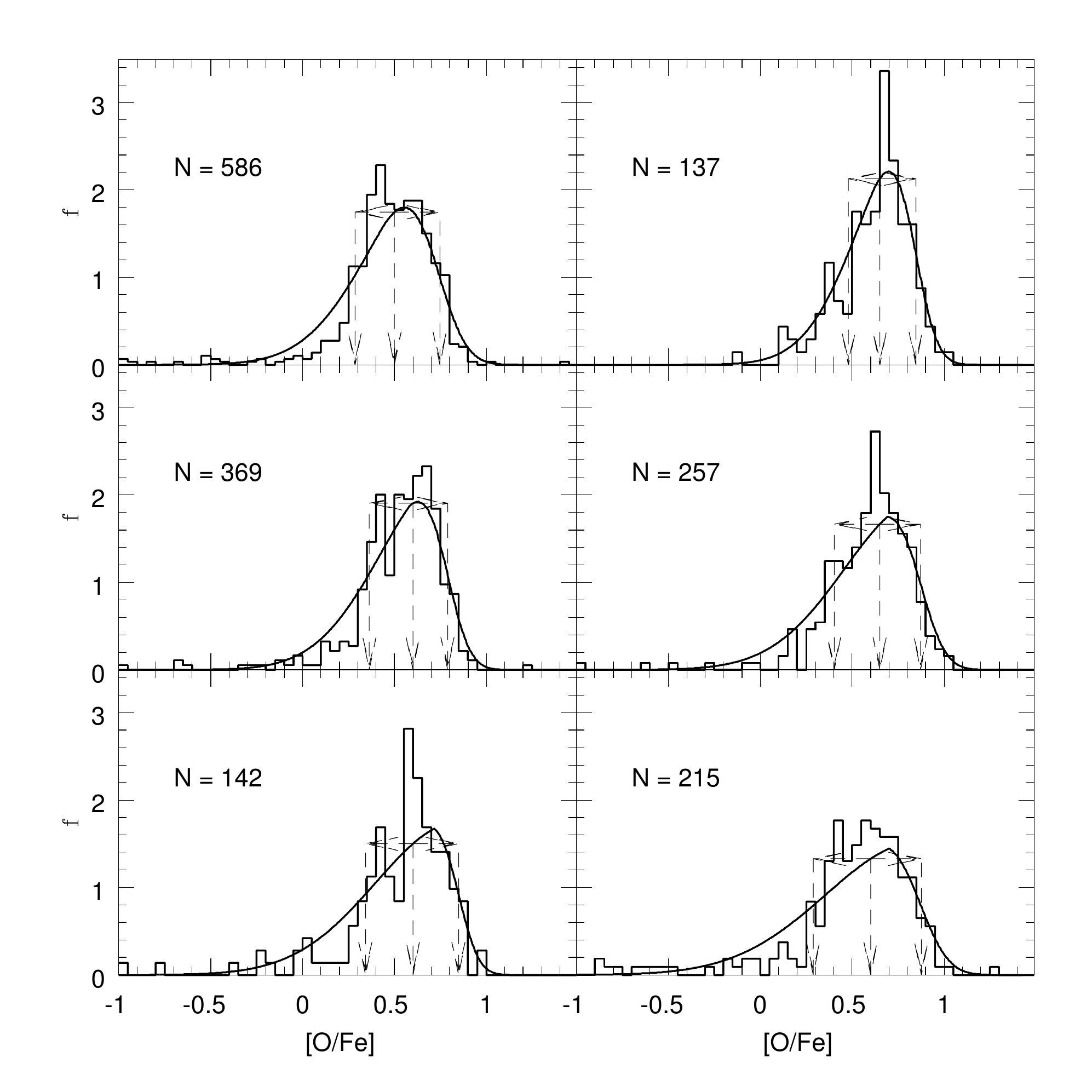}
\caption{(continued) Halo stellar particle $[$O/Fe$]$ distribution for the M$_{\rm tot}$~=~5$\times$10$^{11}$~M$_\odot$ semi--cosmological simulations in Renda~et~al.~(2005b). The 68\%~Confidence~Level range and the number of stellar particles each distribution refers to are also shown.}
\label{appC:sim5e11:fig3}
\end{center}
\end{figure}

\begin{figure}
\begin{center}
\includegraphics[width=1.0\textwidth]{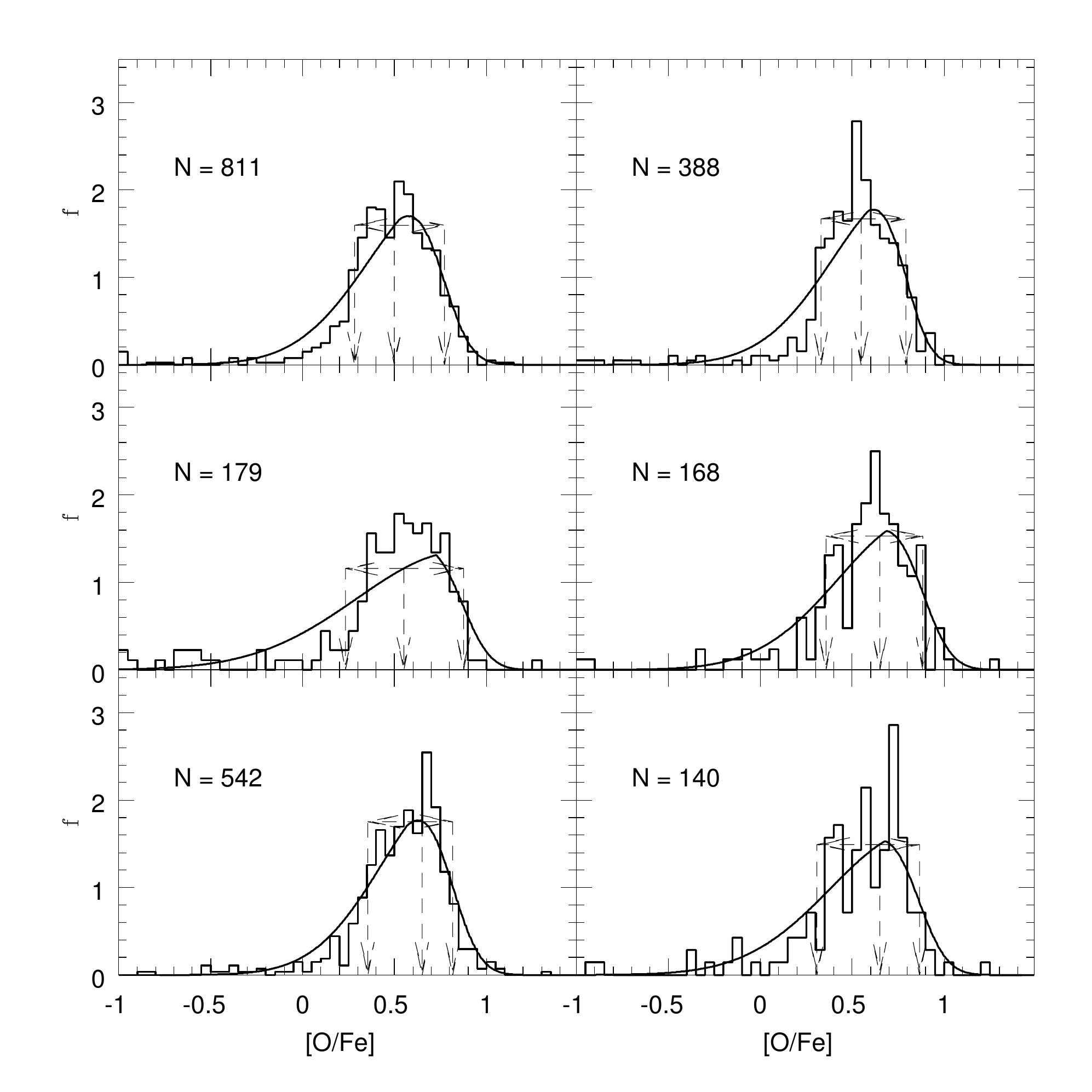}
\caption{(continued) Halo stellar particle $[$O/Fe$]$ distribution for the M$_{\rm tot}$~=~5$\times$10$^{11}$~M$_\odot$ semi--cosmological simulations in Renda~et~al.~(2005b). The 68\%~Confidence~Level range and the number of stellar particles each distribution refers to are also shown.}
\label{appC:sim5e11:fig4}
\end{center}
\end{figure}

\begin{figure}
\begin{center}
\includegraphics[width=1.0\textwidth]{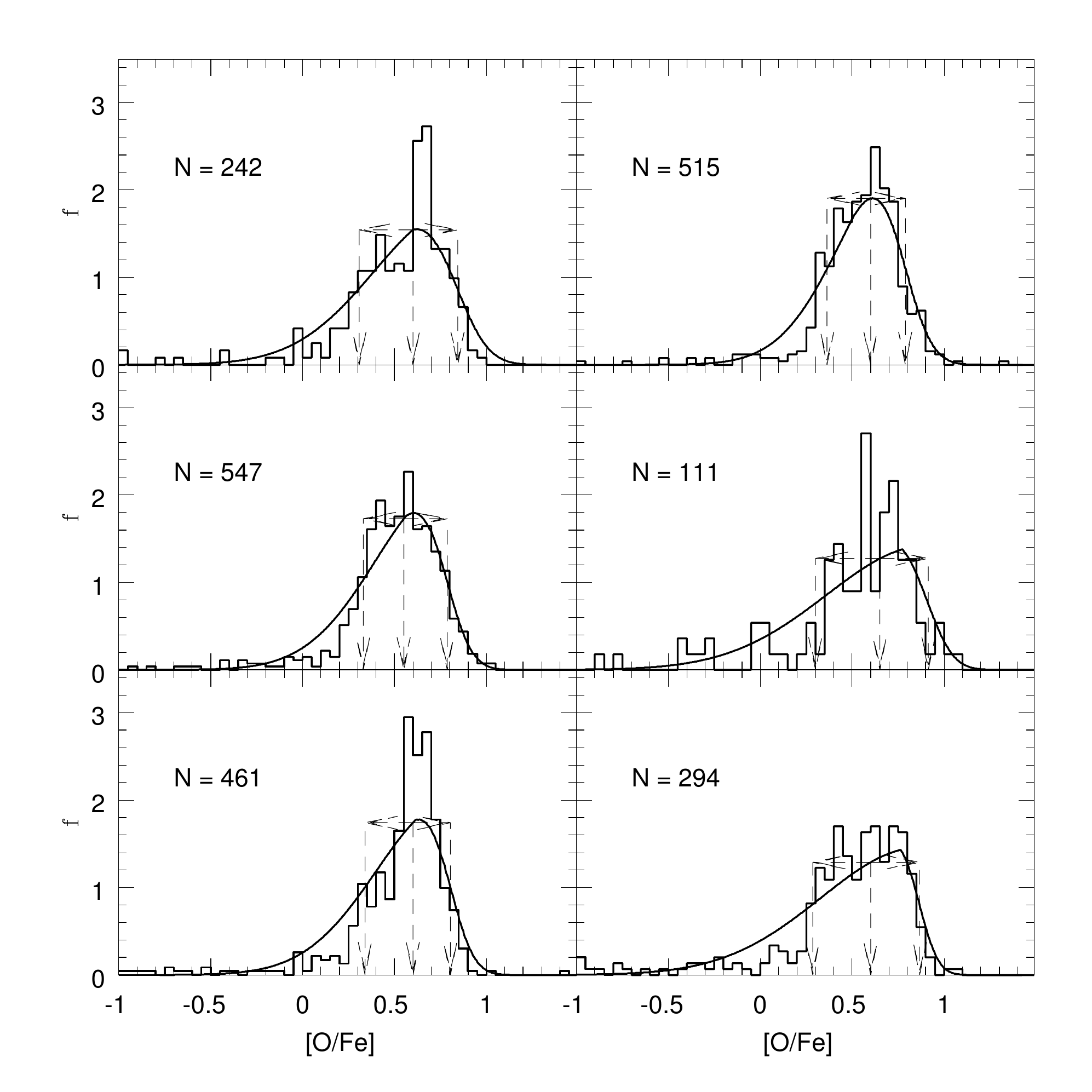}
\caption{Halo stellar particle $[$O/Fe$]$ distribution for the M$_{\rm tot}$~=~1$\times$10$^{12}$~M$_\odot$ semi--cosmological simulations in Renda~et~al.~(2005b). The 68\%~Confidence~Level range and the number of stellar particles each distribution refers to are also shown.}
\label{appC:sim1e12:fig1}
\end{center}
\end{figure}

\begin{figure}
\begin{center}
\includegraphics[width=1.0\textwidth]{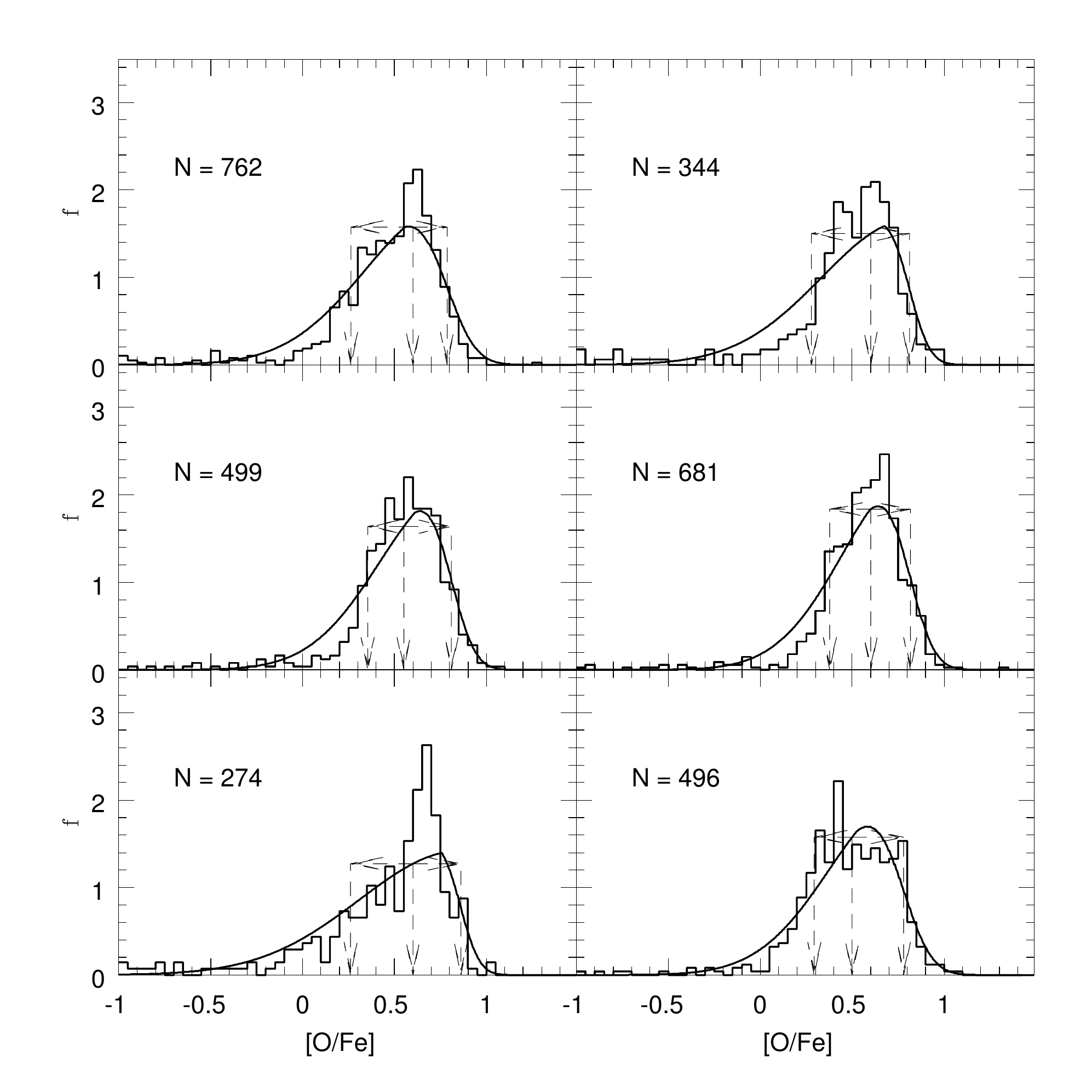}
\caption{(continued) Halo stellar particle $[$O/Fe$]$ distribution for the M$_{\rm tot}$~=~10$^{12}$~M$_\odot$ semi--cosmological simulations in Renda~et~al.~(2005b). The 68\%~Confidence~Level range and the number of stellar particles each distribution refers to are also shown.}
\label{appC:sim1e12:fig2}
\end{center}
\end{figure}

\begin{figure}
\begin{center}
\includegraphics[width=1.0\textwidth]{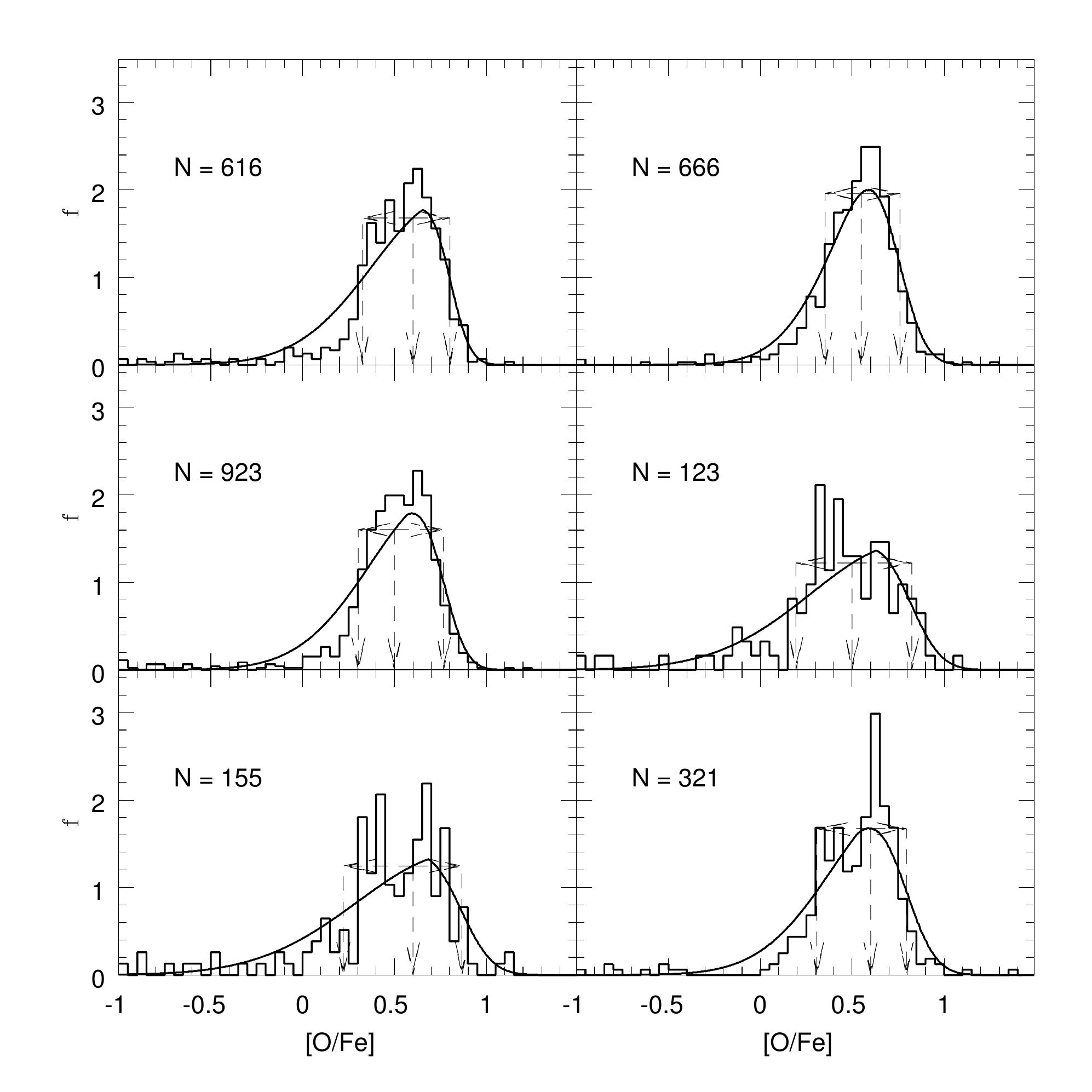}
\caption{(continued) Halo stellar particle $[$O/Fe$]$ distribution for the M$_{\rm tot}$~=~10$^{12}$~M$_\odot$ semi--cosmological simulations in Renda~et~al.~(2005b). The 68\%~Confidence~Level range and the number of stellar particles each distribution refers to are also shown.}
\label{appC:sim1e12:fig3}
\end{center}
\end{figure}

\begin{figure}
\begin{center}
\includegraphics[width=1.0\textwidth]{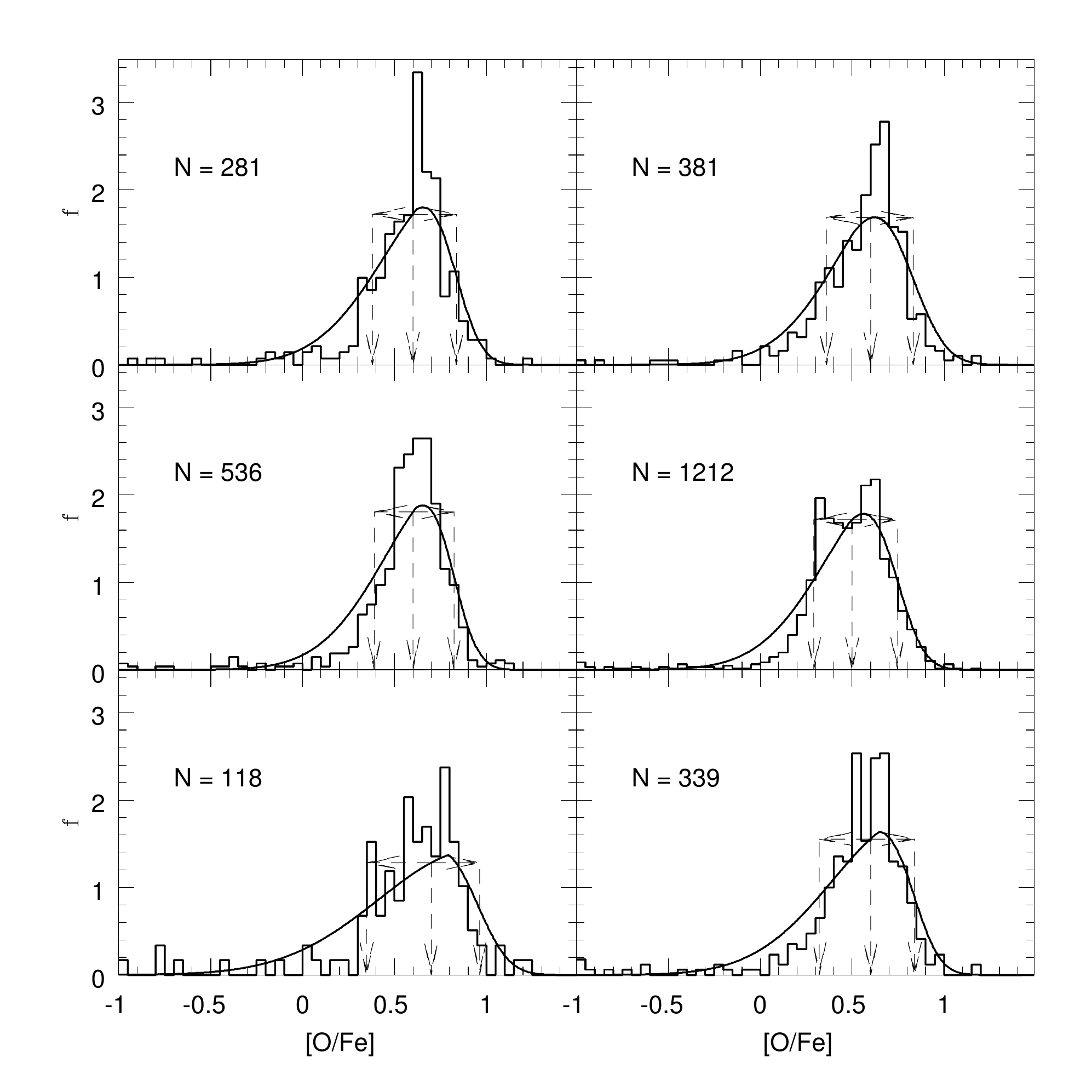}
\caption{(continued) Halo stellar particle $[$O/Fe$]$ distribution for the M$_{\rm tot}$~=~10$^{12}$~M$_\odot$ semi--cosmological simulations in Renda~et~al.~(2005b). The 68\%~Confidence~Level range and the number of stellar particles each distribution refers to are also shown.}
\label{appC:sim1e12:fig4}
\end{center}
\end{figure}

\begin{figure}
\begin{center}
\includegraphics[width=1.0\textwidth]{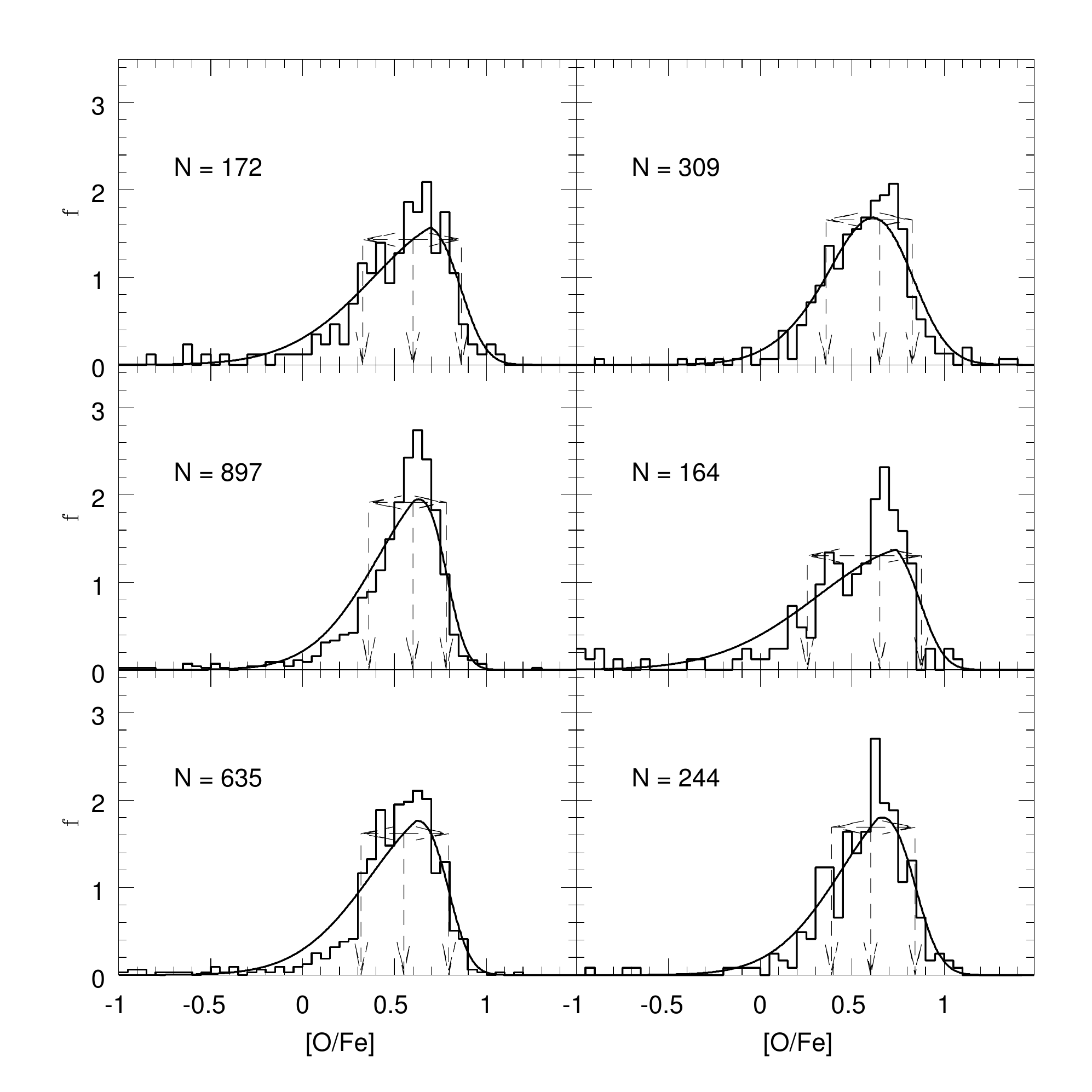}
\caption{(continued) Halo stellar particle $[$O/Fe$]$ distribution for the M$_{\rm tot}$~=~10$^{12}$~M$_\odot$ semi--cosmological simulations in Renda~et~al.~(2005b). The 68\%~Confidence~Level range and the number of stellar particles each distribution refers to are also shown.}
\label{appC:sim1e12:fig5}
\end{center}
\end{figure}

\begin{figure}
\begin{center}
\includegraphics[width=1.0\textwidth]{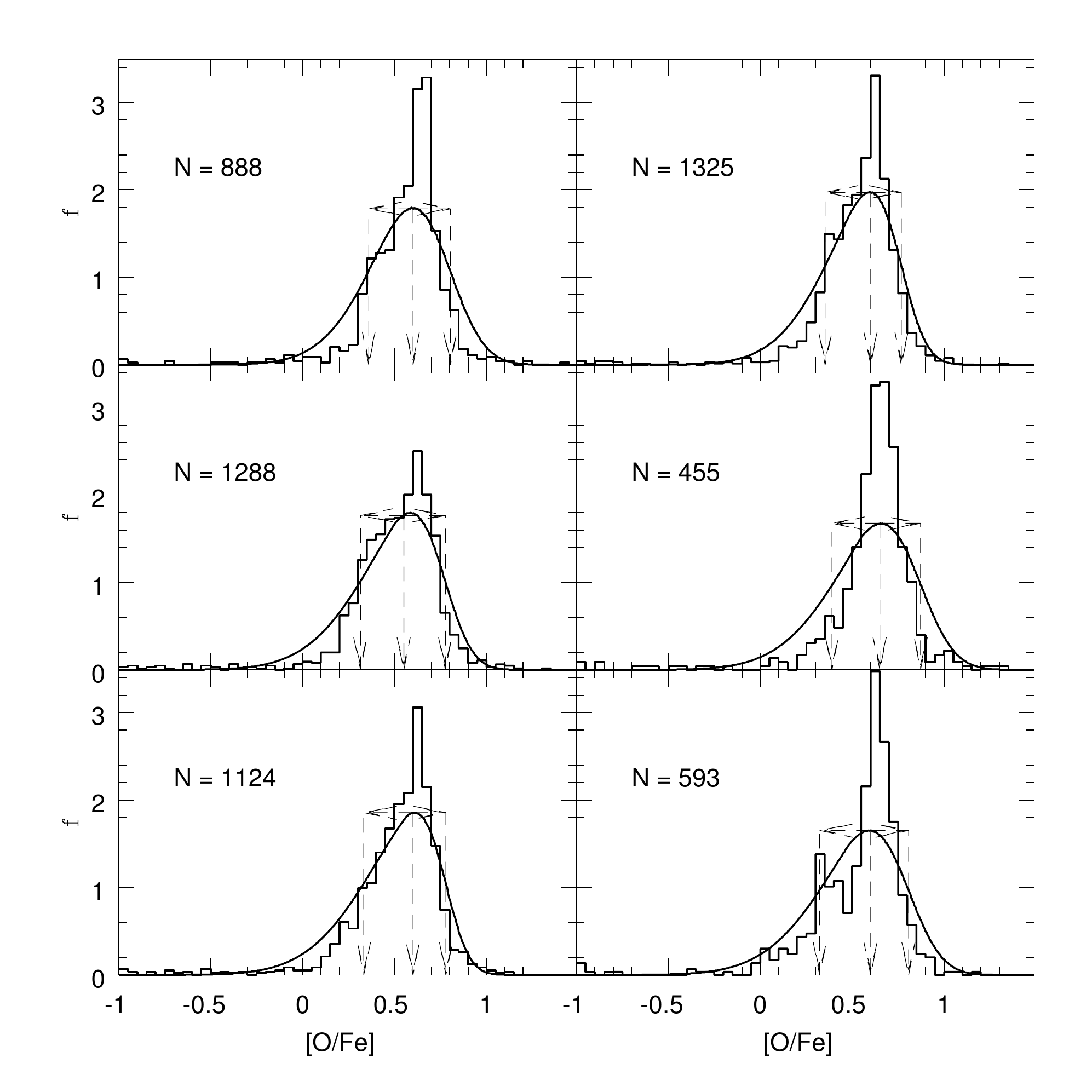}
\caption{Halo stellar particle $[$O/Fe$]$ distribution for the M$_{\rm tot}$~=~5$\times$10$^{12}$~M$_\odot$ semi--cosmological simulations in Renda~et~al.~(2005b). The 68\%~Confidence~Level range and the number of stellar particles each distribution refers to are also shown.}
\label{appC:sim5e12:fig1}
\end{center}
\end{figure}

\begin{figure}
\begin{center}
\includegraphics[width=1.0\textwidth]{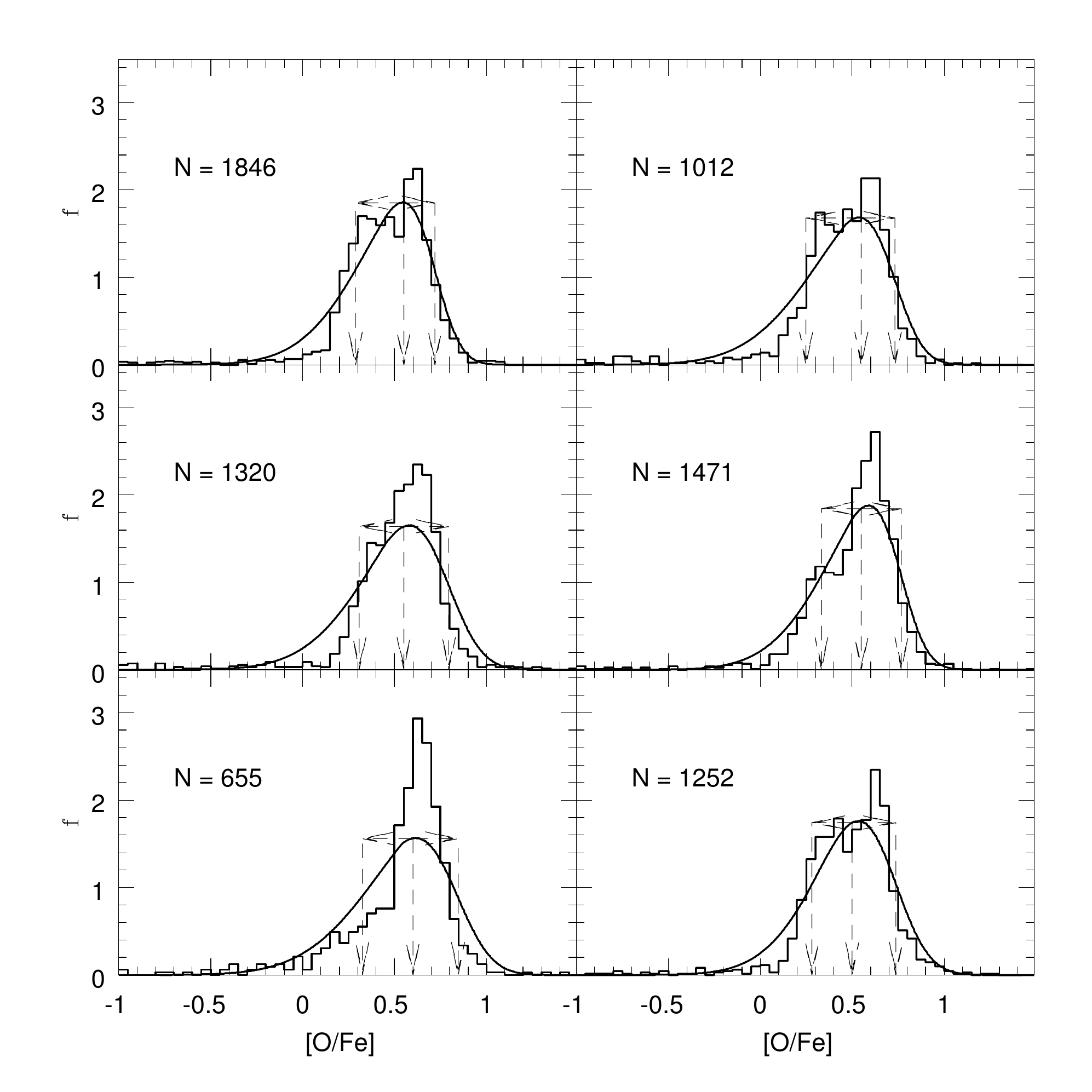}
\caption{(continued) Halo stellar particle $[$O/Fe$]$ distribution for the M$_{\rm tot}$~=~5$\times$10$^{12}$~M$_\odot$ semi--cosmological simulations in Renda~et~al.~(2005b). The 68\%~Confidence~Level range and the number of stellar particles each distribution refers to are also shown.}
\label{appC:sim5e12:fig2}
\end{center}
\end{figure}

\begin{figure}
\begin{center}
\includegraphics[width=1.0\textwidth]{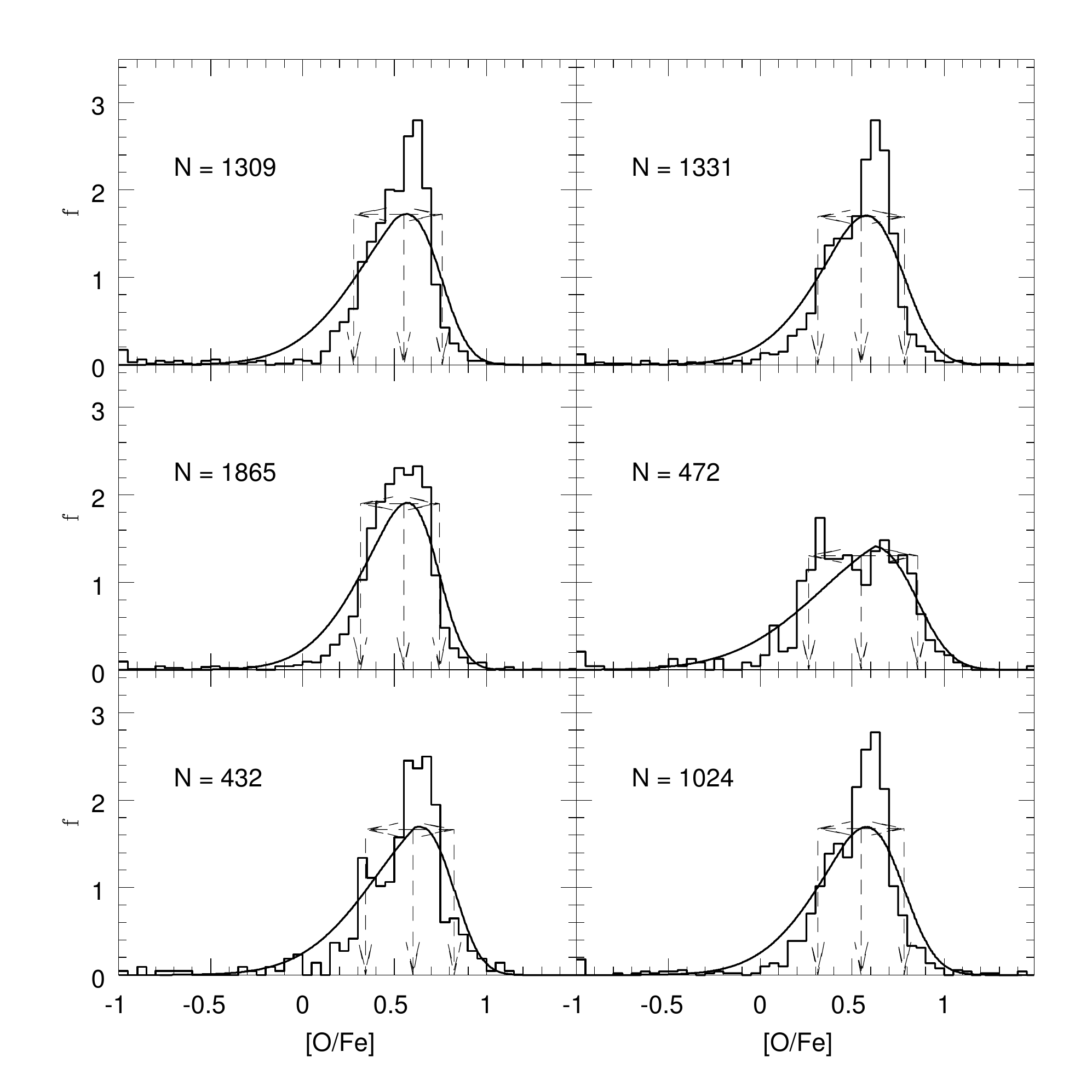}
\caption{(continued) Halo stellar particle $[$O/Fe$]$ distribution for the M$_{\rm tot}$~=~5$\times$10$^{12}$~M$_\odot$ semi--cosmological simulations in Renda~et~al.~(2005b). The 68\%~Confidence~Level range and the number of stellar particles each distribution refers to are also shown.}
\label{appC:sim5e12:fig3}
\end{center}
\end{figure}

\begin{figure}
\begin{center}
\includegraphics[width=1.0\textwidth]{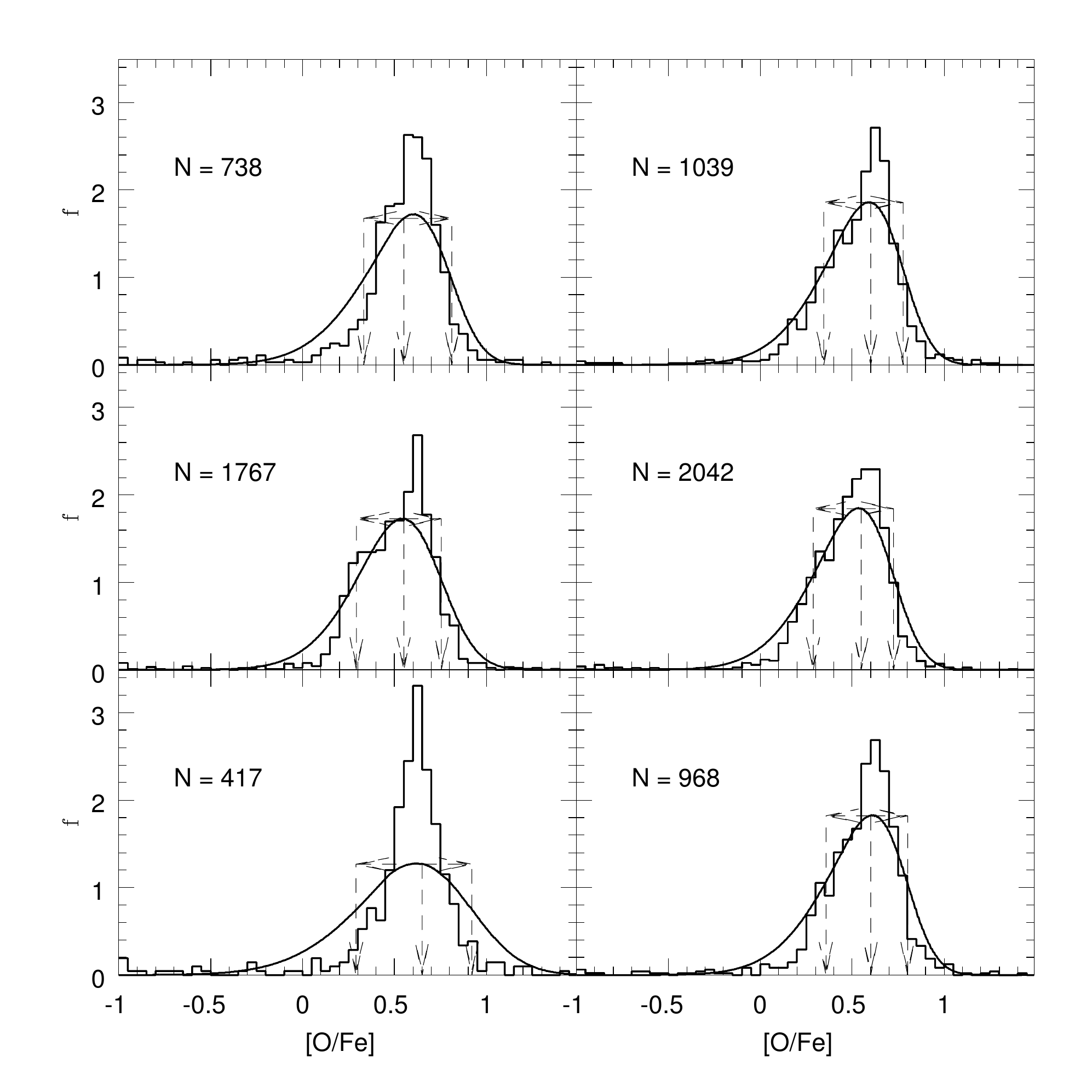}
\caption{(continued) Halo stellar particle $[$O/Fe$]$ distribution for the M$_{\rm tot}$~=~5$\times$10$^{12}$~M$_\odot$ semi--cosmological simulations in Renda~et~al.~(2005b). The 68\%~Confidence~Level range and the number of stellar particles each distribution refers to are also shown.}
\label{appC:sim5e12:fig4}
\end{center}
\end{figure}

\begin{figure}
\begin{center}
\includegraphics[width=1.0\textwidth]{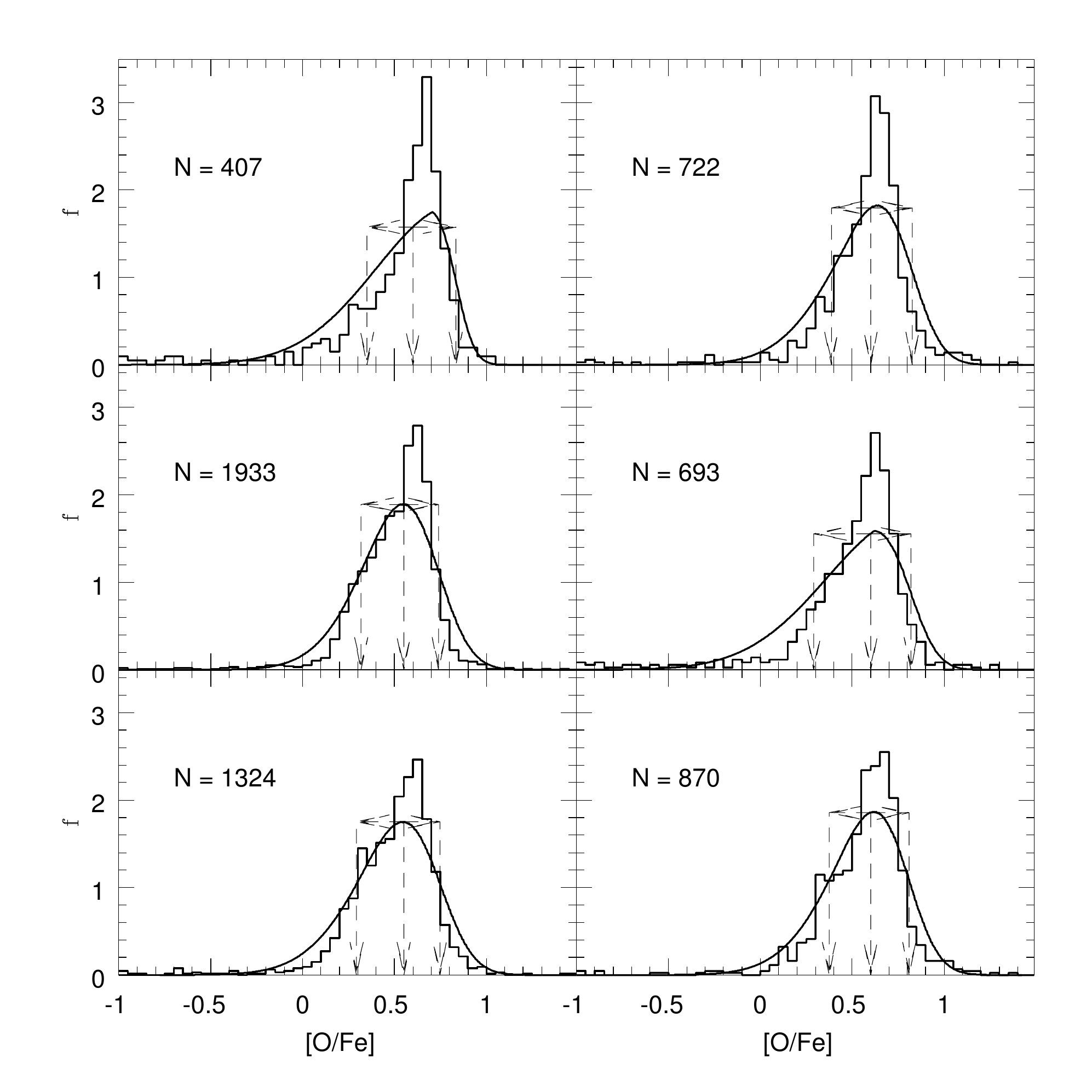}
\caption{(continued) Halo stellar particle $[$O/Fe$]$ distribution for the M$_{\rm tot}$~=~5$\times$10$^{12}$~M$_\odot$ semi--cosmological simulations in Renda~et~al.~(2005b). The 68\%~Confidence~Level range and the number of stellar particles each distribution refers to are also shown.}
\label{appC:sim5e12:fig5}
\end{center}
\end{figure}




\newpage

\begin{center}
\chapter{Stellar Halo Age Distributions}
\label{app:appendixD}
\end{center}

\begin{figure}
\begin{center}
\includegraphics[width=1.0\textwidth]{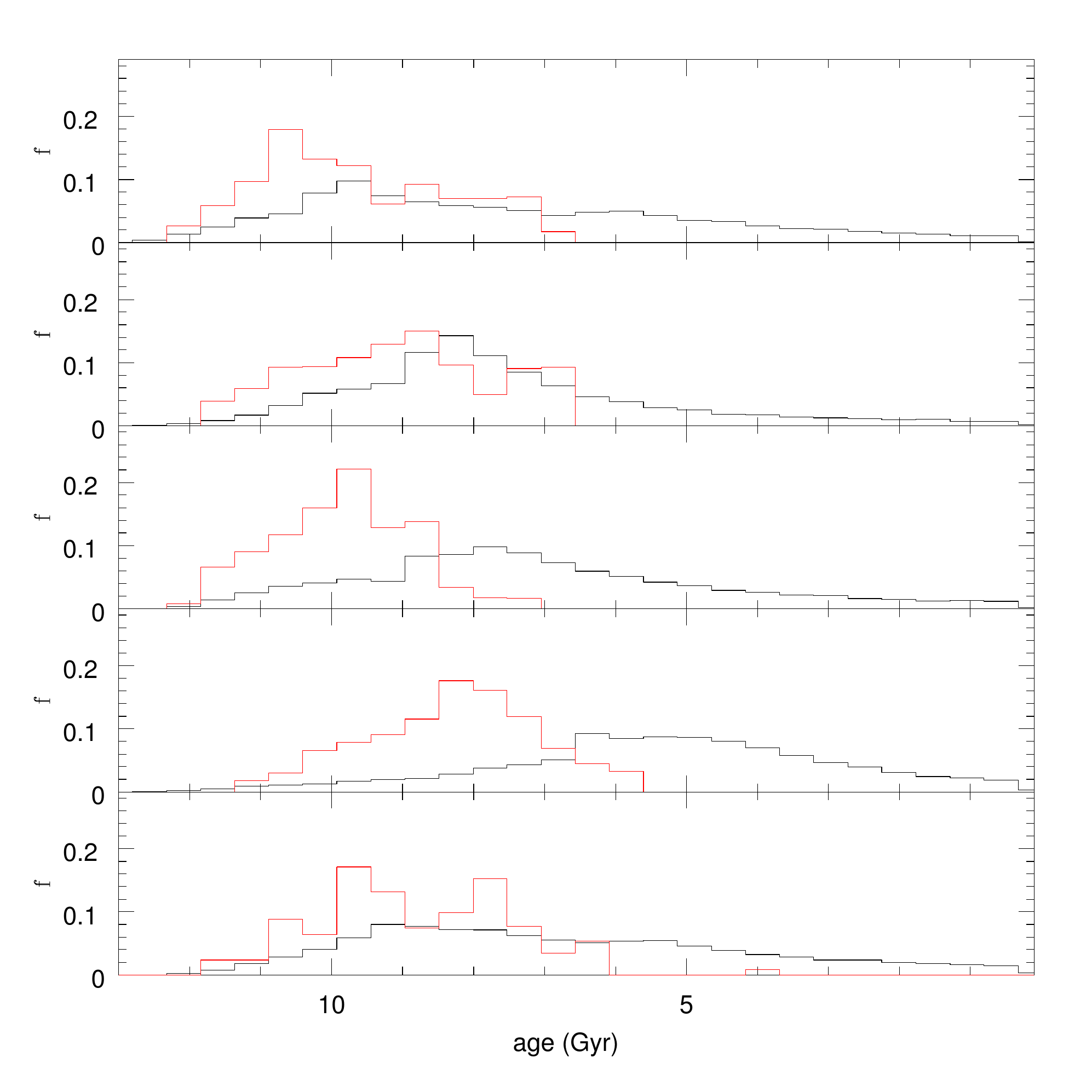}
\caption{Halo (red) and galaxy (black) age distributions for the M$_{\rm tot}$~=~10$^{11}$~M$_\odot$ semi--cosmological simulations in Renda~et~al.~(2005b).}\label{appD:sim1e11:fig}
\end{center}
\end{figure}

\begin{figure}
\begin{center}
\includegraphics[width=1.0\textwidth]{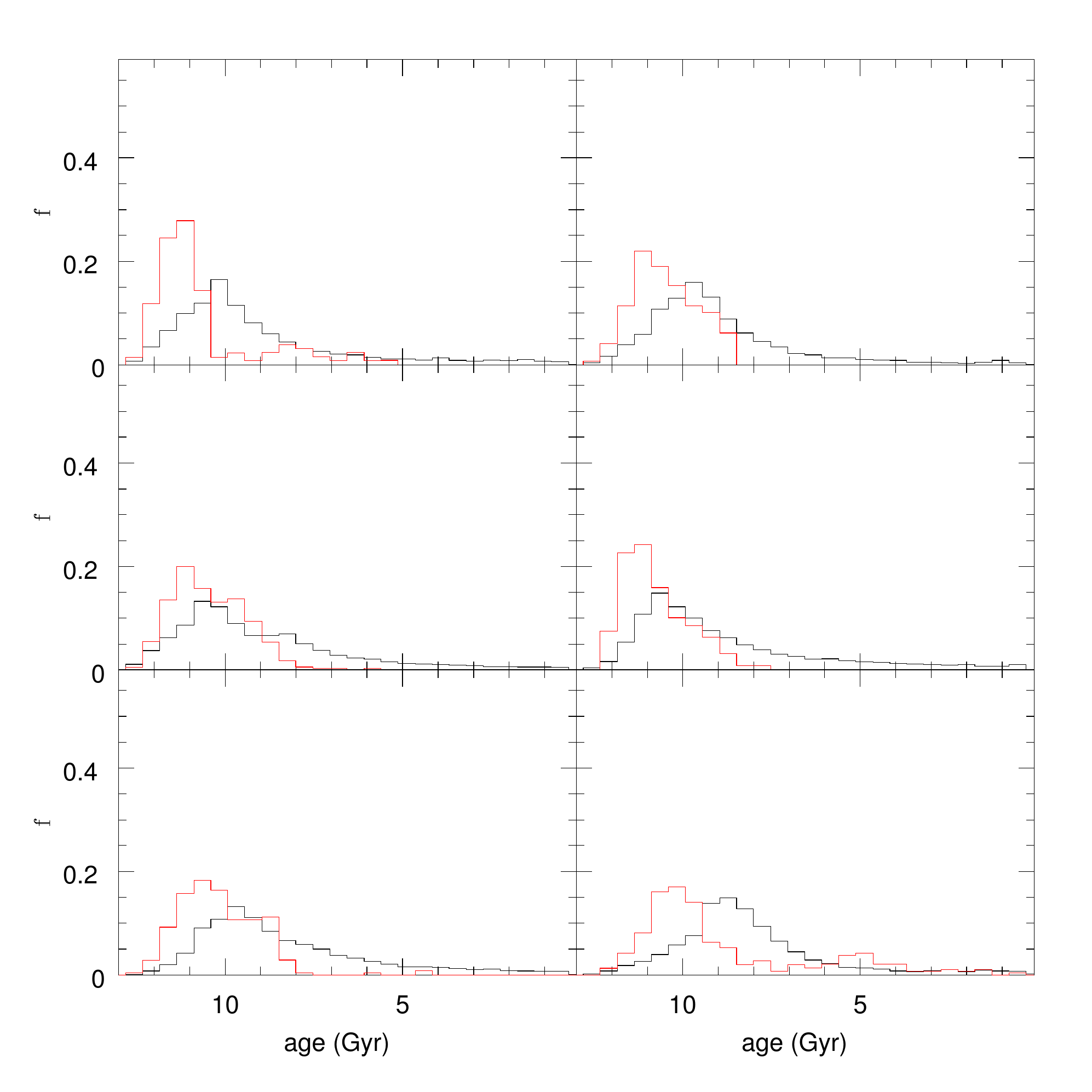}
\caption{Halo (red) and galaxy (black) age distributions for the M$_{\rm tot}$~=~5$\times$10$^{11}$~M$_\odot$ semi--cosmological simulations in Renda~et~al.~(2005b).}\label{appD:sim5e11:fig1}
\end{center}
\end{figure}

\begin{figure}
\begin{center}
\includegraphics[width=1.0\textwidth]{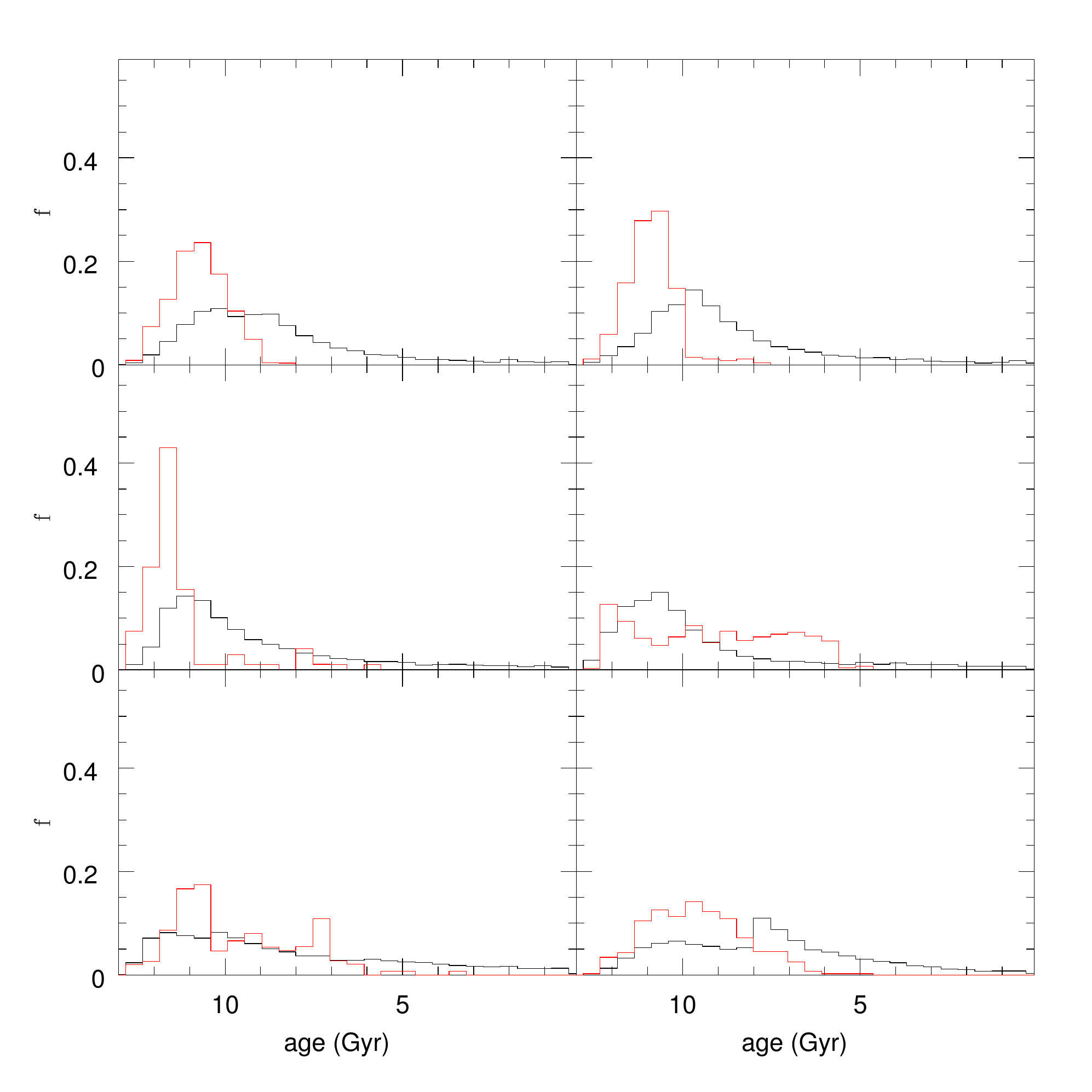}
\caption{(continued) Halo (red) and galaxy (black) age distributions for the M$_{\rm tot}$~=~5$\times$10$^{11}$~M$_\odot$ semi--cosmological simulations in Renda~et~al.~(2005b).}\label{appD:sim5e11:fig2}
\end{center}
\end{figure}

\begin{figure}
\begin{center}
\includegraphics[width=1.0\textwidth]{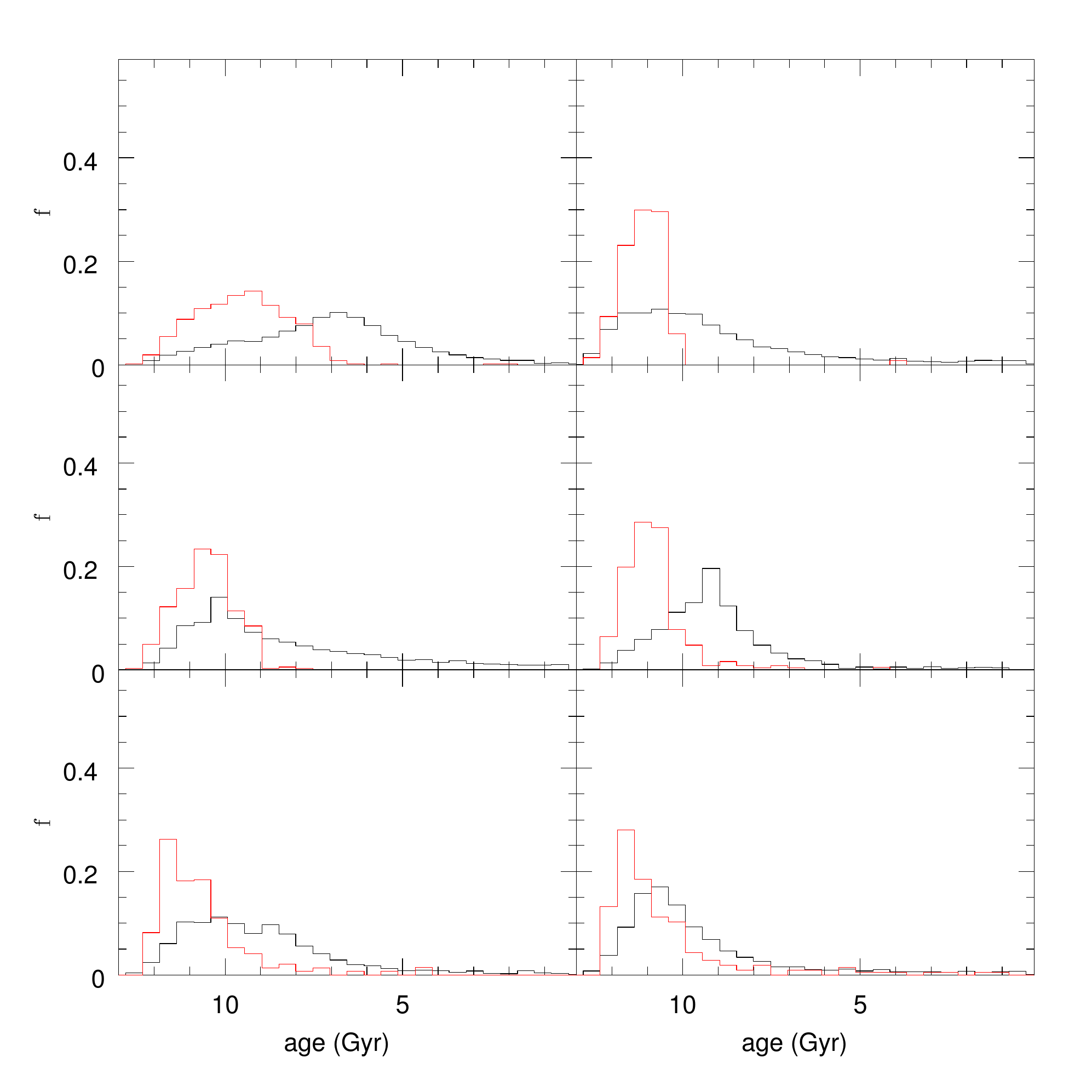}
\caption{(continued) Halo (red) and galaxy (black) age distributions for the M$_{\rm tot}$~=~5$\times$10$^{11}$~M$_\odot$ semi--cosmological simulations in Renda~et~al.~(2005b).}\label{appD:sim5e11:fig3}
\end{center}
\end{figure}

\begin{figure}
\begin{center}
\includegraphics[width=1.0\textwidth]{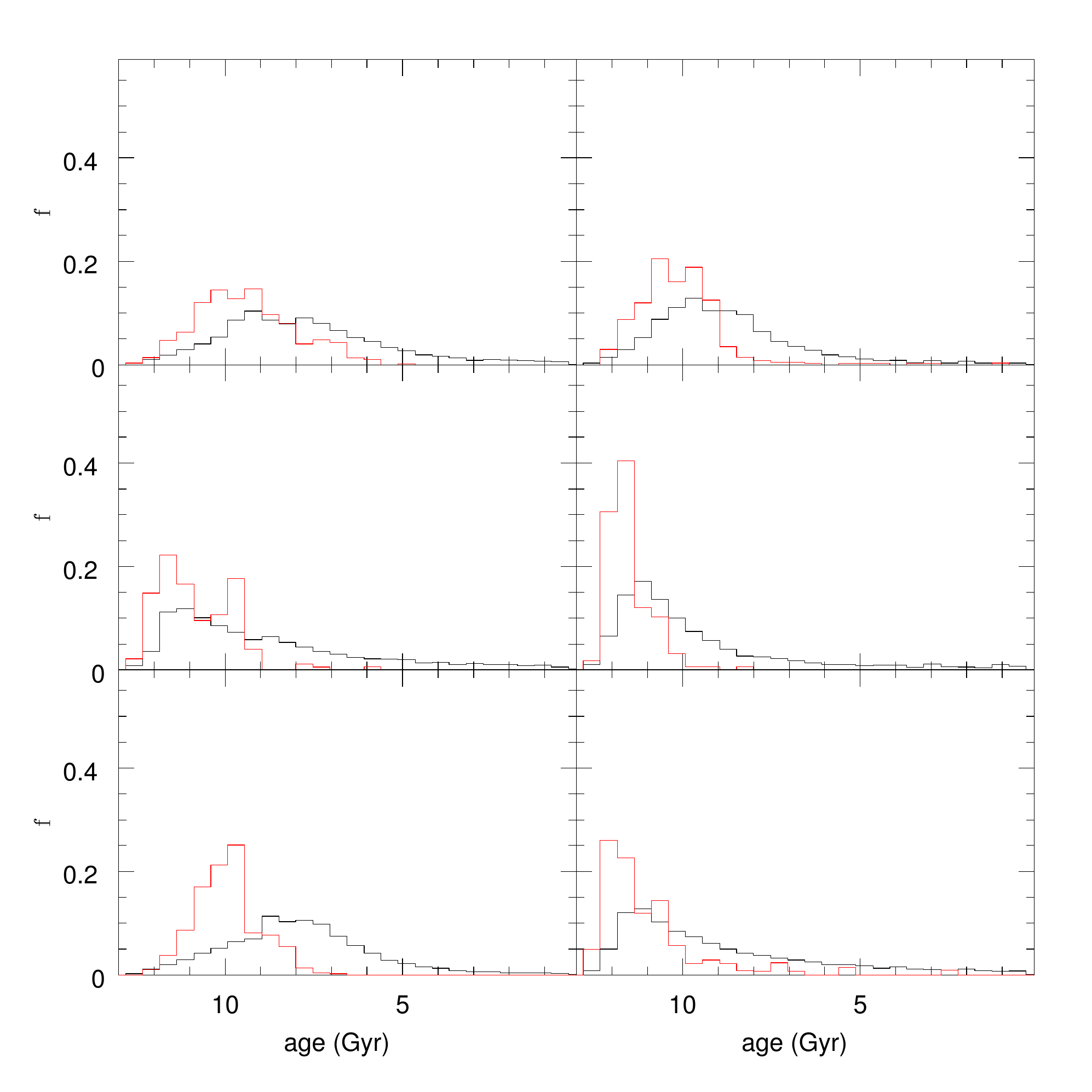}
\caption{(continued) Halo (red) and galaxy (black) age distributions for the M$_{\rm tot}$~=~5$\times$10$^{11}$~M$_\odot$ semi--cosmological simulations in Renda~et~al.~(2005b).}\label{appD:sim5e11:fig4}
\end{center}
\end{figure}

\begin{figure}
\begin{center}
\includegraphics[width=1.0\textwidth]{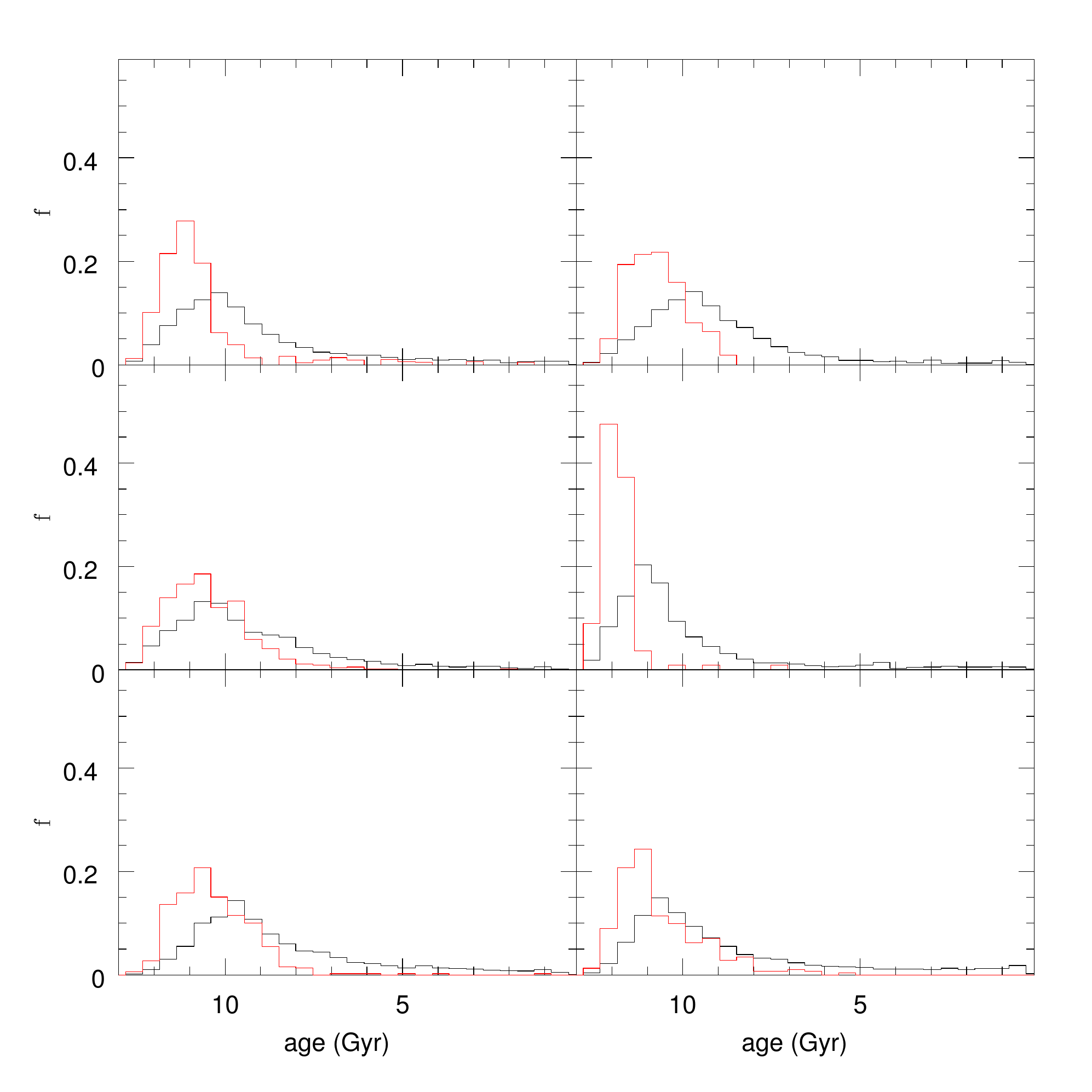}
\caption{Halo (red) and galaxy (black) age distributions for the M$_{\rm tot}$~=~10$^{12}$~M$_\odot$ semi--cosmological simulations in Renda~et~al.~(2005b).}\label{appD:sim1e12:fig1}
\end{center}
\end{figure}

\begin{figure}
\begin{center}
\includegraphics[width=1.0\textwidth]{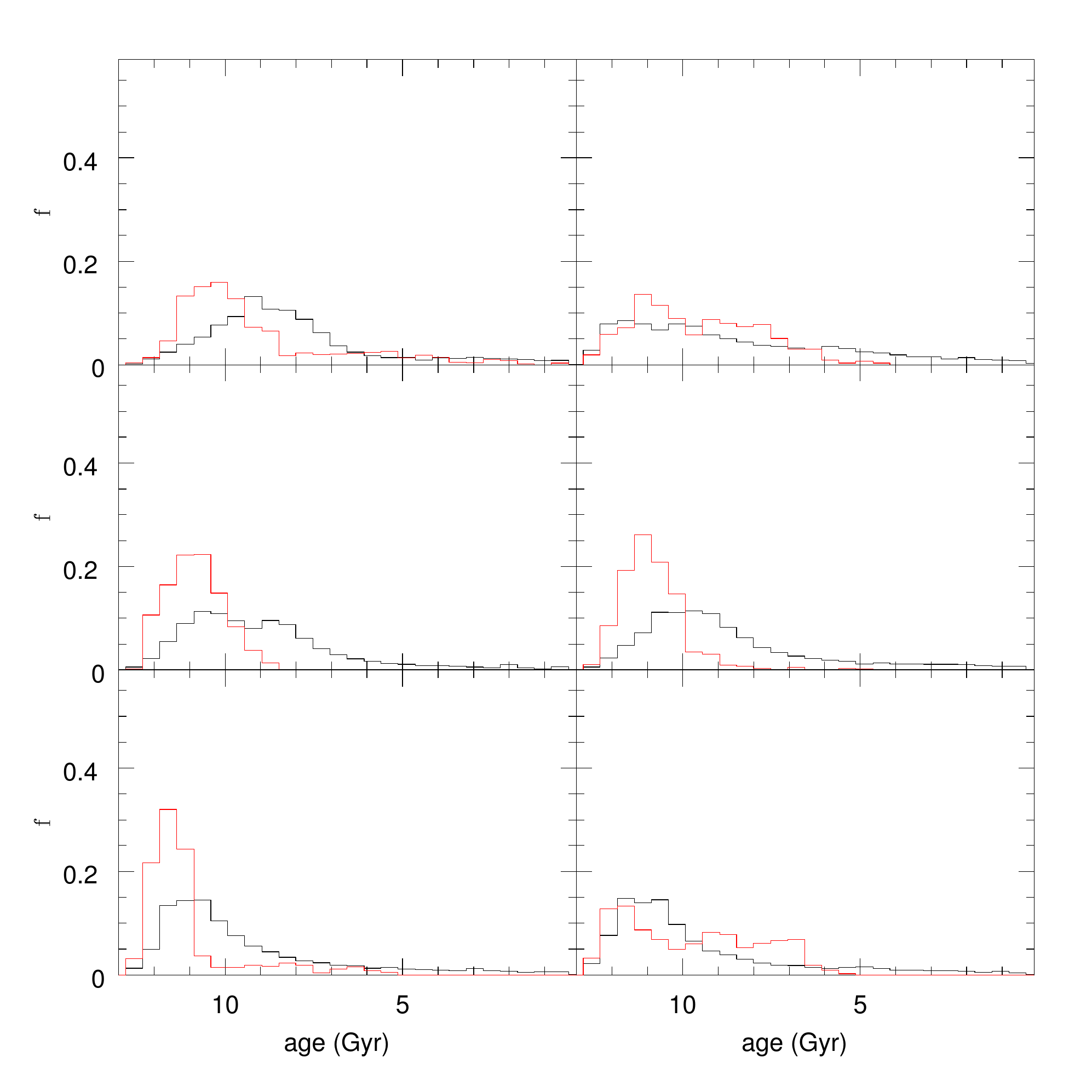}
\caption{(continued) Halo (red) and galaxy (black) age distributions for the M$_{\rm tot}$~=~10$^{12}$~M$_\odot$ semi--cosmological simulations in Renda~et~al.~(2005b).}\label{appD:sim1e12:fig2}
\end{center}
\end{figure}

\begin{figure}
\begin{center}
\includegraphics[width=1.0\textwidth]{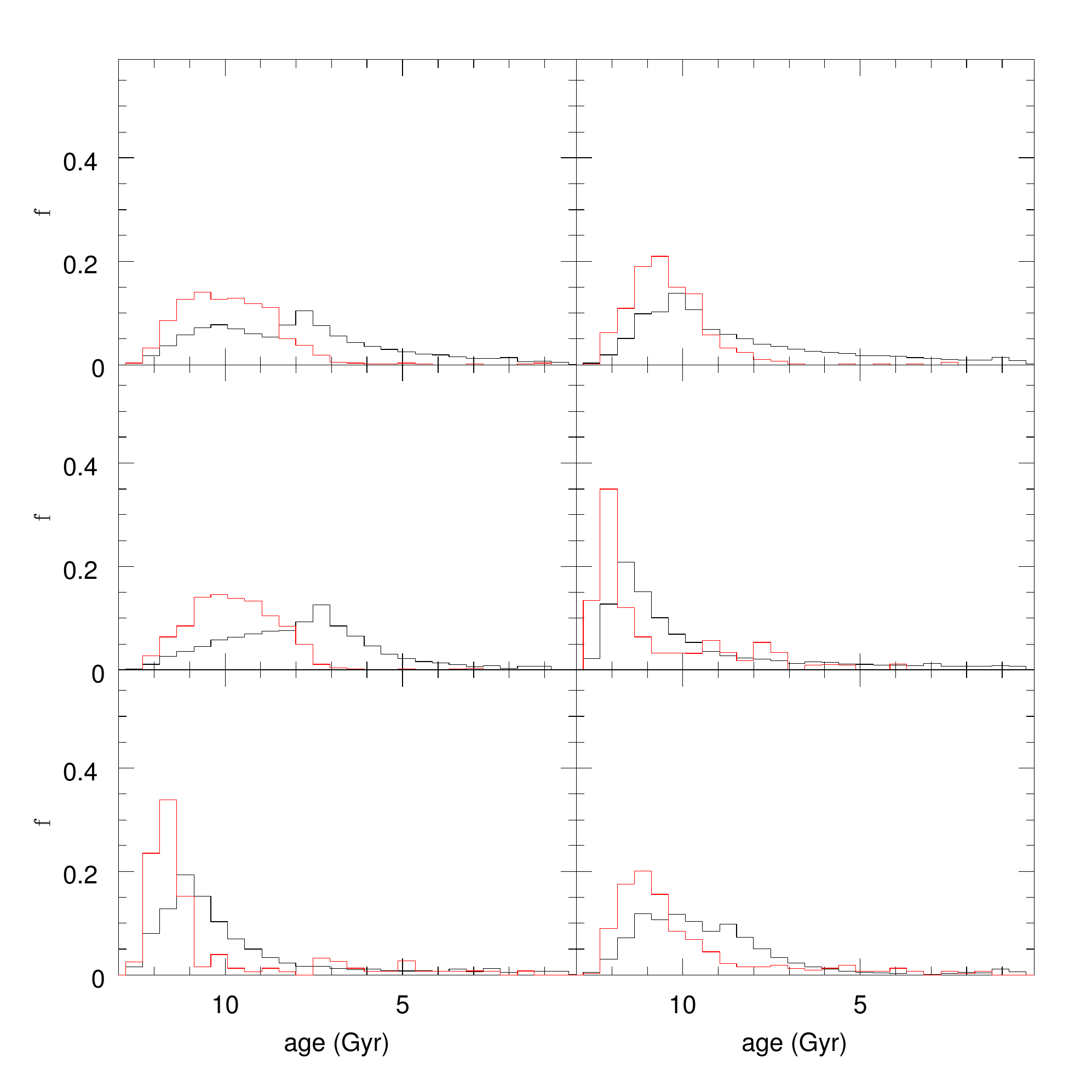}
\caption{(continued) Halo (red) and galaxy (black) age distributions for the M$_{\rm tot}$~=~10$^{12}$~M$_\odot$ semi--cosmological simulations in Renda~et~al.~(2005b).}\label{appD:sim1e12:fig3}
\end{center}
\end{figure}

\begin{figure}
\begin{center}
\includegraphics[width=1.0\textwidth]{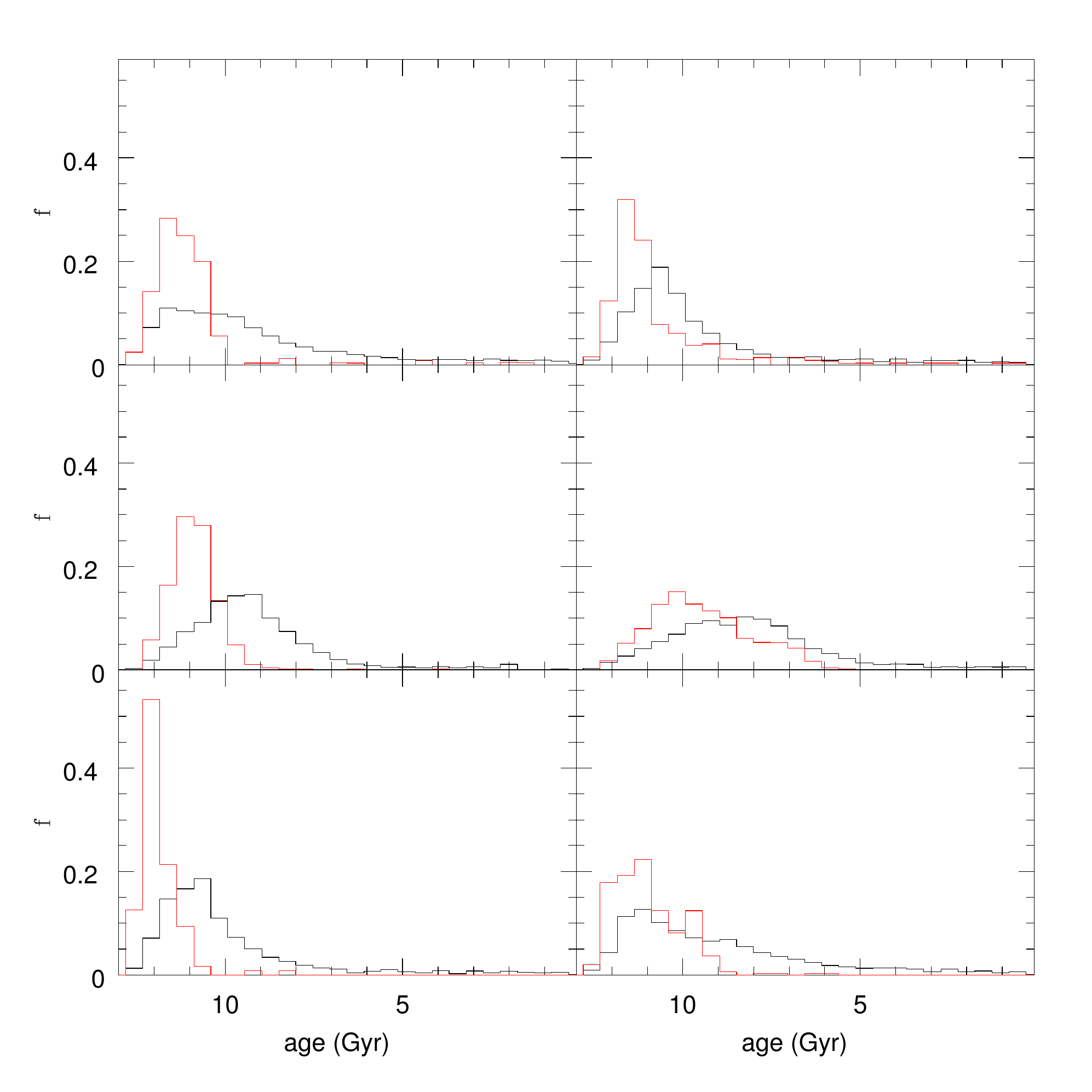}
\caption{(continued) Halo (red) and galaxy (black) age distributions for the M$_{\rm tot}$~=~10$^{12}$~M$_\odot$ semi--cosmological simulations in Renda~et~al.~(2005b).}\label{appD:sim1e12:fig4}
\end{center}
\end{figure}

\begin{figure}
\begin{center}
\includegraphics[width=1.0\textwidth]{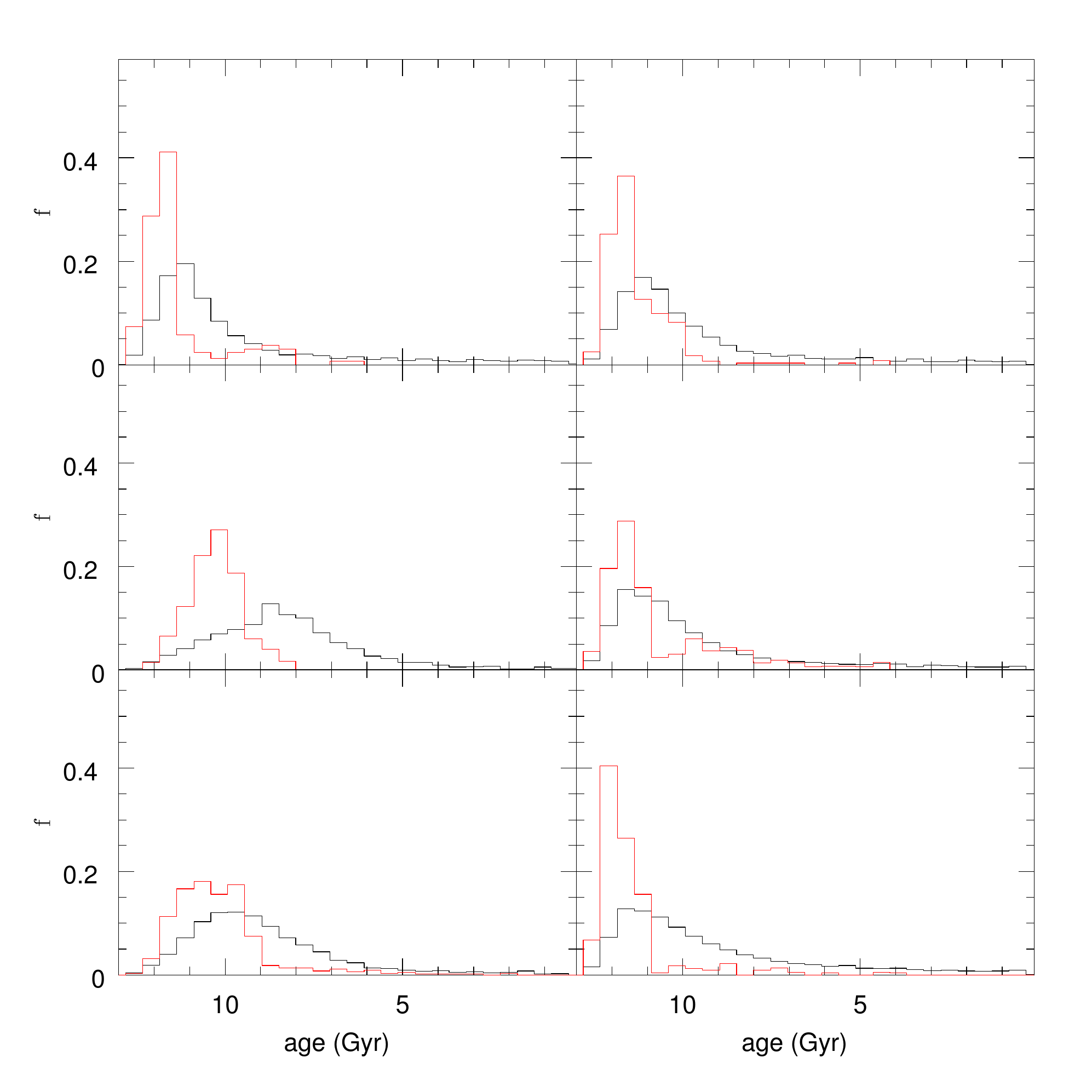}
\caption{(continued) Halo (red) and galaxy (black) age distributions for the M$_{\rm tot}$~=~10$^{12}$~M$_\odot$ semi--cosmological simulations in Renda~et~al.~(2005b).}\label{appD:sim1e12:fig5}
\end{center}
\end{figure}

\begin{figure}
\begin{center}
\includegraphics[width=1.0\textwidth]{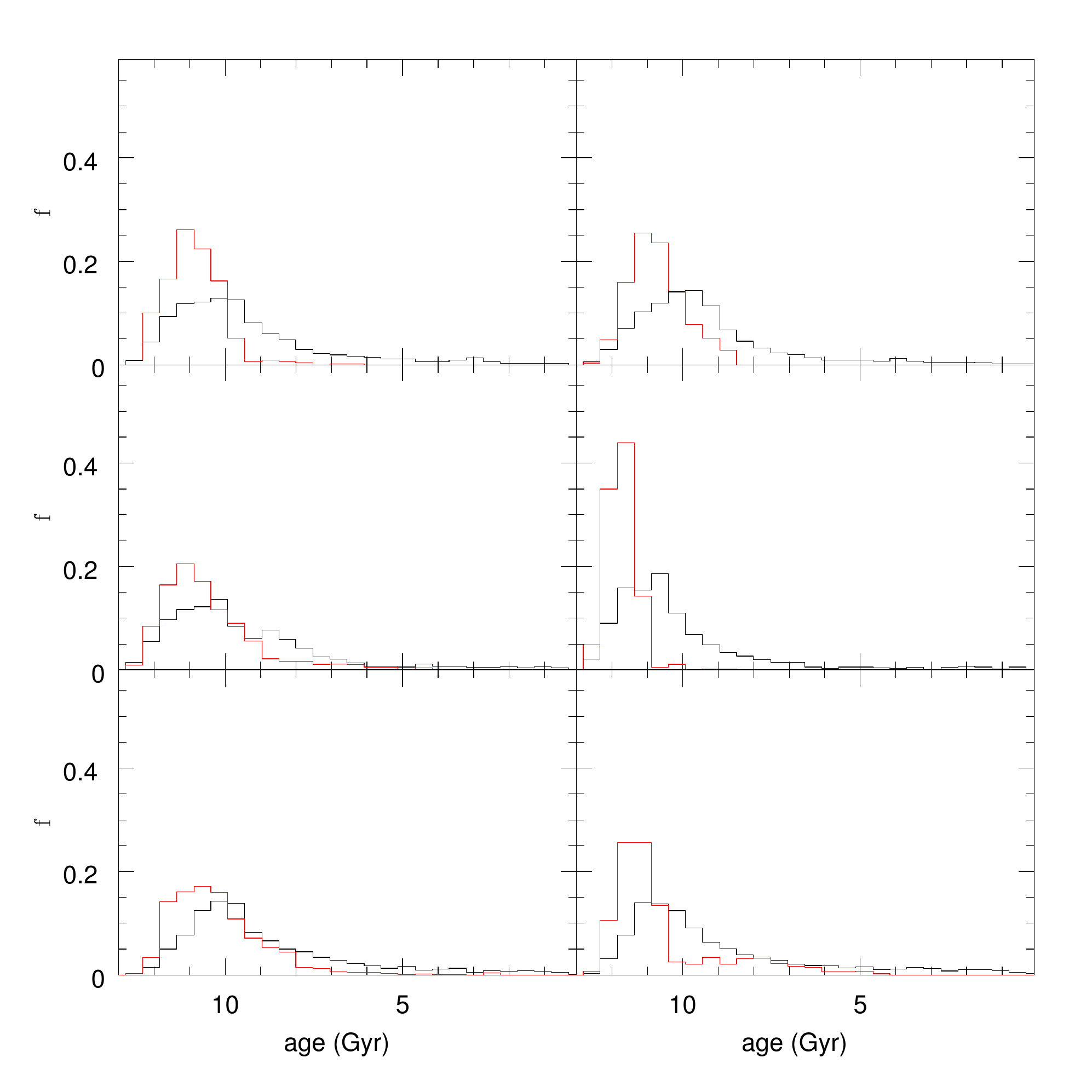}
\caption{Halo (red) and galaxy (black) age distributions for the M$_{\rm tot}$~=~5$\times$10$^{12}$~M$_\odot$ semi--cosmological simulations in Renda~et~al.~(2005b).}\label{appD:sim5e12:fig1}
\end{center}
\end{figure}

\begin{figure}
\begin{center}
\includegraphics[width=1.0\textwidth]{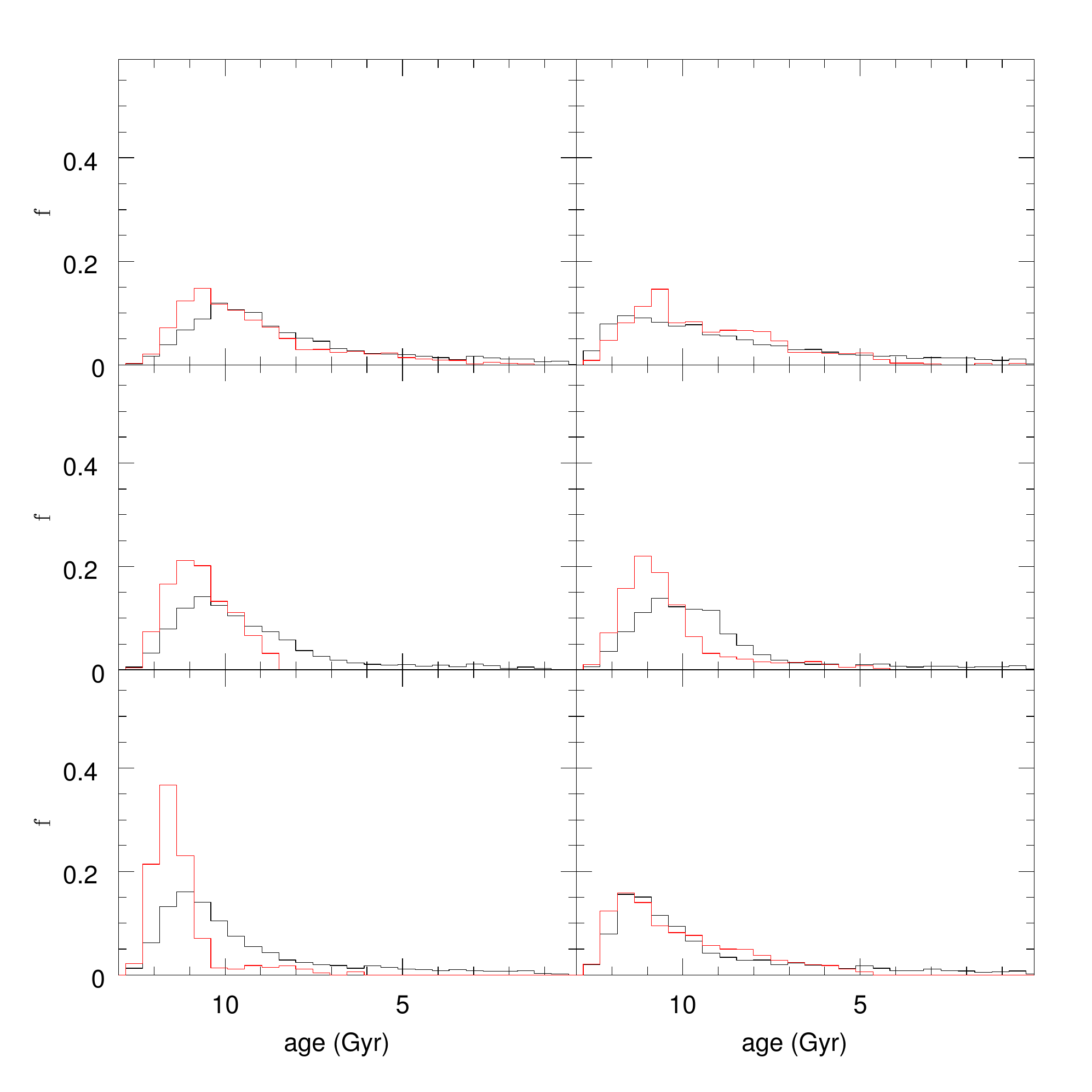}
\caption{(continued) Halo (red) and galaxy (black) age distributions for the M$_{\rm tot}$~=~5$\times$10$^{12}$~M$_\odot$ semi--cosmological simulations in Renda~et~al.~(2005b).}\label{appD:sim5e12:fig2}
\end{center}
\end{figure}

\begin{figure}
\begin{center}
\includegraphics[width=1.0\textwidth]{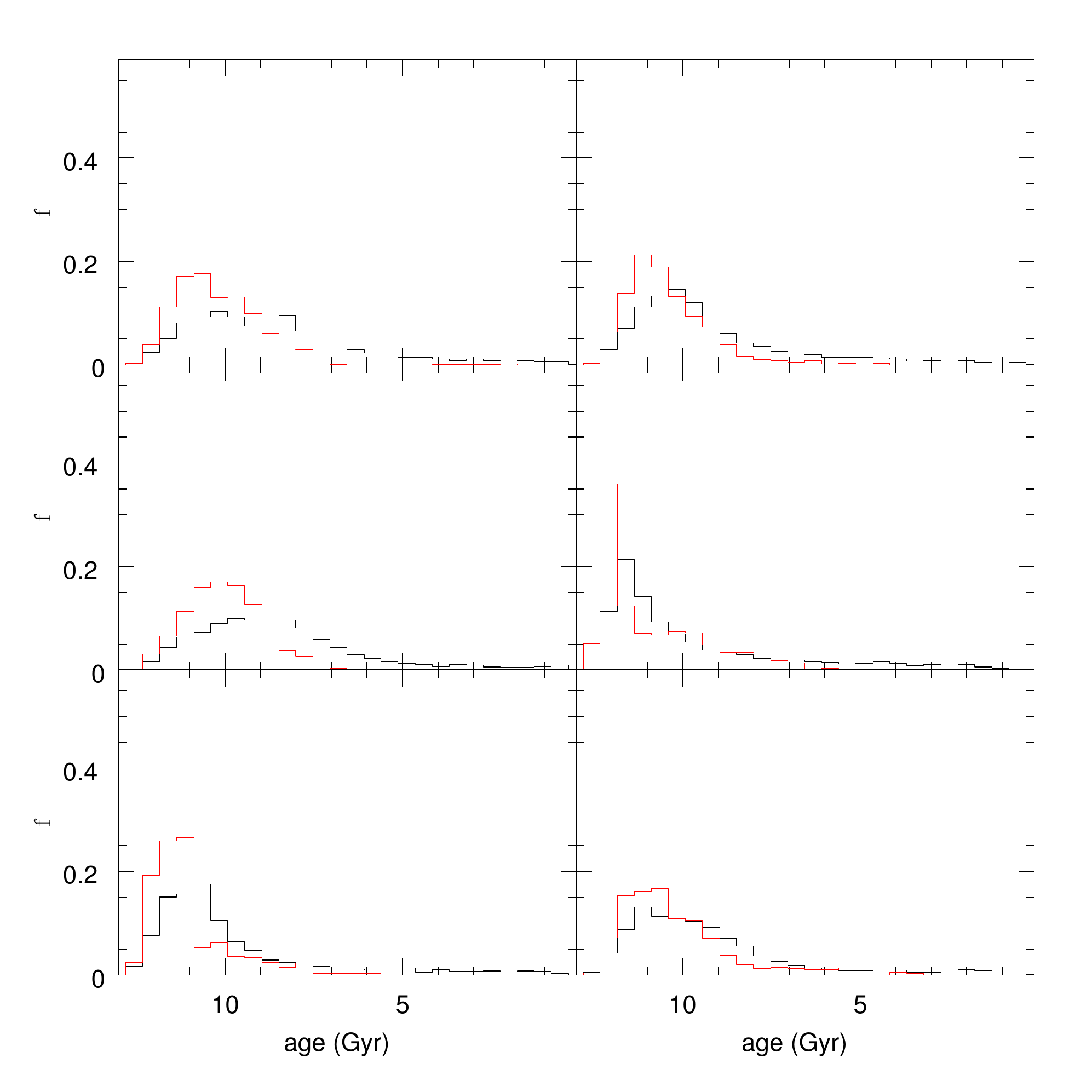}
\caption{(continued) Halo (red) and galaxy (black) age distributions for the M$_{\rm tot}$~=~5$\times$10$^{12}$~M$_\odot$ semi--cosmological simulations in Renda~et~al.~(2005b).}\label{appD:sim5e12:fig3}
\end{center}
\end{figure}

\begin{figure}
\begin{center}
\includegraphics[width=1.0\textwidth]{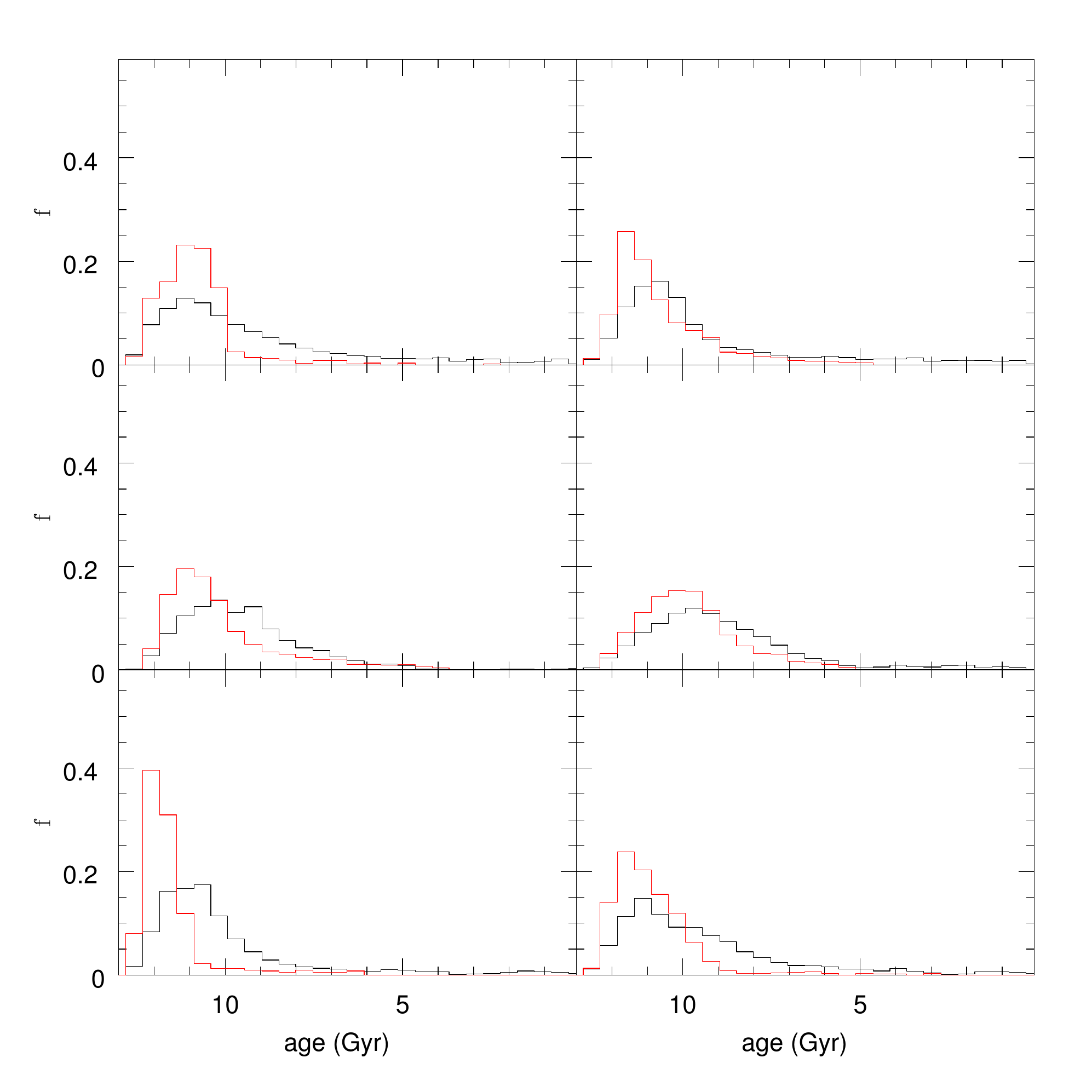}
\caption{(continued) Halo (red) and galaxy (black) age distributions for the M$_{\rm tot}$~=~5$\times$10$^{12}$~M$_\odot$ semi--cosmological simulations in Renda~et~al.~(2005b).}\label{appD:sim5e12:fig4}
\end{center}
\end{figure}

\begin{figure}
\begin{center}
\includegraphics[width=1.0\textwidth]{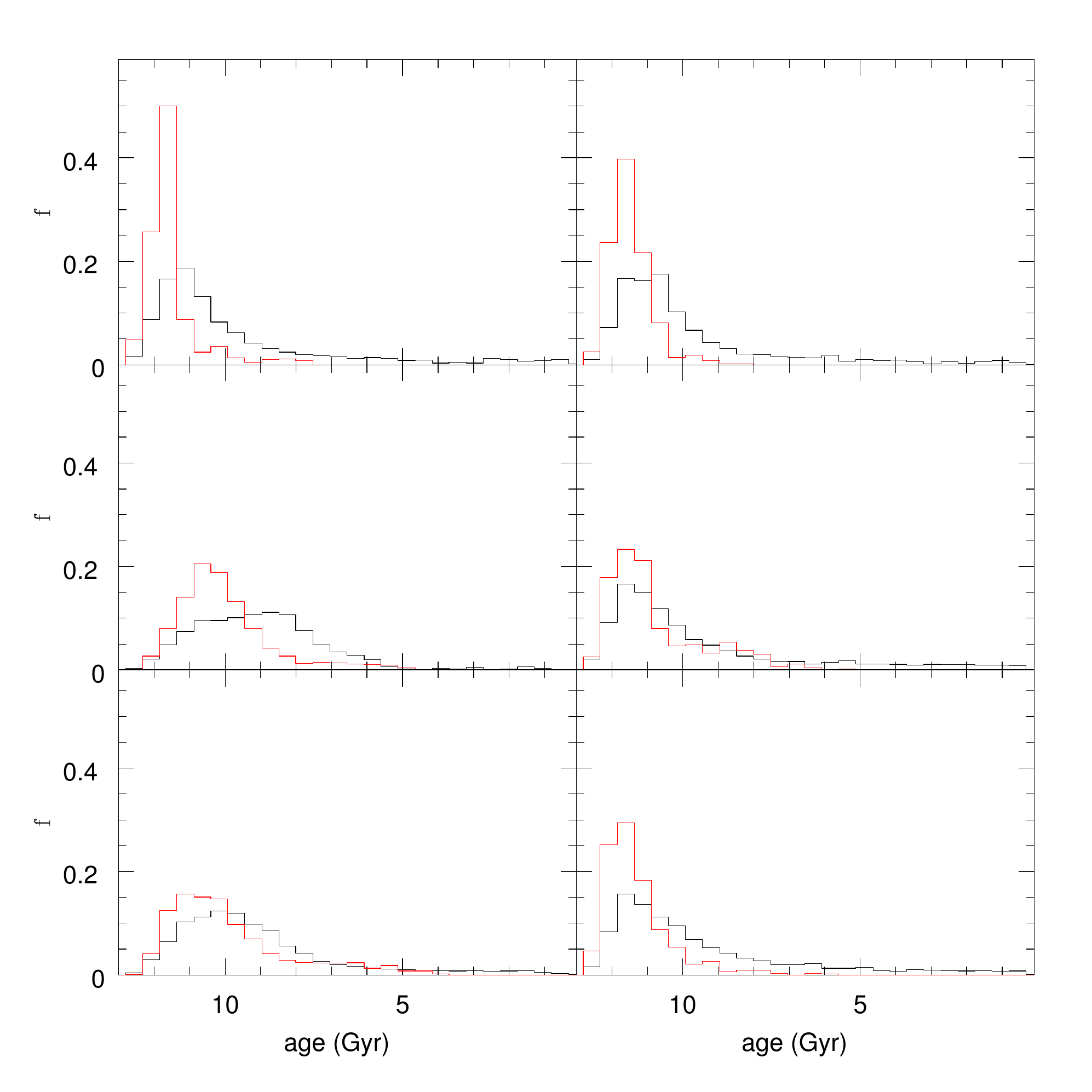}
\caption{(continued) Halo (red) and galaxy (black) age distributions for the M$_{\rm tot}$~=~5$\times$10$^{12}$~M$_\odot$ semi--cosmological simulations in Renda~et~al.~(2005b).}\label{appD:sim5e12:fig5}
\end{center}
\end{figure}




\newpage

\begin{center}
\chapter{Galaxy Rotation Curves}
\label{app:appendixE}
\end{center}

\begin{figure}
\begin{center}
\includegraphics[width=1.0\textwidth]{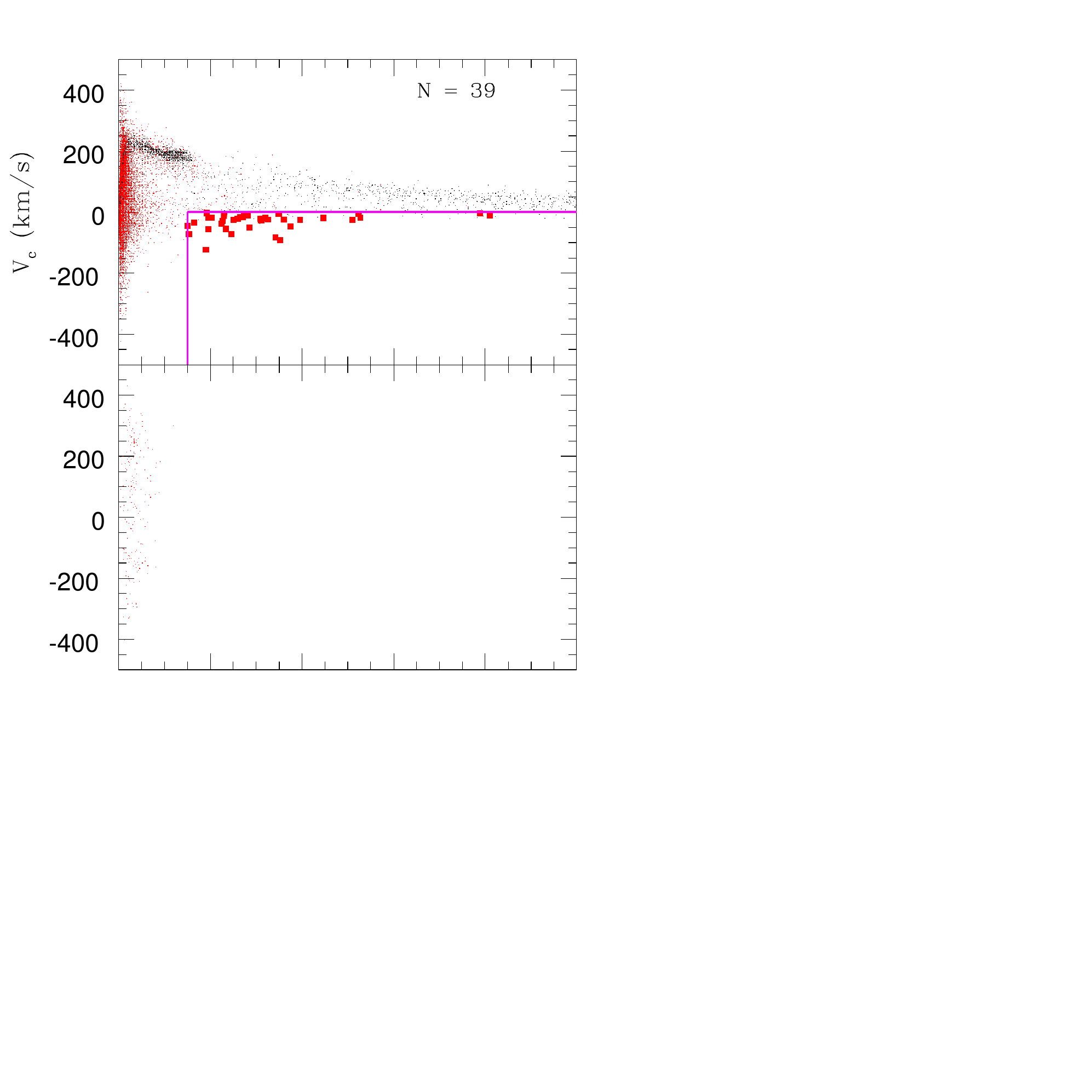}
\caption{Galaxy rotation curves for the M$_{\rm tot}$~=~5$\times$10$^{11}$~M$_\odot$ semi--cosmological simulations in Renda~et~al.~(2005b). Black dots for the gaseous particles. Red dots for the stellar particles; red boxes for the ensemble of stellar particles (enclosed within the magenta frame) which are counter--rotating with v$_{\theta}~<~0$ within each topographical halo at a projected radius R$~>~$15~kpc. The size of such ensemble is also shown in each panel.}
\label{appE:sim5e11:fig1}
\end{center}
\end{figure}

\begin{figure}
\begin{center}
\includegraphics[width=1.0\textwidth]{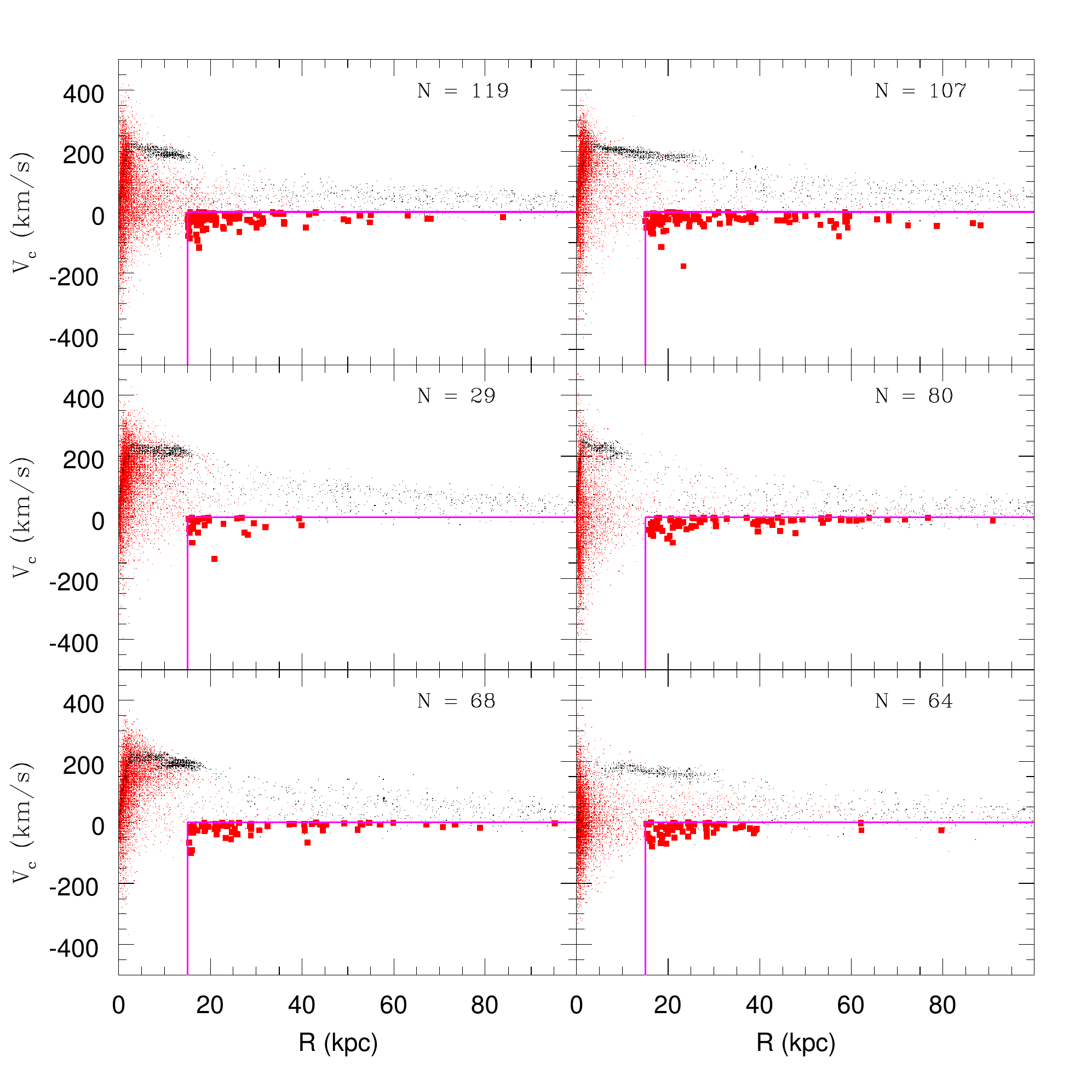}
\caption{(continued) Galaxy rotation curves for the M$_{\rm tot}$~=~5$\times$10$^{11}$~M$_\odot$ semi--cosmological simulations in Renda~et~al.~(2005b). Black dots for the gaseous particles. Red dots for the stellar particles; red boxes for the ensemble of stellar particles (enclosed within the magenta frame) which are counter--rotating with v$_{\theta}~<~0$ within each topographical halo at a projected radius R$~>~$15~kpc. The size of such ensemble is also shown in each panel.}
\label{appE:sim5e11:fig2}
\end{center}
\end{figure}

\begin{figure}
\begin{center}
\includegraphics[width=1.0\textwidth]{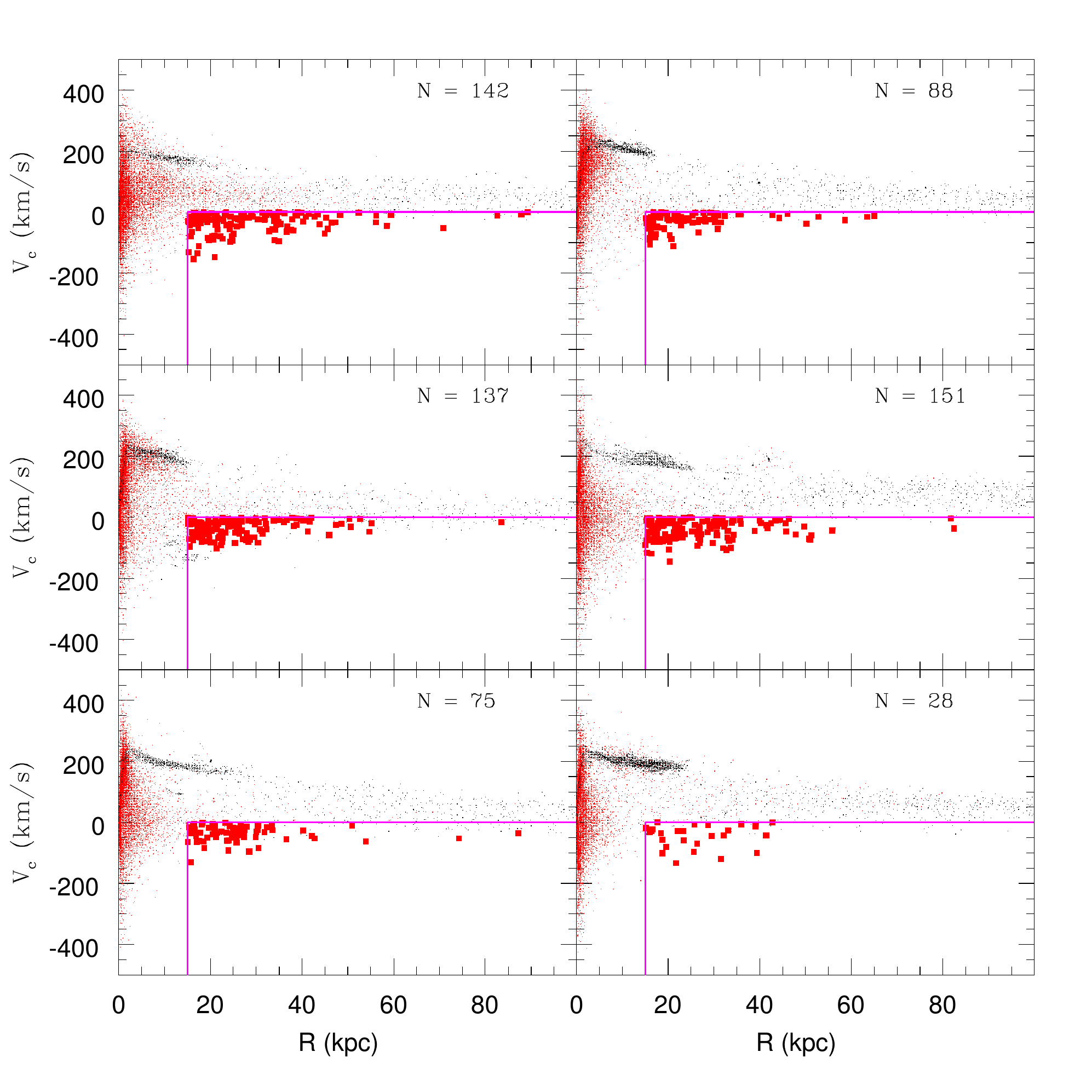}
\caption{(continued) Galaxy rotation curves for the M$_{\rm tot}$~=~5$\times$10$^{11}$~M$_\odot$ semi--cosmological simulations in Renda~et~al.~(2005b). Black dots for the gaseous particles. Red dots for the stellar particles; red boxes for the ensemble of stellar particles (enclosed within the magenta frame) which are counter--rotating with v$_{\theta}~<~0$ within each topographical halo at a projected radius R$~>~$15~kpc. The size of such ensemble is also shown in each panel.}
\label{appE:sim5e11:fig3}
\end{center}
\end{figure}

\begin{figure}
\begin{center}
\includegraphics[width=1.0\textwidth]{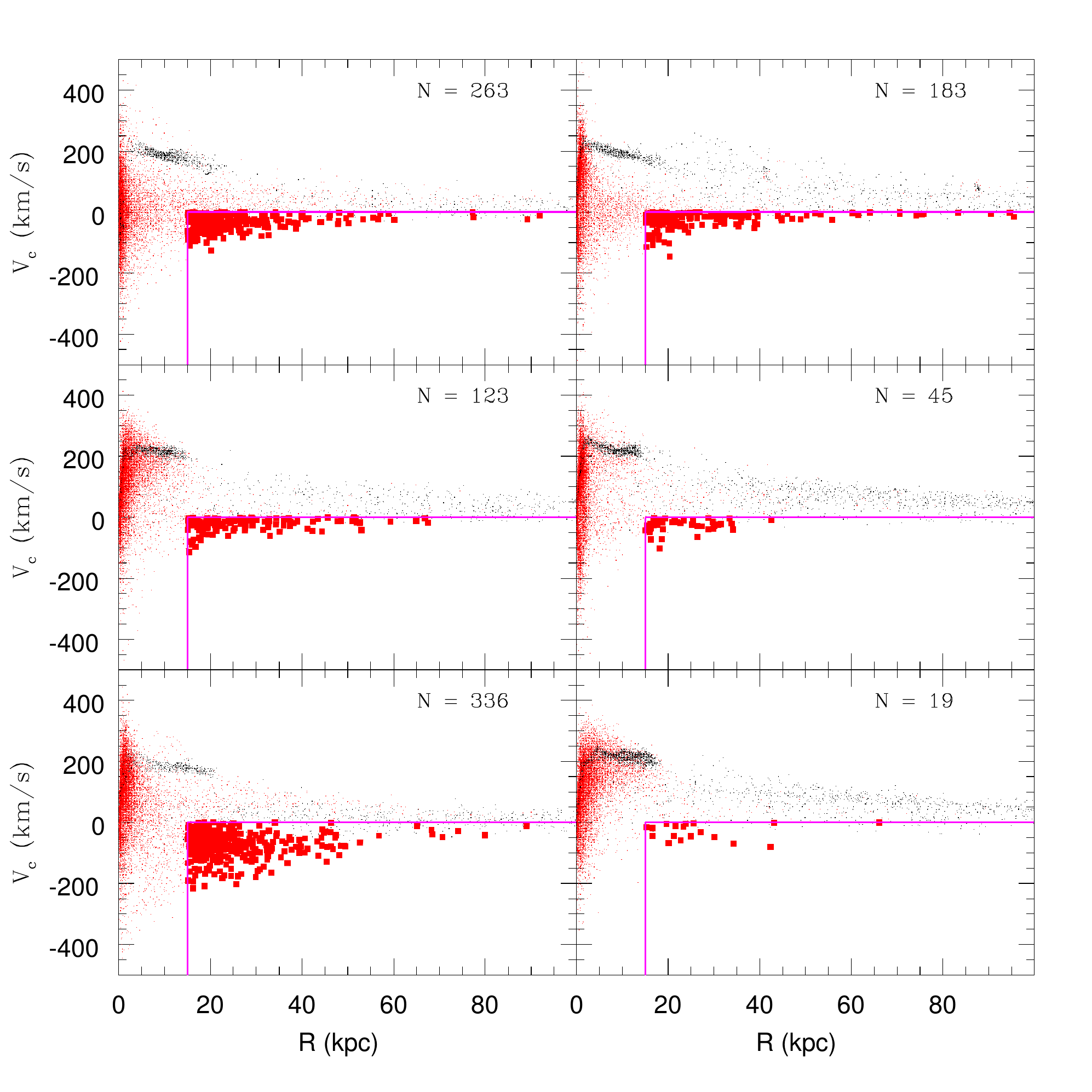}
\caption{(continued) Galaxy rotation curves for the M$_{\rm tot}$~=~5$\times$10$^{11}$~M$_\odot$ semi--cosmological simulations in Renda~et~al.~(2005b). Black dots for the gaseous particles. Red dots for the stellar particles; red boxes for the ensemble of stellar particles (enclosed within the magenta frame) which are counter--rotating with v$_{\theta}~<~0$ within each topographical halo at a projected radius R$~>~$15~kpc. The size of such ensemble is also shown in each panel.}
\label{appE:sim5e11:fig4}
\end{center}
\end{figure}

\begin{figure}
\begin{center}
\includegraphics[width=1.0\textwidth]{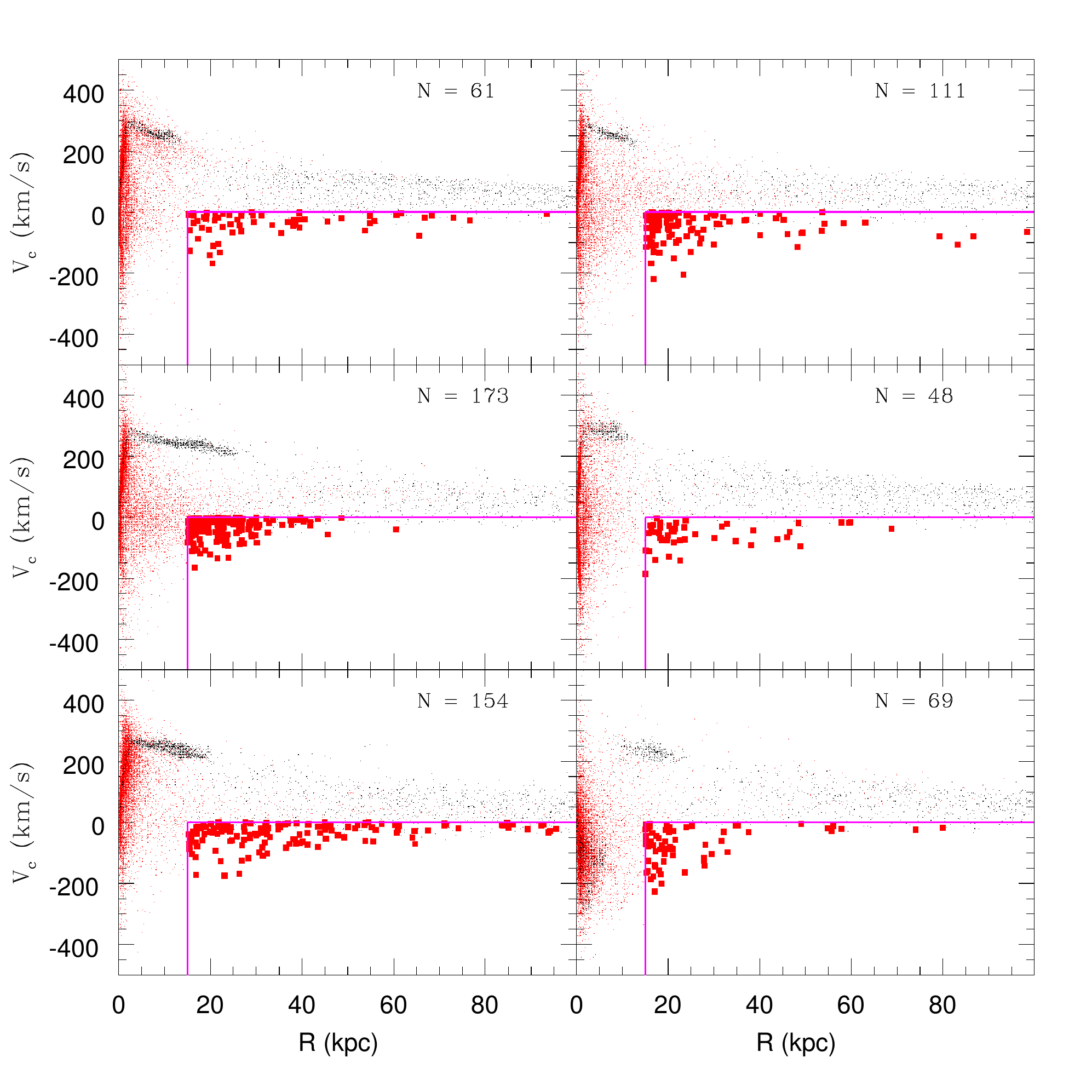}
\caption{Galaxy rotation curves for the M$_{\rm tot}$~=~10$^{12}$~M$_\odot$ semi--cosmological simulations in Renda~et~al.~(2005b). Black dots for the gaseous particles. Red dots for the stellar particles; red boxes for the ensemble of stellar particles (enclosed within the magenta frame) which are counter--rotating with v$_{\theta}~<~0$ within each topographical halo at a projected radius R$~>~$15~kpc. The size of such ensemble is also shown in each panel.}
\label{appE:sim1e12:fig1}
\end{center}
\end{figure}

\begin{figure}
\begin{center}
\includegraphics[width=1.0\textwidth]{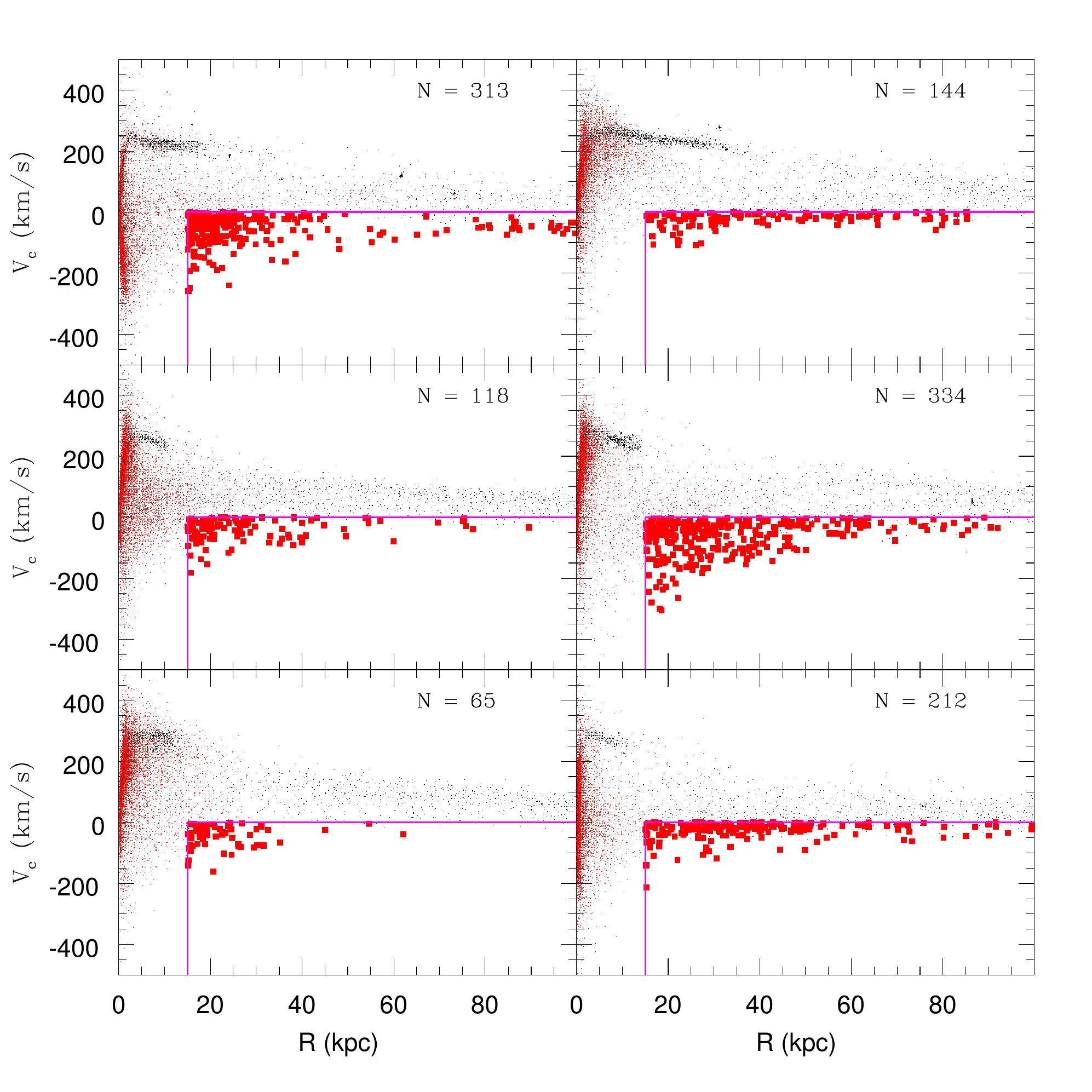}
\caption{(continued) Galaxy rotation curves for the M$_{\rm tot}$~=~10$^{12}$~M$_\odot$ semi--cosmological simulations in Renda~et~al.~(2005b). Black dots for the gaseous particles. Red dots for the stellar particles; red boxes for the ensemble of stellar particles (enclosed within the magenta frame) which are counter--rotating with v$_{\theta}~<~0$ within each topographical halo at a projected radius R$~>~$15~kpc. The size of such ensemble is also shown in each panel.}
\label{appE:sim1e12:fig2}
\end{center}
\end{figure}

\begin{figure}
\begin{center}
\includegraphics[width=1.0\textwidth]{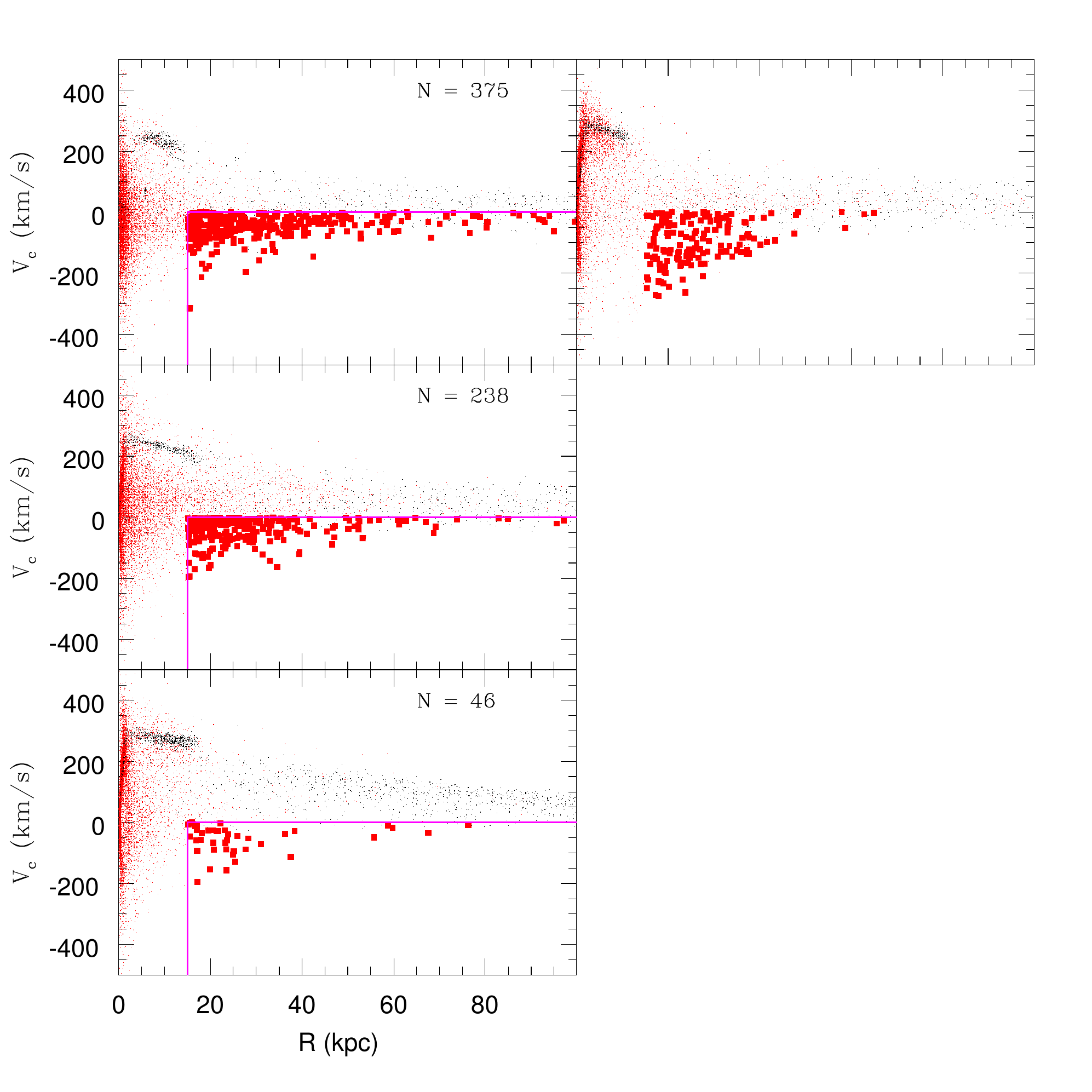}
\caption{(continued) Galaxy rotation curves for the M$_{\rm tot}$~=~10$^{12}$~M$_\odot$ semi--cosmological simulations in Renda~et~al.~(2005b). Black dots for the gaseous particles. Red dots for the stellar particles; red boxes for the ensemble of stellar particles (enclosed within the magenta frame) which are counter--rotating with v$_{\theta}~<~0$ within each topographical halo at a projected radius R$~>~$15~kpc. The size of such ensemble is also shown in each panel.}
\label{appE:sim1e12:fig3}
\end{center}
\end{figure}

\begin{figure}
\begin{center}
\includegraphics[width=1.0\textwidth]{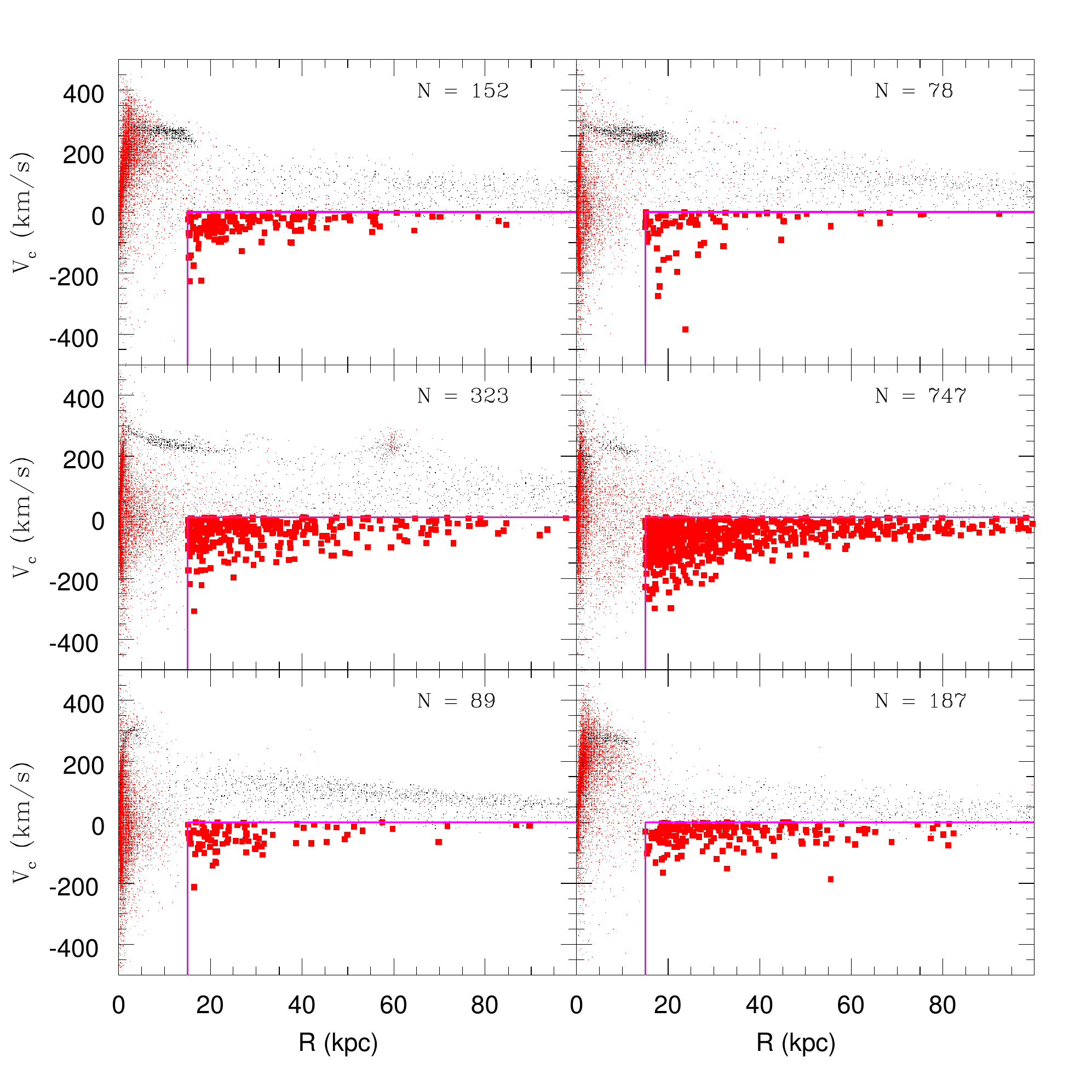}
\caption{(continued) Galaxy rotation curves for the M$_{\rm tot}$~=~10$^{12}$~M$_\odot$ semi--cosmological simulations in Renda~et~al.~(2005b). Black dots for the gaseous particles. Red dots for the stellar particles; red boxes for the ensemble of stellar particles (enclosed within the magenta frame) which are counter--rotating with v$_{\theta}~<~0$ within each topographical halo at a projected radius R$~>~$15~kpc. The size of such ensemble is also shown in each panel.}
\label{appE:sim1e12:fig4}
\end{center}
\end{figure}

\begin{figure}
\begin{center}
\includegraphics[width=1.0\textwidth]{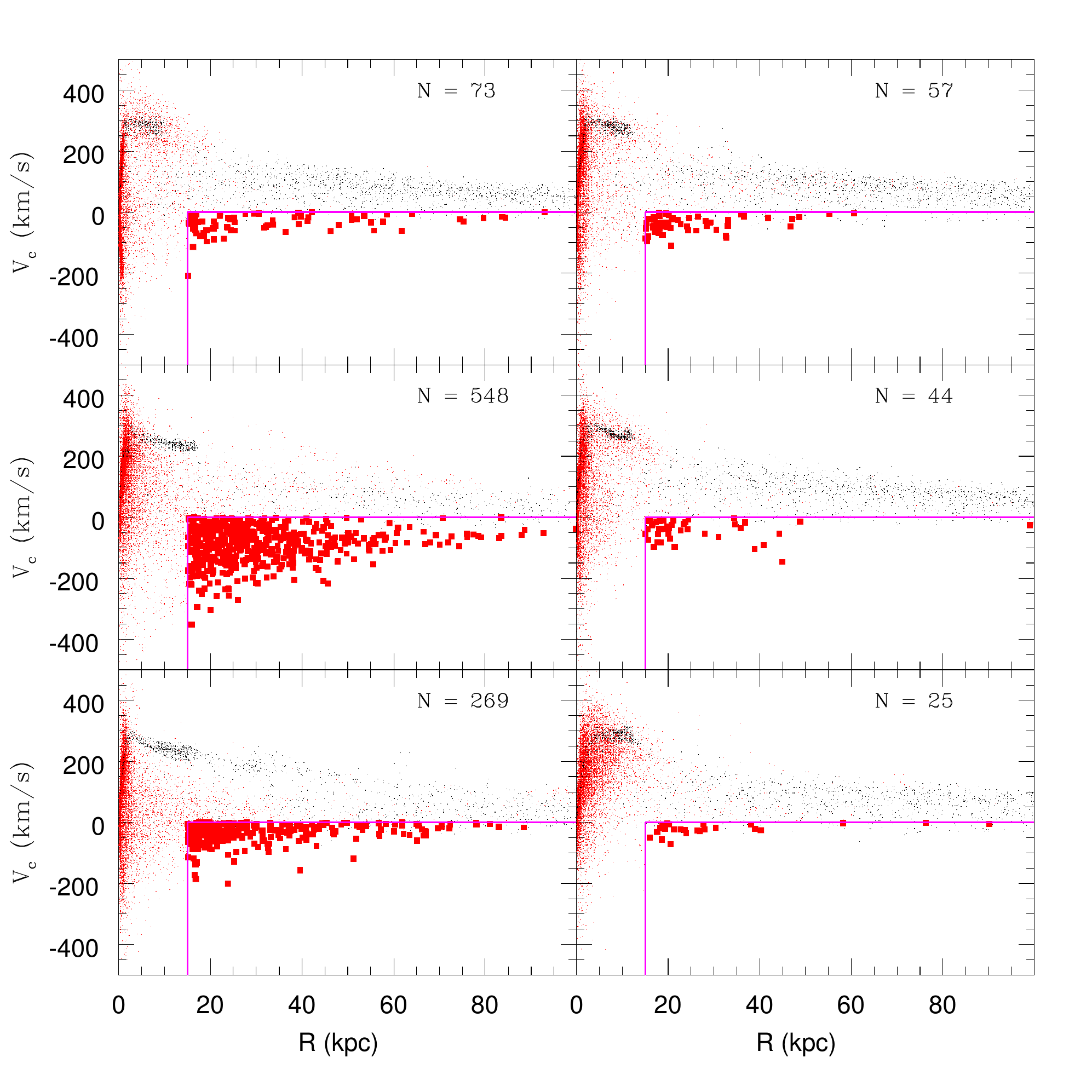}
\caption{(continued) Galaxy rotation curves for the M$_{\rm tot}$~=~10$^{12}$~M$_\odot$ semi--cosmological simulations in Renda~et~al.~(2005b). Black dots for the gaseous particles. Red dots for the stellar particles; red boxes for the ensemble of stellar particles (enclosed within the magenta frame) which are counter--rotating with v$_{\theta}~<~0$ within each topographical halo at a projected radius R$~>~$15~kpc. The size of such ensemble is also shown in each panel.}
\label{appE:sim1e12:fig5}
\end{center}
\end{figure}

\begin{figure}
\begin{center}
\includegraphics[width=1.0\textwidth]{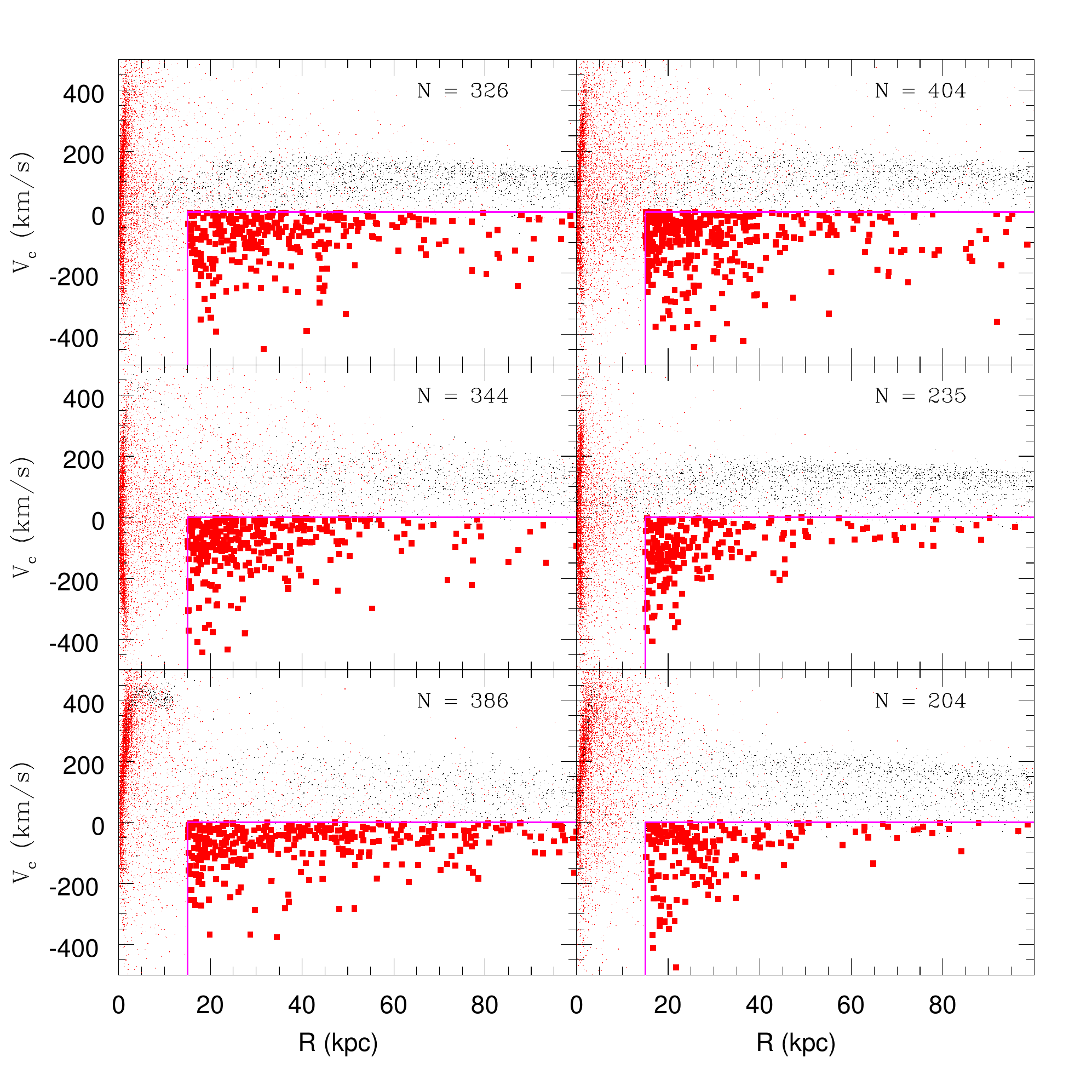}
\caption{Galaxy rotation curves for the M$_{\rm tot}$~=~5$\times$10$^{12}$~M$_\odot$ semi--cosmological simulations in Renda~et~al.~(2005b). Black dots for the gaseous particles. Red dots for the stellar particles; red boxes for the ensemble of stellar particles (enclosed within the magenta frame) which are counter--rotating with v$_{\theta}~<~0$ within each topographical halo at a projected radius R$~>~$15~kpc. The size of such ensemble is also shown in each panel.}
\label{appE:sim5e12:fig1}
\end{center}
\end{figure}

\begin{figure}
\begin{center}
\includegraphics[width=1.0\textwidth]{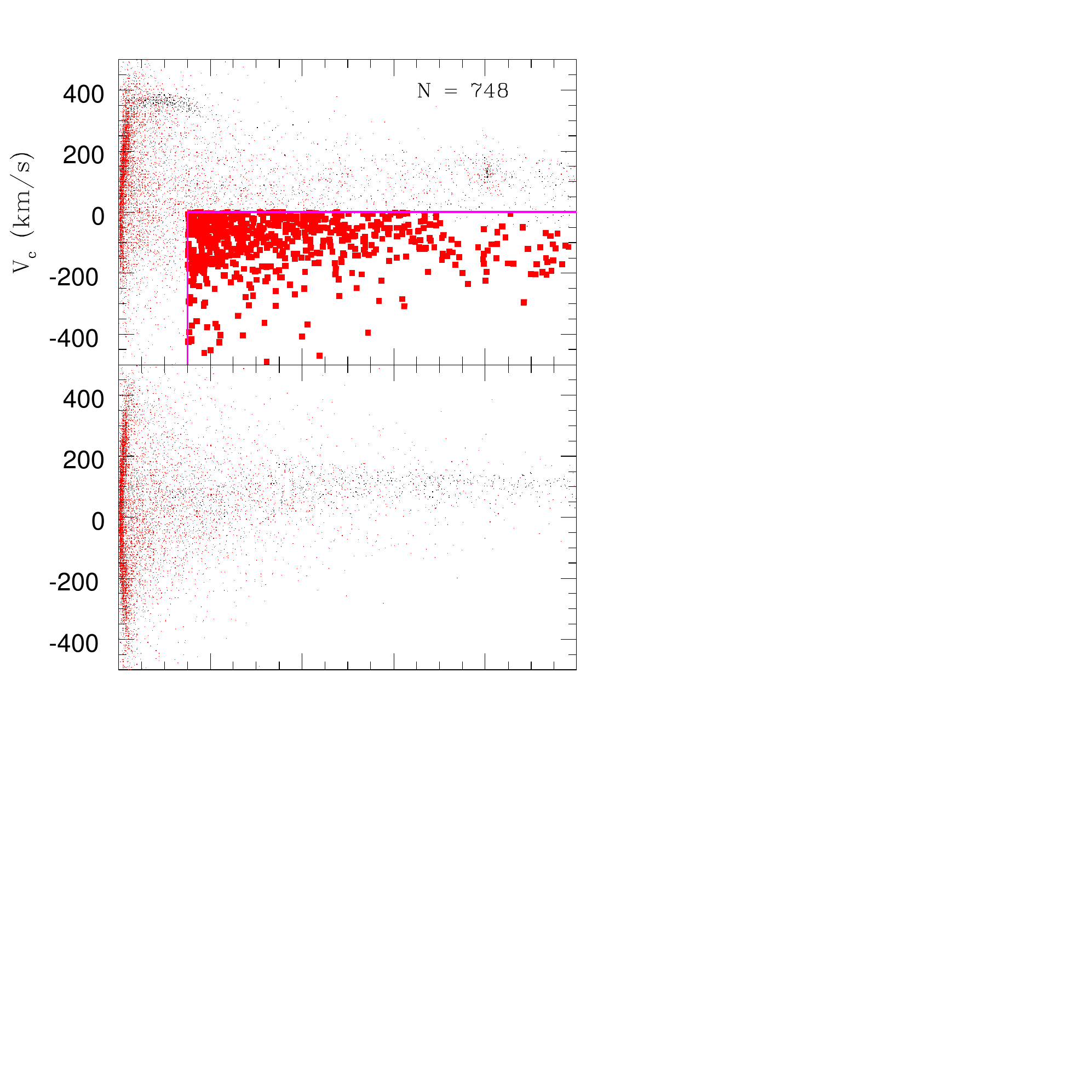}
\caption{(continued) Galaxy rotation curves for the M$_{\rm tot}$~=~5$\times$10$^{12}$~M$_\odot$ semi--cosmological simulations in Renda~et~al.~(2005b). Black dots for the gaseous particles. Red dots for the stellar particles; red boxes for the ensemble of stellar particles (enclosed within the magenta frame) which are counter--rotating with v$_{\theta}~<~0$ within each topographical halo at a projected radius R$~>~$15~kpc. The size of such ensemble is also shown in each panel.}
\label{appE:sim5e12:fig2}
\end{center}
\end{figure}

\begin{figure}
\begin{center}
\includegraphics[width=1.0\textwidth]{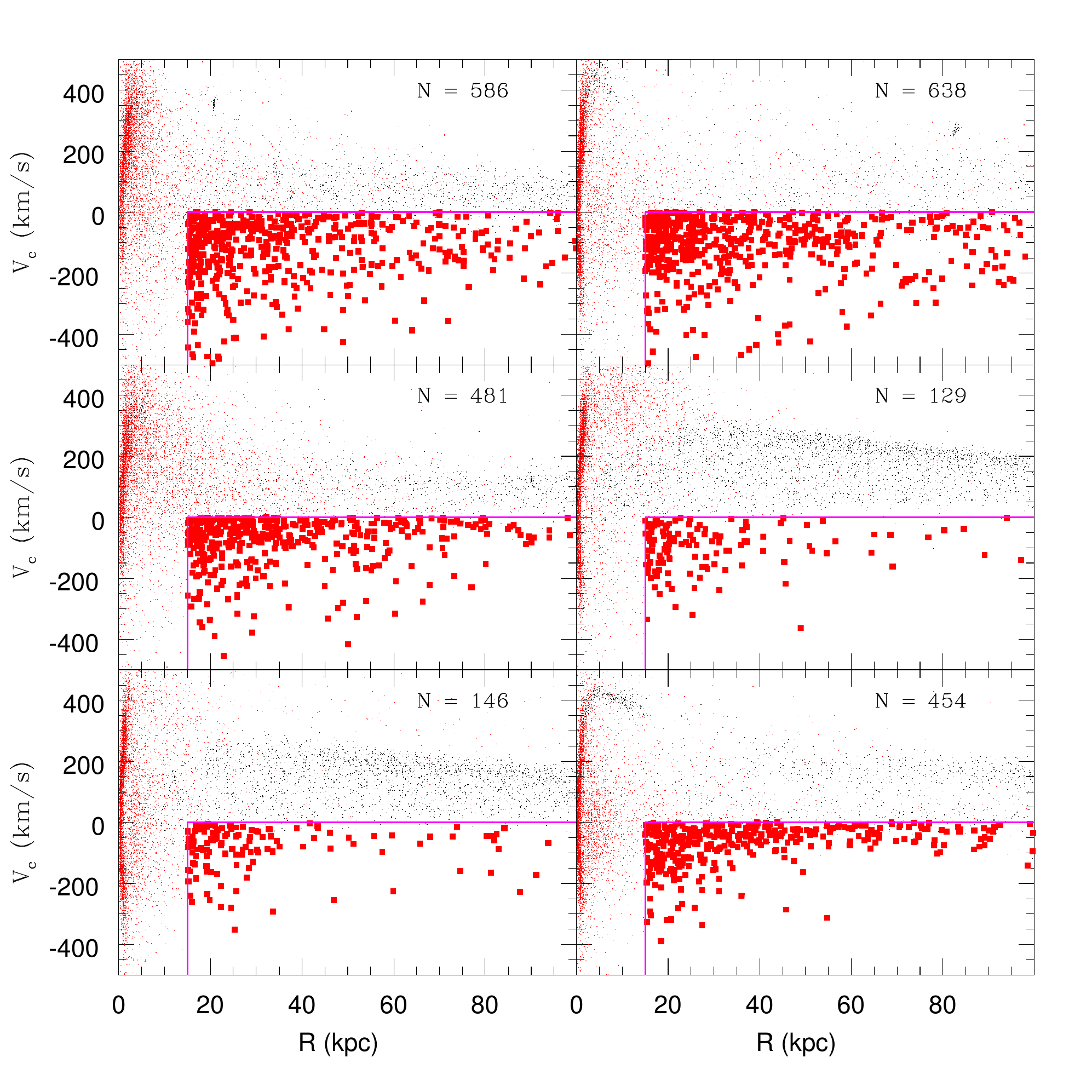}
\caption{(continued) Galaxy rotation curves for the M$_{\rm tot}$~=~5$\times$10$^{12}$~M$_\odot$ semi--cosmological simulations in Renda~et~al.~(2005b). Black dots for the gaseous particles. Red dots for the stellar particles; red boxes for the ensemble of stellar particles (enclosed within the magenta frame) which are counter--rotating with v$_{\theta}~<~0$ within each topographical halo at a projected radius R$~>~$15~kpc. The size of such ensemble is also shown in each panel.}
\label{appE:sim5e12:fig3}
\end{center}
\end{figure}

\begin{figure}
\begin{center}
\includegraphics[width=1.0\textwidth]{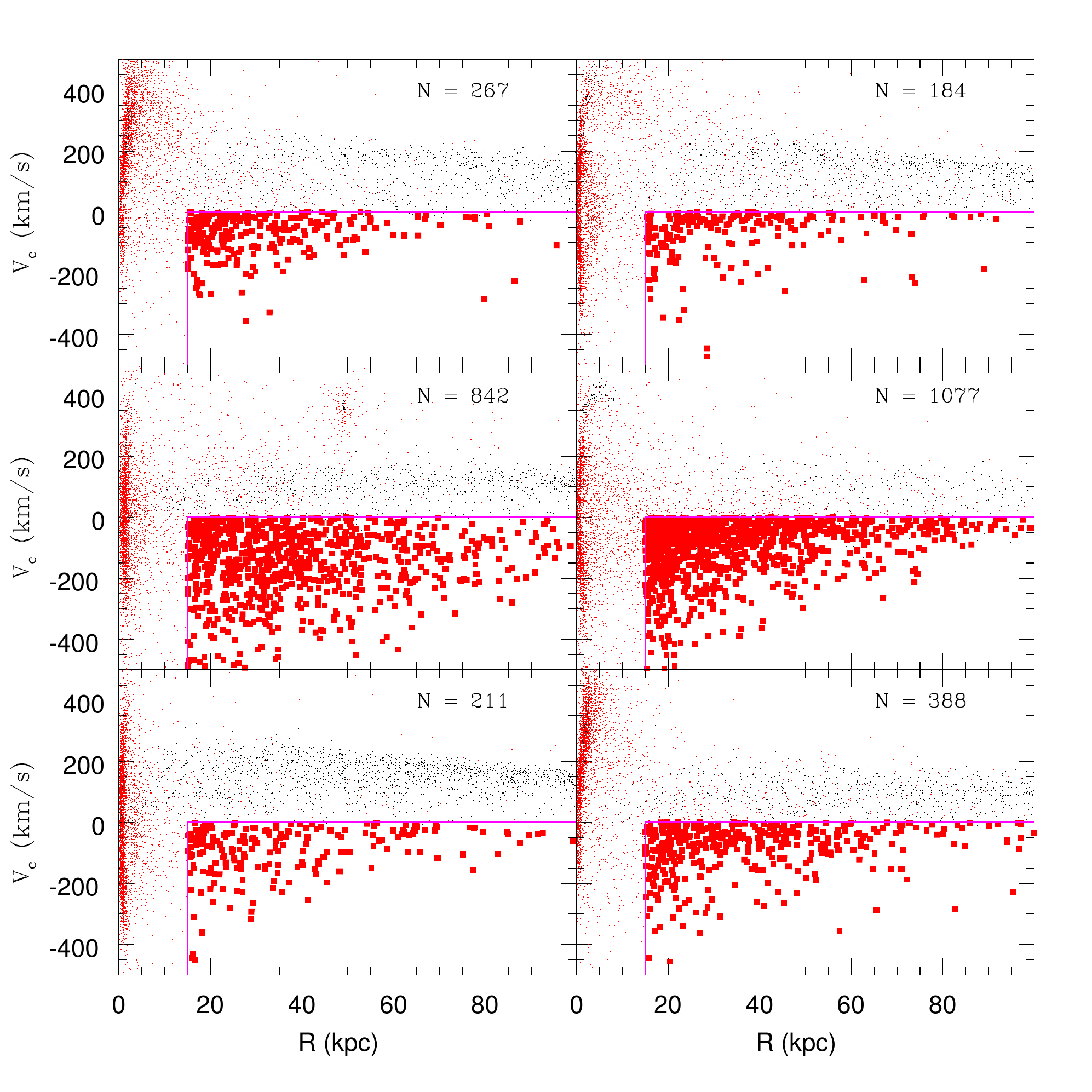}
\caption{(continued) Galaxy rotation curves for the M$_{\rm tot}$~=~5$\times$10$^{12}$~M$_\odot$ semi--cosmological simulations in Renda~et~al.~(2005b). Black dots for the gaseous particles. Red dots for the stellar particles; red boxes for the ensemble of stellar particles (enclosed within the magenta frame) which are counter--rotating with v$_{\theta}~<~0$ within each topographical halo at a projected radius R$~>~$15~kpc. The size of such ensemble is also shown in each panel.}
\label{appE:sim5e12:fig4}
\end{center}
\end{figure}

\begin{figure}
\begin{center}
\includegraphics[width=1.0\textwidth]{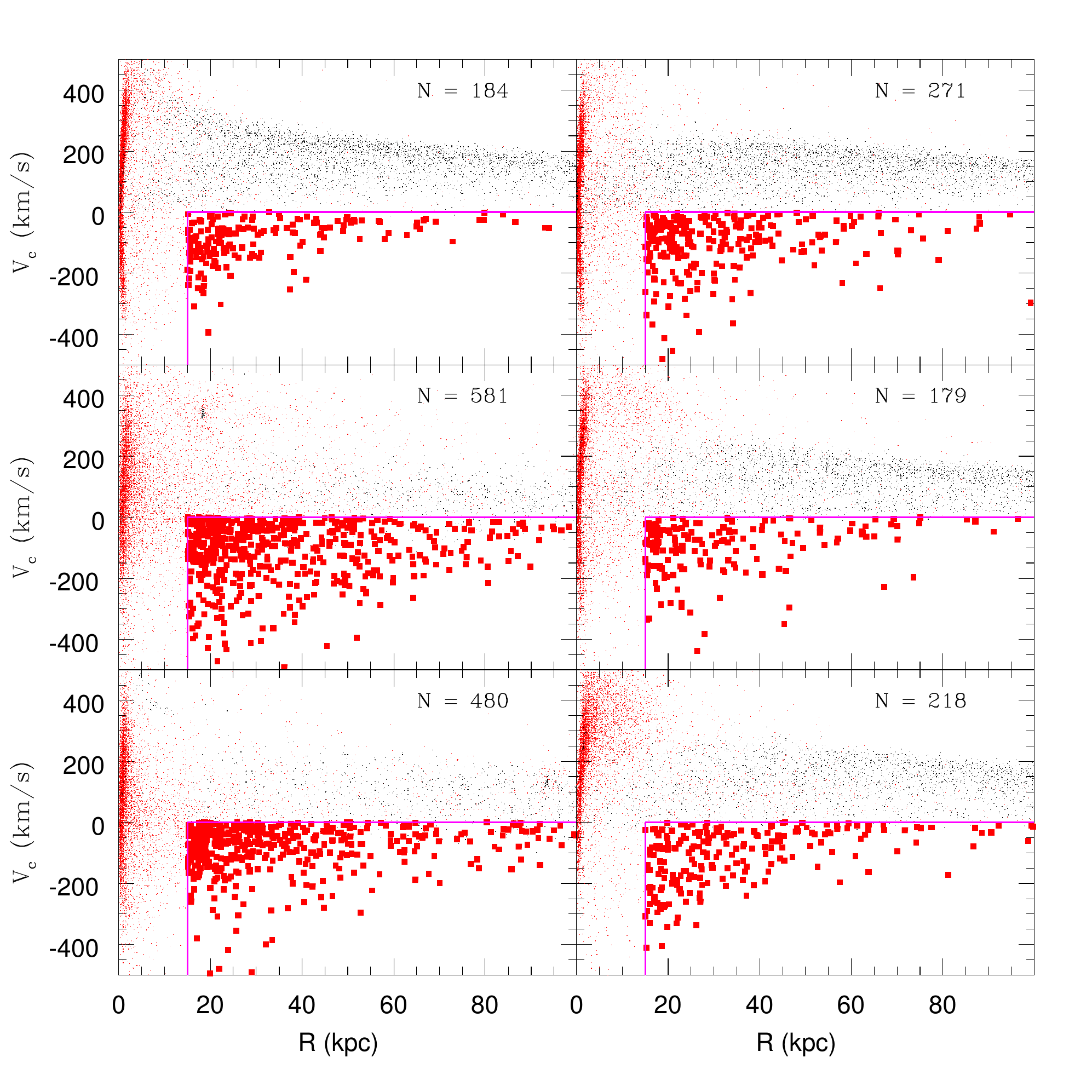}
\caption{(continued) Galaxy rotation curves for the M$_{\rm tot}$~=~5$\times$10$^{12}$~M$_\odot$ semi--cosmological simulations in Renda~et~al.~(2005b). Black dots for the gaseous particles. Red dots for the stellar particles; red boxes for the ensemble of stellar particles (enclosed within the magenta frame) which are counter--rotating with v$_{\theta}~<~0$ within each topographical halo at a projected radius R$~>~$15~kpc. The size of such ensemble is also shown in each panel.}
\label{appE:sim5e12:fig5}
\end{center}
\end{figure}




\newpage

\begin{center}
\chapter[(Once More) Stellar Halo MDF]{(Once More) Stellar Halo Metallicity Distributions}
\label{app:appendixF}
\end{center}

\begin{figure}
\begin{center}
\includegraphics[width=1.0\textwidth]{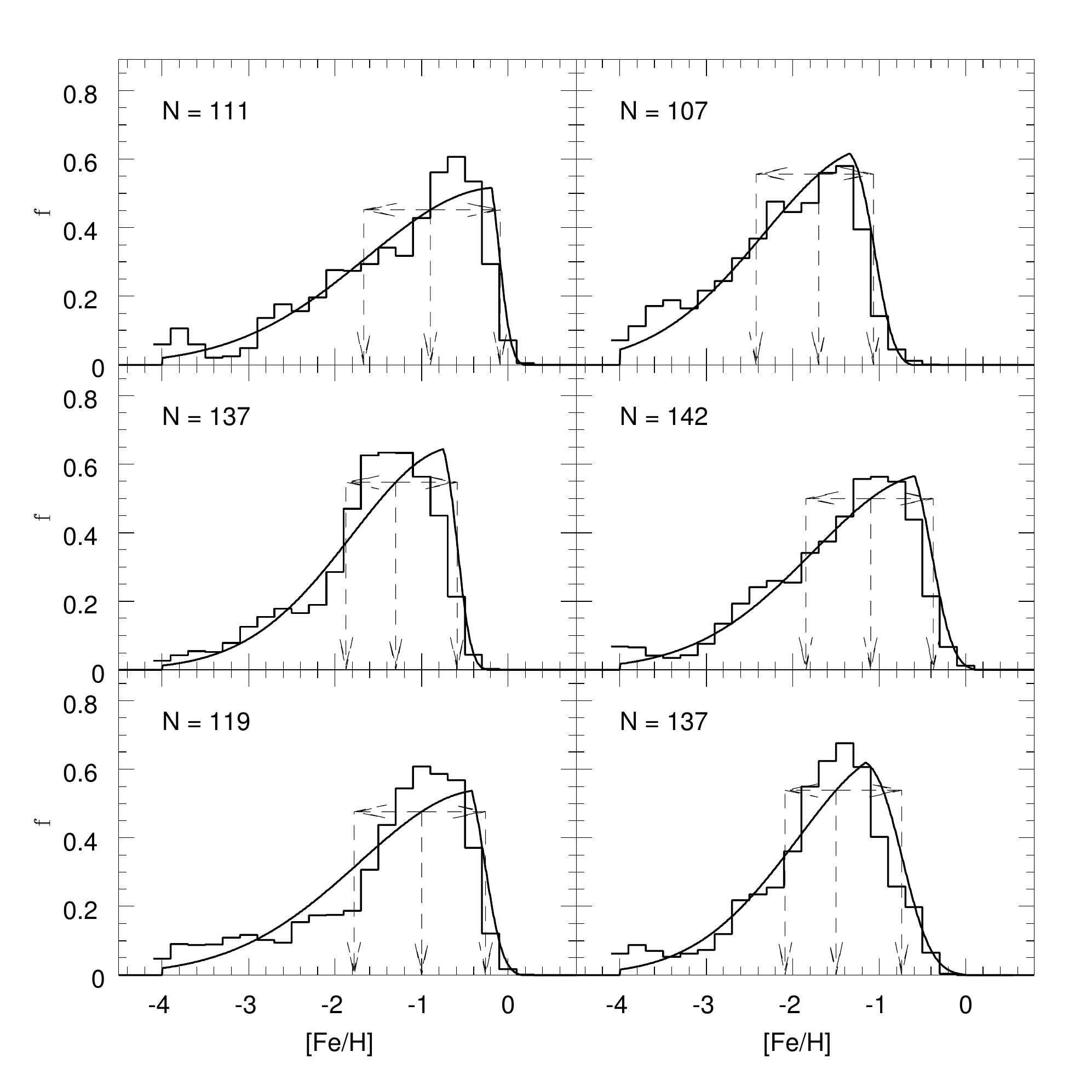}
\caption{Kinematically selected halo MDFs for the M$_{\rm tot}$~=~5$\times$10$^{11}$~M$_\odot$ semi--cosmological simulations in Renda~et~al.~(2005b). The 68\%~Confidence~Level range and the number of stellar particles each MDF relates to are also shown. Each MDF refers to the stellar particles in the simulation which are counter--rotating with v$_{\theta}~<~0$ at a projected radius R$~>~$15~kpc.}
\label{appF:sim5e11:fig1}
\end{center}
\end{figure}

\begin{figure}
\begin{center}
\includegraphics[width=1.0\textwidth]{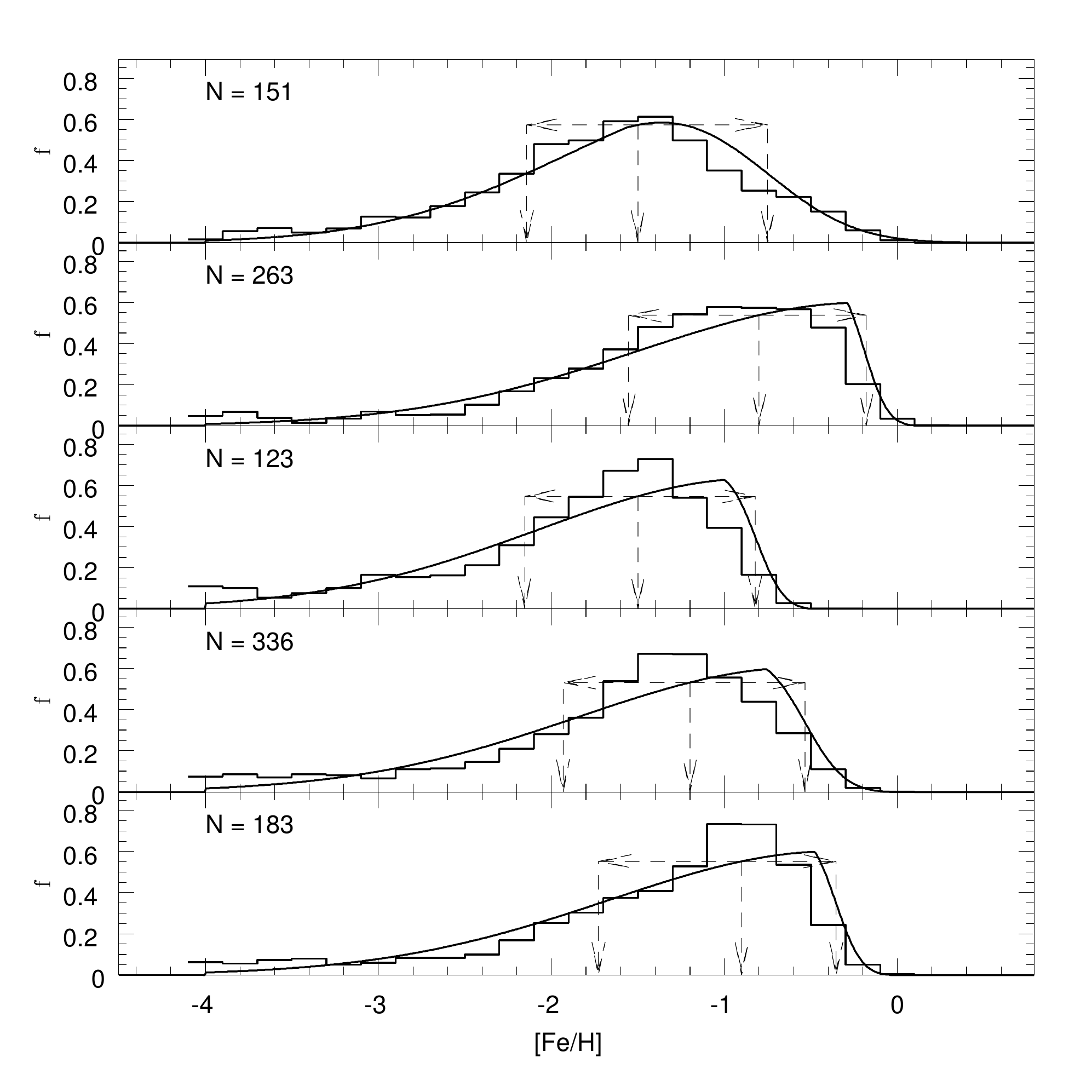}
\caption{(continued) Kinematically selected halo MDFs for the M$_{\rm tot}$~=~5$\times$10$^{11}$~M$_\odot$ semi--cosmological simulations in Renda~et~al.~(2005b). The 68\%~Confidence~Level range and the number of stellar particles each MDF relates to are also shown. Each MDF refers to the stellar particles in the simulation which are counter--rotating with v$_{\theta}~<~0$ at a projected radius R$~>~$15~kpc.}
\label{appF:sim5e11:fig2}
\end{center}
\end{figure}

\begin{figure}
\begin{center}
\includegraphics[width=1.0\textwidth]{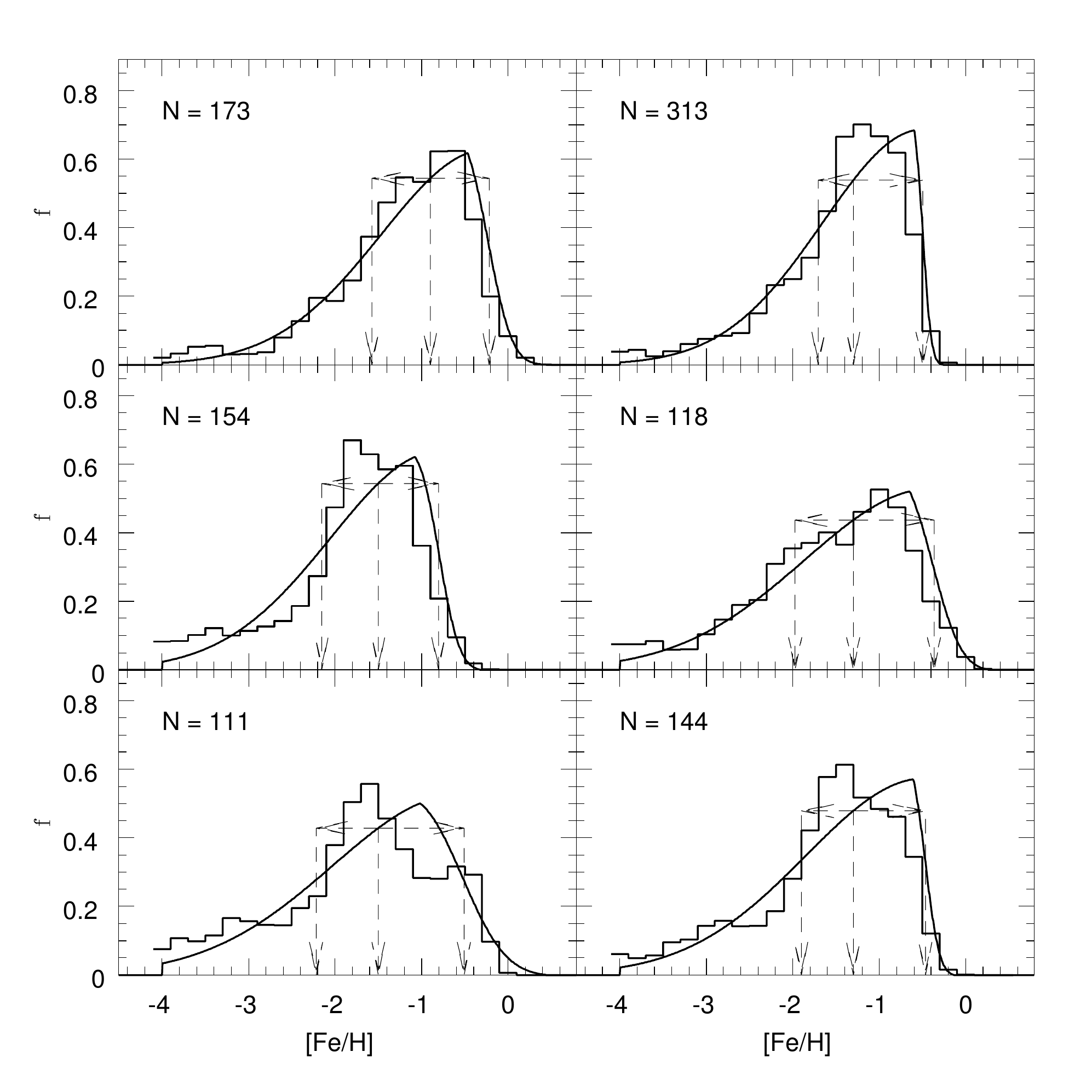}
\caption{Kinematically selected halo MDFs for the M$_{\rm tot}$~=~10$^{12}$~M$_\odot$ semi--cosmological simulations in Renda~et~al.~(2005b). The 68\%~Confidence~Level range and the number of stellar particles each MDF relates to are also shown. Each MDF refers to the stellar particles in the simulation which are counter--rotating with v$_{\theta}~<~0$ at a projected radius R$~>~$15~kpc.}
\label{appF:sim1e12:fig1}
\end{center}
\end{figure}

\begin{figure}
\begin{center}
\includegraphics[width=1.0\textwidth]{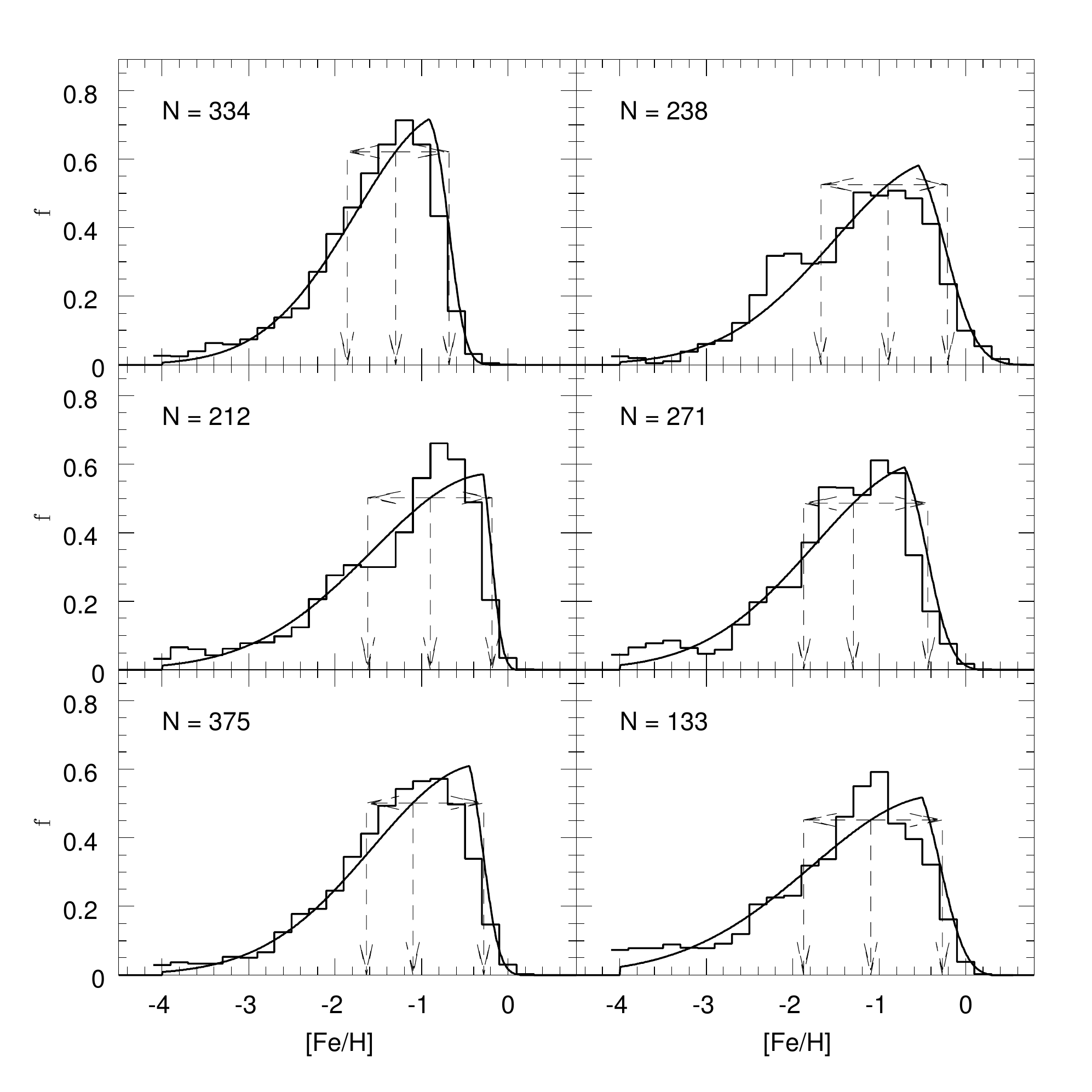}
\caption{(continued) Kinematically selected halo MDFs for the M$_{\rm tot}$~=~10$^{12}$~M$_\odot$ semi--cosmological simulations in Renda~et~al.~(2005b). The 68\%~Confidence~Level range and the number of stellar particles each MDF relates to are also shown. Each MDF refers to the stellar particles in the simulation which are counter--rotating with v$_{\theta}~<~0$ at a projected radius R$~>~$15~kpc.}
\label{appF:sim1e12:fig2}
\end{center}
\end{figure}

\begin{figure}
\begin{center}
\includegraphics[width=1.0\textwidth]{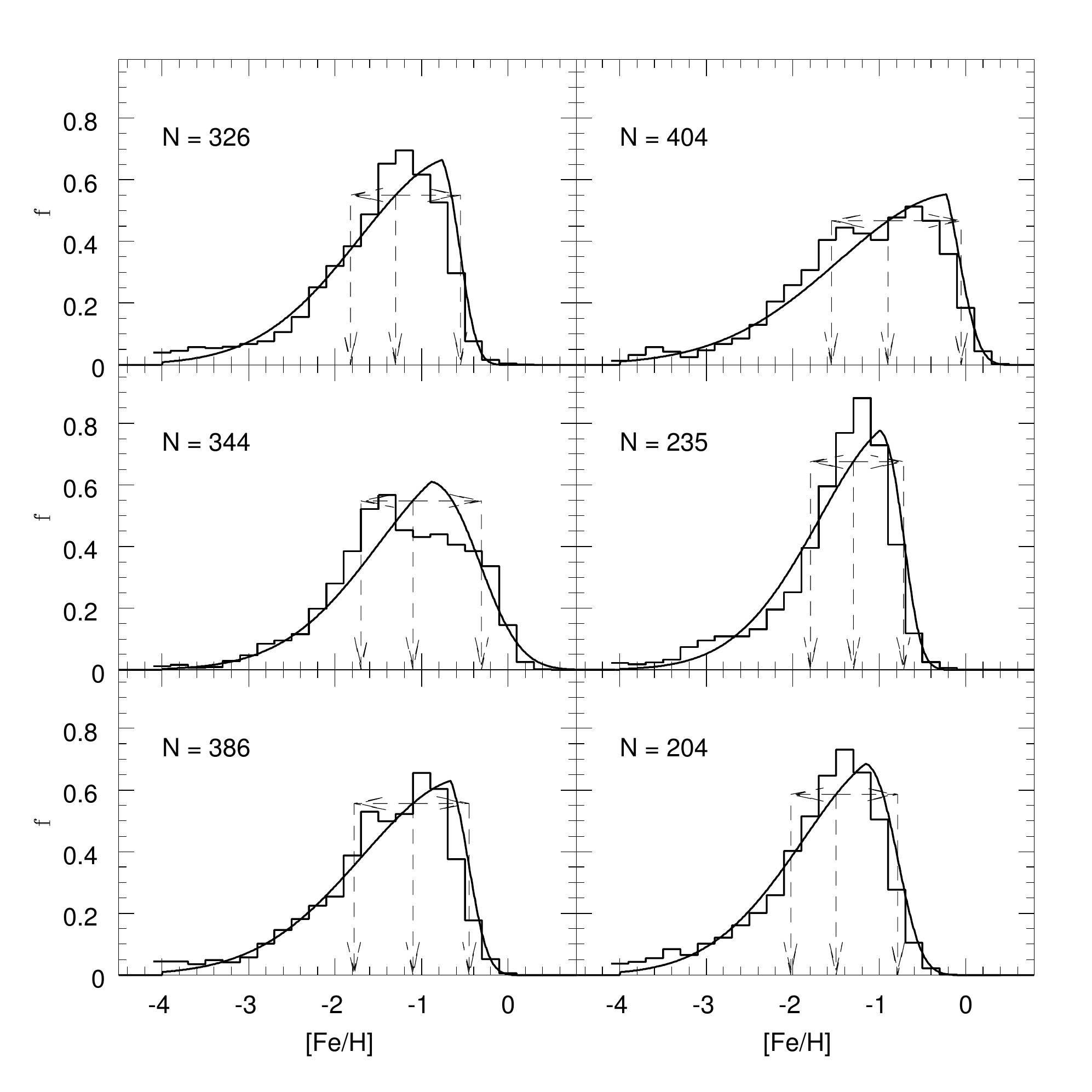}
\caption{Kinematically selected halo MDFs for the M$_{\rm tot}$~=~5$\times$10$^{12}$~M$_\odot$ semi--cosmological simulations in Renda~et~al.~(2005b). The 68\%~Confidence~Level range and the number of stellar particles each MDF relates to are also shown. Each MDF refers to the stellar particles in the simulation which are counter--rotating with v$_{\theta}~<~0$ at a projected radius R$~>~$15~kpc.}
\label{appF:sim5e12:fig1}
\end{center}
\end{figure}

\begin{figure}
\begin{center}
\includegraphics[width=1.0\textwidth]{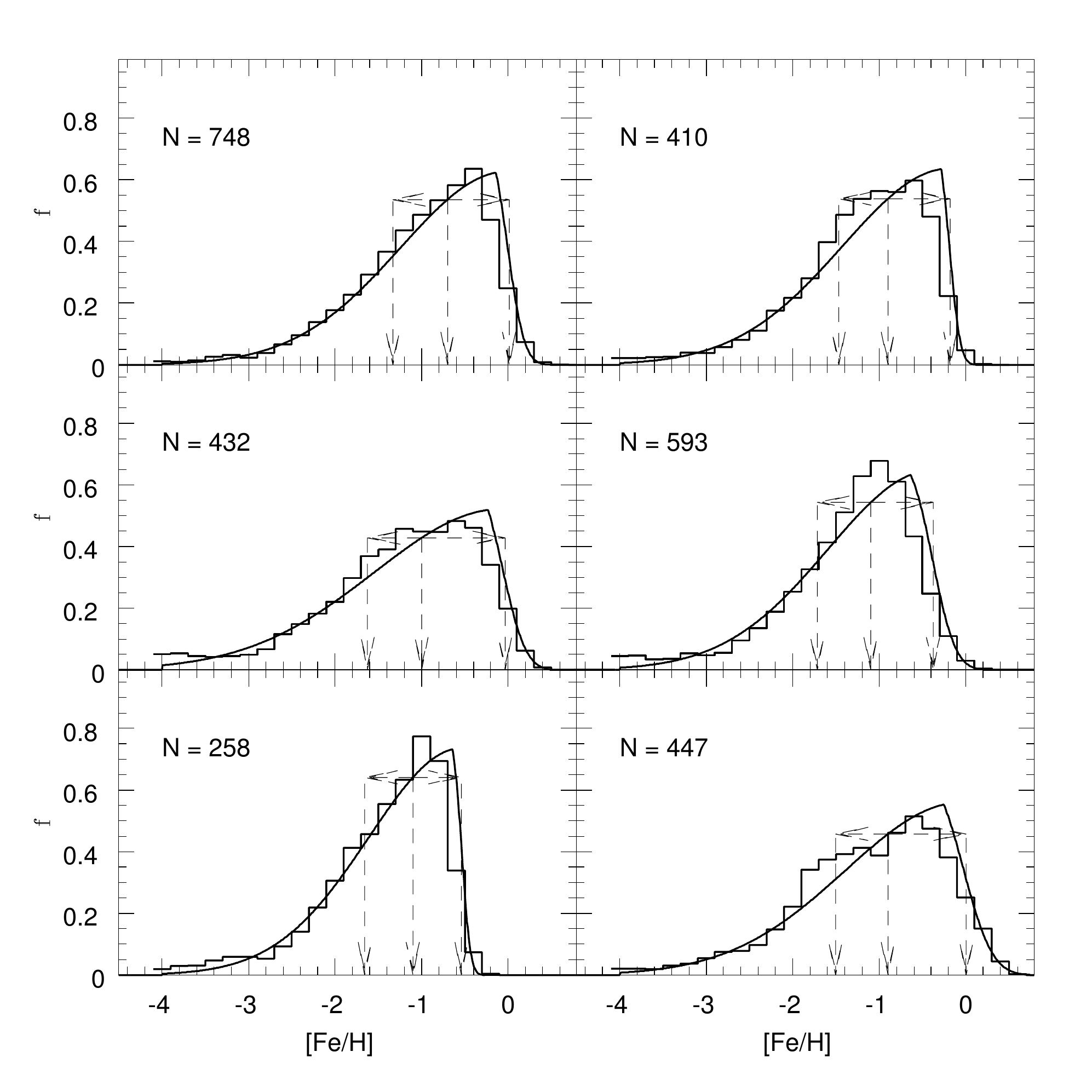}
\caption{(continued) Kinematically selected halo MDFs for the M$_{\rm tot}$~=~5$\times$10$^{12}$~M$_\odot$ semi--cosmological simulations in Renda~et~al.~(2005b). The 68\%~Confidence~Level range and the number of stellar particles each MDF relates to are also shown. Each MDF refers to the stellar particles in the simulation which are counter--rotating with v$_{\theta}~<~0$ at a projected radius R$~>~$15~kpc.}
\label{appF:sim5e12:fig2}
\end{center}
\end{figure}

\begin{figure}
\begin{center}
\includegraphics[width=1.0\textwidth]{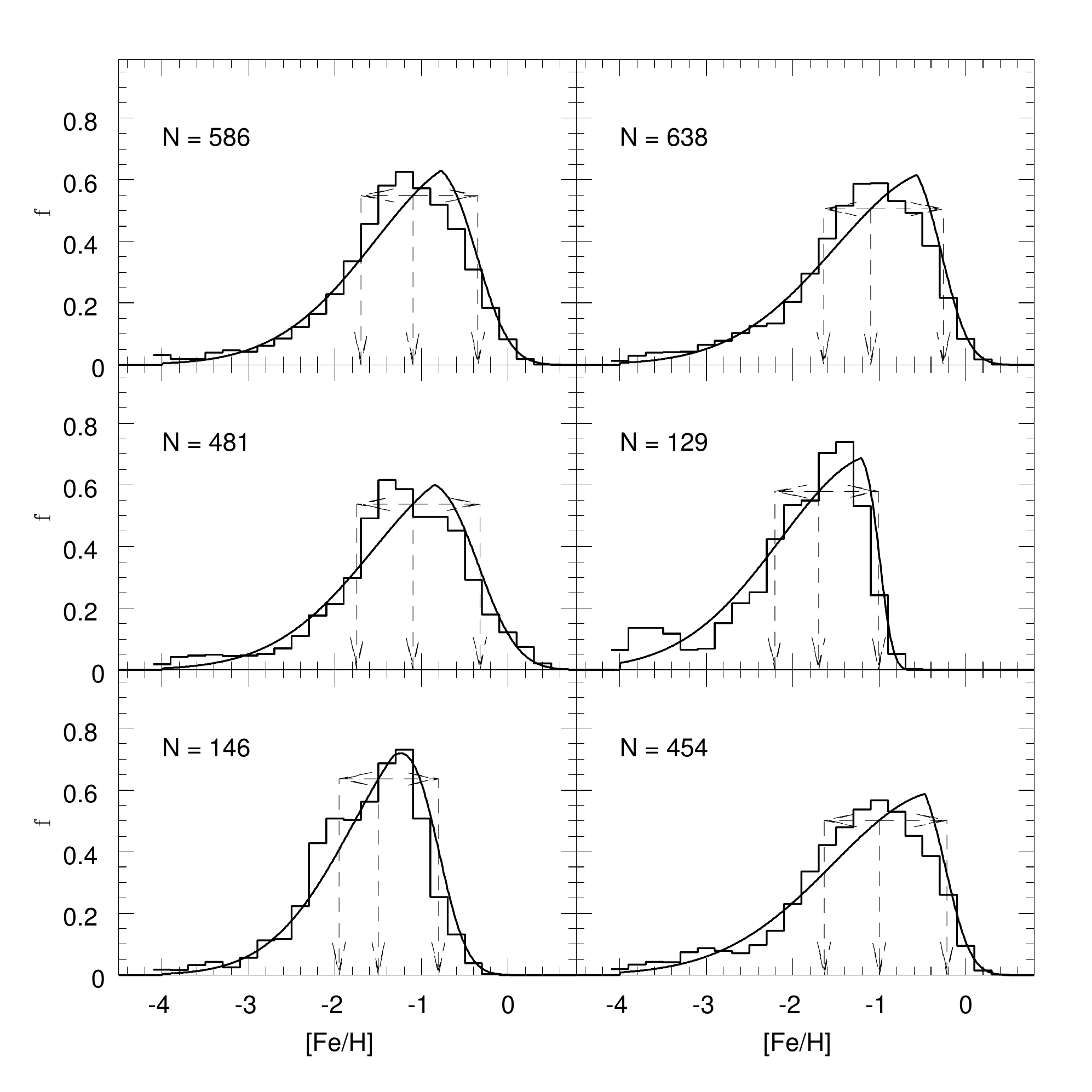}
\caption{(continued) Kinematically selected halo MDFs for the M$_{\rm tot}$~=~5$\times$10$^{12}$~M$_\odot$ semi--cosmological simulations in Renda~et~al.~(2005b). The 68\%~Confidence~Level range and the number of stellar particles each MDF relates to are also shown. Each MDF refers to the stellar particles in the simulation which are counter--rotating with v$_{\theta}~<~0$ at a projected radius R$~>~$15~kpc.}
\label{appF:sim5e12:fig3}
\end{center}
\end{figure}

\begin{figure}
\begin{center}
\includegraphics[width=1.0\textwidth]{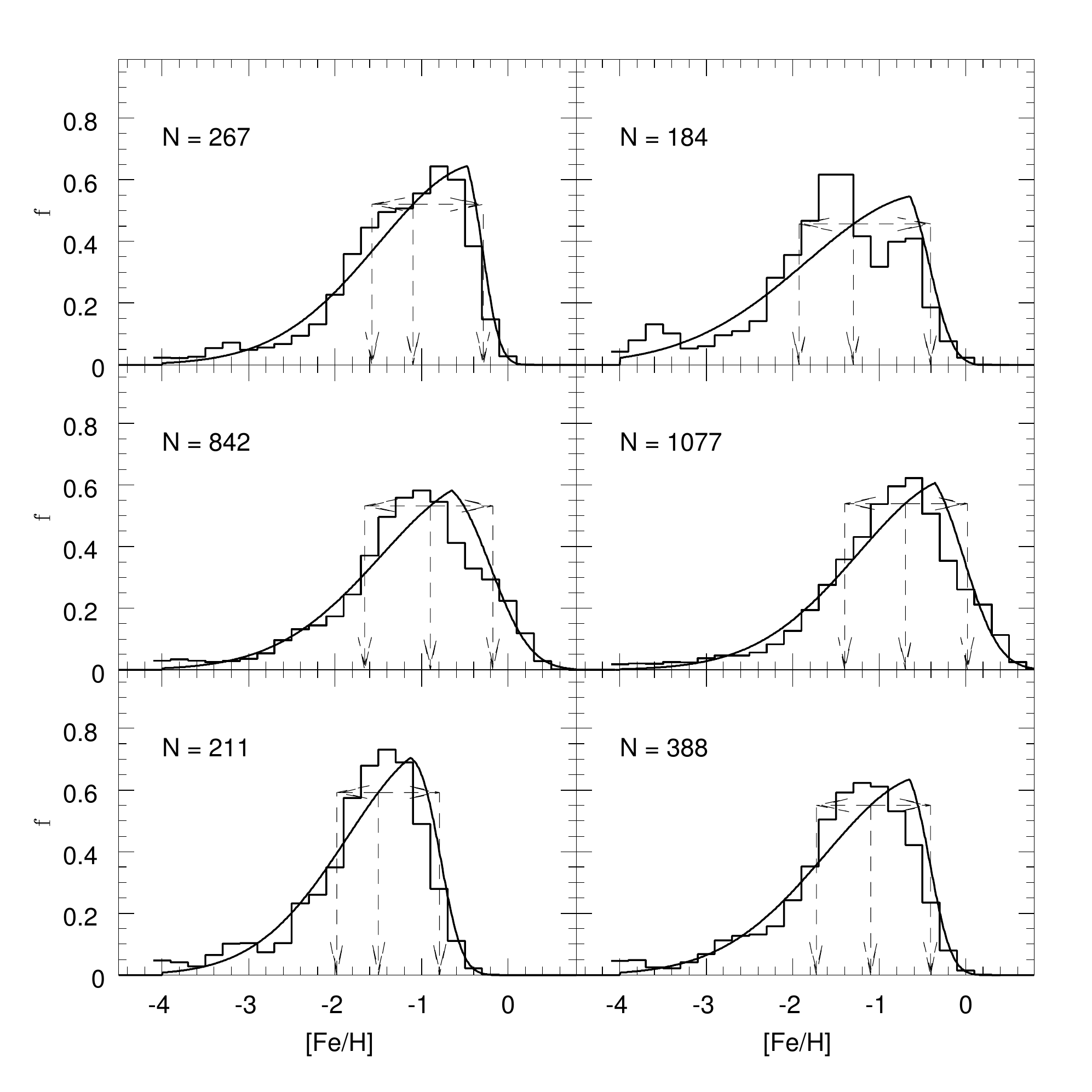}
\caption{(continued) Kinematically selected halo MDFs for the M$_{\rm tot}$~=~5$\times$10$^{12}$~M$_\odot$ semi--cosmological simulations in Renda~et~al.~(2005b). The 68\%~Confidence~Level range and the number of stellar particles each MDF relates to are also shown. Each MDF refers to the stellar particles in the simulation which are counter--rotating with v$_{\theta}~<~0$ at a projected radius R$~>~$15~kpc.}
\label{appF:sim5e12:fig4}
\end{center}
\end{figure}

\begin{figure}
\begin{center}
\includegraphics[width=1.0\textwidth]{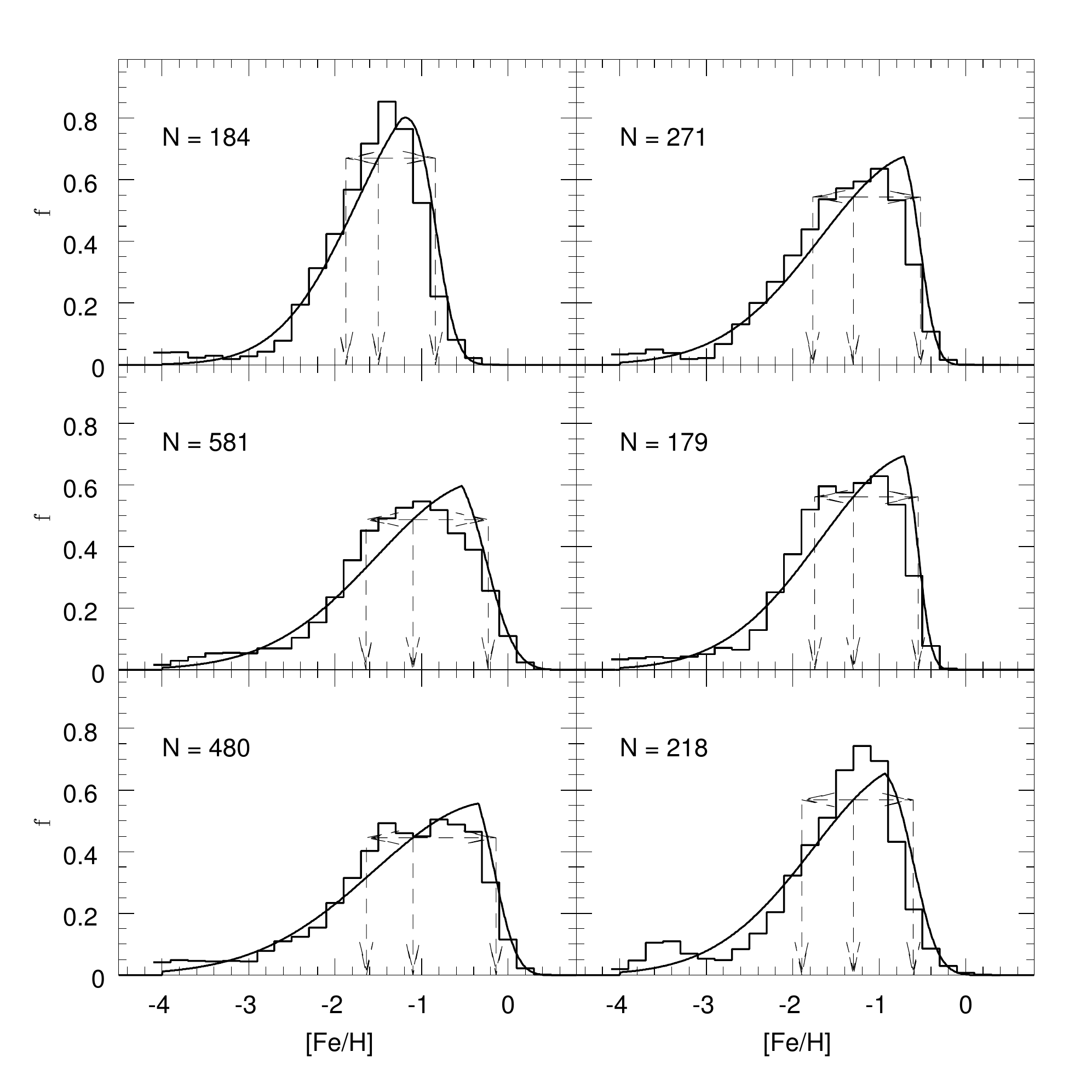}
\caption{(continued) Kinematically selected halo MDFs for the M$_{\rm tot}$~=~5$\times$10$^{12}$~M$_\odot$ semi--cosmological simulations in Renda~et~al.~(2005b). The 68\%~Confidence~Level range and the number of stellar particles each MDF relates to are also shown. Each MDF refers to the stellar particles in the simulation which are counter--rotating with v$_{\theta}~<~0$ at a projected radius R$~>~$15~kpc.}
\label{appF:sim5e12:fig5}
\end{center}
\end{figure}




\newpage 

\begin{center}
\chapter{Stellar Halo (V~-~I) Colour Distributions}
\label{app:appendixG}
\end{center}

\begin{figure}
\begin{center}
\includegraphics[width=1.0\textwidth]{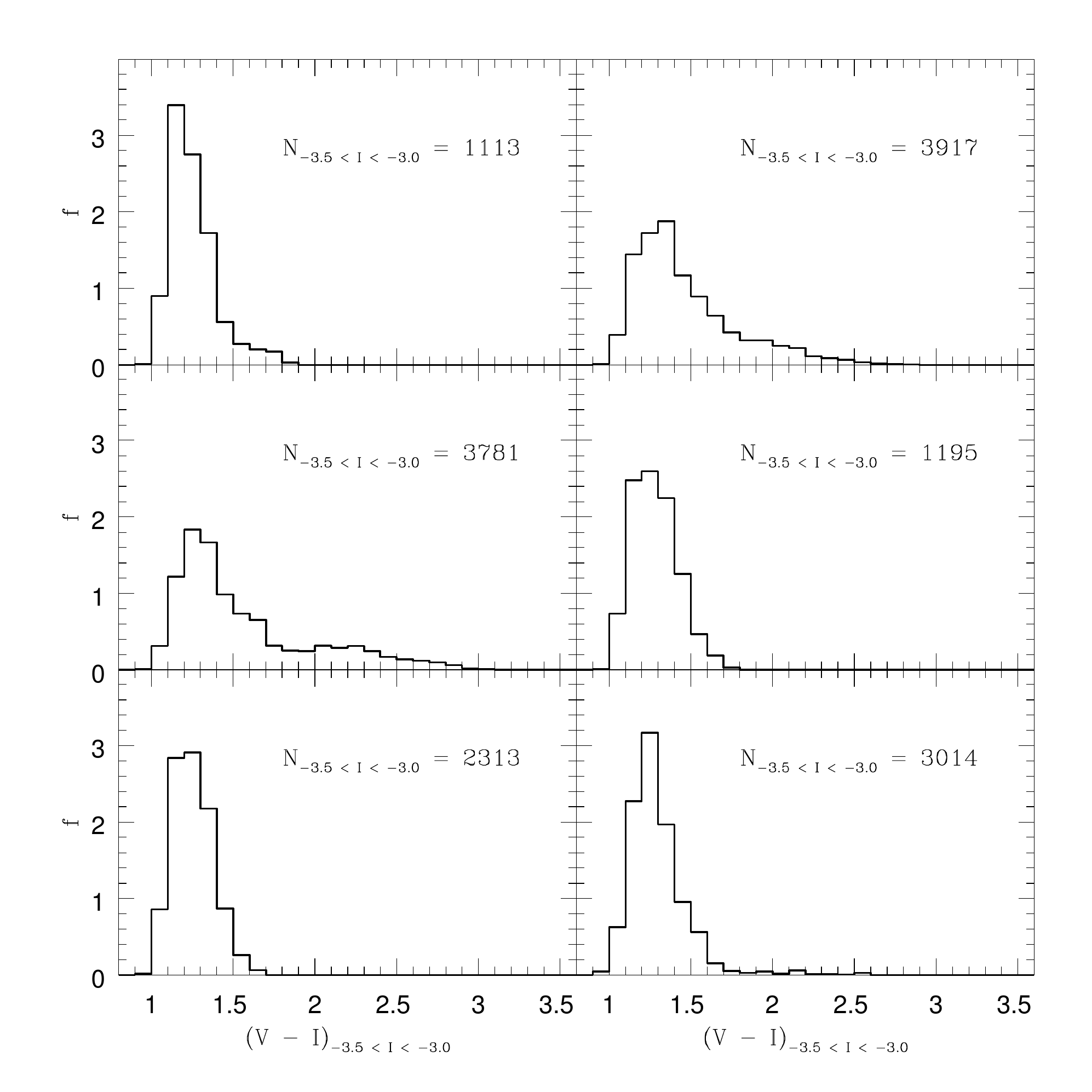}
\caption{Halo synthetic CMD (V~-~I) colour distribution at -3.5~$<$~I~$<$~-3.0 for the M$_{\rm tot}$~=~5$\times$10$^{11}$~M$_\odot$ semi--cosmological simulations in Renda~et~al.~(2005b). Each panel also shows the number of generated stars at -3.5~$<$~I~$<$~-3.0 in each synthetic CMD.}
\label{appG:sim5e11:fig1}
\end{center}
\end{figure}

\begin{figure}
\begin{center}
\includegraphics[width=1.0\textwidth]{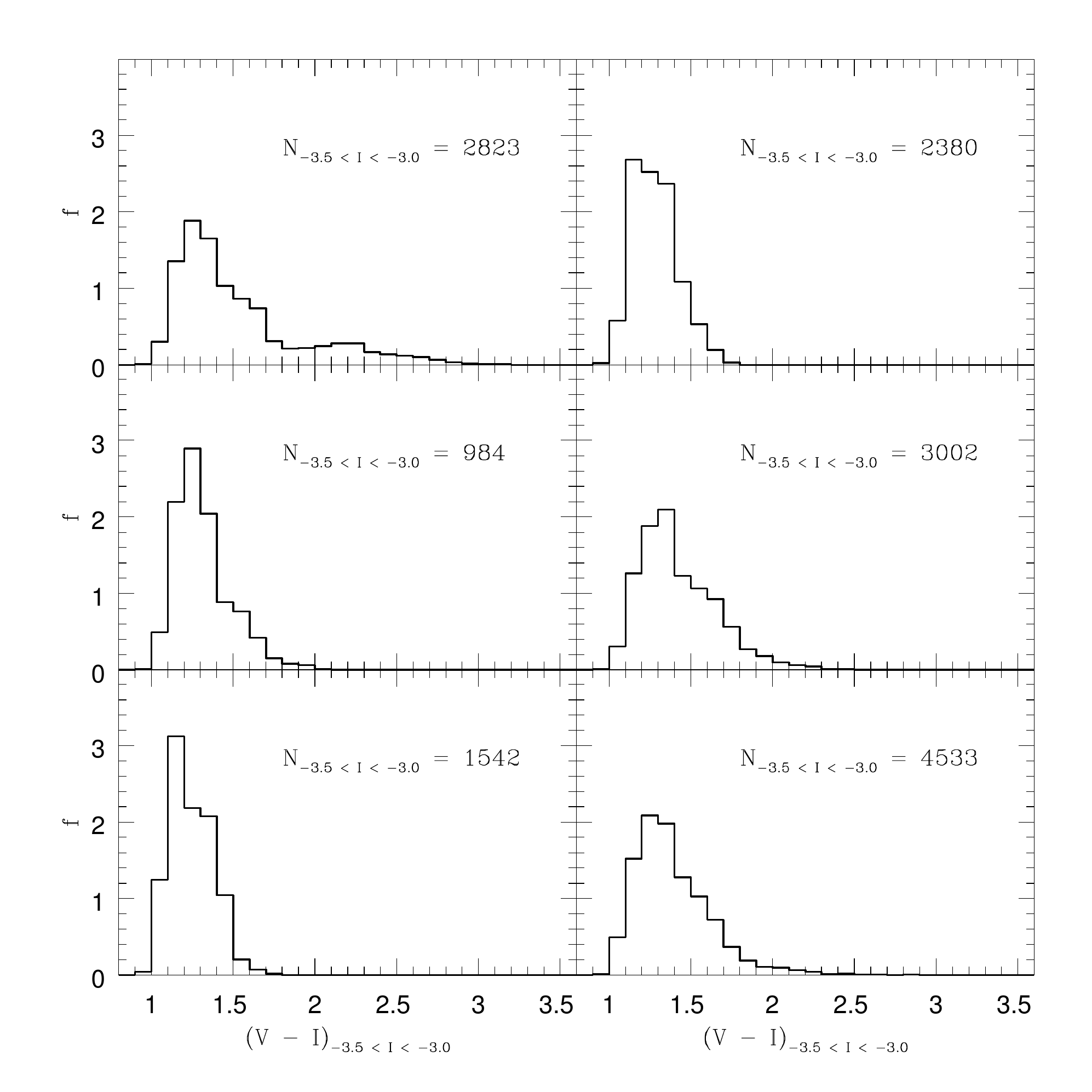}
\caption{(continued) Halo synthetic CMD (V~-~I) colour distribution at -3.5~$<$~I~$<$~-3.0 for the M$_{\rm tot}$~=~5$\times$10$^{11}$~M$_\odot$ semi--cosmological simulations in Renda~et~al.~(2005b). Each panel also shows the number of generated stars at -3.5~$<$~I~$<$~-3.0 in each synthetic CMD.}
\label{appG:sim5e11:fig2}
\end{center}
\end{figure}

\begin{figure}
\begin{center}
\includegraphics[width=1.0\textwidth]{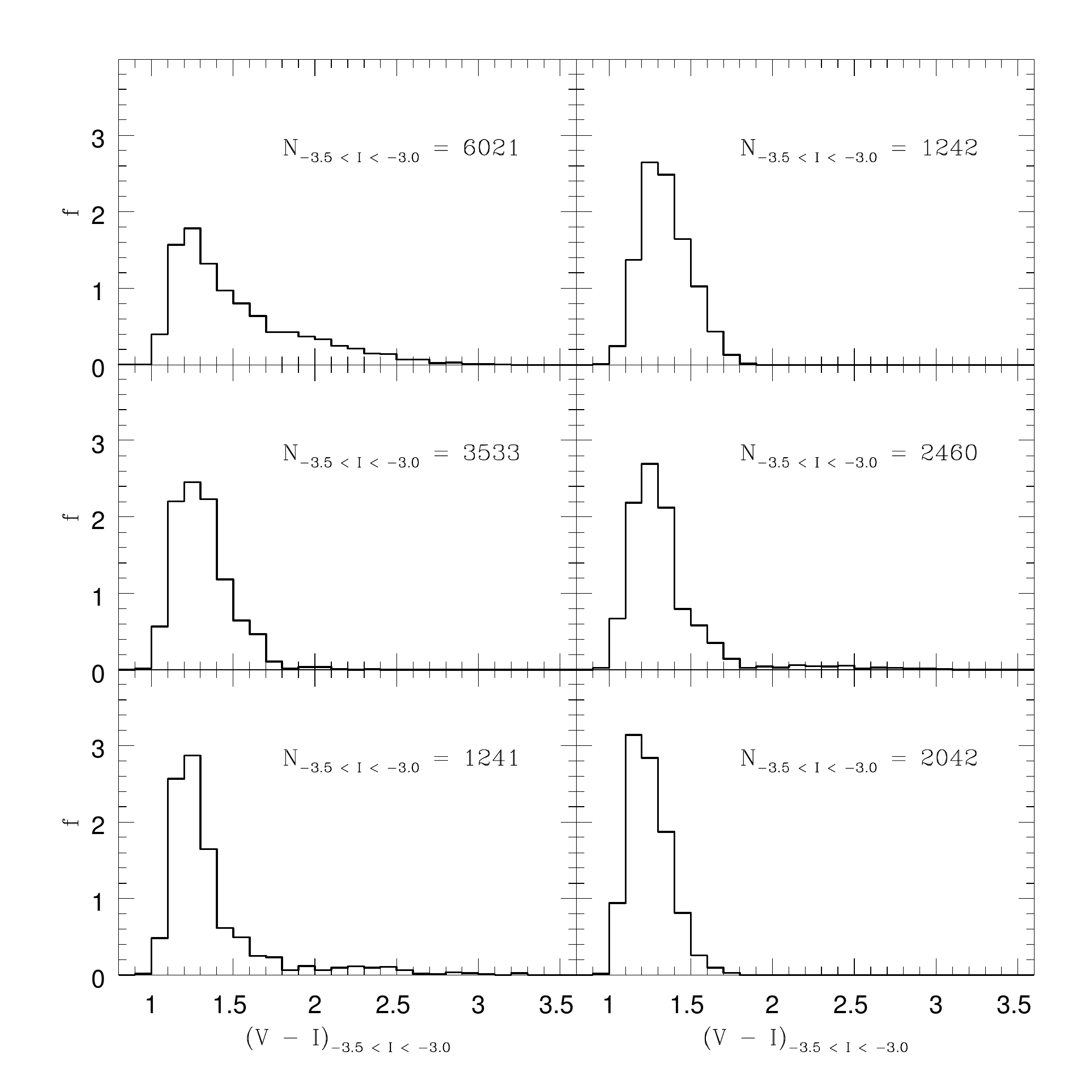}
\caption{(continued) Halo synthetic CMD (V~-~I) colour distribution at -3.5~$<$~I~$<$~-3.0 for the M$_{\rm tot}$~=~5$\times$10$^{11}$~M$_\odot$ semi--cosmological simulations in Renda~et~al.~(2005b). Each panel also shows the number of generated stars at -3.5~$<$~I~$<$~-3.0 in each synthetic CMD.}
\label{appG:sim5e11:fig3}
\end{center}
\end{figure}

\begin{figure}
\begin{center}
\includegraphics[width=1.0\textwidth]{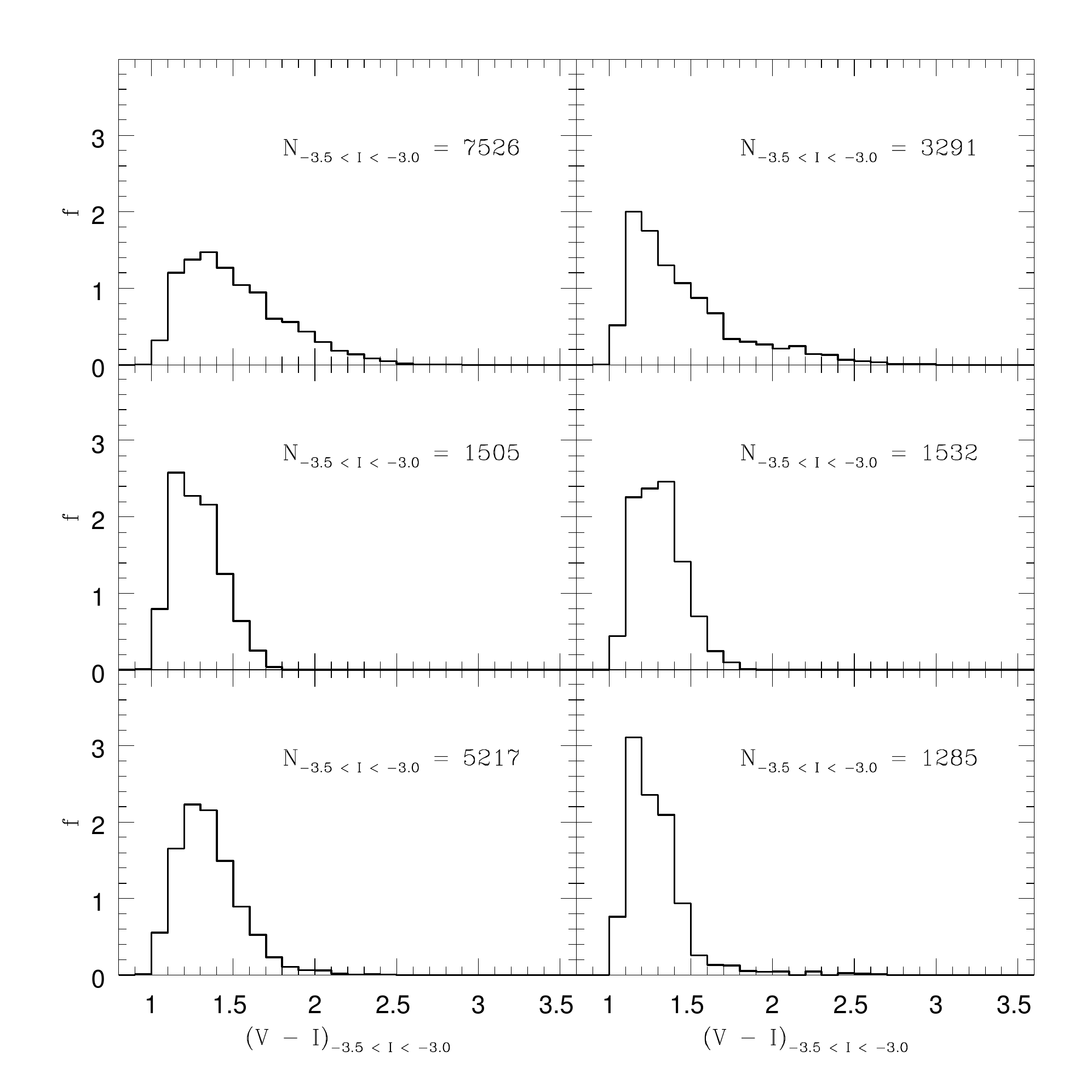}
\caption{(continued) Halo synthetic CMD (V~-~I) colour distribution at -3.5~$<$~I~$<$~-3.0 for the M$_{\rm tot}$~=~5$\times$10$^{11}$~M$_\odot$ semi--cosmological simulations in Renda~et~al.~(2005b). Each panel also shows the number of generated stars at -3.5~$<$~I~$<$~-3.0 in each synthetic CMD.}
\label{appG:sim5e11:fig4}
\end{center}
\end{figure}

\begin{figure}
\begin{center}
\includegraphics[width=1.0\textwidth]{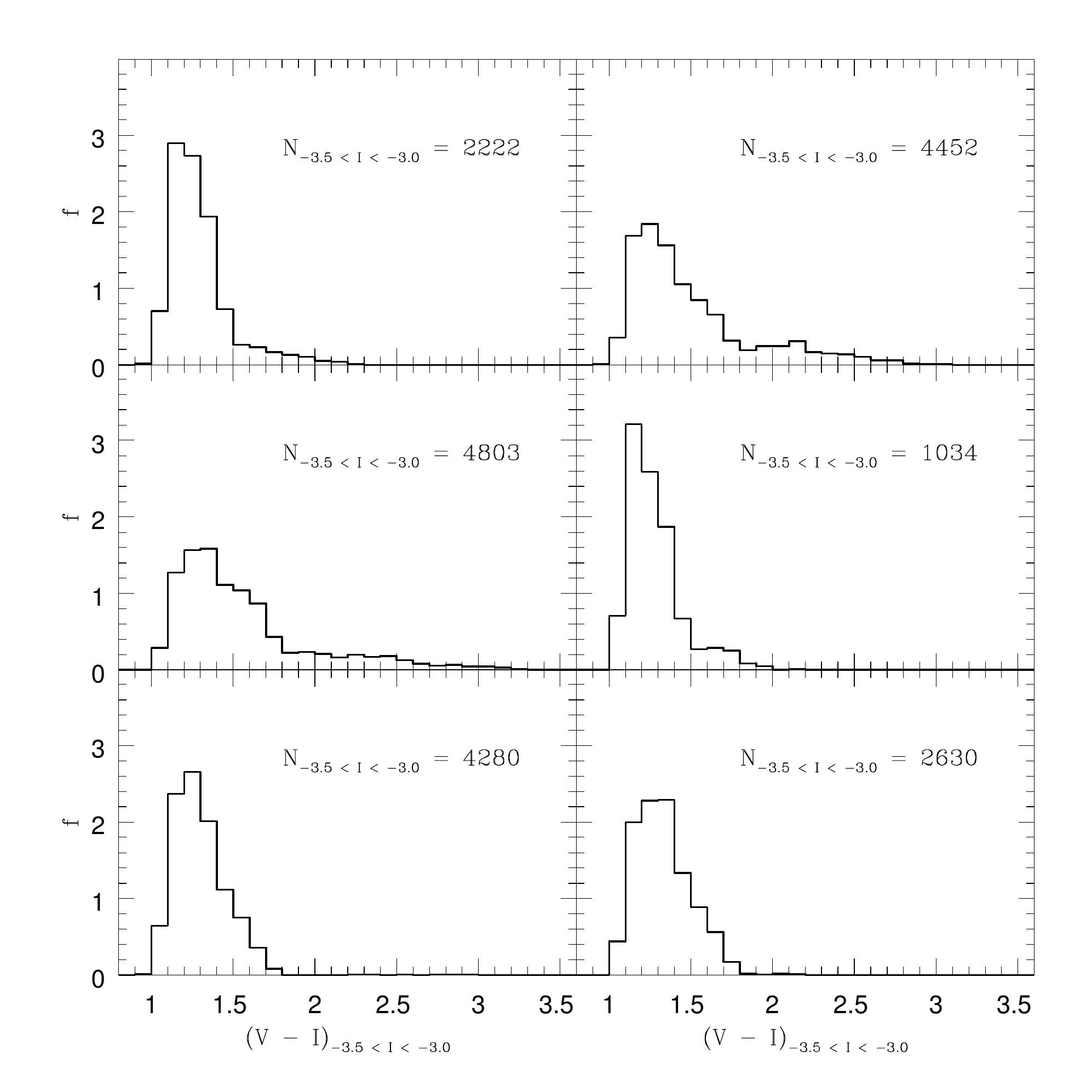}
\caption{Halo synthetic CMD (V~-~I) colour distribution at -3.5~$<$~I~$<$~-3.0 for the M$_{\rm tot}$~=~1$\times$10$^{12}$~M$_\odot$ semi--cosmological simulations in Renda~et~al.~(2005b). Each panel also shows the number of generated stars at -3.5~$<$~I~$<$~-3.0 in each synthetic CMD.}
\label{appG:sim1e12:fig1}
\end{center}
\end{figure}

\begin{figure}
\begin{center}
\includegraphics[width=1.0\textwidth]{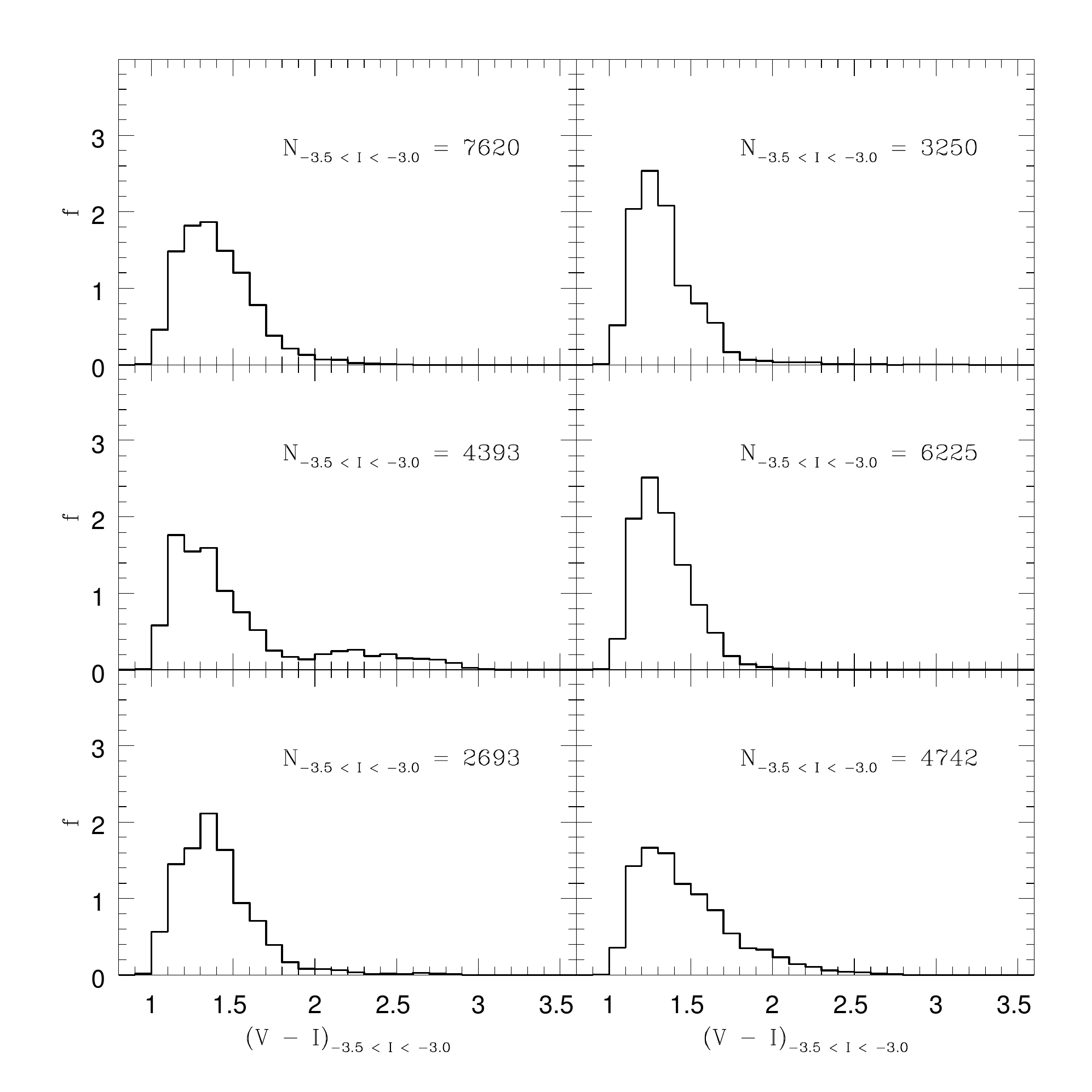}
\caption{(continued) Halo synthetic CMD (V~-~I) colour distribution at -3.5~$<$~I~$<$~-3.0 for the M$_{\rm tot}$~=~10$^{12}$~M$_\odot$ semi--cosmological simulations in Renda~et~al.~(2005b). Each panel also shows the number of generated stars at -3.5~$<$~I~$<$~-3.0 in each synthetic CMD.}
\label{appG:sim1e12:fig2}
\end{center}
\end{figure}

\begin{figure}
\begin{center}
\includegraphics[width=1.0\textwidth]{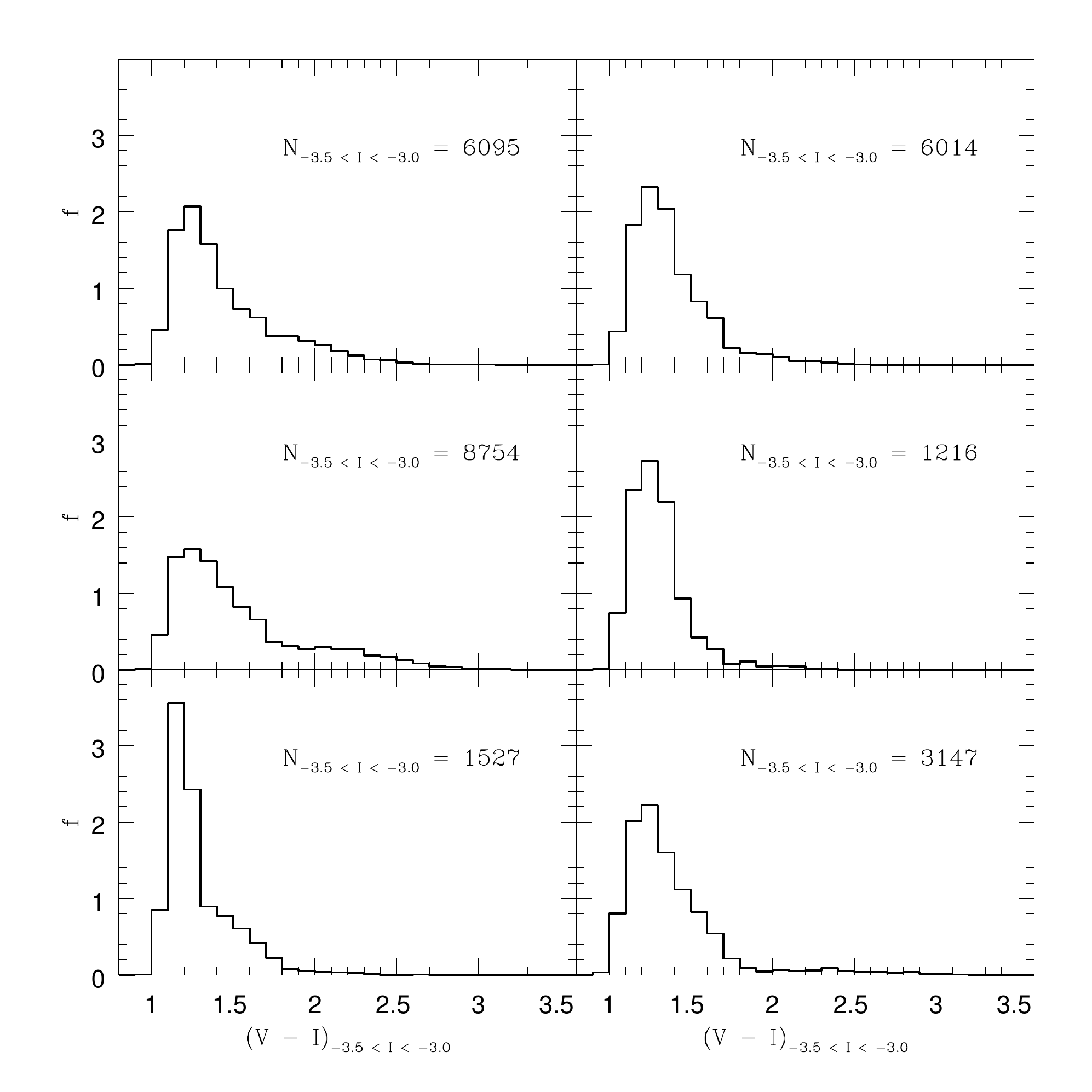}
\caption{(continued) Halo synthetic CMD (V~-~I) colour distribution at -3.5~$<$~I~$<$~-3.0 for the M$_{\rm tot}$~=~10$^{12}$~M$_\odot$ semi--cosmological simulations in Renda~et~al.~(2005b). Each panel also shows the number of generated stars at -3.5~$<$~I~$<$~-3.0 in each synthetic CMD.}
\label{appG:sim1e12:fig3}
\end{center}
\end{figure}

\begin{figure}
\begin{center}
\includegraphics[width=1.0\textwidth]{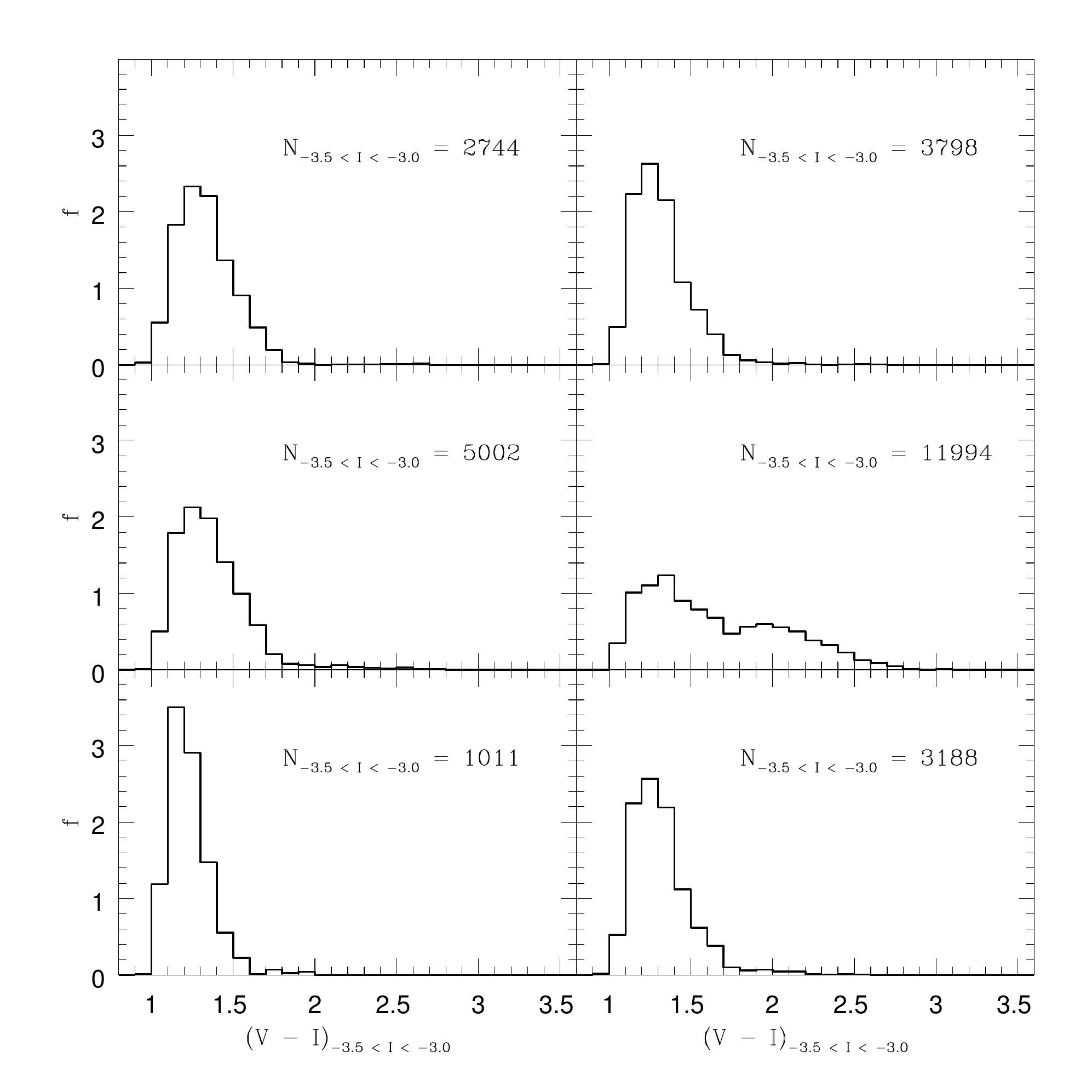}
\caption{(continued) Halo synthetic CMD (V~-~I) colour distribution at -3.5~$<$~I~$<$~-3.0 for the M$_{\rm tot}$~=~10$^{12}$~M$_\odot$ semi--cosmological simulations in Renda~et~al.~(2005b). Each panel also shows the number of generated stars at -3.5~$<$~I~$<$~-3.0 in each synthetic CMD.}
\label{appG:sim1e12:fig4}
\end{center}
\end{figure}

\begin{figure}
\begin{center}
\includegraphics[width=1.0\textwidth]{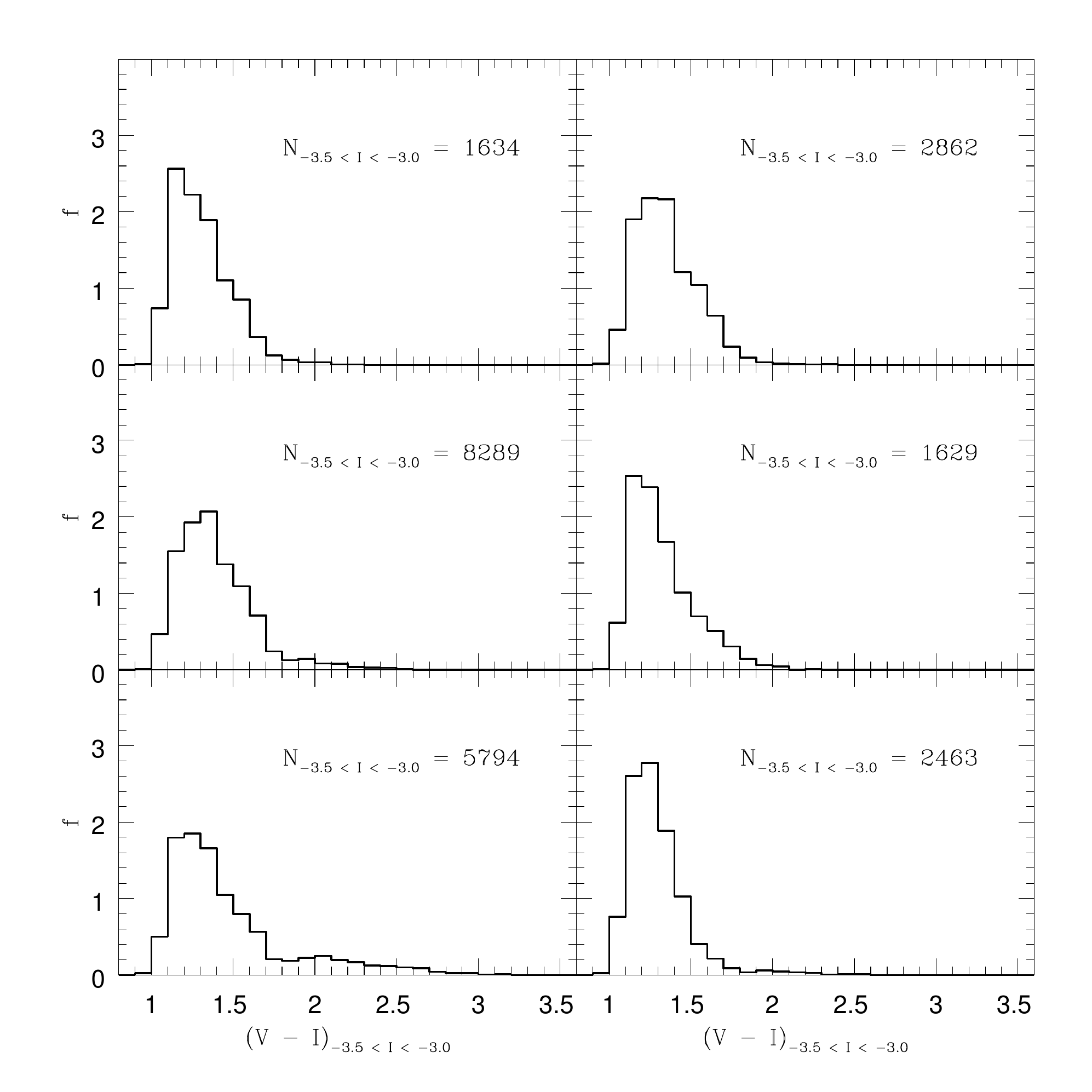}
\caption{(continued) Halo synthetic CMD (V~-~I) colour distribution at -3.5~$<$~I~$<$~-3.0 for the M$_{\rm tot}$~=~10$^{12}$~M$_\odot$ semi--cosmological simulations in Renda~et~al.~(2005b). Each panel also shows the number of generated stars at -3.5~$<$~I~$<$~-3.0 in each synthetic CMD.}
\label{appG:sim1e12:fig5}
\end{center}
\end{figure}

\begin{figure}
\begin{center}
\includegraphics[width=1.0\textwidth]{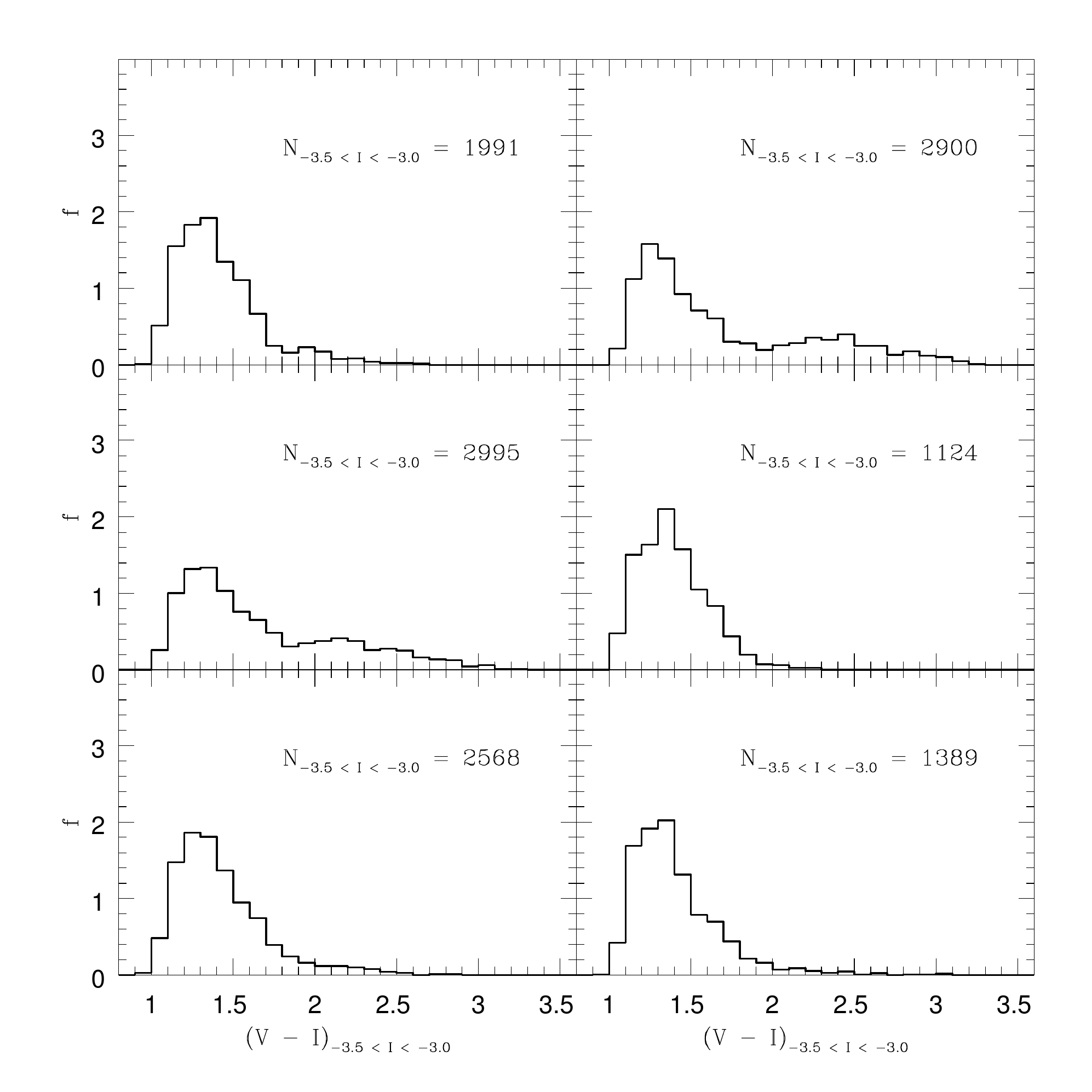}
\caption{Halo synthetic CMD (V~-~I) colour distribution at -3.5~$<$~I~$<$~-3.0 for the M$_{\rm tot}$~=~5$\times$10$^{12}$~M$_\odot$ semi--cosmological simulations in Renda~et~al.~(2005b). Each panel also shows the number of generated stars at -3.5~$<$~I~$<$~-3.0 in each synthetic CMD.}
\label{appG:sim5e12:fig1}
\end{center}
\end{figure}

\begin{figure}
\begin{center}
\includegraphics[width=1.0\textwidth]{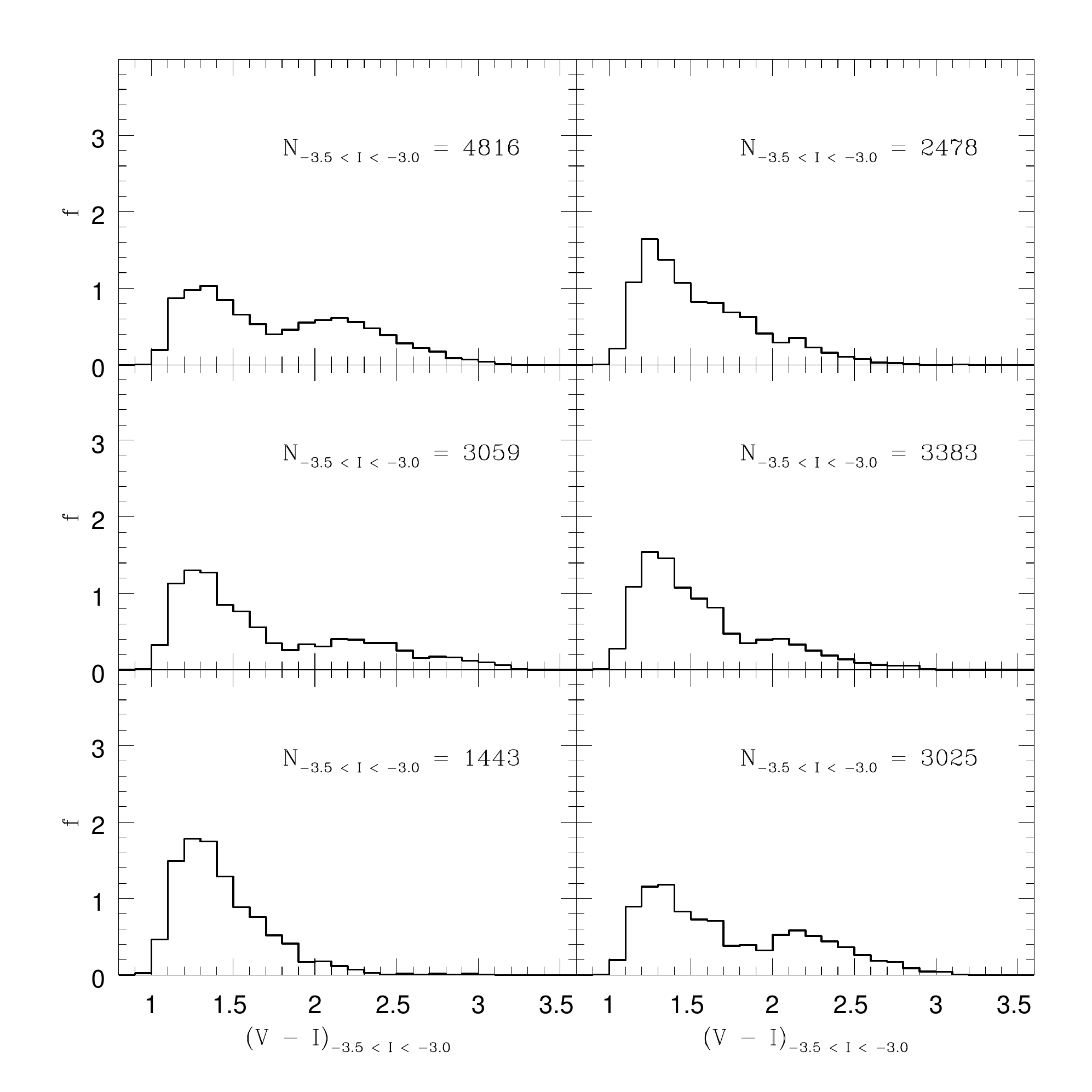}
\caption{(continued) Halo synthetic CMD (V~-~I) colour distribution at -3.5~$<$~I~$<$~-3.0 for the M$_{\rm tot}$~=~5$\times$10$^{12}$~M$_\odot$ semi--cosmological simulations in Renda~et~al.~(2005b). Each panel also shows the number of generated stars at -3.5~$<$~I~$<$~-3.0 in each synthetic CMD.}
\label{appG:sim5e12:fig2}
\end{center}
\end{figure}

\begin{figure}
\begin{center}
\includegraphics[width=1.0\textwidth]{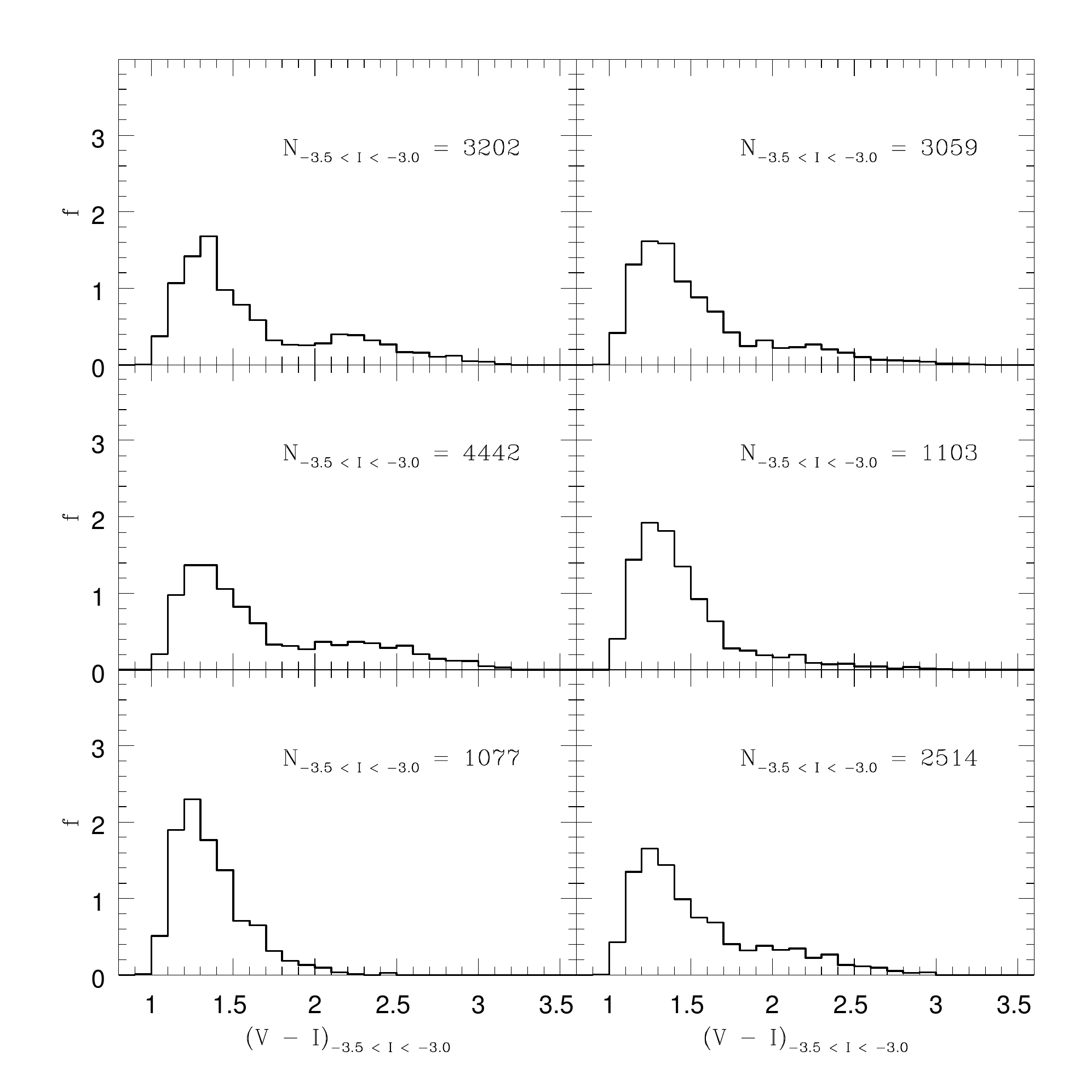}
\caption{(continued) Halo synthetic CMD (V~-~I) colour distribution at -3.5~$<$~I~$<$~-3.0 for the M$_{\rm tot}$~=~5$\times$10$^{12}$~M$_\odot$ semi--cosmological simulations in Renda~et~al.~(2005b). Each panel also shows the number of generated stars at -3.5~$<$~I~$<$~-3.0 in each synthetic CMD.}
\label{appG:sim5e12:fig3}
\end{center}
\end{figure}

\begin{figure}
\begin{center}
\includegraphics[width=1.0\textwidth]{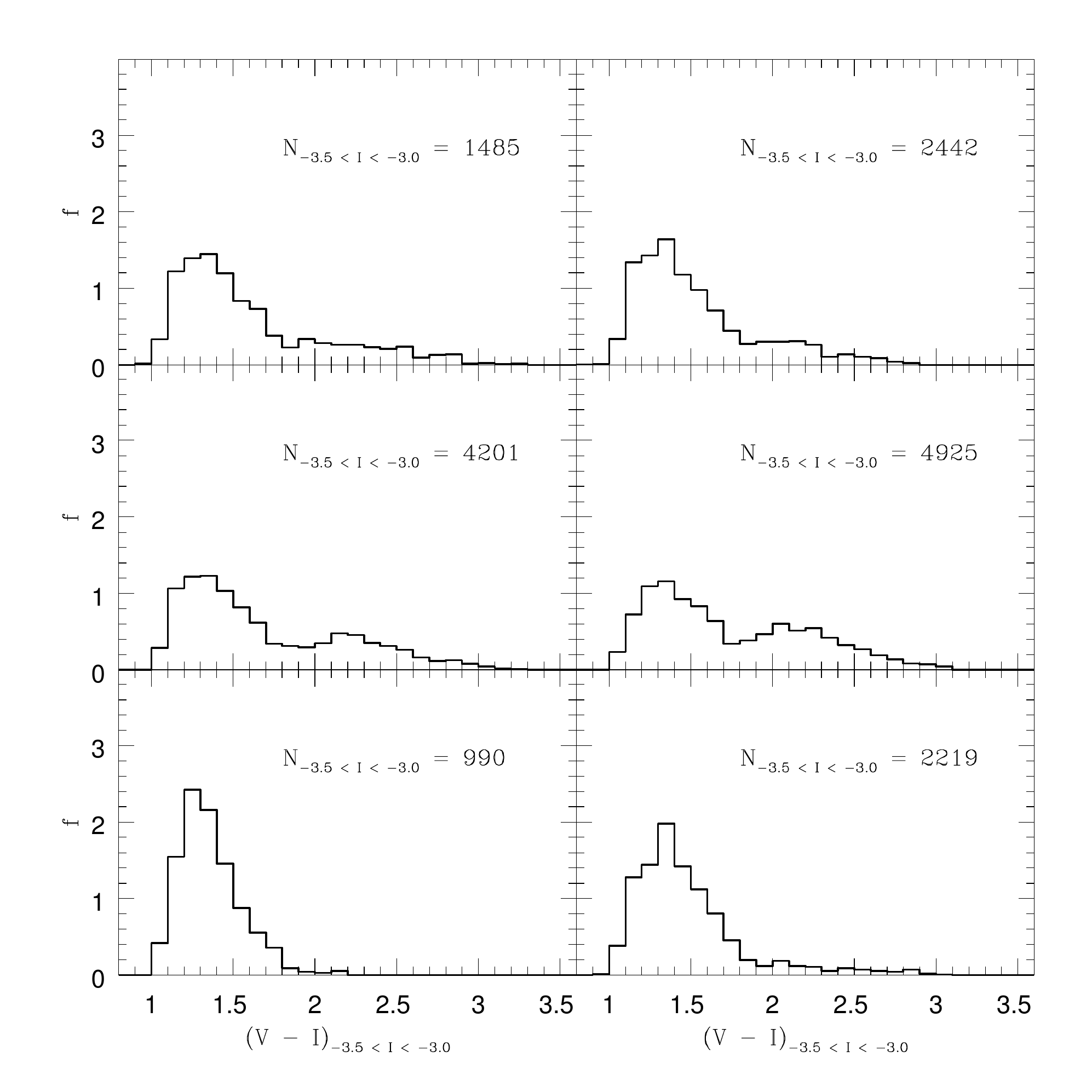}
\caption{(continued) Halo synthetic CMD (V~-~I) colour distribution at -3.5~$<$~I~$<$~-3.0 for the M$_{\rm tot}$~=~5$\times$10$^{12}$~M$_\odot$ semi--cosmological simulations in Renda~et~al.~(2005b). Each panel also shows the number of generated stars at -3.5~$<$~I~$<$~-3.0 in each synthetic CMD.}
\label{appG:sim5e12:fig4}
\end{center}
\end{figure}

\begin{figure}
\begin{center}
\includegraphics[width=1.0\textwidth]{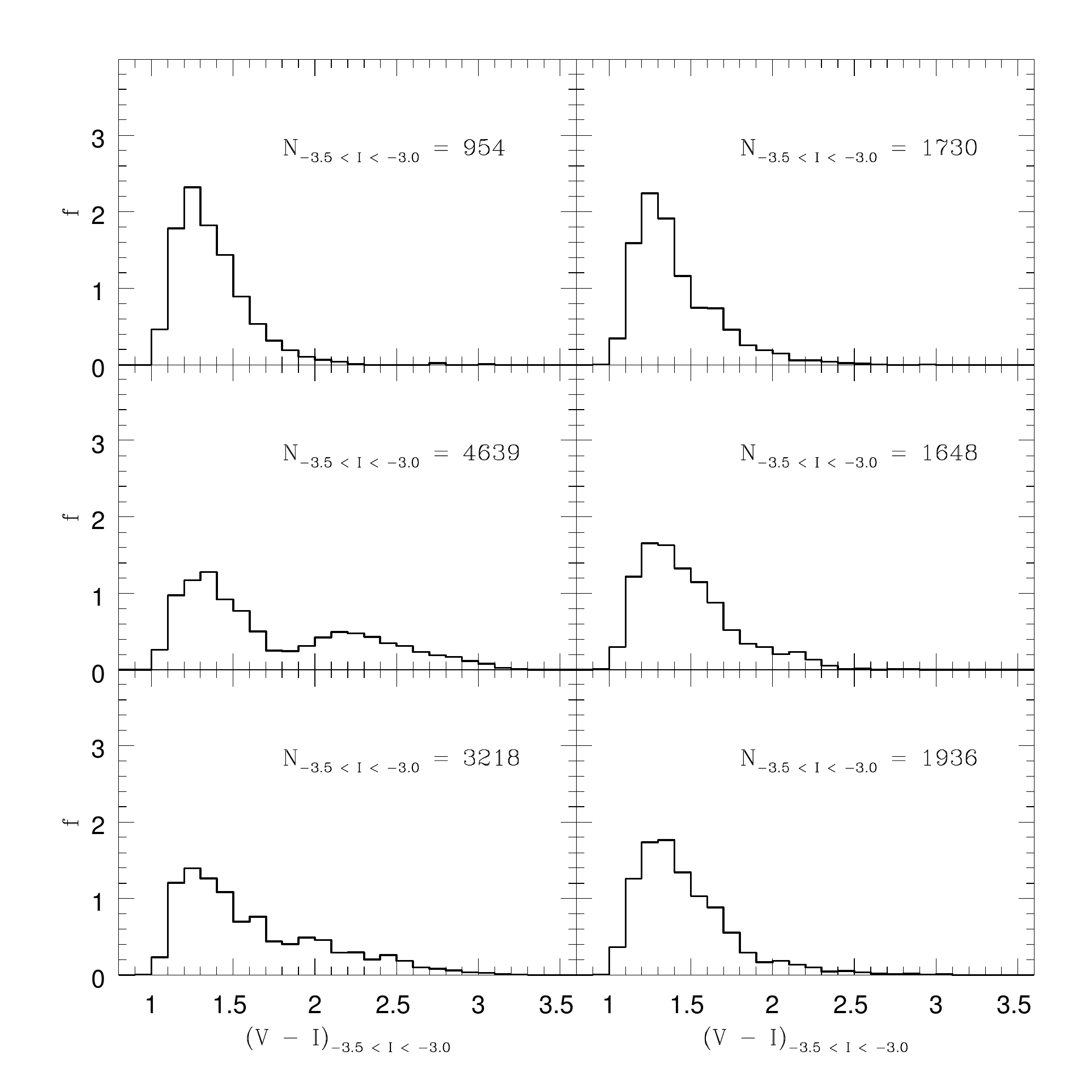}
\caption{(continued) Halo synthetic CMD (V~-~I) colour distribution at -3.5~$<$~I~$<$~-3.0 for the M$_{\rm tot}$~=~5$\times$10$^{12}$~M$_\odot$ semi--cosmological simulations in Renda~et~al.~(2005b). Each panel also shows the number of generated stars at -3.5~$<$~I~$<$~-3.0 in each synthetic CMD.}
\label{appG:sim5e12:fig5}
\end{center}
\end{figure}




\newpage

\begin{center}
\chapter[Halo MDF via Metallicity--Colour]{\LARGE Stellar Halo MDFs\\ via Metallicity--Colour Relationship}
\label{app:appendixH}
\end{center}

\begin{figure}
\begin{center}
\includegraphics[width=1.0\textwidth]{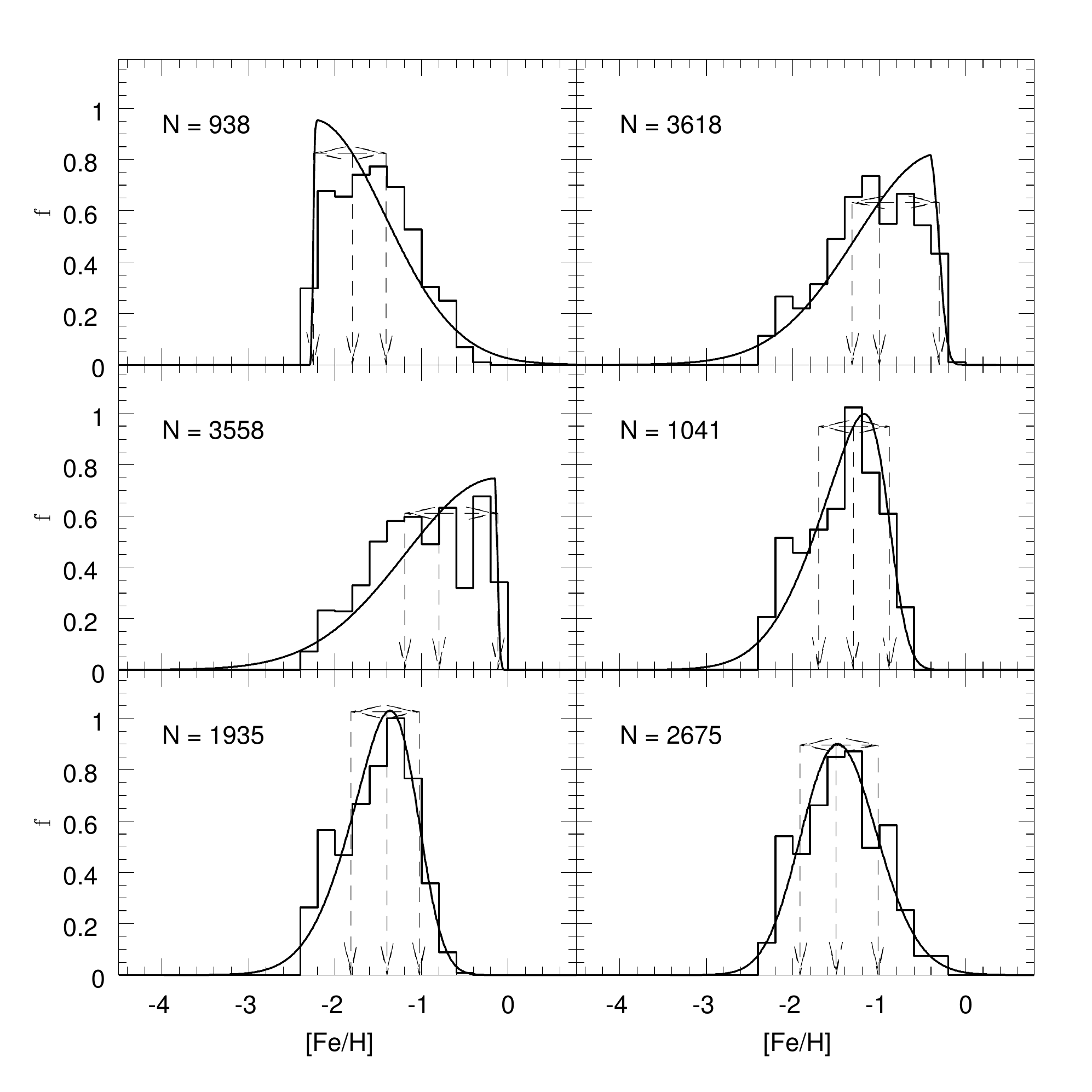}
\caption{Halo MDF via Metallicity--Colour relationship by the same pipeline as in Mouhcine~et~al.~(2005) for the M$_{\rm tot}$~=~5$\times$10$^{11}$~M$_\odot$ semi--cosmological simulations in Renda~et~al.~(2005b). The 68\%~Confidence~Level range and the number of stellar particles each MDF refers to are also shown.}
\label{appH:sim5e11:fig1}
\end{center}
\end{figure}

\begin{figure}
\begin{center}
\includegraphics[width=1.0\textwidth]{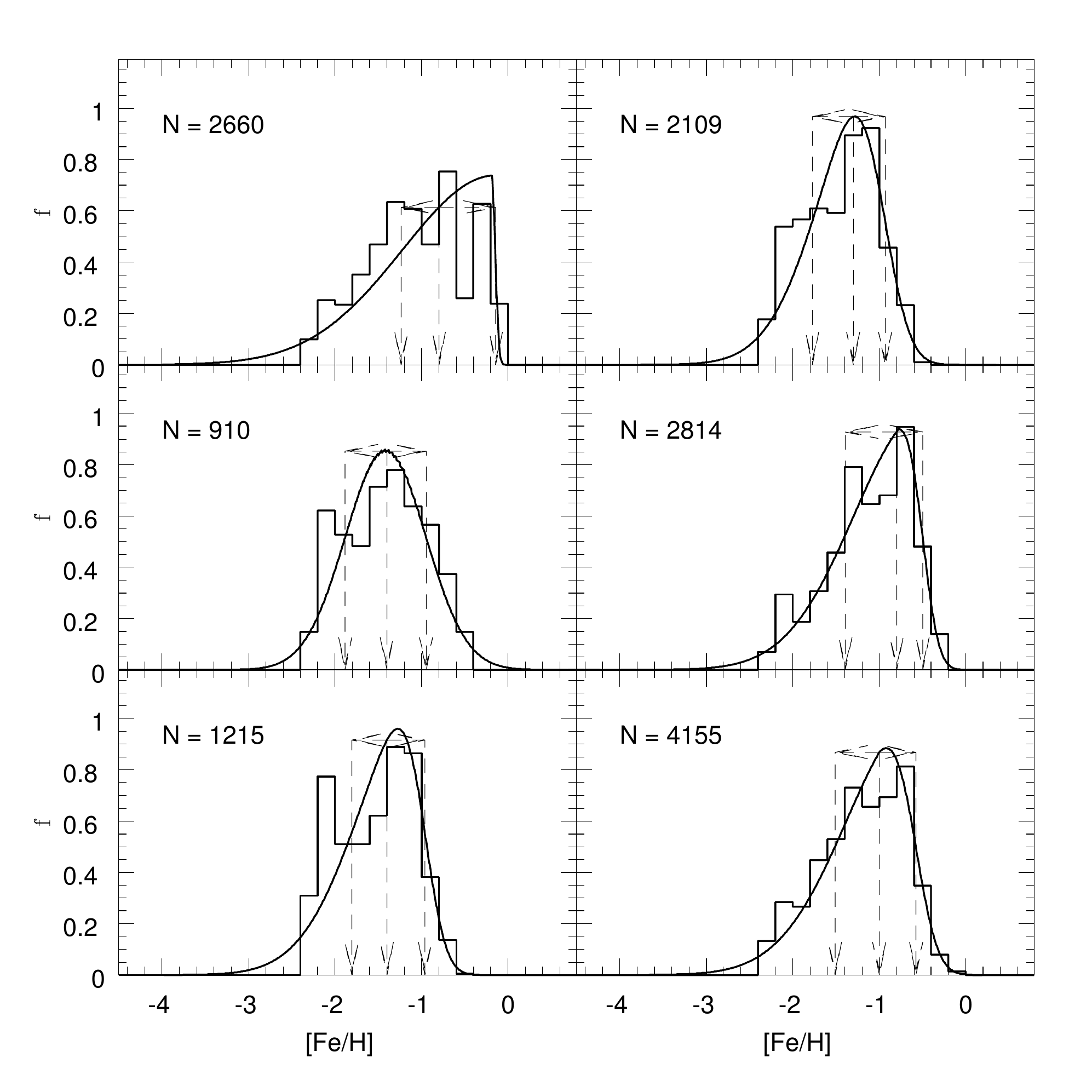}
\caption{(continued) Halo MDF via Metallicity--Colour relationship by the same pipeline as in Mouhcine~et~al.~(2005) for the M$_{\rm tot}$~=~5$\times$10$^{11}$~M$_\odot$ semi--cosmological simulations in Renda~et~al.~(2005b). The 68\%~Confidence~Level range and the number of stellar particles each MDF refers to are also shown.}
\label{appH:sim5e11:fig2}
\end{center}
\end{figure}

\begin{figure}
\begin{center}
\includegraphics[width=1.0\textwidth]{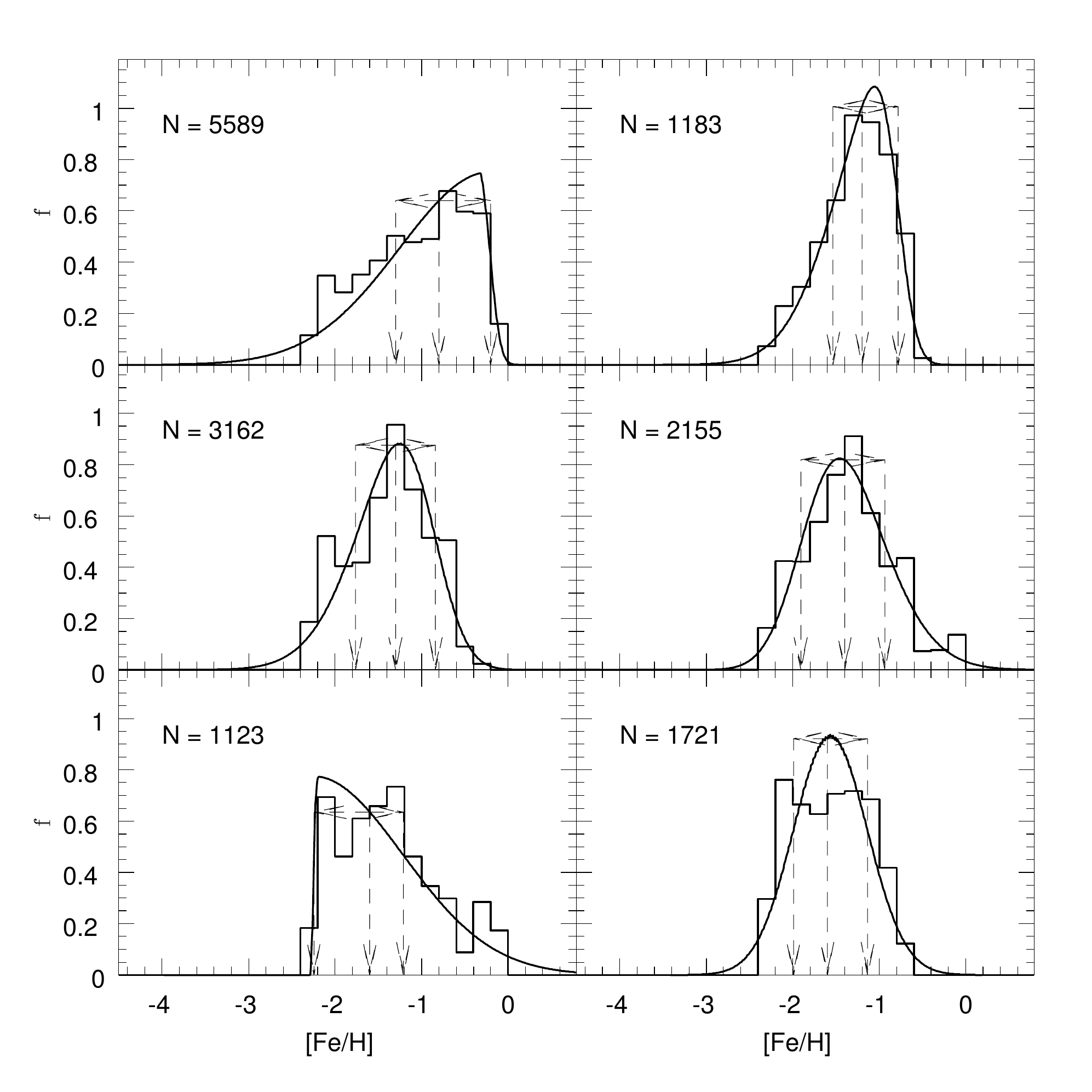}
\caption{(continued) Halo MDF via Metallicity--Colour relationship by the same pipeline as in Mouhcine~et~al.~(2005) for the M$_{\rm tot}$~=~5$\times$10$^{11}$~M$_\odot$ semi--cosmological simulations in Renda~et~al.~(2005b). The 68\%~Confidence~Level range and the number of stellar particles each MDF refers to are also shown.}
\label{appH:sim5e11:fig3}
\end{center}
\end{figure}

\begin{figure}
\begin{center}
\includegraphics[width=1.0\textwidth]{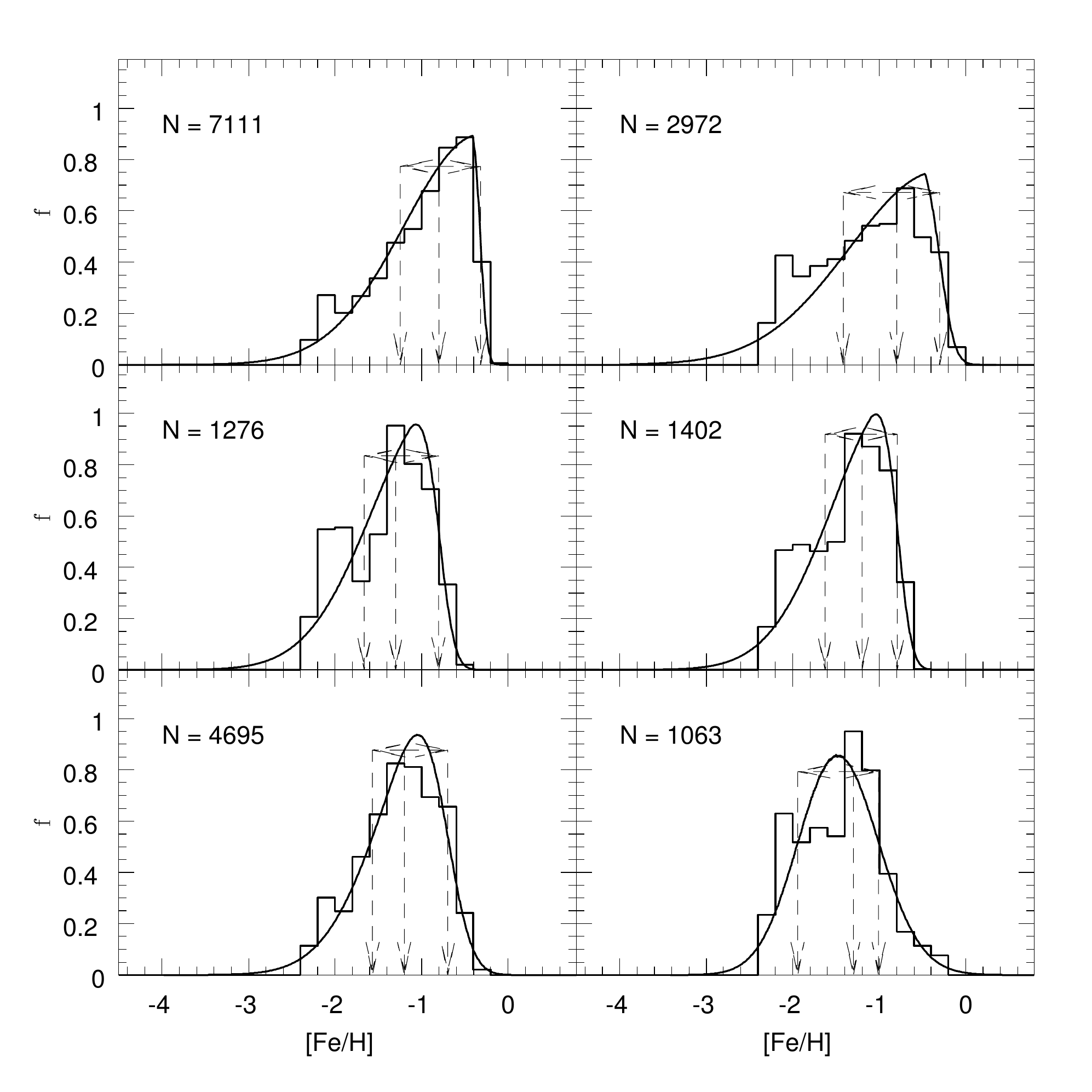}
\caption{(continued) Halo MDF via Metallicity--Colour relationship by the same pipeline as in Mouhcine~et~al.~(2005) for the M$_{\rm tot}$~=~5$\times$10$^{11}$~M$_\odot$ semi--cosmological simulations in Renda~et~al.~(2005b). The 68\%~Confidence~Level range and the number of stellar particles each MDF refers to are also shown.}
\label{appH:sim5e11:fig4}
\end{center}
\end{figure}

\begin{figure}
\begin{center}
\includegraphics[width=1.0\textwidth]{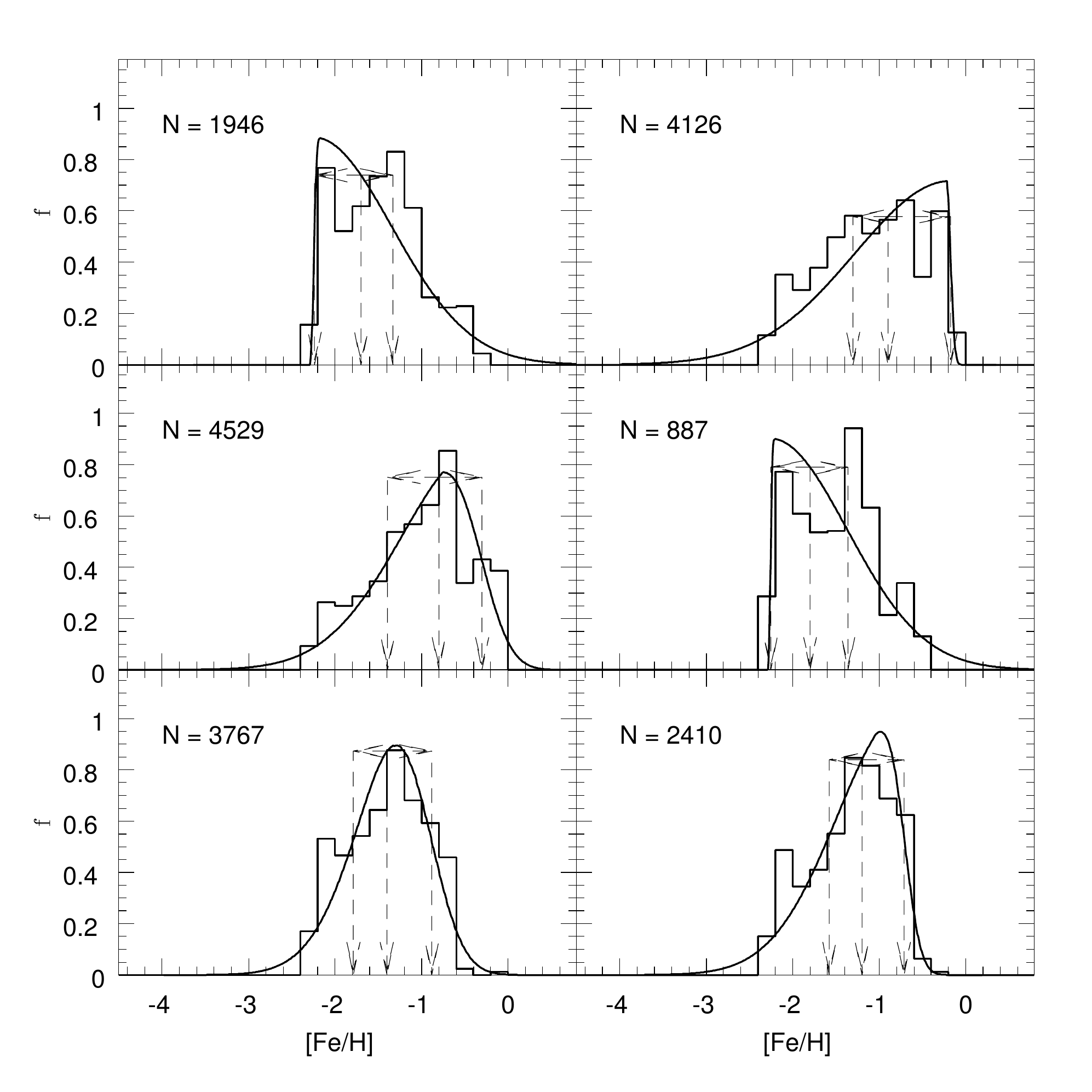}
\caption{Halo MDF via Metallicity--Colour relationship by the same pipeline as in Mouhcine~et~al.~(2005) for the M$_{\rm tot}$~=~10$^{12}$~M$_\odot$ semi--cosmological simulations in Renda~et~al.~(2005b). The 68\%~Confidence~Level range and the number of stellar particles each MDF refers to are also shown.}
\label{appH:sim1e12:fig1}
\end{center}
\end{figure}

\begin{figure}
\begin{center}
\includegraphics[width=1.0\textwidth]{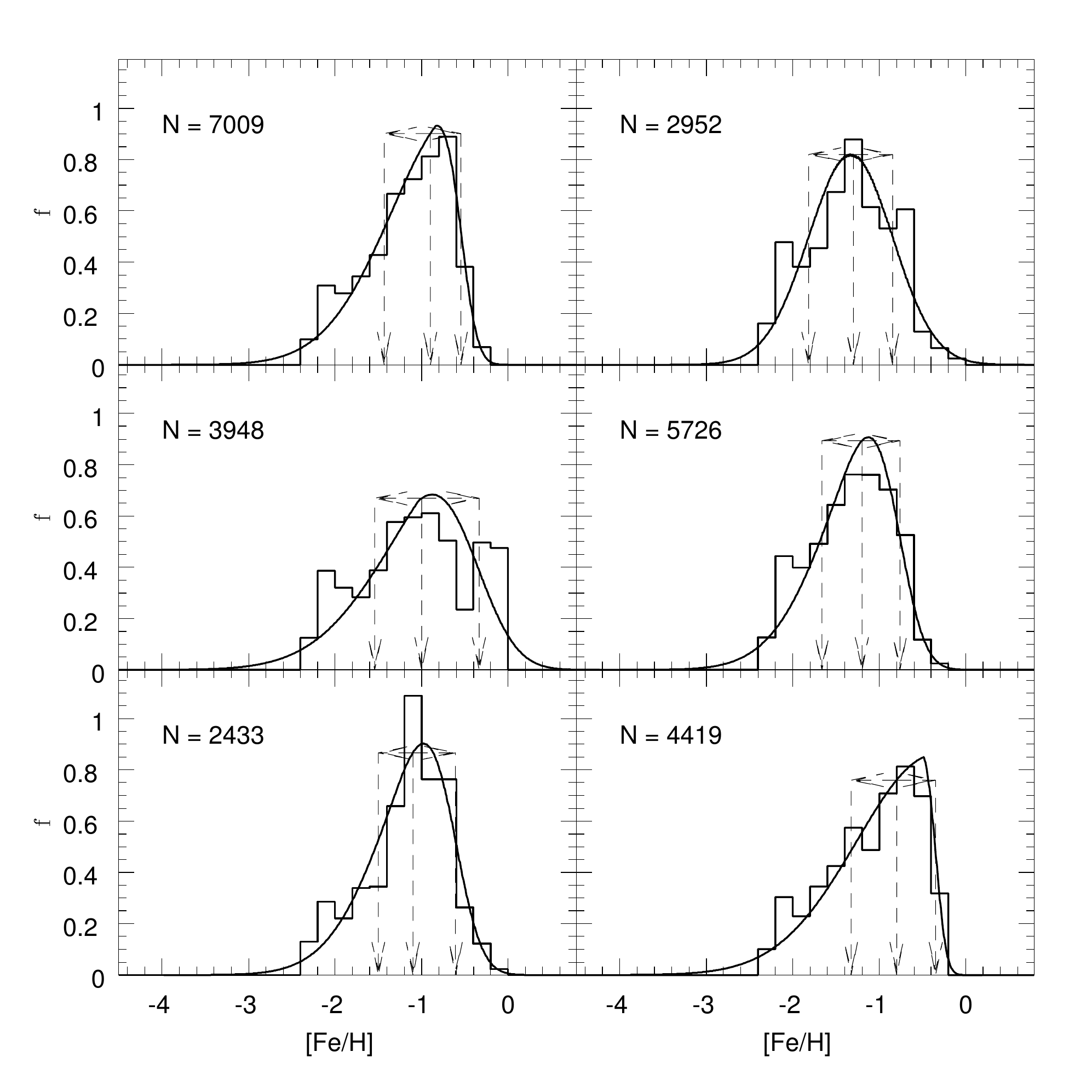}
\caption{(continued) Halo MDF via Metallicity--Colour relationship by the same pipeline as in Mouhcine~et~al.~(2005) for the M$_{\rm tot}$~=~10$^{12}$~M$_\odot$ semi--cosmological simulations in Renda~et~al.~(2005b). The 68\%~Confidence~Level range and the number of stellar particles each MDF refers to are also shown.}
\label{appH:sim1e12:fig2}
\end{center}
\end{figure}

\begin{figure}
\begin{center}
\includegraphics[width=1.0\textwidth]{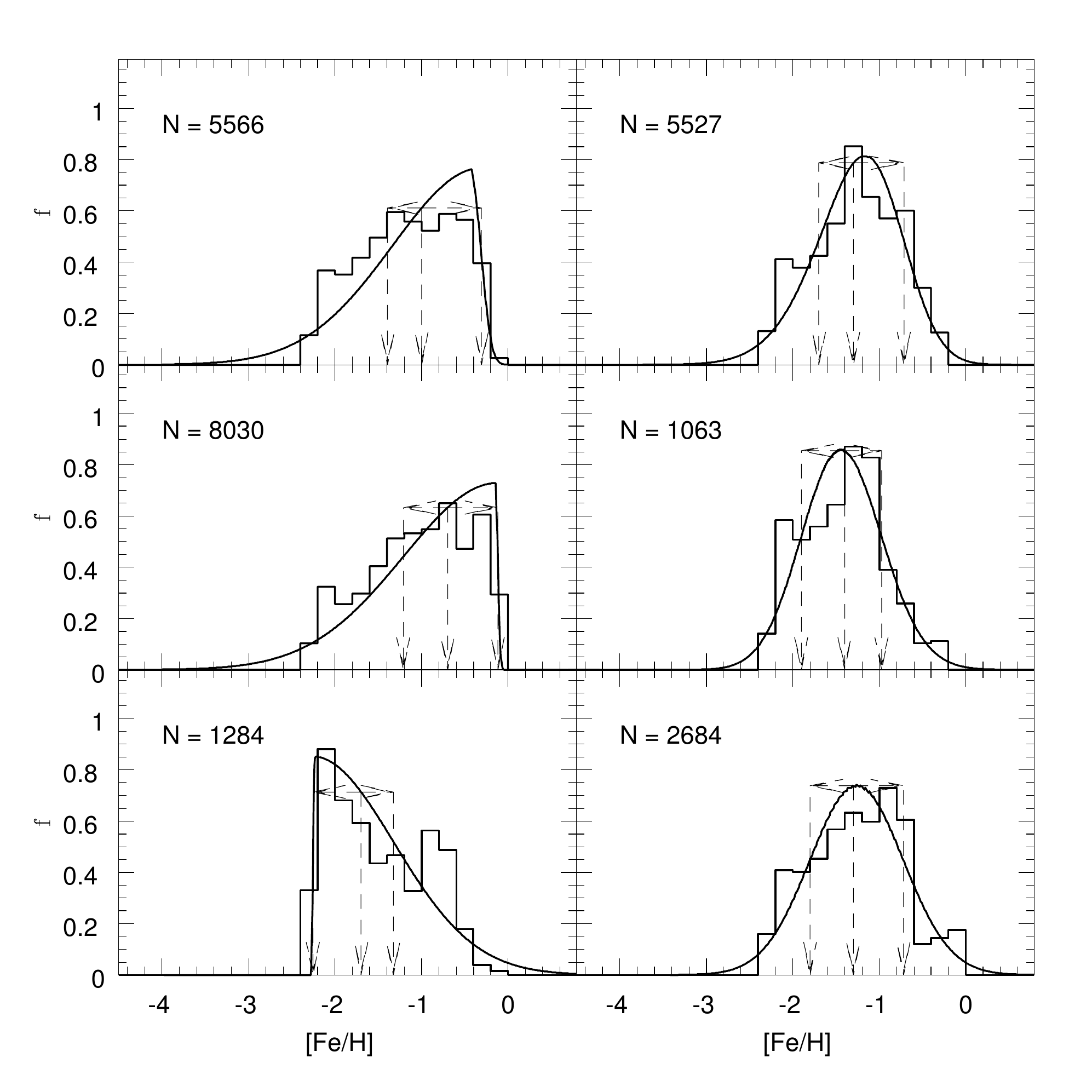}
\caption{(continued) Halo MDF via Metallicity--Colour relationship by the same pipeline as in Mouhcine~et~al.~(2005) for the M$_{\rm tot}$~=~10$^{12}$~M$_\odot$ semi--cosmological simulations in Renda~et~al.~(2005b). The 68\%~Confidence~Level range and the number of stellar particles each MDF refers to are also shown.}
\label{appH:sim1e12:fig3}
\end{center}
\end{figure}

\begin{figure}
\begin{center}
\includegraphics[width=1.0\textwidth]{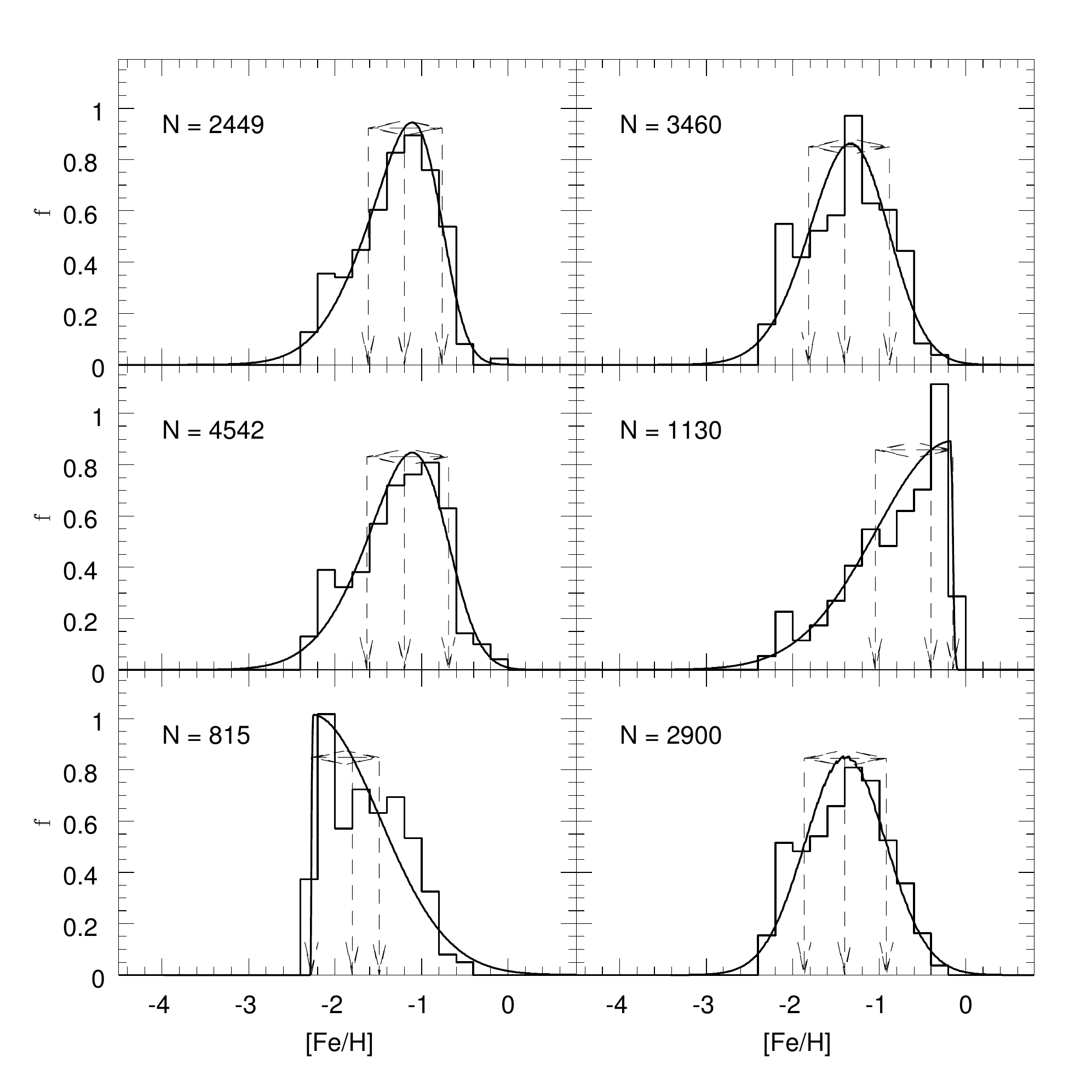}
\caption{(continued) Halo MDF via Metallicity--Colour relationship by the same pipeline as in Mouhcine~et~al.~(2005) for the M$_{\rm tot}$~=~10$^{12}$~M$_\odot$ semi--cosmological simulations in Renda~et~al.~(2005b). The 68\%~Confidence~Level range and the number of stellar particles each MDF refers to are also shown.}
\label{appH:sim1e12:fig4}
\end{center}
\end{figure}

\begin{figure}
\begin{center}
\includegraphics[width=1.0\textwidth]{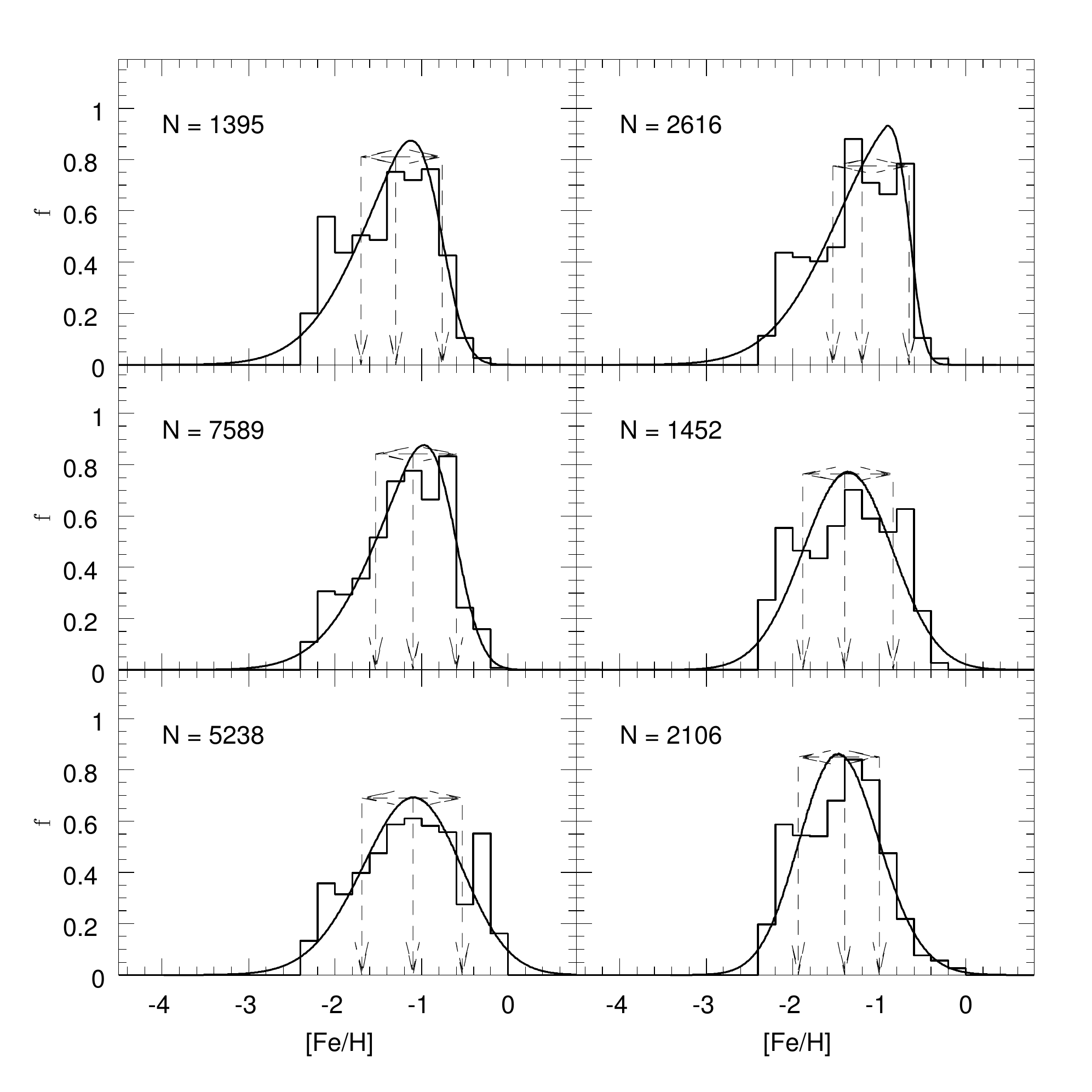}
\caption{(continued) Halo MDF via Metallicity--Colour relationship by the same pipeline as in Mouhcine~et~al.~(2005) for the M$_{\rm tot}$~=~10$^{12}$~M$_\odot$ semi--cosmological simulations in Renda~et~al.~(2005b). The 68\%~Confidence~Level range and the number of stellar particles each MDF refers to are also shown.}
\label{appH:sim1e12:fig5}
\end{center}
\end{figure}

\begin{figure}
\begin{center}
\includegraphics[width=1.0\textwidth]{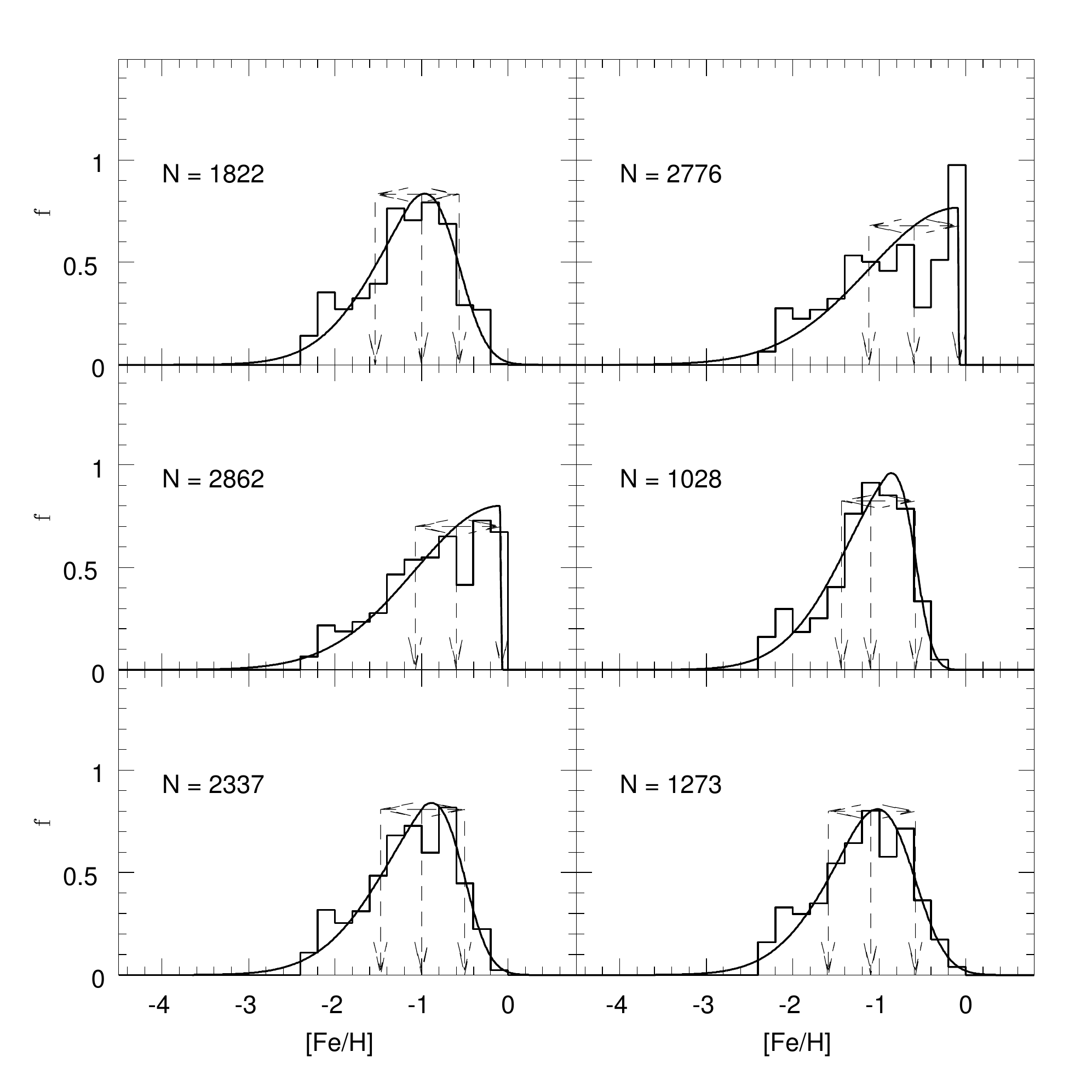}
\caption{Halo MDF via Metallicity--Colour relationship by the same pipeline as in Mouhcine~et~al.~(2005) for the M$_{\rm tot}$~=~5$\times$10$^{12}$~M$_\odot$ semi--cosmological simulations in Renda~et~al.~(2005b). The 68\%~Confidence~Level range and the number of stellar particles each MDF refers to are also shown.}
\label{appH:sim5e12:fig1}
\end{center}
\end{figure}

\begin{figure}
\begin{center}
\includegraphics[width=1.0\textwidth]{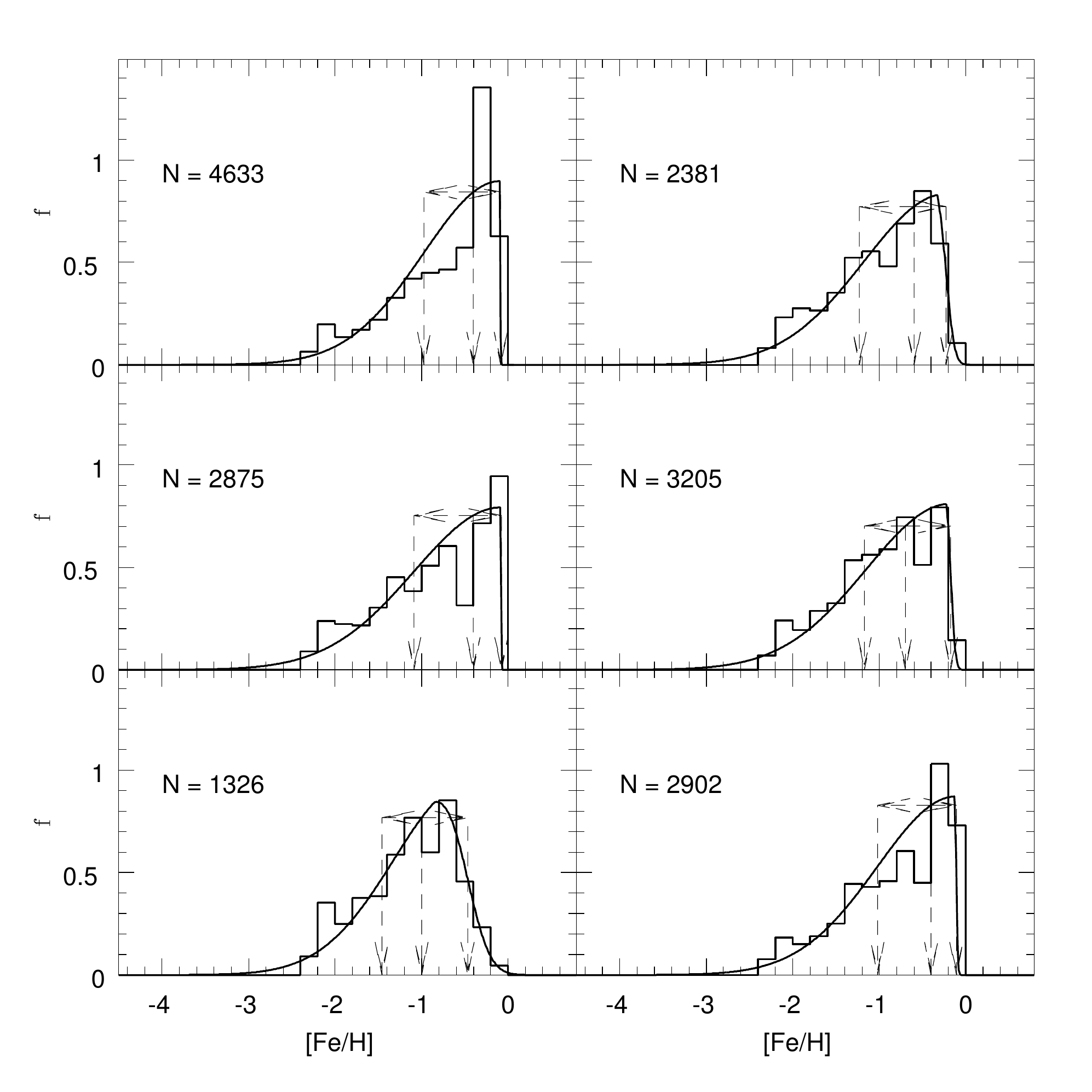}
\caption{(continued) Halo MDF via Metallicity--Colour relationship by the same pipeline as in Mouhcine~et~al.~(2005) for the M$_{\rm tot}$~=~5$\times$10$^{12}$~M$_\odot$ semi--cosmological simulations in Renda~et~al.~(2005b). The 68\%~Confidence~Level range and the number of stellar particles each MDF refers to are also shown.}
\label{appH:sim5e12:fig2}
\end{center}
\end{figure}

\begin{figure}
\begin{center}
\includegraphics[width=1.0\textwidth]{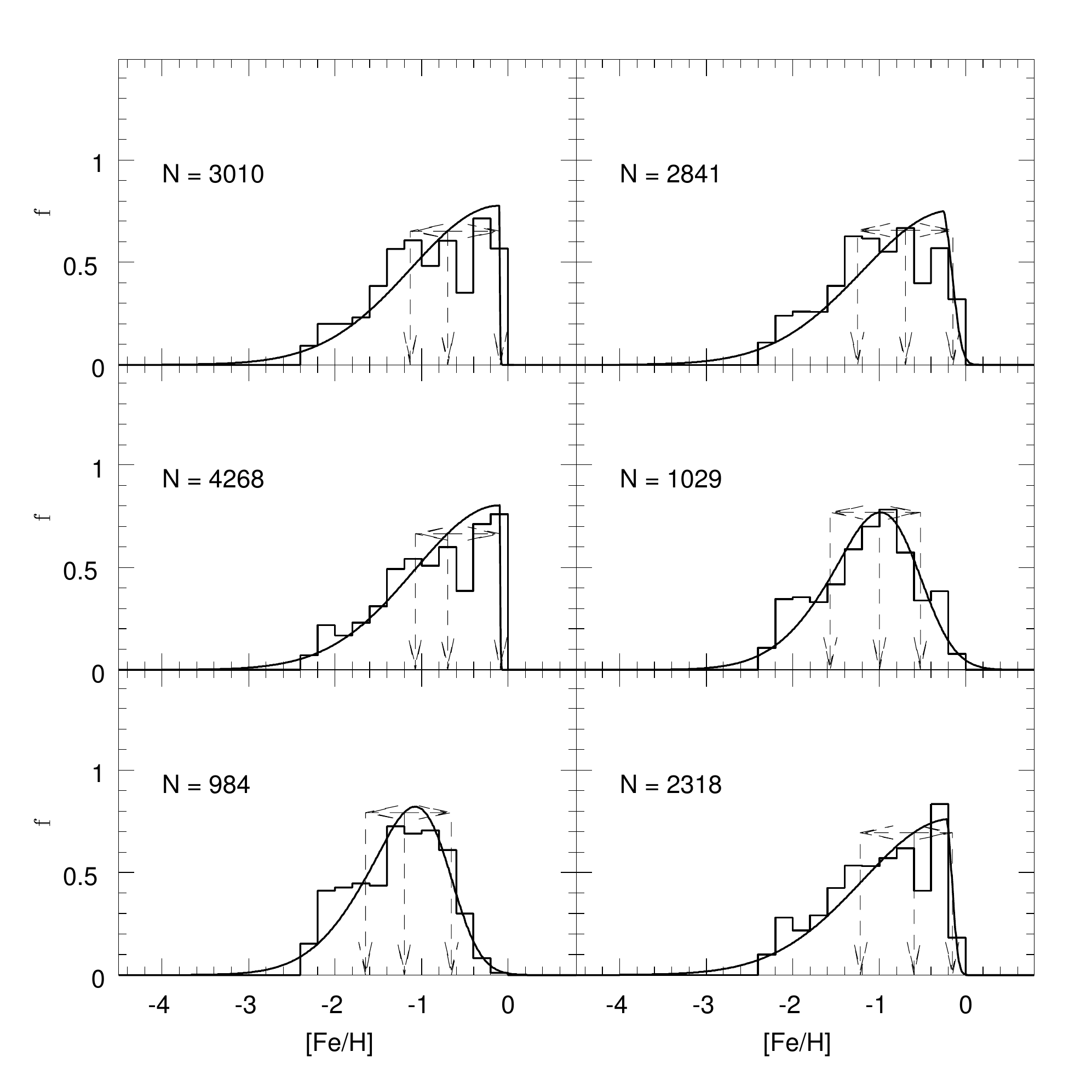}
\caption{(continued) Halo MDF via Metallicity--Colour relationship by the same pipeline as in Mouhcine~et~al.~(2005) for the M$_{\rm tot}$~=~5$\times$10$^{12}$~M$_\odot$ semi--cosmological simulations in Renda~et~al.~(2005b). The 68\%~Confidence~Level range and the number of stellar particles each MDF refers to are also shown.}
\label{appH:sim5e12:fig3}
\end{center}
\end{figure}

\begin{figure}
\begin{center}
\includegraphics[width=1.0\textwidth]{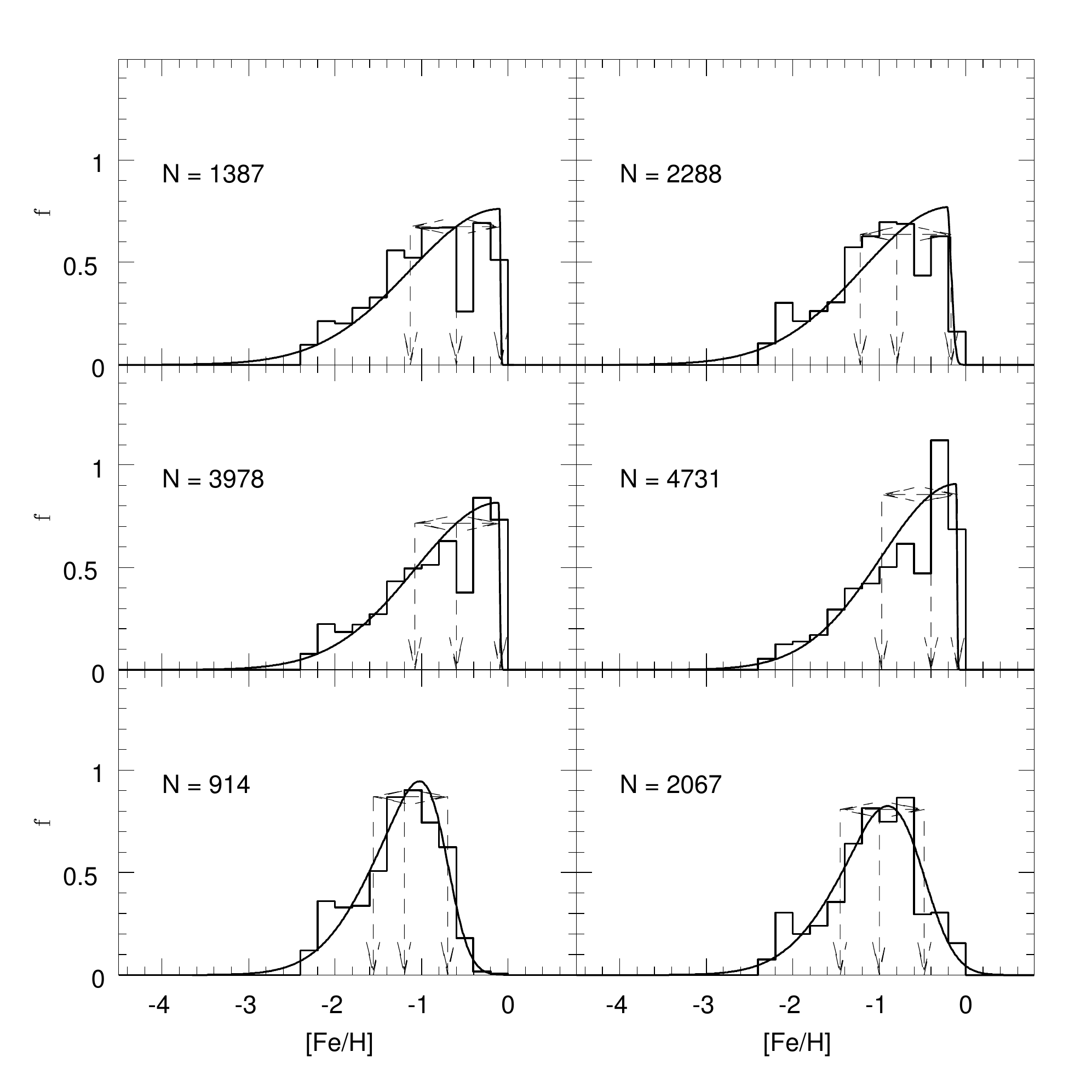}
\caption{(continued) Halo MDF via Metallicity--Colour relationship by the same pipeline as in Mouhcine~et~al.~(2005) for the M$_{\rm tot}$~=~5$\times$10$^{12}$~M$_\odot$ semi--cosmological simulations in Renda~et~al.~(2005b). The 68\%~Confidence~Level range and the number of stellar particles each MDF refers to are also shown.}
\label{appH:sim5e12:fig4}
\end{center}
\end{figure}

\begin{figure}
\begin{center}
\includegraphics[width=1.0\textwidth]{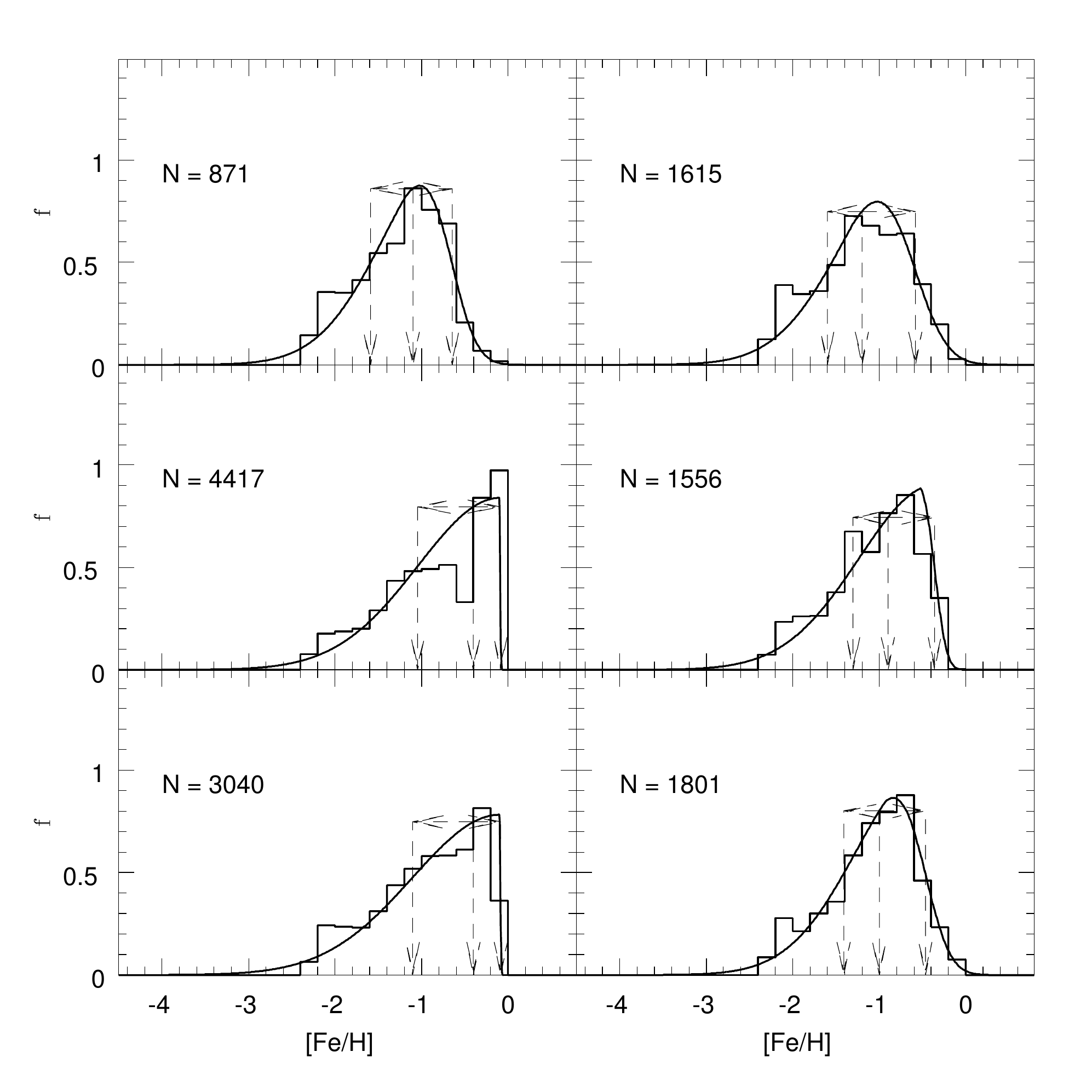}
\caption{(continued) Halo MDF via Metallicity--Colour relationship by the same pipeline as in Mouhcine~et~al.~(2005) for the M$_{\rm tot}$~=~5$\times$10$^{12}$~M$_\odot$ semi--cosmological simulations in Renda~et~al.~(2005b). The 68\%~Confidence~Level range and the number of stellar particles each MDF refers to are also shown.}
\label{appH:sim5e12:fig5}
\end{center}
\end{figure}




\bibliographystyle{mn2e}


\clearpage
\onehalfspacing
\addcontentsline{toc}{chapter}{Publications}

\pagestyle{empty}
\noindent
{\Large\bf Publications}

\vspace{4mm}

\begin{tabular}{l}
\rule[-2mm]{0mm}{3mm}
Refereed Journals\\ 
\hline
\vspace{1mm}\\
\begin{tabular}{ll}
$[$1$]$ & {\bf Renda A.}, Gibson B.K., Mouhcine M., Ibata R.A.,\\
        & Kawata D., Flynn C., Brook C.B.,\\
        & \textsf{Mon.\ Not.\ R.\ Astron.\ Soc.\ Lett.} \textbf{363}, L16 (2005)\\
        &\\
$[$2$]$ & {\bf Renda A.}, Kawata D., Fenner Y., Gibson B.K.,\\
        & \textsf{Mon.\ Not.\ R.\ Astron.\ Soc.} \textbf{356}, 1071 (2005)\\
        &\\
$[$3$]$ & {\bf Renda A.}, Fenner Y., Gibson B.K., Karakas A.I., Lattanzio J.C.,\\ 
        & Campbell S., Chieffi A., Cunha K., Smith V.V.,\\ 
        & \textsf{Mon.\ Not.\ R.\ Astron.\ Soc.} \textbf{354}, 575 (2004)\\
        &\\
\end{tabular}\\
Proceedings\\ 
\hline
\vspace{1mm}\\
\begin{tabular}{ll}
$[$1$]$ & {\bf Renda A.}, Fenner Y., Gibson B.K., Karakas A.I., Lattanzio J.C.,\\ 
        & Campbell S., Chieffi A., Cunha K., Smith V.V.,\\
        & \textsf{Nuclear\ Physics\ A} \textbf{758}, 324 (2005),\\
        & Buchmann L., Comyn M., Thomson J. Editors,\\
        & Proceedings of The 8th Int. Symposium on Nuclei in the Cosmos,\\ 
        & Vancouver, British Columbia, Canada, 2004\\
        &\\
\end{tabular}
\end{tabular}\\

\begin{tabular}{l}
\rule[-2mm]{0mm}{3mm}
Poster Presentations\\ 
\hline
\vspace{1mm}\\
\begin{tabular}{ll}
$[$1$]$ & {\bf Renda A.}, Gibson B.K., Mouhcine M., Ibata R.A.,\\
        & Kawata D., Flynn C., Brook C.B.,\\ 
        & {\it{The Stellar Halo Metallicity-Luminosity Relationship for Spiral Galaxies}},\\
        & ``The Fabulous Destiny of Galaxies: Bridging Past and Present''\\ 
        & Laboratoire d'Astrophysique de Marseille 5th Int. Cosmology Conf.,\\ 
        & Marseille, France, 2005\\
        &\\
$[$2$]$ & {\bf Renda A.}, Fenner Y., Gibson B.K., Karakas A.I., Lattanzio J.C.,\\ 
        & Campbell S., Chieffi A., Cunha K., Smith V.V.,\\
        & {\it{The Evolution of Fluorine in Galactic Systems}},\\
        & The 8th Int. Symposium on Nuclei in the Cosmos,\\ 
        & Vancouver, British Columbia, Canada, 2004\\
        &\\
\end{tabular}\\
Colloquia\\
\hline
\vspace{1mm}\\
\begin{tabular}{ll}
$[$1$]$ & {\it{The Stellar Halo Metallicity-Luminosity Relationship for Spiral Galaxies}},\\
        & held June 9, 2005, at L'Observatoire Astronomique de Strasbourg,\\
        & Strasbourg, France\\
        &\\
$[$2$]$ & {\it{The Stellar Halo Metallicity-Luminosity Relationship for Spiral Galaxies}},\\
        & held May 31, 2005, at The Obs. of the Carnegie Inst. of Washington,\\
        & Pasadena, Cal., U.S.A.\\
        &\\
\end{tabular}
\end{tabular}



\end{document}